\documentclass[twocolumn]{aastex62}
\usepackage{newtxtext,newtxmath}

\usepackage[T1]{fontenc}
\usepackage{ae,aecompl}
\usepackage{graphicx}	
\usepackage{amsmath}	
\usepackage{amssymb}	

\usepackage{ltxtable}
\usepackage{placeins}
\usepackage{longtable}
\usepackage{tabularx} 
\usepackage{tablefootnote}
\usepackage{pifont}
\newcommand{\cmark}{\ding{51}}%
\newcommand{\xmark}{\ding{55}}%

\newcommand{\ovi}{O\,{\sc VI}}

\newcommand{\ha}{H\,$\alpha$} 
\newcommand{\hbeta}{H\,$\beta$}

\newcommand{\helium}{He\,{\sc I}}
\newcommand{\heliumb}{He\,{\sc II}}

\newcommand{\nitrogen}{[N\,{\sc II}]}

\newcommand{\oxygeniii}{[O\,{\sc III}]}

\newcommand{\ironiv}{[Fe\,{\sc IV}]}
\newcommand{\ironvii}{[Fe\,{\sc VII}]}

\newcommand{\kriptoiii}{[Kr\,{\sc III}]}

\newcommand{\carboni}{C\,{\sc I}}

\newcommand{\DD}{D$^{\prime}$}
\newcommand{\degree}{$^{\circ}$}
\def\vhel{\ifmmode{V_{{\rm HEL}}}\else{$V_{{\rm HEL}}$}\fi}
\def\vsys{\ifmmode{V_{\rm sys}}\else{$V_{\rm sys}$}\fi}
\def\kms{\ifmmode{~{\rm km\,s}^{-1}}\else{~km~s$^{-1}$}\fi}
\def\vlsr{\ifmmode{v_{\rm lsr}}\else{$v_{\rm lsr}$}\fi}

\received{January 1, 2018}
\revised{January 7, 2018}
\accepted{\today}
\submitjournal{ApJ}

\shorttitle{A census of symbiotic stars}
\shortauthors{Akras et al.}

\begin{document}

\title{A census of symbiotic stars in the 2MASS, {\it WISE} and {\it Gaia} surveys}

\correspondingauthor{Stavros Akras}
\email{stavrosakras@on.br, akras@astro.ufrj.br}

\author{Stavros Akras}
\affil{Observat\'orio Nacional/MCTIC, Rua Gen. Jos\'{e} Cristino, 77, 20921-400, Rio de Janeiro, Brazil}
\affil{Observat\'orio do Valongo, Universidade Federal do Rio de Janeiro, Ladeira Pedro Antonio 43, 20080-090, Rio de Janeiro, Brazil}

\author{Lizette Guzman-Ramirez}
\affiliation{Leiden Observatory, Leiden University, Niels Bohrweg 2, 2333 CA Leiden, The Netherlands}
\affiliation{European Southern Observatory, Alonso de C\'ordova 3107, Casilla 19001, Santiago, Chile}

\author{Marcelo L. Leal-Ferreira}
\affiliation{Leiden Observatory, Leiden University, Niels Bohrweg 2, 2333 CA Leiden, Netherlands}
\affiliation{Argelander-Institut f\"{u}r Astronomie, Universit\"{a}t Bonn, Auf dem H\"{u}gel 71, D-53121, Bonn, Germany}

\author{Gerardo Ramos-Larios}
\affiliation{Instituto de Astronom\'ia y Meteorolog\'ia, Av. Vallarta No. 2602, Col. Arcos Vallarta, C.P. 44130 Guadalajara, Jalisco, Mexico}

\begin{abstract}

We present a new census of Galactic and extragalactic symbiotic stars (SySts). This compilation contains 323 known 
and 87 candidate SySts. Of the confirmed SySts, 257 are Galactic and 66 extragalactic. The spectral energy distributions (SEDs) of 
348 sources have been constructed using 2MASS and AllWISE data. Regarding the Galactic SySts, 74\% are S-types, 13\% D and 3.5\% \DD. 
S-types show an SED peak between 0.8 and 1.7~$\mu$m, whereas D-type show a peak at longer wavelengths between 2 and 4~$\mu$m. \DD-type, 
on the other hand, display a nearly flat profile. {\it Gaia} distances and effective temperatures are also presented. 
According to their {\it Gaia} distances, S-type are found to be members of both thin and thick Galactic disk populations, 
while S$+$IR- and D-types are mainly thin disk sources. {\it Gaia} temperatures show a reasonable agreement with the 
temperatures derived from SEDs within their uncertainties. A new census of the \ovi\ $\lambda$6830 Raman-scattered line in SySts is also 
presented. From a sample of 298 SySts with available optical spectra, 55\% are found to emit the line. No significant preference 
is found among the different types. The report of the \ovi\ $\lambda$6830 Raman-scattered line in non-SySts is also discussed as well as 
the correlation between the Raman-scattered \ovi\ line and X-ray emission. We conclude that the presence of the \ovi\ Raman-scattered line 
still provides a strong criterion for identifying a source as a SySt.

\end{abstract}

\keywords{catalogs - stars: binaries: symbiotic - stars: fundamental parameters - (stars:) white dwarfs - (ISM): dust, extinction
}


\section{Introduction} \label{sec:intro}

Symbiotic stars (SySts) are interacting, wide binary systems consisting of a red giant or a supergiant star that transfers 
matter to a much hotter companion, usually a white dwarf (WD), which can also be a neutron star. In the case of 
a white dwarfs as a primary, the secondary can be either a red giant or an asymptotic giant branch (AGB) star, and they are categorized 
as WD-symbiotics, whereas in the case of a neutron star as a primary, the secondary can be either a giant, AGB star, or a supergiant, 
and they are categorized as symbiotic X-ray binaries (Masetti et al. 2006a; Luna et al. 2013). In this paper, we refer to 
both groups as SySts.

The optical spectrum of SySts consists of both absorption features due to the photosphere of the cool companion (e.g. TiO, VO, C$_2$, CN) 
as well as a number of high-ionization lines (e.g. \heliumb\ $\lambda$4686, \ironvii\ $\lambda\lambda$5727,6087, \ovi\ $\lambda$6830),  
low-/intermediate-ionization lines (e.g. \nitrogen\ $\lambda\lambda$6548,6584, \oxygeniii\ $\lambda\lambda$4959, 5007) and bright 	
Balmer lines (e.g. \ha, \hbeta) due to the presence of a luminous and hot WD. Despite their spectra resembling those of planetary 
nebulae (PNe), they are not considered as genuine PNe. PNe are systems in which the WD ionizes the material that the same star expelled 
during the AGB phase, while in case of SySts the material that the WD ionizes comes from the secondary.

The most common criteria to classify a source as a SySt are the following: (i) the presence of strong \heliumb\ $\lambda$4686 and \ha\ 
lines as well as emission lines from high-excitation ions (e.g. \ironvii\ $\lambda\lambda$5727,6087), (ii) the presence of the absorption 
features TiO, VO and CN associated with the photosphere of the cool companion, and (iii) the presence of the \ovi\ Raman-scattered lines 
centered at 6830 and 7088\AA\ (e.g. Kenyon 1986; Mikolajewska et al. 1997; Belczy\'{n}ski et al. 2000)

The two broad \ovi\ $\lambda\lambda$6830,7088 lines usually seen in SySts are interpreted as the result of the Raman-scattering of 
the UV \ovi\ $\lambda$1032 and $\lambda$1038 resonance lines by neutral hydrogen (Nussbaumer et al. 1989; Schmid 1989). 
Even before the identification of these two lines, Allen (1980) and Schmid \& Schild (1994) had pointed out that 50\% or more of 
SySts exhibit the \ovi\ $\lambda$6830 line in their spectrum. The \ovi\ $\lambda\lambda$6830,7088 Raman-scattered lines have been proven 
to be a powerful tool for studying jets and accretion disks in SySts (e.g. Lee \& Kang 2007; Heo \& Lee 2015; Heo et al. 2016). Besides the spectroscopic observations for searching SySts, \ovi\ $\lambda$6830 imaging polarimetry can also become a very efficient method for discovering new SySts without follow-up spectroscopic observations or additional emission line images (Akras 2017)

SySts are classified into two main categories based on their near-infrared data (Allen \& Glass 1974; Webster \& Allen 1975): 
(i) those with a near-IR color temperature of $\sim$3000-4000~K, which is attributed to the 
temperature of a K, M, or G spectral-type giant (stellar or S-type SySts),\footnote{S-type SySts are also divided into two subgroups: 
(i) the yellow SySts with a K or G giant companion  and (ii) the red SySts with an M-type giant companion (Schmid \& Nussbaumer 1993; 
Frankowski \& Jorissen 2007)}, and (ii) those with a near-IR color temperature around 700-1000~K, indicating a warm dusty 
circumstellar envelope surrounding a more evolved AGB star, usually a Mira variable (dusty or D-type SySts). From the point of view of their spectral 
energy distributions (SEDs), S-type and D-type SySts have a peak in their SED profiles at 1-2 and 5-15~$\mu$m, 
respectively (Ivison et al. 1995).

Allen (1982) added a third type of SySt, namely \DD-type, in order to separate those with SED profiles that peak at even longer wavelengths 
than the normal D-type between 20 and 30~$\mu$m. The true nature of \DD-type SySts 
is still controversial. According to Allen (1982), the cool companions of \DD-type SySts are either a K or a G spectral type giant and 
their spectra exhibit a mid-IR excess due to the presence of a dusty component  with temperatures lower than those of D-type SySts. 
On the other hand, Kenyon, Fernandez-Castro \& Stencel (1988) claimed that this group of SySts have far-IR 
colors that resemble those of compact planetary nebulae (see also Corradi \& Schwartz 1997; Pereira, Smith \& Cunha 2005).

SySts are considered as potential progenitors of type~Ia supernova (SNe Ia) due to the large amount of mass 
that WDs accrete from the winds of the cool companions, resulting in them exceeding the Chandrasekhar mass (1.4 M$\odot$) and exploding 
as a SN Ia (Munari \& Renzini 1992; Han \& Podsiadlowski 2004; Di Stefano 2010; Dilday et al. 2012).
Consequently, the interest in SySts has been gradually increasing, and many attempts have been made to discover new members in this 
class of objects either in our galaxy or nearby galaxies in the Local Group (e.g. Miszalski et al. 2014; Mikolajewska et al. 2014).

SySts are also important X-ray sources. The origin of X-ray emission in SySts is manifold: (i) the thermonuclear activity 
on the hot WDs, (ii) the colliding winds, and/or (iii) the accretion disk. Muerset, Wolff \& Jordan (1997) identified 16 SySts as 
supersoft X-ray sources based on {\it ROSAT} observations. The authors proposed a classification scheme based only on their 
X-ray emission, dividing SySts into three groups: (a) the supersoft X-ray sources with energies $\le$0.4~keV, likely emitted directly from the white dwarf ($\alpha$-type), (b) soft X-ray objects that exhibit a peak at 0.8~kev and maximum energies up to 2.4~keV, 
likely originating from a hot, shocked gas where the stellar winds collide ($\beta$-type) and 
(c) objects with a non-thermal emission and energies higher than 2.4~keV ($\gamma$-type) due to the accretion 
of mass onto a neutron star.

New X-ray observations have revealed SySts with very hard X-ray emission ($>$10~keV) (Chernyakova et al. 2005; 
Tueller et al. 2005). Recently, Luna and collaborators presented new {\it Swift} X-ray Telescope (XRT) data 
of nine SySts, and they refined the previous X-ray classification scheme by adding a new type, namely the $\delta$-type (Luna et al. 2013). 
All these SySts have very hard X-ray thermal emission with energies higher than 2.4~keV likely originating from the inner regions of 
an accretion disk. Overall, the number of SySts with X-ray emission is increasing, and the current 
number is 44. Seven out of 44 are classified as $\alpha$-type, 12 as $\beta$-type, 9 as $\gamma$-type, 8 as $\delta$-type, and 8 as $\beta$/$\delta$-type (Luna et al. 2013 and references therein; Mukai et al. 2016, Wheatley, Mukai \& de Martino 2003, Nu\~{n}ez et al. 2016).

Over the last 20 years many new SySts have been discovered in our Galaxy and nearby galaxies. We thus decided to census the general 
population of known and candidate SySts, present their photometric data from the Two Micron All Sky Survey 
(2MASS, Cutri et al. 2003; Skrutskie et al. 2006) and the {\it Wide-field Infrared Survey Explorer} ({\it WISE}; Wright et al. 2010, Cutri et al. 2014) 
all-sky near-IR surveys and provide their classification in the scheme of S/D/\DD-types from a wider spectral range (1-22~$\mu$m).

The paper is organized as follows: the sample selection, the description of the census, and the cross-matching between the 2MASS and AllWISE 
catalogs are presented in Sect. 2. The SED profiles, the blackbody fitting, the temperature and distance using {\it Gaia} ( Gaia Collaboration 
et al. 2016, 2018), and the classification of SySts are described in Sect 3. We present an updated census of the \ovi\ $\lambda$6830 Raman-scattered line in SySts in Sect. 4. We also discuss the report of the \ovi\ Raman-scattered line in non-SySts. In Sect. 5, we explore the 
link of the \ovi\ Raman-scattered lines with the metallicity and X-ray emission. We end up with our conclusions in Sect. 6.

\section{Sample Selection}

The most comprehensive compilation of SySts was published by Belczy\'{n}ski and collaborators in 2000 (Belczy\'{n}ski et al. 2000, hereafter Bel2000). 
Their catalog includes all of the known Galactic and extragalactic SySts (188), as well as a number of candidate SySts (30). 
Before that catalog, two more had been already published by Allen (1984) and by Kenyon (1986). 

The total number of known SySts has been continuously increasing since 2000. 
Most of the discoveries have been made by two independent groups:
(a) Miszalski, Mikolajewska, and collaborators and (b) the INT Photometric \ha\ Survey (IPHAS collaboration; Drew et al. 2005, 
see Table~\ref{table1} in Appendix~A for references). 

Our new census of SySts lists 323 known and 87 candidate SySts. From the known 
SySts, 257 are Galactic and 66 extragalactic. This corresponds to a $\sim$45\% increase in the population of Galactic SySts and 
a $\sim$350\% increase in the population of extragalactic ones since the last SySt catalog (Table~\ref{table1} in Appendix~A) 

Aside from the most commonly and widely used criteria presented in Bel2000, there are a number of SySts identified based on their strong 
UV excess and the characteristic spectrum of a red giant (e.g. SU Lyn; Mukai et al. 2016), or their strong blue-violet continuum ({\it U-B} and 
{\it B-V}), and their H$\alpha$ line profile and orbital periods $>$600 days (Hen 4-18 and Hen 4-121; Van Eck \& Jorissen 2002). Mukai et al. (2016) 
argued that the classical criteria for identifying SySts (e.g. Bel2000) are biased toward systems with luminous WDs and that the 
true population is underestimated.

\begin{figure*}
\centering
\includegraphics[scale=0.335]{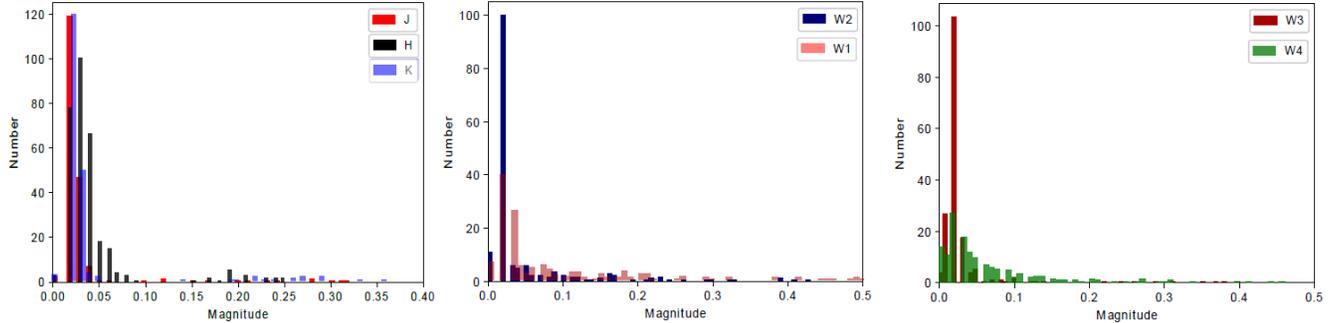}
\caption{Histogram of the photometric errors for the 2MASS and {\it WISE} bands.} 
\label{fig3newa}
\end{figure*}

Despite all these new discoveries, the total number of SySts still remains very low compared to the expected 
number of SySts in our galaxy (3$\times$10$^3$, Allen 1984; 3$\times$10$^4$, Kenyon et al. 1993; 3$\times$10$^5$,
Munari \& Renzini 1992; 4$\times$10$^5$, Magrini et al. 2003; 1.2-15$\times$10$^3$, L\"{u} et al.n 2006). Nevertheless, 
the number of SySts is expected to significantly increase over the next few years due to ongoing surveys like VPHAS+ of the southern 
Galactic plane and Bulge (Drew et al. 2014), J-PAS (Benitez et al. 2014)/J-PLUS (Cenarro et al. 2018, submitted) of the northern hemisphere;  
and S-PLUS of the southern hemisphere (Mendes de Oliveira C., et al. 2018, in preparation); as well as private survey programs (Miszalski et al. 2014; 
Mikolajewska et al. 2014)

\begin{figure}
\centering
\includegraphics[scale=0.35]{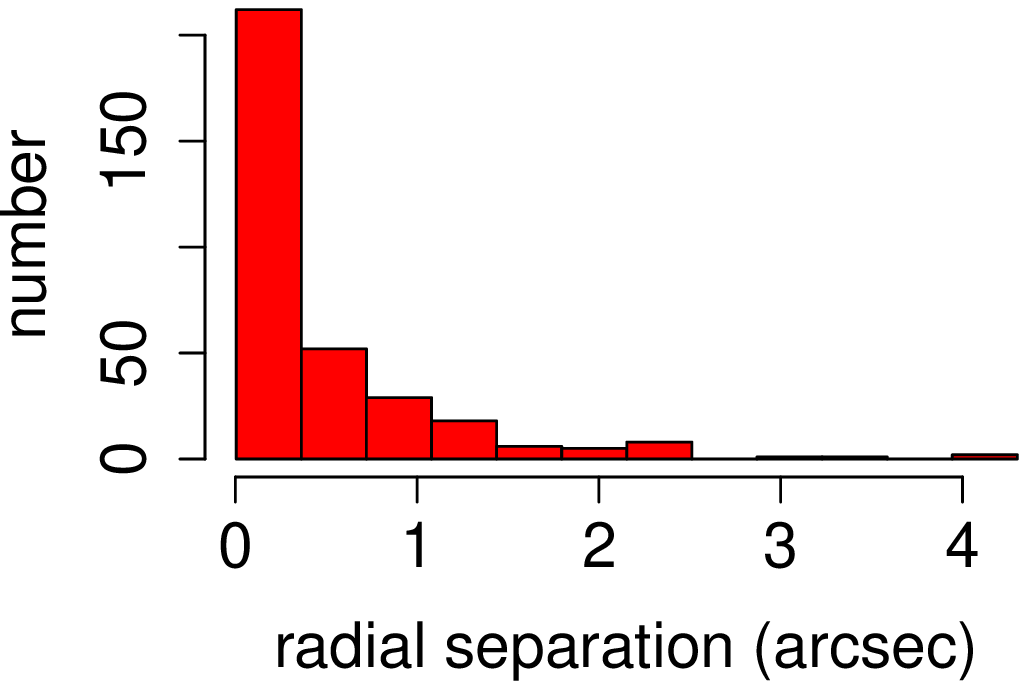}
\includegraphics[scale=0.35]{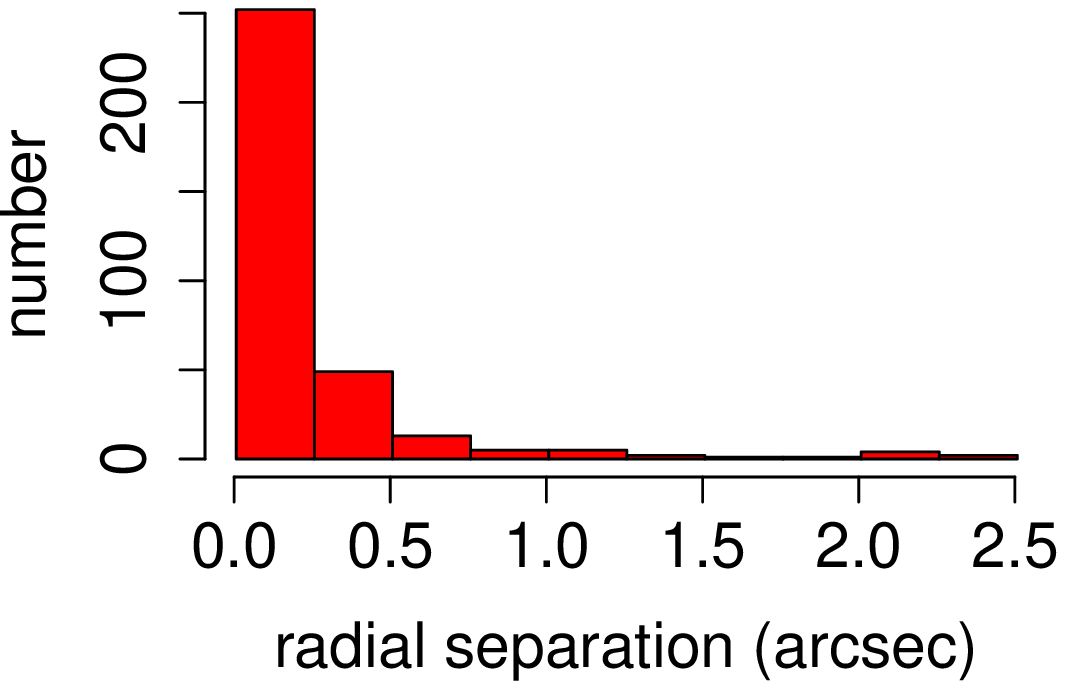}
\caption{Radial distance separation between the positions of our list and those of AllWISE catalog (left panel) and between 
the {\it WISE} positions and the associated 2MASS counterparts (right panel). }
\label{fig3newb}
\end{figure}

2MASS data have been widely used to study SySts as well as to divide them into the S and D types (e.g Allen \& Glass 1974; 
Phillips 2007; Corradi et al. 2008). However, the peak of the SEDs of D-type SySts is suggested to occur at wavelengths between 
5 and 15~$\mu$m (Ivison et al. 1995), longer than the wavelength range covered by 2MASS (between 1 and 2.16~$\mu$m). 
The {\it WISE} survey, on the other hand, covers these longer wavelengths (between 3.6 and 22.1~$\mu$m), and it is coherent to construct 
and study the SED profiles of SySts using both surveys. Yet, it is well known that SySts exhibit strong flux variations. In Section 3, we discuss the impact of possible flux variations between the two surveys because of the different epochs in which they were carried out. 

The first classification of SySts into S and D types by Allen and Glass (1974) was made based on observations in the {\it J}, {\it H}, {\it K$_{\rm s}$} and 
{\it L} bands. The latter is centered at 3.5~$\mu$m, very close to the {\it W1} band of the {\it WISE} survey. These authors mentioned that 
the {\it H-J} color index alone is not a reliable criterion for classifying SySts into S and D types but the {\it K$_{\rm s}$--L} 
index is. S type SySts are found to exhibit {\it K$_{\rm s}$--L}$<$0.9, while D type SySts exhibit {\it K$_{\rm s}$--L}$>$1.3. Photometric 
data from longer wavelengths are crucial for studying these objects and may provide a more robust classification.

Table~\ref{table2} in Appendix~A presents our list of SySts. The first column lists the name of SySts ordered by R.A. The second and third columns give the coordinates (R.A. and decl.) in the J2000.0 epoch of SySts. The rest of the columns list the 2MASS 
(from the All-Sky Catalog of Point Sources; Cutri et al. 2003) and {\it WISE} (from the AllWISE catalog; Cutri et al. 2014) photometric 
magnitudes and associated errors (except for the upper limit values).

The histogram of the photometric errors for all of the bands is illustrated in Figure~\ref{fig3newa}. The vast majority of the 2MASS 
measurements have photometric errors lower than 0.1~mag, while those from the AllWISE catalog are lower than 0.17~mag. In particular, 
95\%, 92\% and 93\% of SySts in the {\it J}, {\it H}, and {\it K$_{\rm s}$} bands, respectively, have photometric errors lower than 0.075~mag. 
The {\it WISE} measurements show higher photometric errors compared to the 2MASS data. Only 70\%, 80\%, 93\% and 
70\% of SySts have photometric errors lower than 0.075~mag in the {\it W1}, {\it W2}, {\it W3} and {\it W4} bands, respectively. 
We also find that the {\it W1} and {\it W4} bands show systematically higher errors than the {\it W2} and {\it W3} bands.

All SySts have been cross-matched with the 2MASS and AllWISE catalogs assuming a matching radius of six arcsec due to 
the resolution of the {\it WISE} survey (Wright et al. 2010). The radial distance separation between the positions of the sources in our list and the 
AllWISE catalog as well as that between the AllWISE catalog and the associated 2MASS counterparts are presented in Figure~\ref{fig3newb}. 
For the majority of the SySts, the matching between the catalogs is very good. In particular, 88\% of all SySts in our list 
have a radial distance separation with AllWISE lower then 1 arcsec and 94\% per cent lower than 2 arcsec, 
while 96\% have a radial distance separation between the AllWISE and 2MASS counterparts lower than 1 arcsec and 98\% lower than 2 arcsec.

\section{Spectral energy distributions}

\begin{figure*}
\centering
\begin{center}
\includegraphics[scale=0.24]{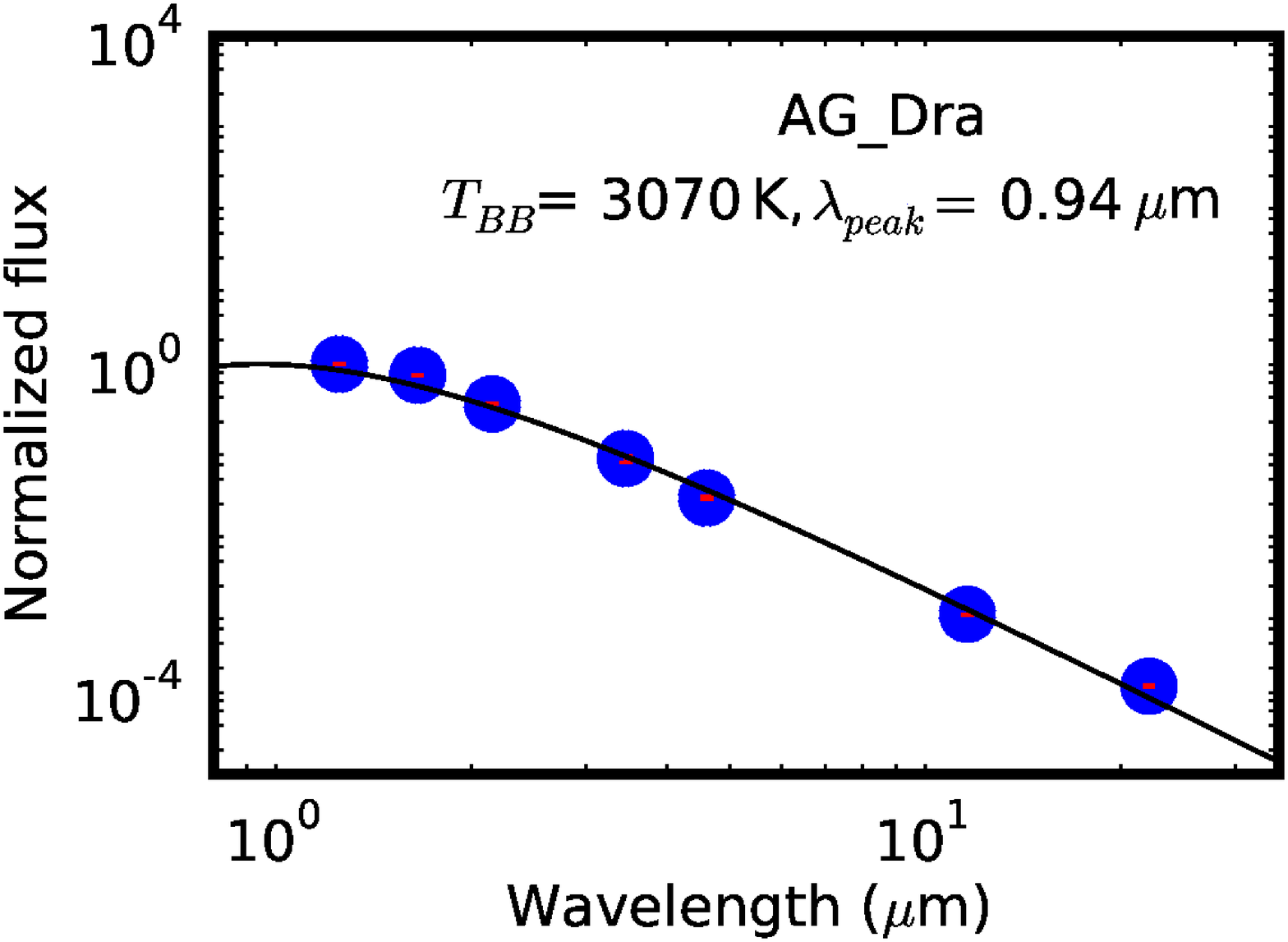}
\includegraphics[scale=0.24]{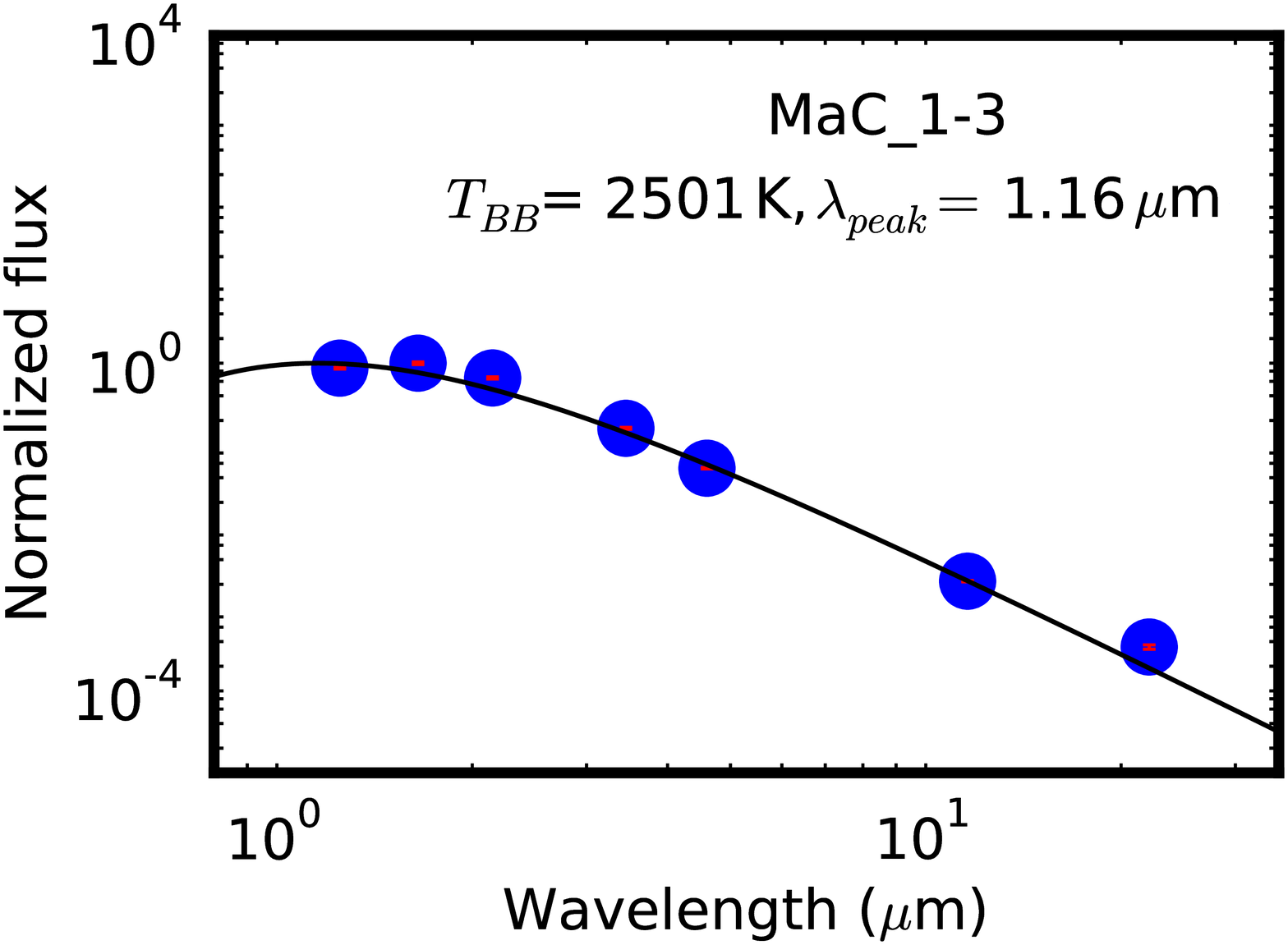}
\includegraphics[scale=0.24]{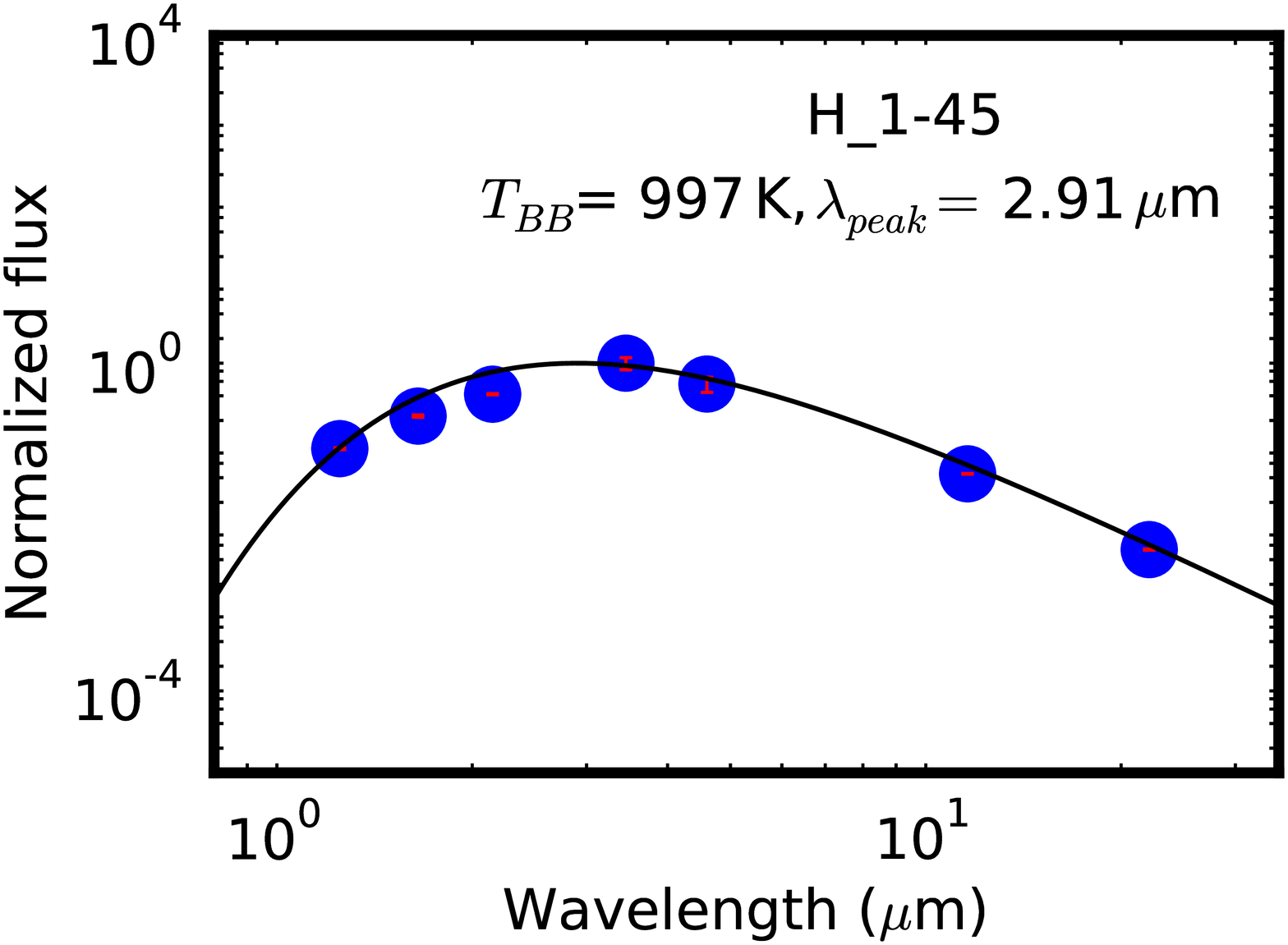}
\includegraphics[scale=0.24]{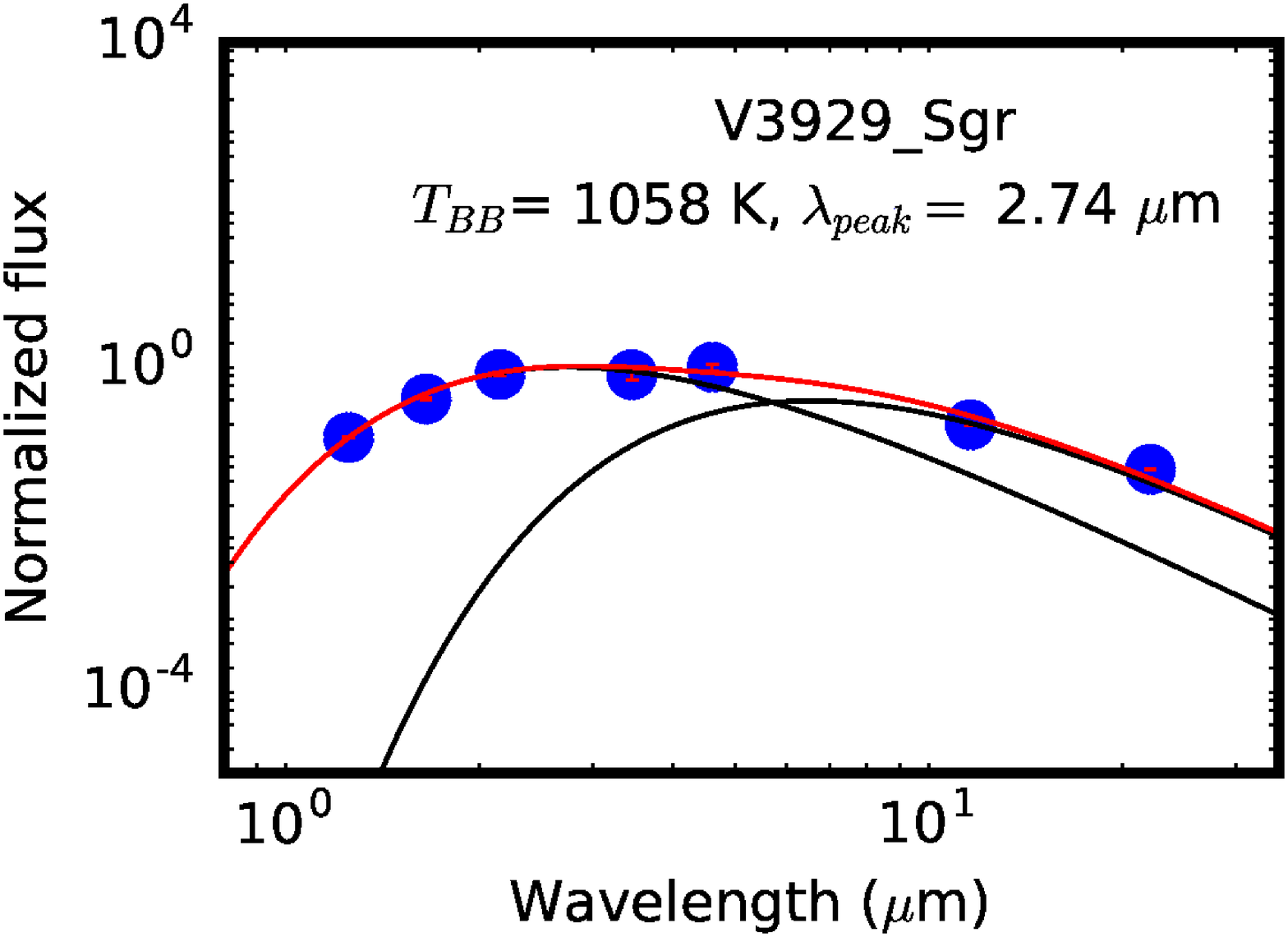}
\includegraphics[scale=0.24]{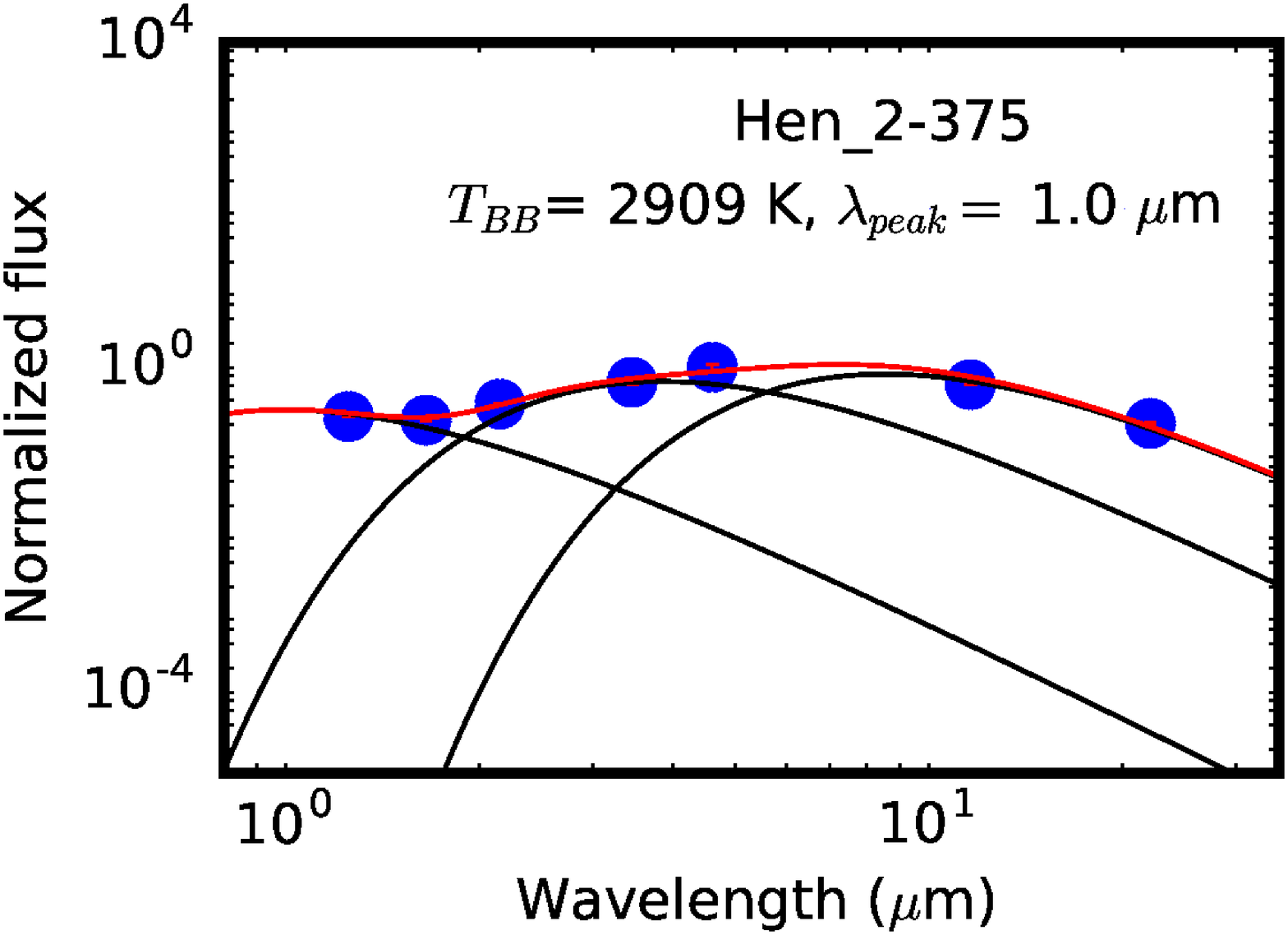}
\includegraphics[scale=0.24]{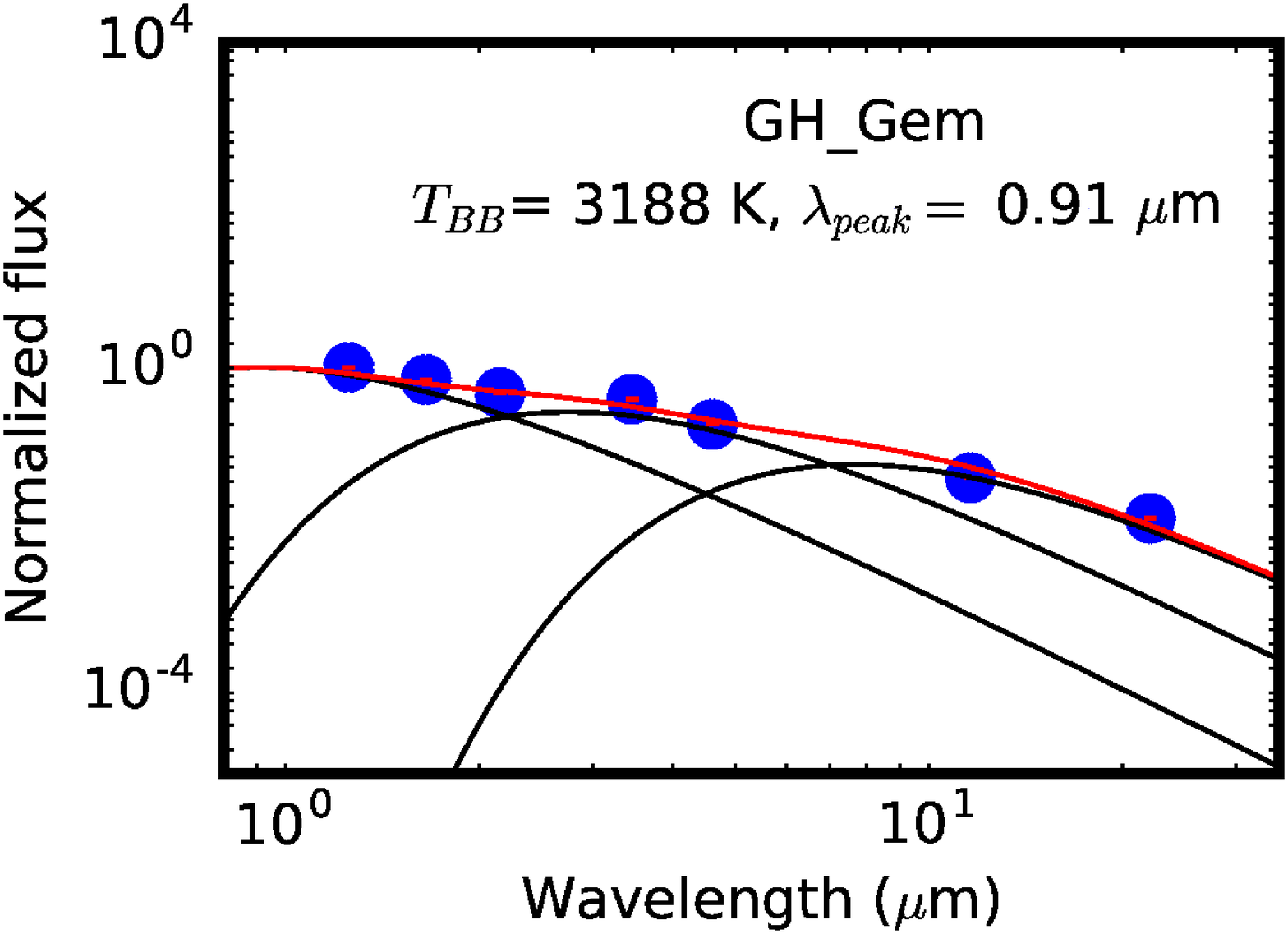}
\includegraphics[scale=0.24]{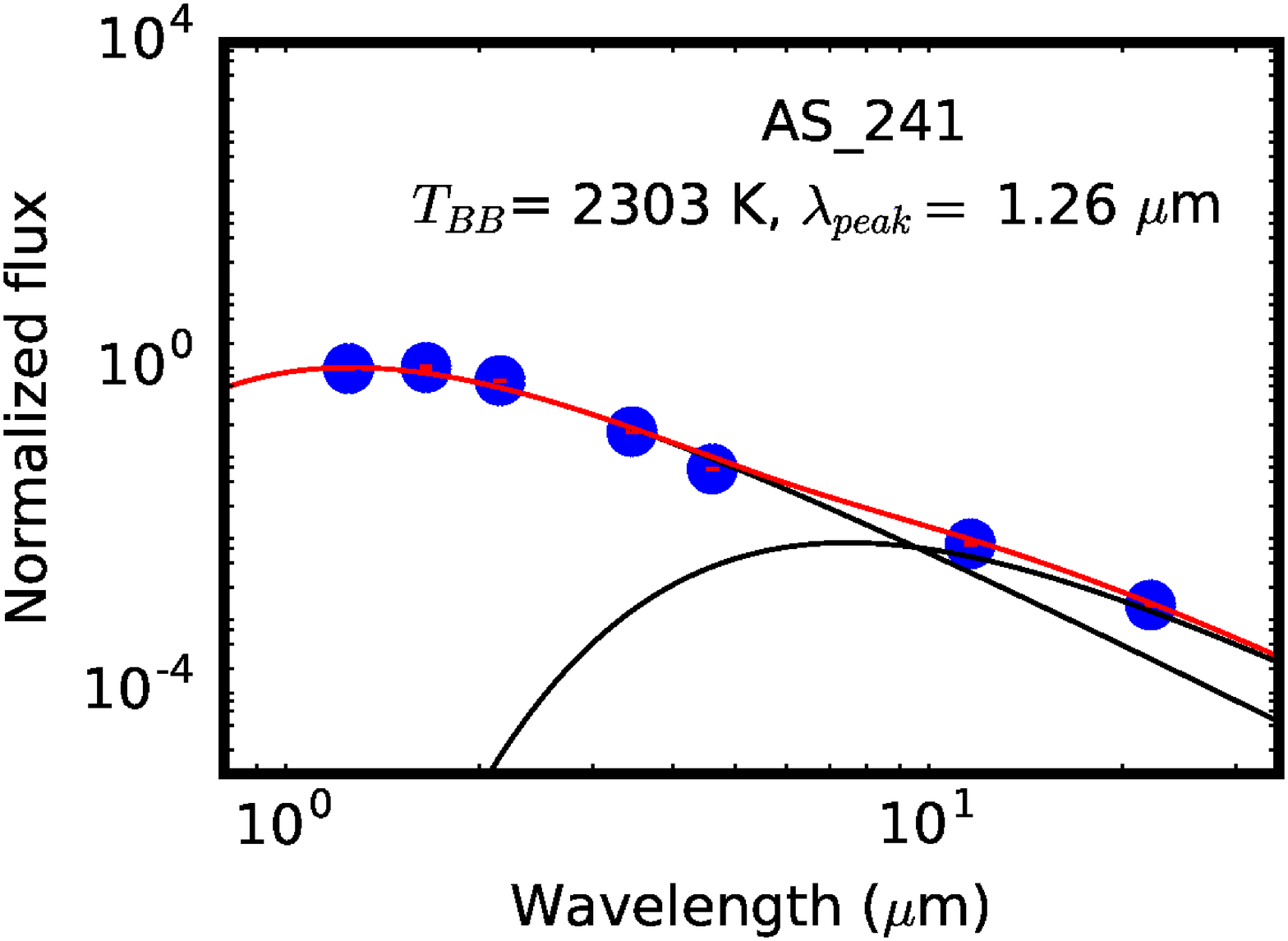}
\includegraphics[scale=0.24]{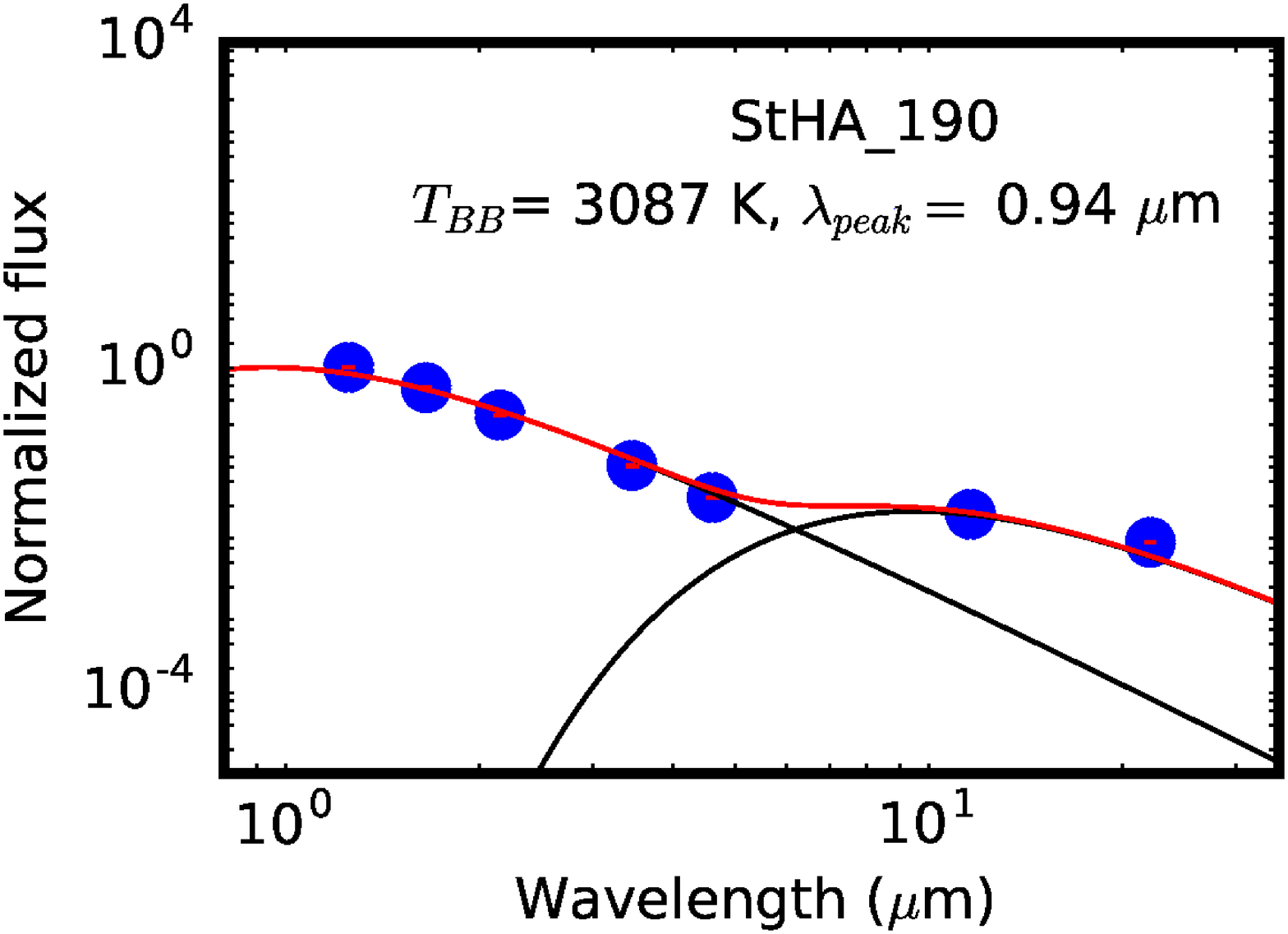}
\caption{Examples of normalized SED profiles for two S- (first line), two D- (second line), two S$+$IR- (third line), and 
two \DD-type SySts (fourth line). The blue dots correspond to the photometric data from the 2MASS and {\it WISE} surveys, the black lines to 
the BB models for each component, and the red line to the final fit only for cases of multicomponent SEDs. The errors of the data are 
portrayed by the vertical red bars, but in most cases are very small. (Figure~set~\ref{fig4} presents the SED profiles for all SySts, 
Figures~3.1--3.348).}
\label{fig4}
\end{center}
\end{figure*}

Due to the fact that the S- and D-type SySts show a peak in their SEDs between 1 and 15~$\mu$m, we constructed only the infrared 
SEDs of the SySts from 1 to 22~$\mu$m. To fit the 2MASS and AllWISE photometric data, we considered blackbody (BB) models.
This approximation is adequate for obtaining a rough estimate of the temperature of the cool companion in S-type SySts as well 
as the temperature of the dust shells in D-type SySts. But not all of the SED profiles can be fitted assuming one BB model.
It is known that some D-type SySts exhibit two dust components (Angeloni et al. 2010).

\begin{figure*}
\includegraphics[scale=0.2425]{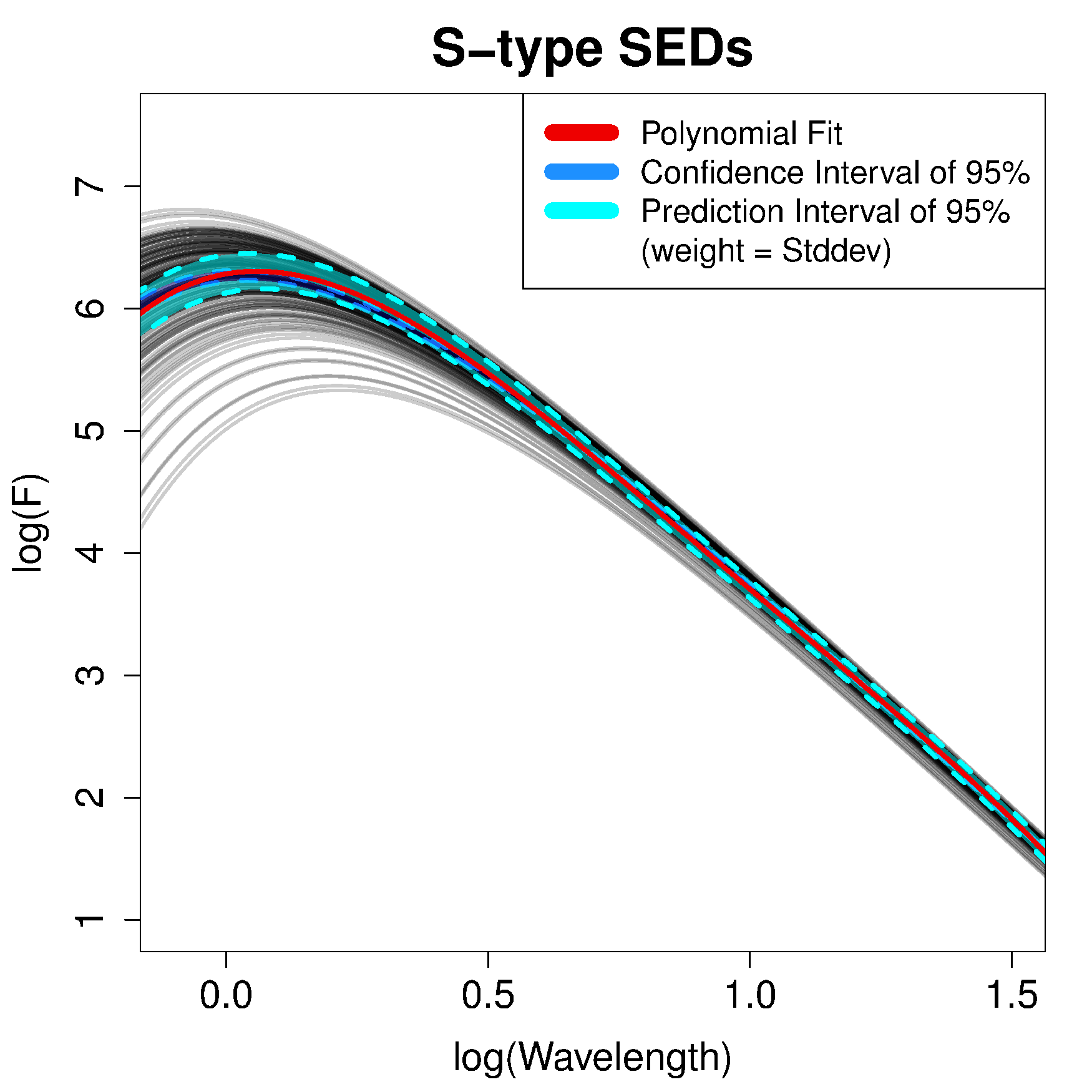}
\includegraphics[scale=0.2425]{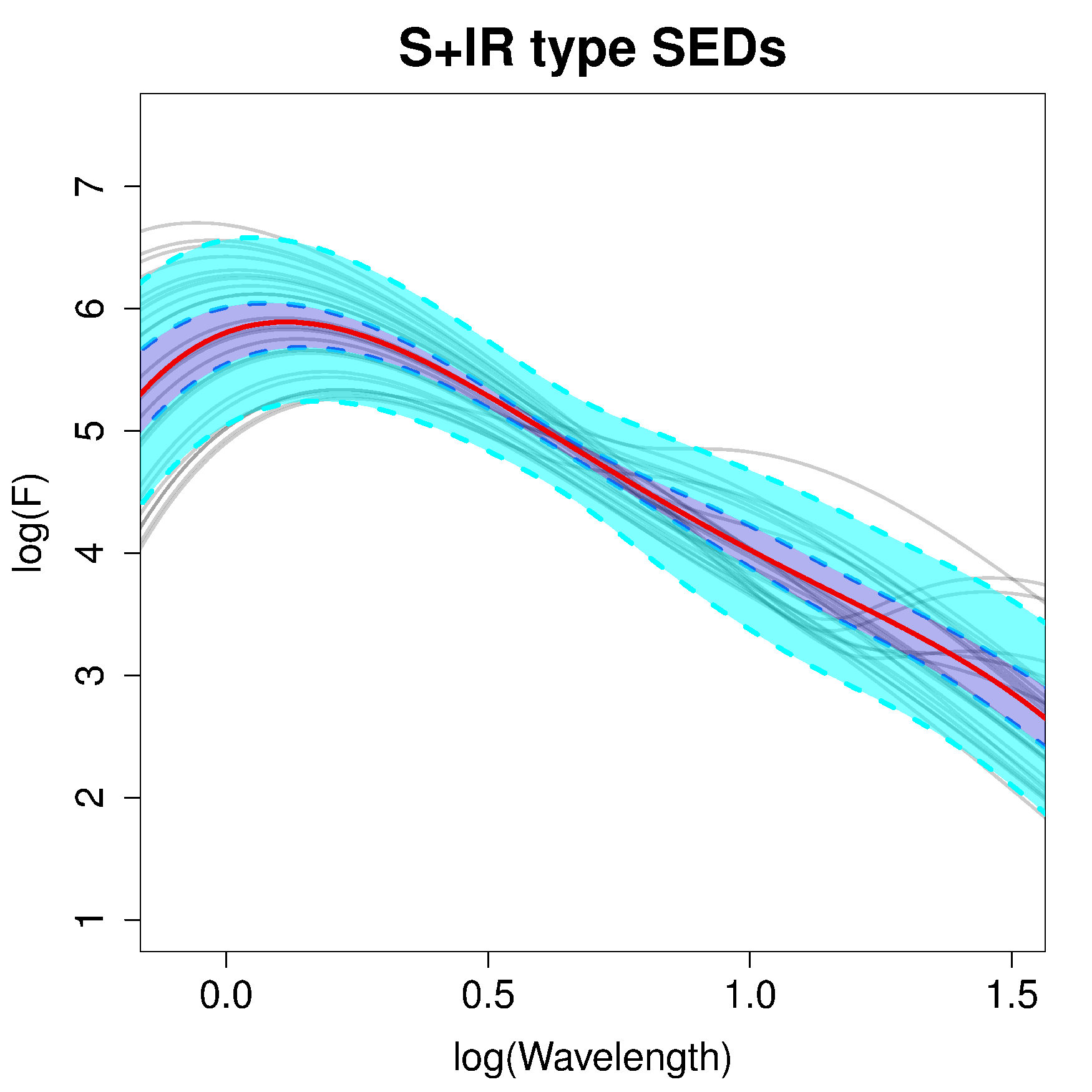}
\includegraphics[scale=0.2425]{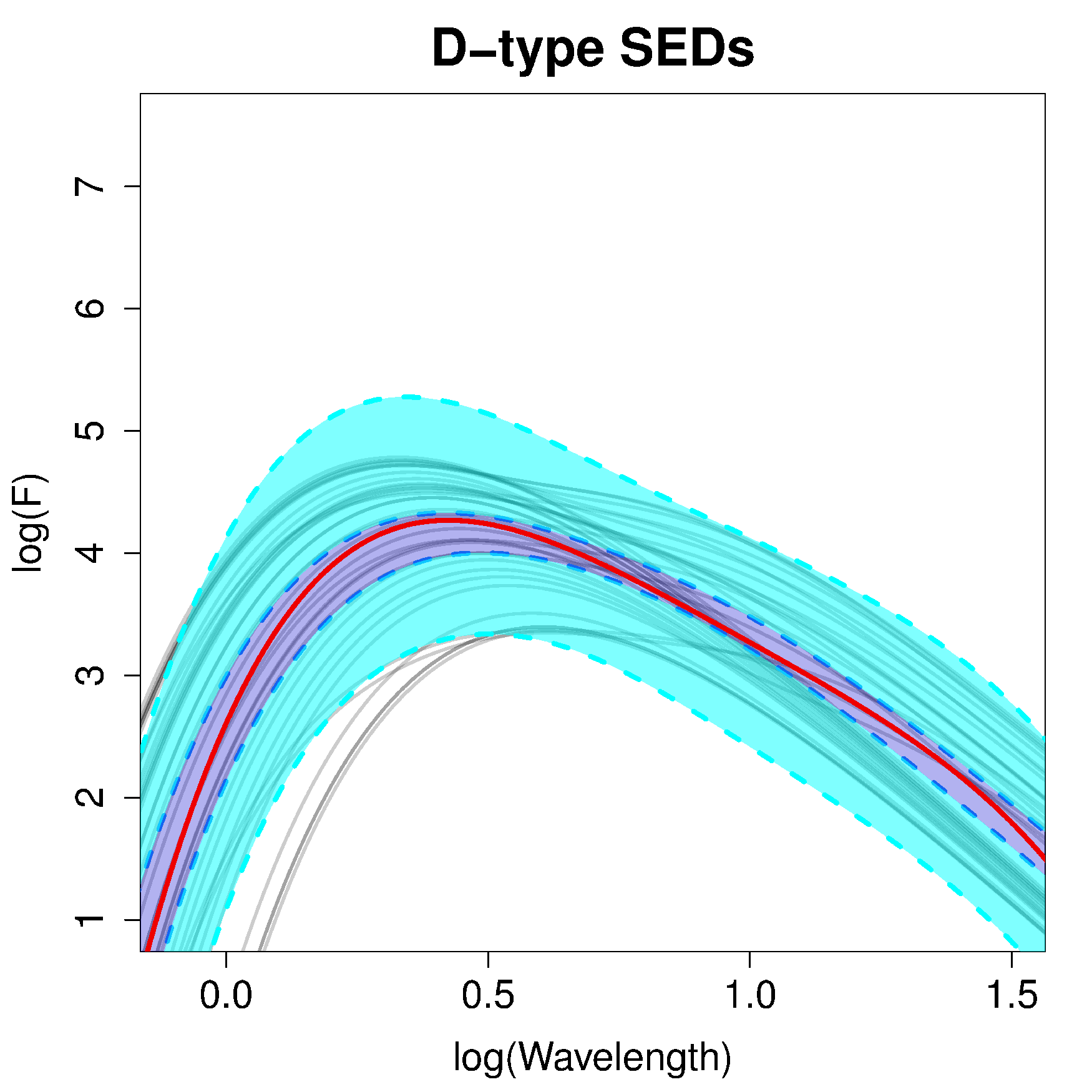}
\includegraphics[scale=0.2425]{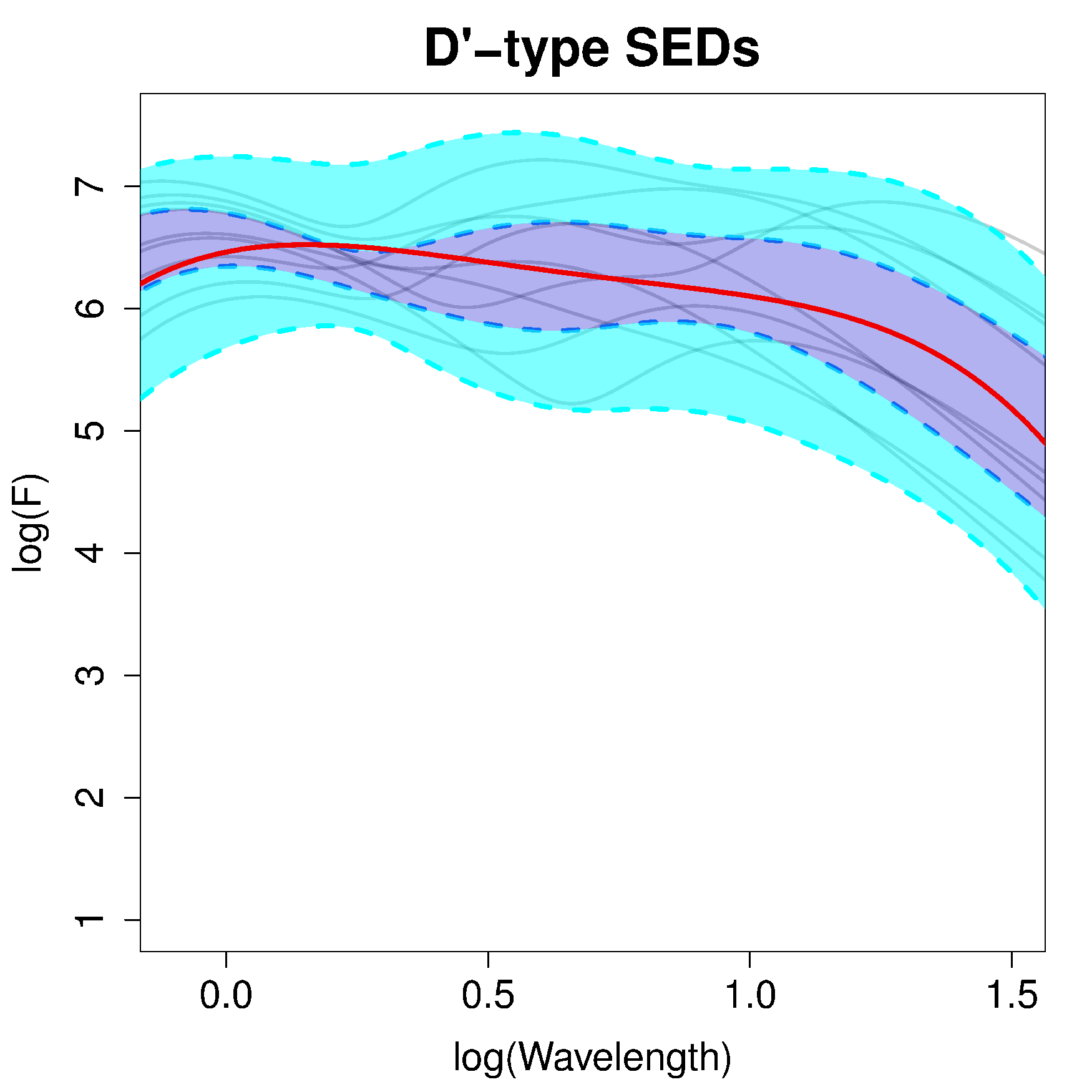}
\caption{SED flux density profiles of all the S-, S$+$IR-, D- and \DD-type SySts (gray lines). The typical SED profile for each type has also been determined with a confidence interval of 95\% (blue dark) and prediction interval of 95\% (cyan) derived with weights inversely proportional to the standard deviation. The red lines give the typical SED profile for each type by applying a polynomial fitting.}
\label{SEDsAll}
\end{figure*}

The majority of SySts are well fit by assuming only one BB model. But surprisingly, we found a number of SySts 
with an S-type SED profile and an infrared excess in the mid-infrared regime (see Fig.~\ref{fig4}~and~\ref{SEDsAll}). 
In the case where a BB model does not fit well all the data points well, we fit the data again taking into account only the 
first five data points (2MASS, {\it W1}, and {\it W2}) in the calculation of the reduced $\chi^2$. If the reduced $\chi^2$  of the second 
fitting is smaller than the first one and the visual inspection of the SED fittings indicates a possible excess at the {\it W3} and {\it W4} 
points, we apply the second BB model to fit the last two points.

All of these SySts, except for four display clear excesses at both {\it W3} and {\it W4} bands, which ensures the presence of the second dust 
component (see Fig.~\ref{fig4}). The low {\it W3} and {\it W4} photometric errors make the detection of the dust component more 
confident. Upper limits on the {\it W3} and/or {\it W4} data points are not considered as possible excesses. The likelihood of false detections 
is small but not negligible, especially for the cases (4 out of 27) in which the excess is detected only in the {\it W4} band.

Regarding the \DD-type SySts, three BB models (or two BB models for SySts with upper limit data) are used in order to reproduce their 
flat SED profiles, one for the cold giant and two for the dust components (Figure~\ref{fig4}). The temperatures of the BB models were 
varied until a good fitting was obtained by visually inspection of the SED fittings. The small number of data points used to fit each model makes 
the fittings for \DD-type SySts less precise, although the difference in the final SED profiles between \DD-type SySts and other 
types is still perceptible.

For a more detailed multiwavelength modeling of SySts' SEDs, we refer the reader to Skopal (2005) and Skopal (2015a,b,c). 
Only a handful of SySts have been modeled in such detail so far. In order to perform a similar study for all SySts, 
one would need data that are not available.

It should be noted that any BB model can adequately describe the temperature of a dust shell. 
However, the effective temperature (T$_{\rm eff}$) of red giants shows a significant departure from the BB temperatures (T$_{\rm BB}$), 
and further corrections is needed (see \S~3.1). 

The photometric data are not corrected for the interstellar extinction, which can alter the derived SED temperatures. However, 
it is very difficult to disentangle the interstellar extinction from the intrinsic extinction of SySts (e.g. Phillips 2007). 

The interstellar color excess, {\it E(B-V)}, has been estimated for several SySts using either the 2200\AA\ feature using UV data 
(Baratta \& Viotti 1983; Sahade et al. 1984; Parimucha \& Va\v{n}ko 2006) or recombination lines in the optical (de Freitas Pacheco \& Costa 
1992; Costa \& de Freitas Pacheco 1994; Pereira 1995; Mikolajewska, Acker \& Stenholm 1997 among others). The former method gives 
{\it E(B-V)} values in the range of 0.2-0.67, while the latter gives values up to $\sim$2~mag. The difference in {\it E(B-V)} estimates 
between the two methods is significant. There are also cases in which the {\it E(B-V)} differ by more than 0.5~mag despite using
the same method. 

SEDs for different {\it E(B-V)} from 0.1 to 1.5 (or {\it A$_{\rm V}$} from 3.1 to 4.65~mag, assuming {\it R$_{\rm V}$}=3.1 and the interstellar 
extinction law from Cardelli, et al. 1989) were simulated in order to verify the effect of interstellar extinction 
on our SED temperatures. The final effective temperatures of the cold companions in SySts (corrected for the blackbody approximation) 
differ by $\sim$240~K (620~K) for an extinction {\it A$_{\rm V}$}=3~mag (6~mag). For the SySt with the highest {\it E(B-V)} value of 0.67 
(RS~Oph) in the list from Parimucha \& Va\v{n}ko (2006), the temperature of the cold companion is 3460~K before correcting for the 
interstellar extinction and 3610~K after the correction. This difference of 150~K is within the overall error of our temperature 
estimates. For {\it E(B-V)}$>$1.6 the temperature difference exceeds the error of our temperature estimates (450~K). From the sample of 
67 southern SySts in Mikolajewska et al. (1997), there are 13 SySts with {\it E(B-V)}$>$1.5. It is, thus, very difficult to determine 
the effect of interstellar extinction in SySts, and we decided to provide temperatures without correcting for the interstellar extinction.

Figure~\ref{fig4} portrays the normalized flux SED profiles for two examples each of S-, D-, and \DD-type SySts as well as 
a fourth type of SySt that we call S$+$IR (see below). The peak of the SED profile for each SySt is normalized to one. The cool red giant 
clearly dominates the SED of the S-type, whereas dust emission dominates that of D-type SySts. In the case of V3929~Sgr, 
two dust components are required in order to reproduce the total SED profile in agreement with the results 
from Angeloni et al. (2010). In \DD-type SySts, both components, the red giant and the dust shells, contribute to the total SED, 
resulting in a nearly flat profile. The common characteristic of \DD-type SySts is the need for three BB components to fit the entire 
SED profile. D- and \DD-type SySts seem to compose two different groups, which is not unambiguous when only 2MASS data are used. 
Figure~set~\ref{fig4} presents all of the SED profiles.

In Figure~\ref{SEDsAll}, we present all of the flux density SEDs superimposed together in four individual plots. 
The SED fits of all known SySts are shown with black lines with 80\% transparency so that one can easily notice the high density of 
S-type SySts close to its corresponding prediction interval. We derived polynomial fits of fourth degree to define the typical SED 
representation of each type. These are indicated by the red solid lines in the figures. The polynomial fits were given as input in the 
"predict" function in R \citep{R} so that the prediction interval of the corresponding SED type could be calculated. We assumed 
weights inversely proportional to the standard deviation of the SEDs when deriving the prediction interval. They are shown in shaded 
cyan in the figures and represent the regions where the SEDs of 95\% of new observations are expected to occur. We also show 
in shaded blue the confidence intervals (where the mean of new observations is expected to be) of each type in the figures. They were 
derived with the R function "t.test", and a confidence level of 95\% was used.

\begin{figure*}
\includegraphics[scale=0.325]{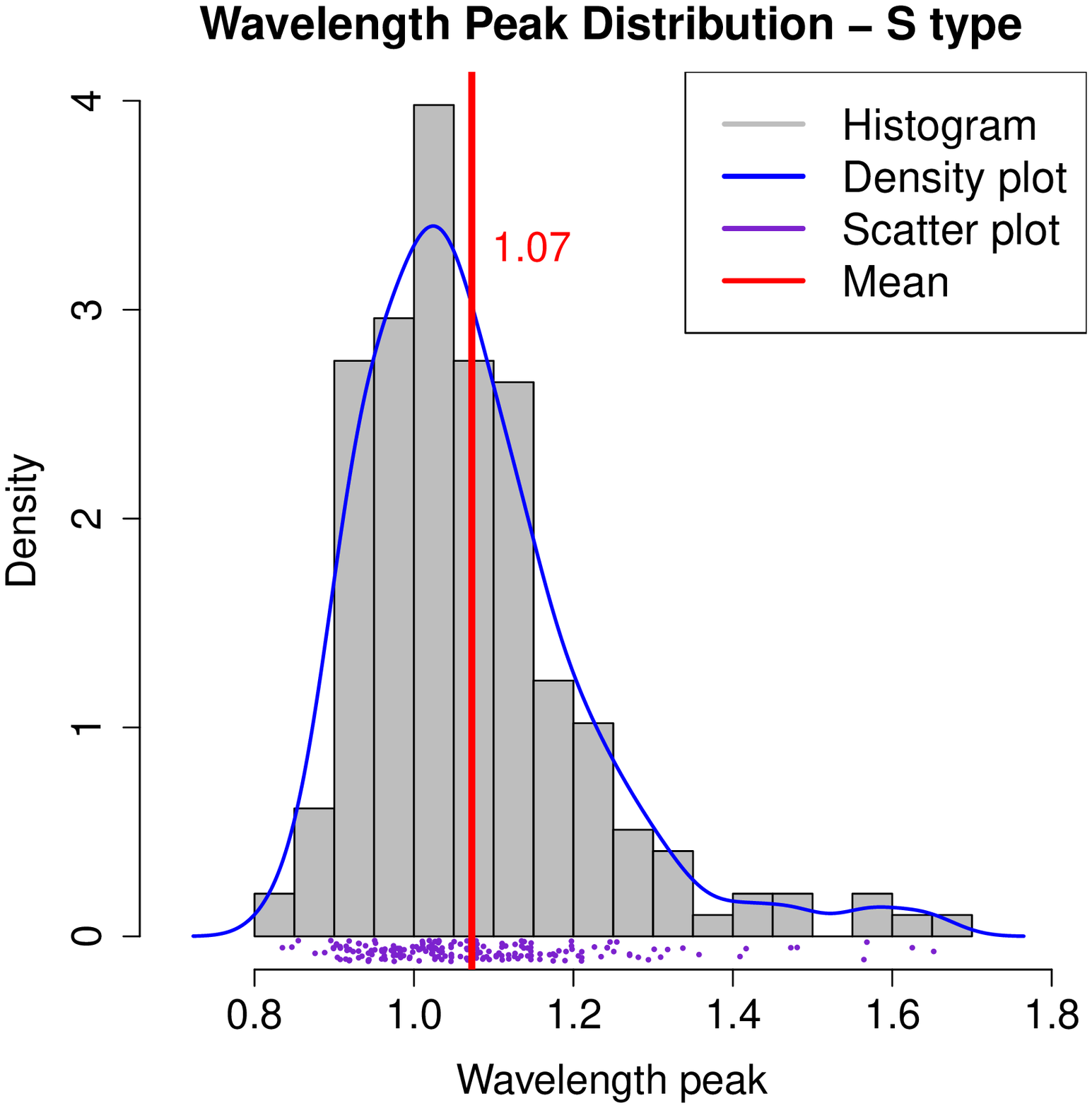}
\includegraphics[scale=0.325]{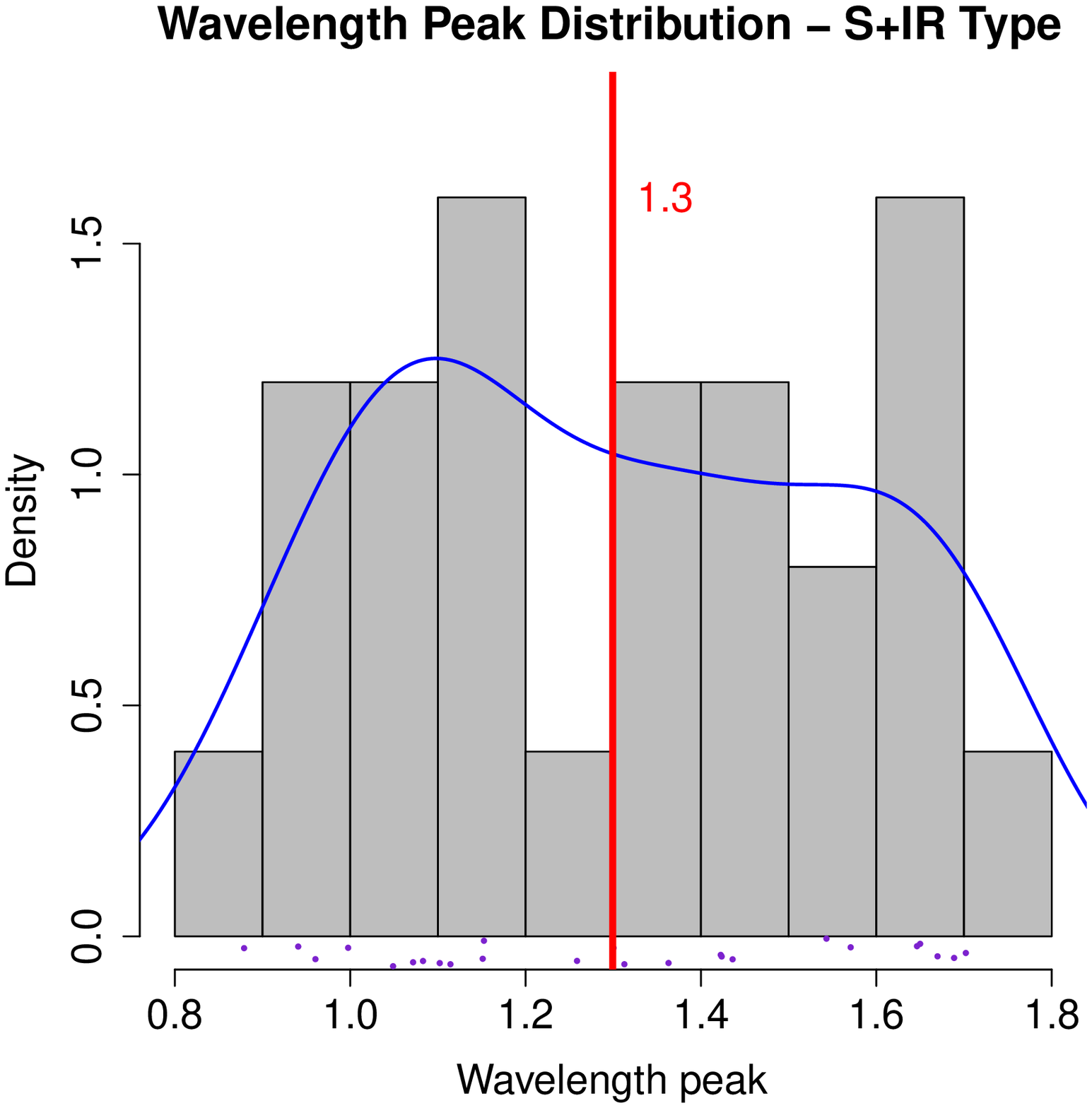}
\includegraphics[scale=0.325]{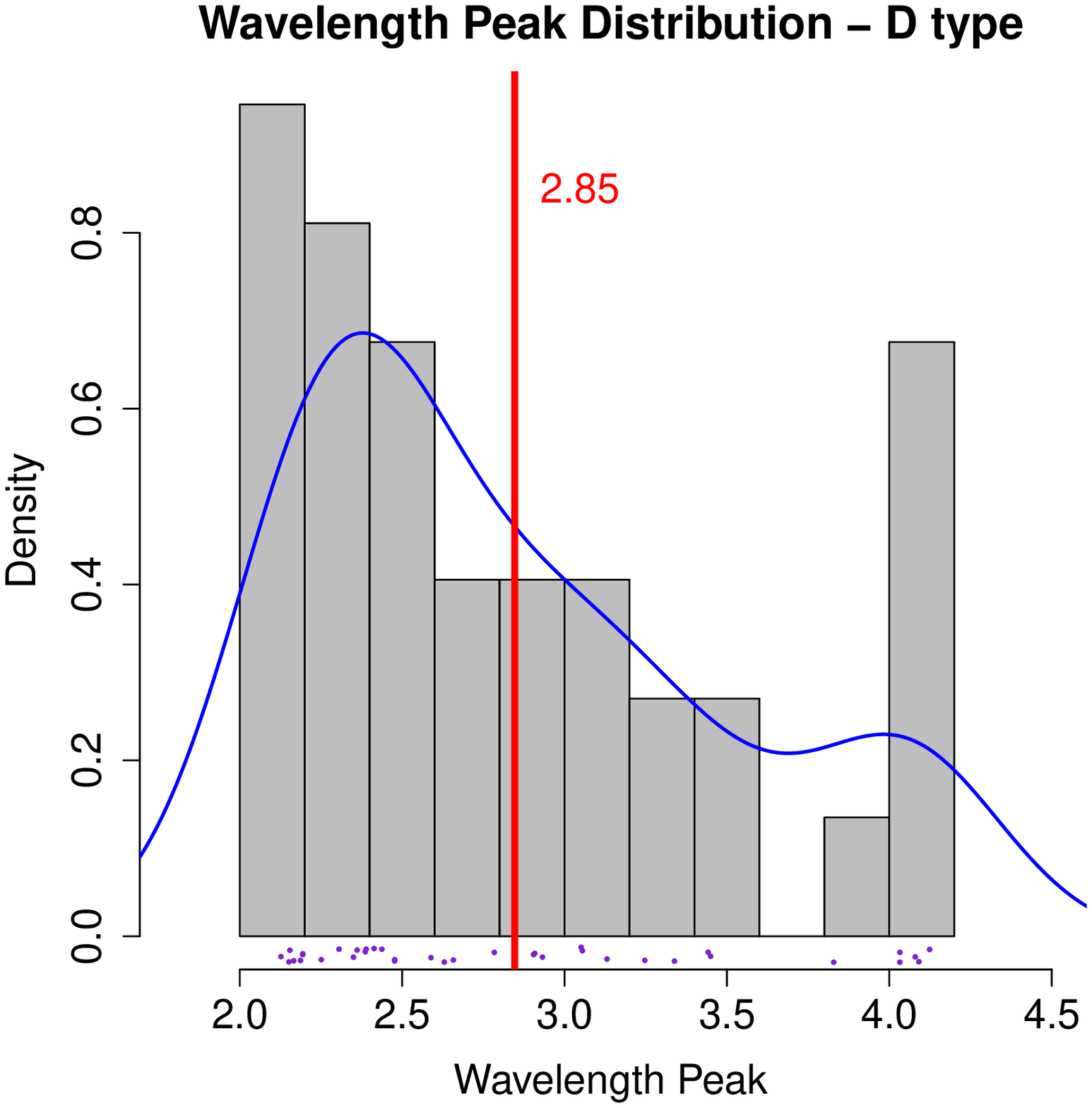}
\caption{Histograms of the wavelengths at which the SED profiles for each type of SySt peaks, except for that of \DD-type. 
The vertical red lines indicate the mean values for each type, the purple points portray the scattering of the sources, 
while the blues lines represent the density distribution.}
\label{fig6}
\end{figure*}

Generally, the SED profiles of S-type SySts exhibit a peak intensity at wavelengths between 0.8 and 1.7~$\mu$m with a high occurrence at 
1.0-1.1~$\mu$m, while the D-type SySts have an SED peak at longer wavelengths between 2 and 4~$\mu$m, with a high occurrence between 
2 and 2.5~$\mu$m (Figure~\ref{fig6}). As for the new S$+$IR type, they are well spread out along the wavelength range between 0.9 and 
1.7~$\mu$m (Figure~\ref{fig6})

Table~\ref{table3new} in Appendix~A lists all information obtained from the SED fitting. The first column gives the name of SySts in the same 
order as in Table~\ref{table2} (Appendix~A). The second and third columns give the old (from the literature) and the new classification 
(this work), respectively. The fourth column gives the temperature of the cold companions for the S-, S$+$IR- and \DD-type SySts while the 
fifth column lists the temperature of the dust shells for the S$+$IR-, D- and \DD-type SySts. The sixth column gives the effective temperature 
of the cold companions corrected for the blackbody approximation (see Section~3.1). The seventh and eighth columns provide the stellar effective temperatures of the secondary companions from the second {\it Gaia} data release (Gaia Collaboration et al. 2018) and their geometrical distances estimated from {\it Gaia} parallaxes (Bailer-Jones et al. 2018). The ninth column gives the wavelength at which the stellar component or 
the brightest dust shell shows a radiation peak. Finally, the last three columns provide information on the detection of the \ovi\ Raman-scattered line, X-ray emission, and their respective references.  

The vast majority of the SEDs are very well fitted, providing a good classification, and only around 20 of them may 
have a poor fitting ($<$7\%). R~Aqr, for instance, is one example with a new classification. In particular, Contini 
\& Formiggini (2003) have pointed out that the cool companion of R~Aqr is a Mira variable star surrounded by a dust envelope
with a temperature of 1000~K. Our SED profile indicates an S-type SySt, but the upper limit values for the 
{\it W1}, {\it W2} and {\it W3} bands makes its classification unreliable. Another example is AS~245, which clearly exhibits an 
S-type SED profile rather than of a D-type SySt with a Mira companion. Gromadzki et al. (2009) also reclassified this SySt as 
an S-type, which means that the old classification of SySts in the S/D scheme is not always correct.

Overall, 74\% of the known SySts are classified as S-types, 13\% as D-types, and only 3.5\% as 
\DD-type. Moreover, only 26 SySts have obtained a new classification in this work. Fourteen S-types and two D-types are now
classified as S$+$IR-types, three D-types are classified S-types and four as \DD-types, two S-types have been reclassified as D- and \DD-types, 
and finally, one \DD-type is now classified as D-type.

As mentioned before, it is known that SySts exhibit large flux variations due to the interaction between the evolved red giant and the 
hot WD. The mass transfer from the red giant to the WD is the main cause of these variations together with several other
phenomena like orbital motion, dust obscuration, and radial pulsations of the Mira companions in D-type or semi-regular pulsations 
of the red giants in S-type (see e.g. Whitelock 1987; Henden \& Munari 2000, 2001, 2006; Mikolajewska 2001; Gromadzki et al. 2009, 2013, 
among others). 

Multi-epoch photometric studies have been carried out either in the optical wavelengths or the near-infrared regime(e.g. Lorenzetti et al. 1985; 
Henden \& Munari 2000, 2001, 2006; Mikolajewska 2001; Gromadzki et al. 2009, 2013; Jurkic \& Kotnik-Karuza 2012). 
All of these studies have shown that flux variations can have a maximum amplitude up to 6-7~mag in the optical regime but significantly 
smaller at longer wavelengths. CH Cyg is a notable example with a large amplitude variations  of up to 4~mag in the {\it U}-band but 
significantly lower in the near-infrared regime of 0.5~mag (Munari et al. 1996).

Therefore, it is necessary to examine how much these flux variations between the 2MASS and AllWISE data (due to the different epochs of the two 
surveys) affect the SED fitting and the resulting temperatures. For this exercise, we used an S- and a D-type SySt as test objects. 
We inferred a fluctuation in their photometric data and then we fitted their new data sets assuming six different scenarios: 
(1) 2MASS in the maximum brightness and {\it WISE} in the minimum, (2) 2MASS in the minimum brightness and {\it WISE} in the maximum, (3) 
2MASS in the maximum brightness and {\it WISE} without any variation (4) 2MASS in the maximum brightness and {\it WISE} without any variation, 
(5) 2MASS without any variation and {\it WISE} in the minimum brightness, and (6) 2MASS without any variation and {\it WISE} in the minimum brightness. 
This was repeated for four different amplitude variations (see Appendix~B). We then repeated the whole procedure for four D-type SySts, but 
in this case we used their observed amplitude variations taken from Gromadzki et al. (2009).

The main result from this exercise is that flux variations up to 2~mag do not significantly affect the SED fitting or the classification of SySts 
(see Appendix~B). Moreover, by examining closely all of these artificial SEDs, we came to the conclusion that a poor fitting on the {\it W1} and 
{\it W2} bands or the 2MASS data may indicate some flux variability between the data. Overall, flux variations between the 2MASS and {\it WISE} data 
result in temperature variations of the order of 300-400~K which is comparable to the uncertainties of our estimates.

Only a handful of SySts exhibit noticeable signs of such variations in their SEDs (e.g. 2MASS~J17391715-3546593, 356.04+03.20, AS~245, 
H~2-34, PN~H~2-5, RT~Cru, SMP~LMC~88, UV~Aur, BI~Cru, Hen~2-127, AS~221, Hen~2-139, K~3-9, RR~Tel, V347~Nor, V835~Cen, 354.98-02.87).

Nevertheless, a good agreement is found with previous studies. For instance, RR~Tel is a known and well-studied variable SySt 
with maximum amplitudes up to 3.47~mag in {\it J}, 2.84~mag in {\it H}, 2.05~mag in {\it K} and 1.29~mag in {\it L} magnitudes 
(Gromadzki et al. 2009). Nevertheless, our T$_{\rm dust}$=1258~K agrees with the temperature derived from a more detailed SED study between 
1200 and 1350~K (Jurkic \& Kotnik-Karuza 2012). The S-type SySt CH Cyg is another example for which our T$_{\rm eff}$=3063~K agrees with the 
value reported by Hinkle, Fekel \& Joyce (2009).

Regarding the S$+$IR-type SySts, the excess at 11 and 22 $\mu$m cannot be attributed to possible flux variations between the 2MASS and {\it WISE} 
observations since the {\it W1} and {\it W2} bands do not present any significant deviation from the BB fitting.

Only 9 out of 42 D-type SySts have a new classification type (two as S$+$IR-, three as S-, and four as \DD-type). 
The presence of a Mira variable star is not directly confirmed in all these systems. For instance, LAMOST~J202629.80+423652.019 is proposed 
to be a D-type based only on some optical emission lines (Li et al. 2015). EF Aql, on the other hand, shows variability with amplitudes of 
around 2~mag and a period of 330~days (Margon et al. 2016) but its {\it J-H} and {\it H-K} color indices are not consistent with a D-type classification (Corradi et al. 2008; Rodriguez-Flores et al. 2014; Clyne et al. 2015).

\begin{figure}
\includegraphics[scale=0.60]{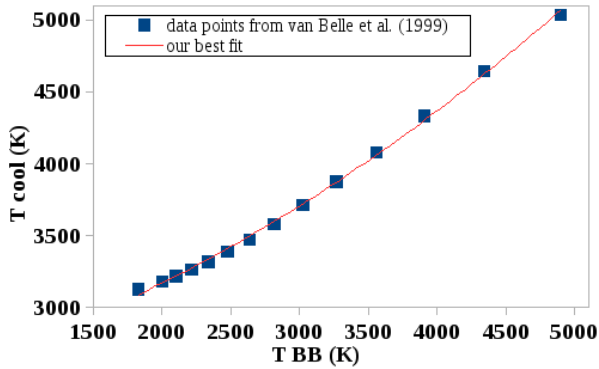}
\caption{T$_{\rm BB}$ vs. T$_{\rm cool}$ plot. The data points have been obtained from van Belle et al. (1999). T$_{\rm BB}$ 
corresponds to the stellar temperature assuming a blackbody approximation and T$_{\rm cool}$ to the observed effective temperatures 
for spectra from M7 to G7. The red line corresponds to our best fit of the data assuming a polynomial function of second degree. }
\label{fig7}
\end{figure}

Multiwavelength modeling of the SySts' SED profiles have shown that during the supersoft X-ray phase or days/weeks after an outburst event, 
the major contributor to the optical and near-IR spectral wavelength regimes is the nebular emission (Skopal 2015a, b, c). The duration 
of this active phase corresponds to a small part of the SySts' lifetime, and it is not expected to affect our SED fits.

\begin{figure*}
\begin{center}
\includegraphics[scale=0.45]{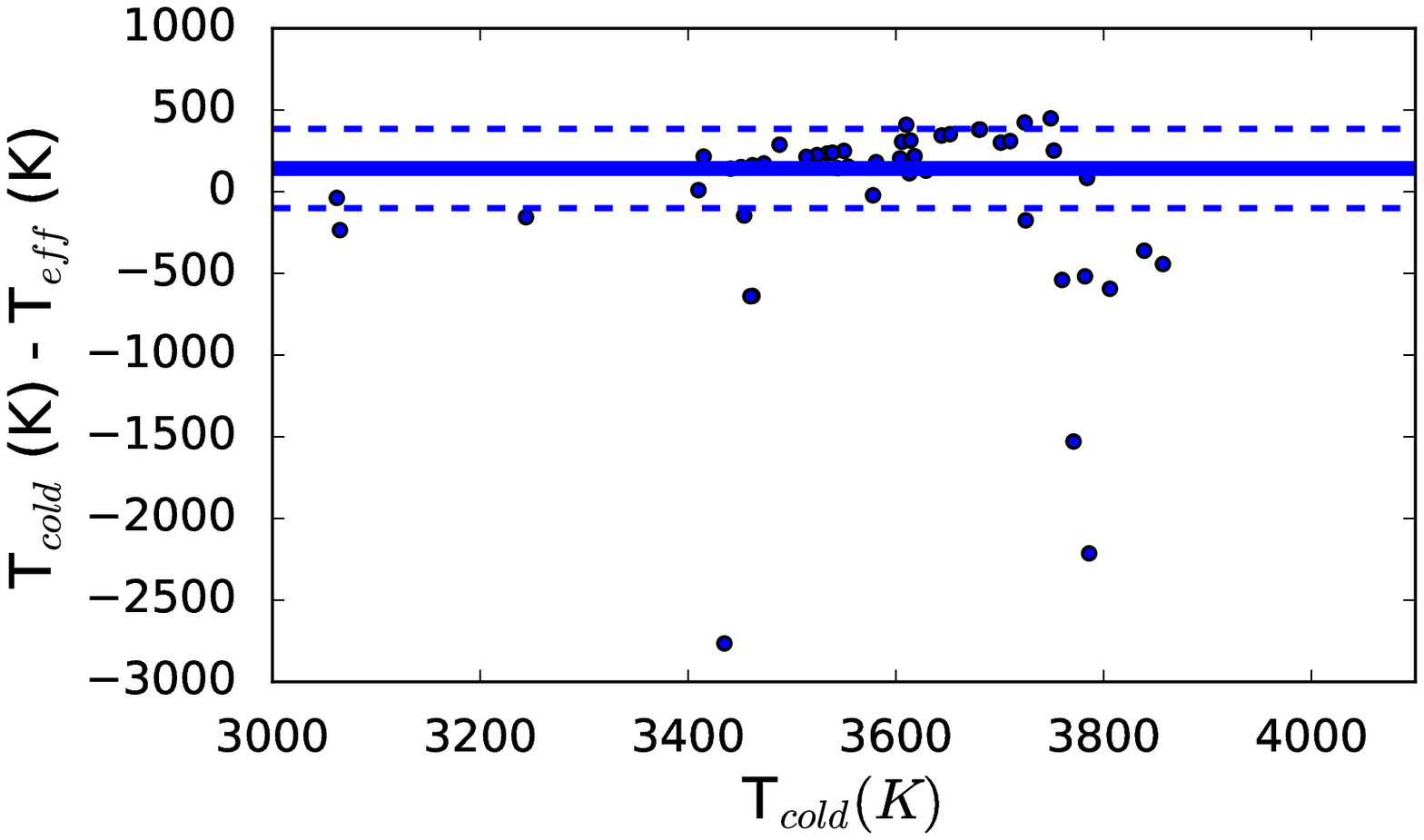}
\includegraphics[scale=0.45]{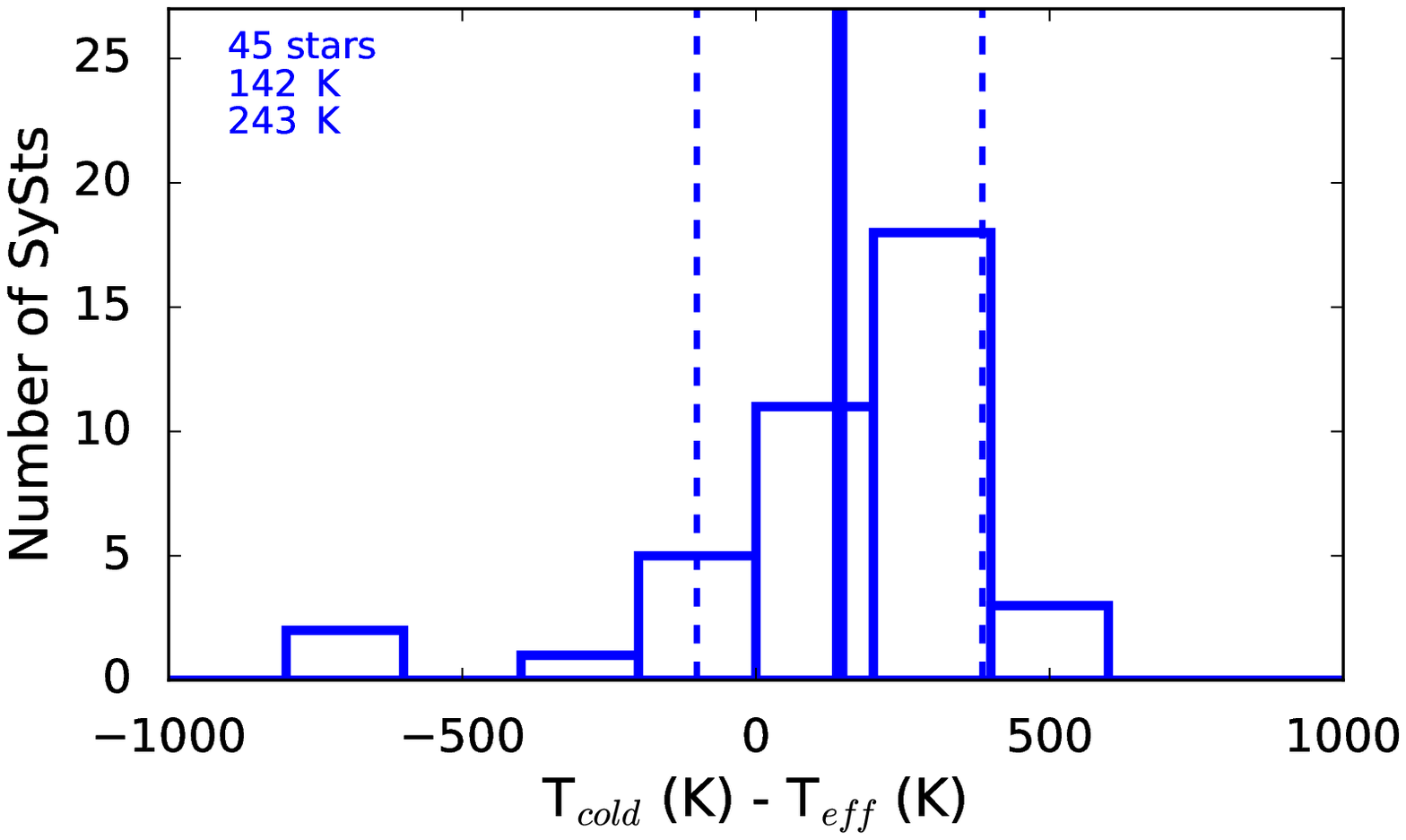}
\includegraphics[scale=0.45]{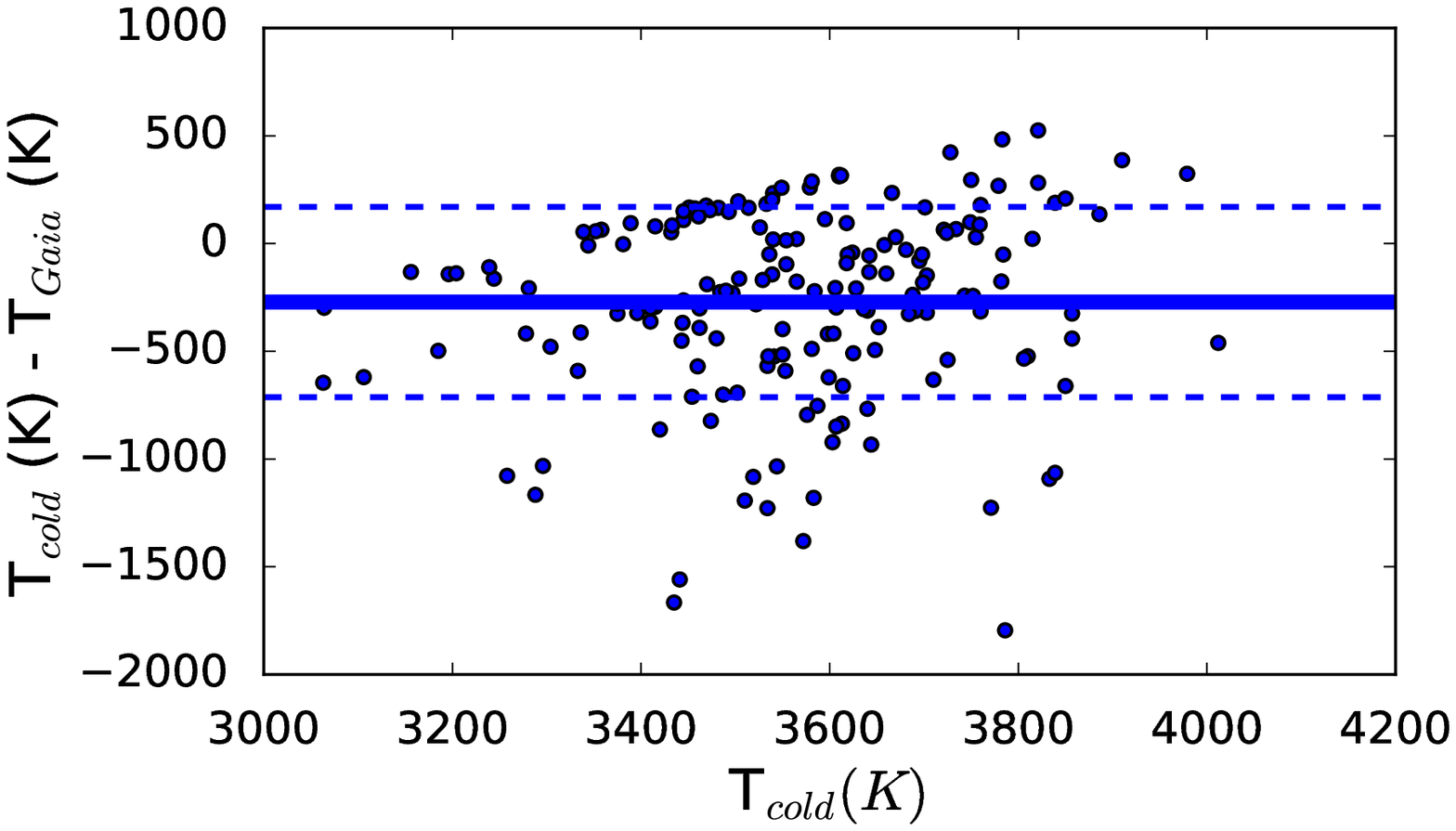}
\includegraphics[scale=0.45]{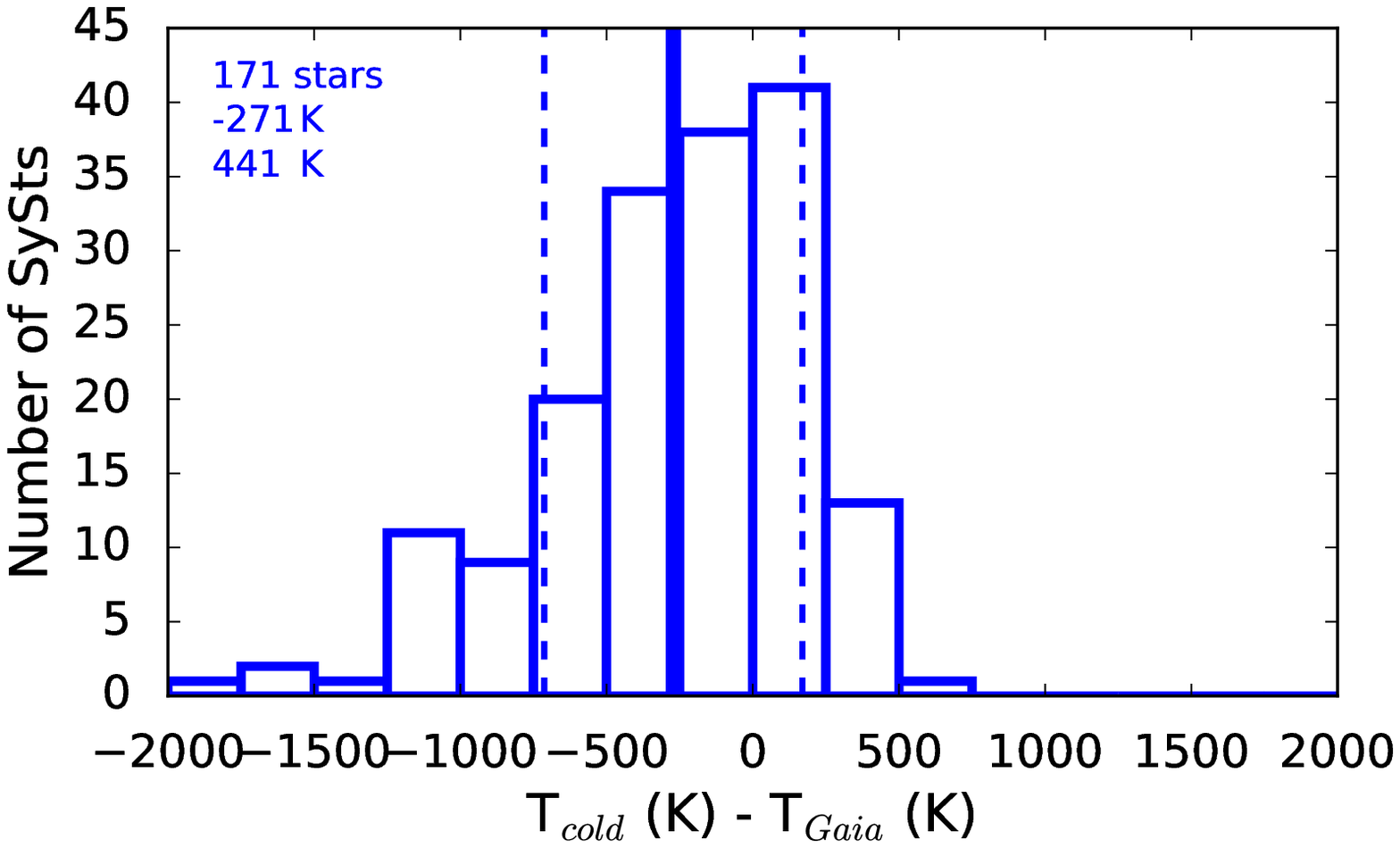}
\caption{Left panels: difference between our temperature estimates (T$_{\rm cool}$) and the effective temperatures taken from the literature 
(T$_{\rm eff}$) for 48 SySts and those derived by {\it Gaia} (T$_{\rm Gaia}$) for 171 SySts as a function of (T$_{\rm cool}$). Right panels: 
histograms of the temperature difference between T$_{\rm cool}$ and T$_{\rm eff}$ (the three sources with high temperature difference are 
excluded from this plot), and T$_{\rm cool}$ and T$_{\rm Gaia}$. The solid lines are the average of the temperature difference, while the 
dashed lines represent the 1$\sigma$ of the distributions.} 
\label{fig20}
\end{center}
\end{figure*}

\subsection{T$_{\rm eff}$ vs. T$_{\rm BB}$}

As mentioned above, it is well known that T$_{\rm BB}$ does not provide the correct T$_{\rm eff}$. Extensive studies on 
the relation between the T$_{\rm eff}$ and T$_{\rm BB}$ of K, M and G giant stars have been performed by several authors 
(Dyck et al. 1996; Van Belle et al. 1999; Alonso et al. 2000; Houdashelt et al. 2000; Tej \& Chandrasekhar 2000, VandenBerg \& Clem 2003). 
T$_{\rm BB}$ shows a significant departure from T$_{\rm eff}$ for red giants with T$_{\rm eff}<$4000~K, and this difference 
becomes more important as T$_{\rm eff}$ becomes lower (see Figure~2 in van Belle et al. 1999). The discrepancy between the 
T$_{\rm eff}$ and T$_{\rm BB}$ is attributed to the presence of absorption features such as $^{12}$CO(2,0) at 2.29~$\mu$m or 
MgH and TiO bands (van Belle et al. 1999 and references therein).

Given that our T$_{\rm BB}$ range between 2200 and 4000~K, it is clear that the difference between T$_{\rm BB}$ and T$_{\rm eff}$ is 
significant for most SySts, and it has to be taken into consideration. To convert T$_{\rm BB}$ into T$_{\rm eff}$, 
we took the values from Table~8 in van Belle et al. (1999) and applied a polynomial function of second degree to fit the data 
(Figure~\ref{fig7}). This relation allows us to convert the T$_{\rm BB}$ of red giants to a more reliable T$_{\rm eff}$. 
Hereafter, we refer to the effective temperature of the cool companions obtained from our best fit as T$_{\rm cool}$. 

The data yield the relation $T_{\rm cool}=6.22*10^{-5}*T_{\rm BB}^{2}+0.23*T_{\rm BB}$+2468~(K) with {\it R}$^2$ (the goodness of fit) 
equals to 0.9986 (Figure~\ref{fig7}). The T$_{\rm cool}$ values of red giants for all SySts are given in Table~\ref{table3new} in Appendix~A. 
The error of T$_{\rm cool}$ is estimated between 250 and 450~K due to the average standard deviation of the observed 
T$_{\rm eff}$ of 250~K (van Belle et al. 1999) and the uncertainty of our fitting.

Before we continue to the analysis of the temperatures for each type of SySts, it is necessary to compare our estimates with
those from previous spectroscopic studies (T$_{\rm eff}$, Table~\ref{table8}). The upper left panel in Figure~\ref{fig20} displays the 
difference between T$_{\rm cool}$ and (T$_{\rm eff}$). The blue solid line represents the average difference between the two temperatures 
(excluding three stars with T$_{\rm eff}>$4400~K), while the blue dashed lines correspond to the 1$\sigma$ deviation of the distribution. 
The three stars with high temperature belong to the rare group of yellow SySts with a K or G spectral-type giant. There are seven more sources 
with a temperature difference of around 500~K, which is comparable with our temperature error of 450~K. The upper right panel in 
Figure~\ref{fig20} illustrates the histogram of the temperature difference without including the three extreme cases.

Given that our SED fitting is restricted only to the IR regime (2MASS and {\it WISE} data) while other authors have used a broader spectral 
wavelength range from X-ray to radio (Skopal 2005, Angeloni et al. 2007), it is worth to comparing our temperatures with those derived  
from these studies. Again, a good agreement between the two temperatures is found. The temperature difference is around 550~K. This verifies that the analysis of SEDs in this work provides a reliable classification as well as temperature within the uncertainties.

\begin{figure}
\includegraphics[scale=0.42]{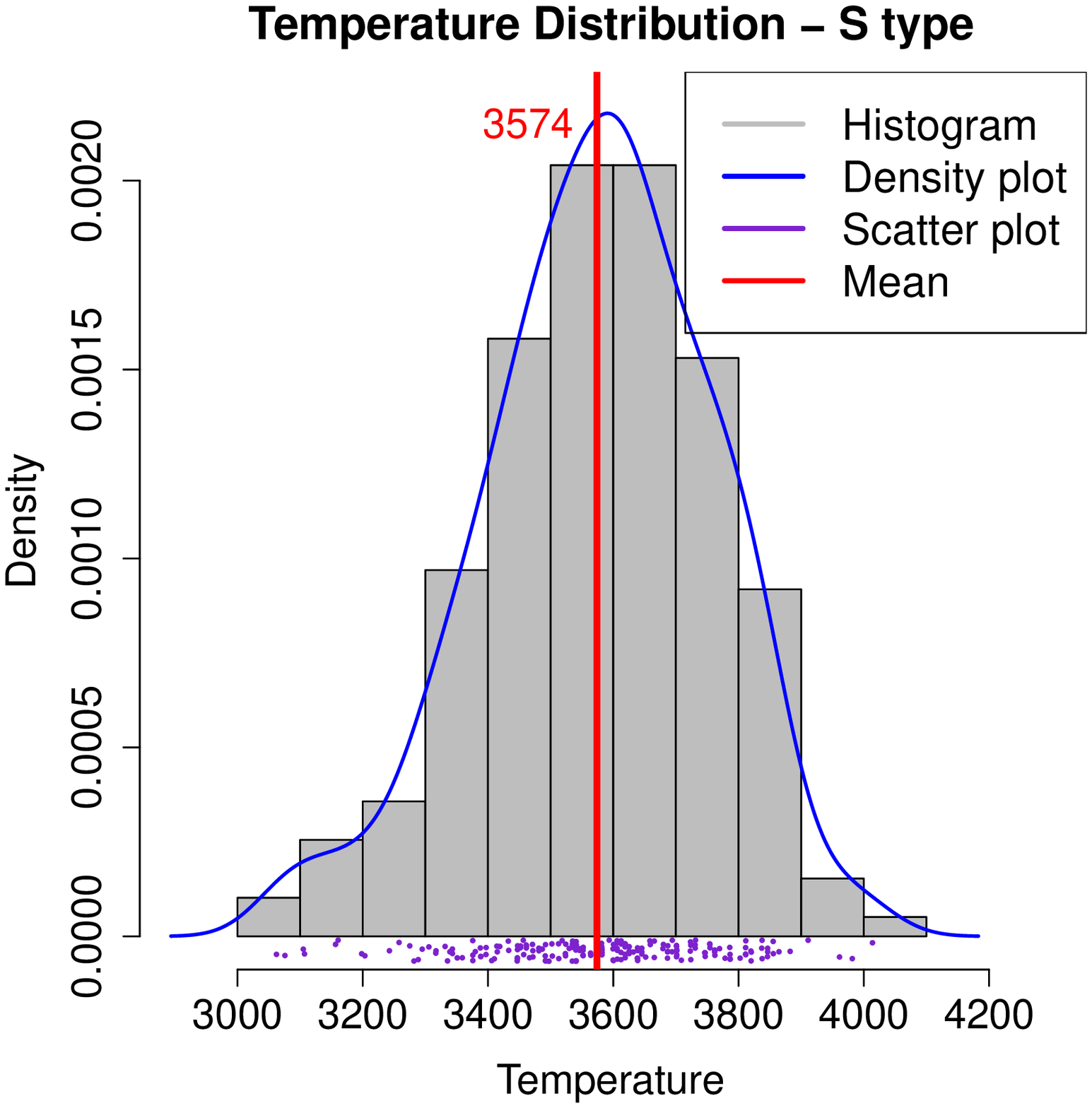}
\includegraphics[scale=0.42]{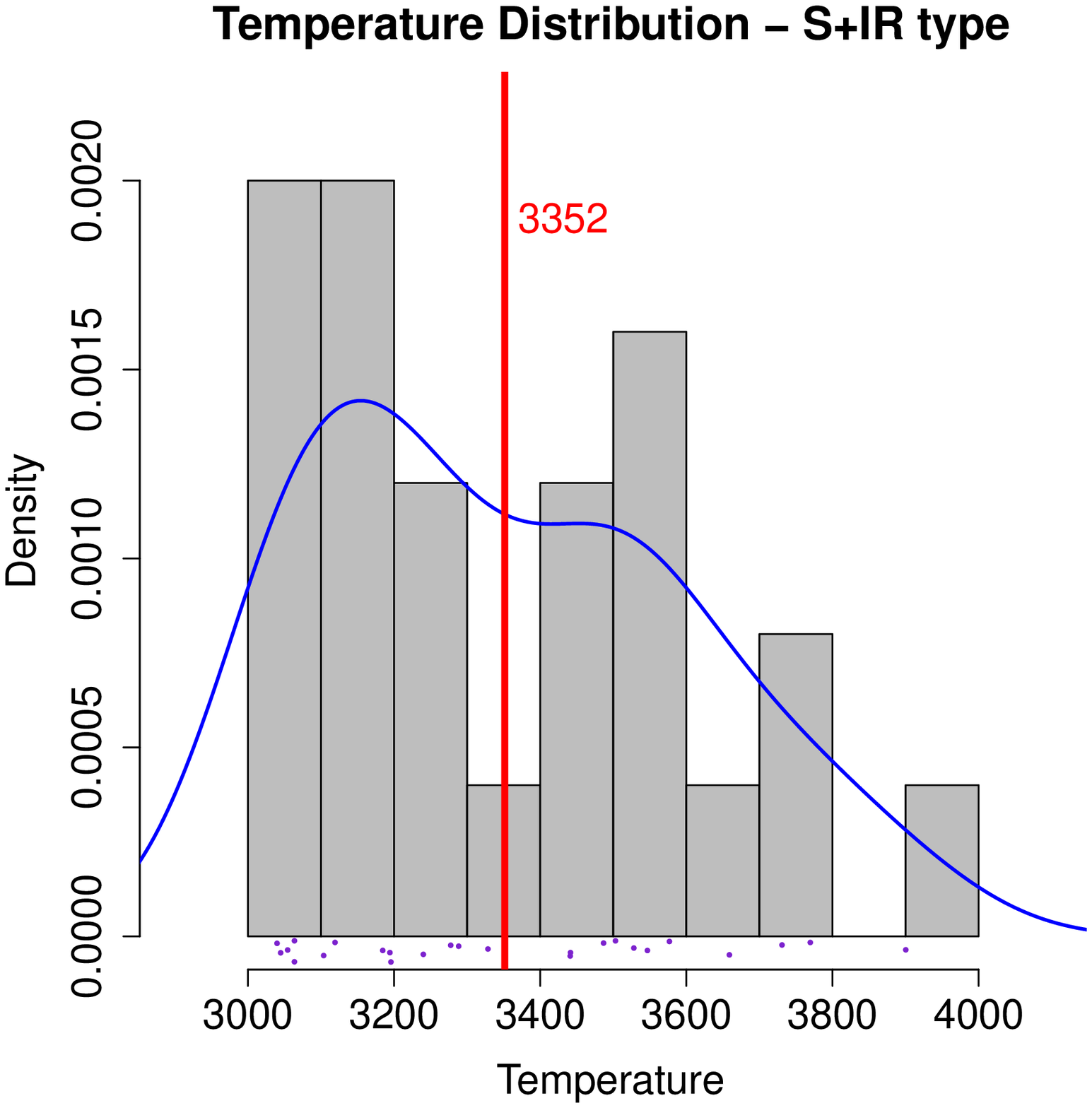}
\caption{Histogram of the T$_{\rm cool}$ for the S-type (upper panel) and for the S$+$IR-type SySts (lower panel). 
The colored lines are the same as those in Figure~\ref{fig6}.}
\label{fig8}
\end{figure}

To go one step further, we also compared the temperatures of the cold companions obtained from Skopal (2005) with the equivalent 
observed values from Table~\ref{table8}. A lower temperature difference of 200~K is found. This better matching is not a result of the 
broader spectral wavelength range used for the SED fitting, but the use of a more robust stellar atmosphere model for the red giants, 
considering T$_{\rm eff}$ and $\theta_{g}$ as free parameters rather than a blackbody.

In Figure~\ref{fig8}, we present the histograms of the T$_{\rm cool}$ for the red giants in S- (upper panel) and S$+$IR-type SySts 
(lower panel). The temperatures range from 3000 to 4100~K which is consistent with red giants of spectral types K, M, and G. 
The vast majority of the cool companions in S- and S$+$IR-type are M-type giants, whereas only a few are likely K- or G-type giants. 
The average temperatures for S- and S$+$IR-type SySts are 3574 and 3352~K, respectively.



\begin{table*}
\centering
\caption{Effective Temperatures (T$_{\rm eff}$) of red stars in SySts}
\label{table8}
\begin{tabular}{llllllllll}
\hline 
\hline
Name & T$_{\rm eff}$ & Type & \ovi\    & References. & Name & T$_{\rm eff}$ & Type & \ovi\    & References.  \\ 
     &   (K)         &      & 6830\AA\ &      &      & (K)           &      & 6830\AA\ &\\   			
\hline 
AG Dra         & 4300 &  S      & \cmark & 1,2,14   &   AE Ara         & 3300 &  S      & \xmark & 8,9,14\\
BD-21 3873     & 4300 &  S      & \xmark & 3,2,14   &   SS~73 96        & 3400 &  S      & \xmark & 8\\
LT Del         & 4400 &  S      & \xmark & 4,2,14   &   AS 270         & 3300 &  S      & \xmark & 8\\
HD 330036      & 6200 &  \DD    & \xmark & 5        &   Y Cra		   & 3300 &  S      & \cmark & 8\\
AS 201         & 6000 &  \DD    & \xmark & 5        &   Hen 2-374      & 3300 &  S      & \cmark & 8\\
StHA 190       & 5300 &  S$+$IR & \xmark & 6        &   Hen 3-1761	   & 3300 &  S      & \xmark & 8\\
CH Cyg         & 3100 &  S      & \xmark & 7        &   WRAY 15-1470   & 3600 &  S      & \xmark & 10\\
CH Cyg	       & 2600 &  S      & \xmark & 14       &   Hen 3-1341     & 3500 &  S      & \cmark & 10\\
BX Mon         & 3400 &  S      & \xmark & 8,9      &   PN Th 3-29     & 3200 &  S      & \cmark & 10\\
V694 Mon       & 3300 &  S      & \xmark & 8        &   V2416 Sgr      & 3300 &  S      & \cmark & 10\\
Hen 3-461      & 3200 &  S      &  $-$   & 8        &   V615 Sgr       & 3300 &  S      & \xmark & 10 \\
SY Mus         & 3400 &  S      & \cmark & 8        &   AS 281         & 3300 &  S      & \cmark & 10\\
Hen 2-87       & 3300 &  S$+$IR & \cmark & 8        &   V2756 Sgr      & 3500 &  S      & \cmark & 10\\
Hen 3-828      & 3300 &  S      & \cmark & 8        &   V2905 Sgr      & 3600 &  S      & $-$ & 10\\
CD-36 8436     & 3300 &  S      & $-$    & 8        &   AR Pav         & 3400 &  S      & \cmark & 10,14\\
RW Hya         & 3700 &  S      & \xmark & 8,14     &   V3804 Sgr      & 3300 &  S$+$IR & \xmark & 10\\
Hen 3-916      & 3400 &  S      & \cmark & 8        &   V4018 Sgr      & 3500 &  S      & \xmark & 10\\
Hen 3-1092     & 3300 &  S      & \cmark & 8        &   V919 Sgr       & 3400 &  S      & \xmark & 10\\
WRAY 16-202    & 3300 &  S      & $-$    & 8        &   CD-43 14304    & 3900 &  S      & \cmark & 10,14\\
Hen 3-1213     & 4100 &  S      & \xmark & 8,2      &   CD-43 14304    & 4300 &  S      & \cmark & 2\\  
Hen 2-173      & 3400 &  S      & \cmark & 8        &   Hen 3-863      & 4300 &  S      & \xmark & 2\\
KX Tra         & 3300 &  S      & \cmark & 8,9      &   StHA 176       & 4200 &  S      & \xmark & 2\\
CL Sco         & 3400 &  S      & \xmark & 8,9      &   V2116 Oph      & 3400 &  S      & \xmark & 11\\
V455 Sco       & 3200 &  S$+$IR & \cmark & 8        &   RS Oph         & 4100 &  S      & \cmark & 12,13\\
M1-21          & 3300 &  S      & \cmark & 8        &   T CrB          & 3500 &  S      & \xmark & 12,14\\ 
RT Ser         & 3300 &  S      & \cmark & 8        &                  &      &         &        &    \\
\hline
\hline
\end{tabular}
\medskip{}
\begin{flushleft}
References: 1.Smith et al. (1996), 2. Pereira \& Roig (2009), 3. Smith et al. (1997), 4. Pereira et al.(1998), 5. Pereira et al. (2005), 6. Smith et al. (2001), 7. Schmidt et al. (2006), 8. Galan et al. (2016), 9. Galan (2015), 10. Galan et al. (2017), 11. Hinkle et al. (2006), 12. Wallerstein et al. (2008), 
13. Pavlenko et al. (2008), 14. Skopal(2005).\\ 
\end{flushleft}
\end{table*}

\subsection{SySts in Gaia}

\subsubsection{Stellar temperatures}
The recent second {\it Gaia} data release provides the stellar temperature for 161 million stars and luminosity for 77 millions stars (Andrae et al. 2018; Gaia Collaboration et al. 2018). {\it Gaia} temperatures are estimated by training a machine learning algorithm (Priam algorithm)\footnote{T$_{\rm eff}$ is estimated assuming zero extinction, due to the strong degeneracy between T$_{\rm eff}$ and A$_{\rm G}$.} using various samples covering a temperature range from 3000 to 10,000~K (part of DR2; Andrae et al. 2018). 
The typical error of the temperature is around 324~K, but it becomes as high as 550~K in the case of cold stars (T$<$4000~K) or hot stars 
(T$>$8000~K). Given that the majority of the cold companions in SySts have an M or K spectral types 
with temperature lower than 4000~K (expect the rare group of yellow SySts), {\it Gaia} temperatures estimates for SySts have errors from 324 
to 550~K (Andrae et al. 2018). 

In the case of colder stars, like the Mira companions in D-type SySts with T$<$3000~K, {\it Gaia} temperatures are even more uncertain, 
and we have decided to not present them in this work. Table~\ref{table3new} (Appendix~A) lists the stellar effective temperatures of the cold companions for 171 Galactic SySts.  

A reasonable agreement within the uncertainties between our and {\it Gaia} temperatures is found. The difference between the estimates is presented 
in the lower left panel in Figure~\ref{fig20}. Temperature estimates of only 17\% (or 10\%) SySts are inconsistent, 
and all of them have T$_{\rm Gaia}>$ 4400~K. This may imply the presence of a G-type giant in these SySts. The average temperature difference is of the order of 271~K with a standard deviation of 441~K (lower right panel in Figure~\ref{fig20}). One can see that our 
T$_{\rm{cool}}$ are in better agreement with the spectroscopic estimates T$_{\rm{eff}}$ rather than with those from {\it Gaia}. 
A comparison between {\it Gaia} temperatures and those from the literature (Table~\ref{table8}) shows an agreement within the uncertainties. The majority of SySts have a temperature difference lower than 600~K which is comparable with the {\it Gaia} uncertainty. 

Regarding the luminosity and radius parameters provided by {\it Gaia} DR2 for the cold companions of SySts, we decided not to present 
them in this work. Unfortunately, only 35 SySts have fractional parallax uncertainty $\sigma\omega/\omega>$0.2 (i.e. signal-to-noise $<$5), 
as it is recommended by {\it Gaia} (see Appendix B in Andrae et al. 2018) and therefore reliable luminosity and radius estimates 
(Andrae et al. 2018).

\subsubsection{Geometrical distances}

Besides the stellar parameters, {\it Gaia} DR2 also provides parallaxes for more than a billion of stars (Gaia Collaboration et al. 2018).
The geometrical distances of 1.33 billion sources (without take into account the stellar type, photometry or extinction) are derived using a probabilistic approach with the advantage of providing lower and higher bounds (within $\pm$1$\sigma$) either for sources with high 
fractional parallax uncertainties or negative parallaxes (Bailer-Jones et al. 2018).

In Table~\ref{table3new} (Appendix~A), we list the distances of 193 SySts (155 S-, 20 S$+$IR-, and 18 D-types) together with their lower 
and higher values. At this point, it is worth mentioning that of all the sources in {\it Gaia} DR2 are considered to be single stars while SySts 
are known to be binary systems. This results in some problematic astrometric solutions like negative parallaxes because of the poor fitting 
of the single-star parallax model (Arenou et al. 2018; Lindegren et al. 2018; Luri et al. 2018). The $astrometric$~$excess$~$noise<$~1~mas 
criterion was used to select only the sources with the most reliable parallax estimates (see Lindegren et al. 2018).

\begin{figure}
\begin{center}
\includegraphics[scale=0.5]{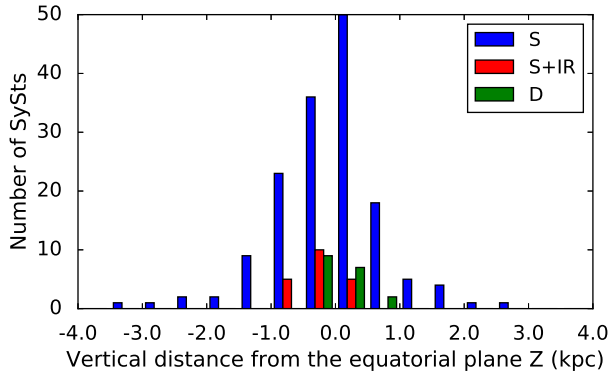}
\caption{Distribution of the vertical distances (Z) from the galactic plane in parsecs for S-, S$+$IR- and D-type SySts. }
\label{figZdistance}
\end{center}
\end{figure}

Fig.~\ref{figZdistance} presents the distribution of the vertical distance from the Galactic plane (Z) 
for all the SySts in our list. The distance of the Sun from the Galactic center is assumed to be 8.0$\pm$0.5 kpc (Reid 1993). 
S-type SySts show a clear Gaussian distribution around the Galactic plane extending up to 3~kpc, while S-IR and D-type SySts are more concentrated 
in the Galactic plane (Z$<$1~kpc). S-type SySts appear to be members of the Galactic disk, both thin and thick (scale heights of 
0.25--0.35~kpc and 0.86$\pm$0.2~kpc, respectively; Ojha 2001 and references therein), while S$+$IR- and D-types are mainly members of the thin 
Galactic disk. 

A comparison of the Z distance distributions between the S-type and the S$+$IR-/D-type SySts shows that very few S$+$IR- or D-type are expected 
to be found at Z$>$1~kpc. In particular, the standard deviation (1$\sigma$) of the S-, S$+$IR- and D-type distributions is found to be 
0.97, 0.37, and 0.28~kpc, respectively. The uncertainty of the standard deviations is around 15\%-20\%  given the error of Z distances 
between a few percent and up to 30\%. This implies that most of the S$+$IR- and D-type (more than two-thirds) belong to the thin 
Galactic disk while the S-type are members of both disk populations. Further study of the chemical composition of these systems may provide 
more information about their formation and evolution as well as their link with the Galactic disk populations. It should also be noted that 
the current population of known Galactic SySts may be biased towards the Galactic disk as the majority of them have been discovered from surveys 
focusing on the Galactic disk ($|d|<$10~\degree; Munari \& Renzini 1992).

Distances for a number of SySts have also been estimated before the release of {\it Gaia} DR2, and for most of them, a reasonable agreement is found 
within the uncertainties. For instance, the distance of RW Hya has been estimated to be either 0.6-0.68 kpc (Allen 1980; 
Muerset  et al. 1991) or 1.23-1.33~kpc (Pereira et al. 2017). Its {\it Gaia} distance of 1217~pc with a lower and an upper bound of 1134 and 1313~pc, respectively, is consistent with the more recent estimate. The current list of SySts distances provides more accurate values than previous studies and 
allow us to explore their spatial distribution in the Milky Way, and it will be very useful for many studies in this field.

\subsection{S-type SySts}

The majority of SySts are classified as S-type, and their SEDs show a peak between 0.8 and 1.7~$\mu$m with a mean value of 1.07~$\mu$m
whereas there is a small but significant number of S-type SySts with a peak either down to 0.7 or up to 1.8~$\mu$m 
(Figures~\ref{fig4}, \ref{SEDsAll}, \ref{fig6} and Table~\ref{table3new} in Appendix~A). The temperature distribution of the cool companions 
in S-type SySts demonstrates that the majority of the SySts have a temperature between 3400 and 3800~K, which corresponds to M1--M5 
spectral types with a peak at M5 (Figure~\ref{fig8}). This is in very good agreement with the distribution reported in previous 
studies (Medina-Tanco \& Steiner 1995; Muerset \& Schmid 1999). Only a small percentage of the S-type SySts population contains 
a K or G spectral-type giant, i.e. yellow SySts.

Interestingly, Galactic S-type SySts are found to contain a normal red giant (M or K), but there are a number of S-type SySts in the SMC 
and LMC that have been found to contain a Mira companion (Muerset et al. 1996). This finding is attributed to the lower metallicity of 
these two galaxies compared to our Galaxy. In low-metallicity environments like the SMC and LMC, the critical mass of a star to become 
a carbon star is lower than that in the Milky way (Marigo et al. 2013). Stars with masses between 2-3~M$\odot$ become carbon stars 
due to the third dredge-up after only a few thermal pulses or, in other words, during the early AGB phase (Marigo et al. 2013). 
The mass-loss rate at this early phase is not high enough to form a dusty shell and give a D-type SED profile.

\subsection{S$+$IR-type SySts}
Besides the typical S-type SySts, we also find a statistically important number of S-type SySts with an infrared excess in 
the 11.6 and/or 22.1~$\mu$m bands (Figs.~\ref{fig4}, \ref{SEDsAll}, and \ref{fig6}, and Table~\ref{table3new} in Appendix~A). We name this new 
type of SySt \lq\lq S$+$IR-type". Twenty-seven SySts in our list are classified as S$+$IR-types. Seven were previously classified as D-types, 
18 as S-types, one as a \DD-type, and one without a previous classification. 

The SED profiles of S$+$IR-type SySts demonstrate a peak at longer wavelengths (1.3~$\mu$m) compared to S-type SySts (1.07$\mu$m), indicating 
the presence of companions with lower temperature. For all the cases of S$+$IR-type SySts, two BB models were used 
to reproduce the total SED. This IR excess may be an indication of the existence of a dusty shell, colder temperatures compared to those of 
D-types, the presence of strong amorphous silicate emission bands at $\sim$10 and $\sim$18~$\mu$m, or an accretion disk around the WD component. 

Interestingly, all SySts, except two very bright cases (R~Aqr and CH~Cyg), with available {\it ISO} data presented by Angeloni et al. (2007), 
exhibit an infrared excess in the {\it W3} and {\it W4} bands and at the same time show strong amorphous silicate emission 
bands at $\sim$10 and $\sim$18~$\mu$m. However, only one has been classified as an S$+$IR-type (candidate V627~Cas), while all of the 
remaining ones are classified as D-type SySts. Therefore, the IR excess observed in S$+$IR-type SySts might not be associated with the 
presence of amorphous silicate emission bands.  

The second possible scenario for the IR excess in S$+$IR-type SySts is the presence of a dusty disk that formed around the WD. The presence of a 
slowly expanding disk that has been proposed in order to explain the two-temperature components of high orbital inclination SySts 
during the active phase (Skopal 2006). 

A third scenario of a much colder and tenuous dusty shell, similar to those in D-types, is also possible. In this case, the cold 
companion in S$+$IR-type SySts may have just become an AGB star. Because of its very low mass-loss rate in the early AGB phase, the stars do not 
have sufficient mass to form a dust shell (i.e. the shell that would obscure the star and result in a D-type SED profile). 
This scenario requires colder companions compared to the red giants in S-type SySts, being in a more evolved phase. This is 
exactly what the temperature distributions display (Figure~\ref{fig8}). There is a significant percentage of S$+$IR-type SySts 
in which the companions have temperatures between 3000 and 3200~K, while the number of S-type SySts in the same temperature range 
is almost negligible. 

We searched in the literature for further information on the S$+$IR-type SySts and variabilities with periods 
between 100 and 400~days and amplitude pulsations between 0.5 and 2~mags have been reported for the several of them (e.g. 
EF~Aql; SS 73~17; H 2-38; SS 73~122; V366~Cas, Whitelock et al. 1983; Whitelock 1987; Pojma\'{n}ski 2002; Matsunaga, Fukushi \& Nakada 
2005; Watson, Henden \& Price 2006; Richards et al. 2012; Soszy\'{n}ski et al., 2013; Samus' et al. 2017). The pulsation
periods and amplitudes from the S$+$IR-type SySts appear to be lower compared to those of Mira stars in D-type SySts. 
R~Aqr has been misclassified as San $+$IR-type due to its uncertain photometric magnitudes while its high amplitude variations ($>$2~mags) 
are consistent with Mira stars.

Given that the pulsation period in the AGB phase increases with time (e.g. Vassiliadis \& Wood 1993), it is possible that the giant companions in 
S$+$IR-type SySts are in a less evolved phase (e.g. early AGB) than the Mira in D-types (thermal pulsating in the TP AGB phase). 
The intense mass-loss rate during the end of the TP-AGB phase is responsible for the formation of the dusty shell in D-type SySts 
(e.g. Muerset et al. 1996), while in the case of S$+$IR-type, the companion has just entered in the AGB phase (early AGB) and its mass-loss 
rate is not high enough to form a similar dusty shell with high infrared excess as in D-types. 
In the early AGB phase, the mass-loss rate is lower compared to the more evolved TP-AGB phase (e.g. Schild 1989; 
Vassiliadis \& Wood 1993, Rosenfield 2014), resulting in a lower infrared excess. The {\it K$_{\rm s}$--}[12] and {\it K$_{\rm s}$--}[22] 
color indices (or equivalently {\it K$_{\rm s}$--W3} and {\it K$_{\rm s}$--W4}) are two widely used indicators of the dust mass loss 
in giants (e.g. Guandalini et al. 2006; Whitelock 2006; Uttenthaler 2013, Akras et al. 2017). The mean values of {\it K$_{\rm s}$--W3} 
and {\it K$_{\rm s}$--W4} for each type of SySts are estimated. We find that both color indices increase from 0.68 and 1.40 
(the standard deviation (SD) is found to be 0.48 and 0.82, respectively) for S-type, to 2.42 and 3.83 (SD of 1.27 and 1.12, 
respectively) for S$+$IR-types, to 4.54 and 5.61 (SD of 1.01 and 1.24, respectively) for D-types and to 
6.63 and 8.79 (SD of 1.31 and 1.71, respectively) for \DD-types. This indicates that the red giants in 
S$+$IR-type exhibit a higher mass-loss rate than their counterparts in S-type and lower than those in D-type. Hence, S$+$IR-type SySts likely 
represent a transition phase from S-type to D-type SySts (see also Medina-Tanco \& Steiner 1995).

Gromadzki et al. (2009) have claimed that AS~245 has a semi-regular star rather than a Mira, due to the low amplitudes and its certain 
pulsation period. From our analysis, AS~245 has been classified as an S-type, and 
not as an S$+$IR-type. However, a low excess at the {\it W4} band is barely seen (see Fig.~A3). Moreover, we calculate its 
T$_{\rm eff}$=3162~K, which is lower than the average temperature of the red giants in S-type SySts by 400~K (Figs.~\ref{fig8}). 
This may indicate a more evolved companion, although not as evolved as Mira stars. The {\it K$_{\rm s}$--W3} and {\it K$_{\rm s}$--W4} color indices of this SySt are found to be 1.58 and 2.38, respectively, between the values of S- and S$+$IR-type.

\subsection{D-type SySts}

From the remaining SySts, 13\% of the known SySts are classified as D-types (Table~\ref{table3new} in Appendix~A). 
The SEDs of D-type SySts show a peak between 2 and 4~$\mu$m with a mean value of 2.85~$\mu$m (Figure.~\ref{fig4}-\ref{fig6}). 
This peak range is shorter than the one reported by Ivison et al. (1995). 

An important number of D-type SySts is found for which two BB models are required in order to fit the whole SED profile. 
The average values of the dust temperatures for these two dusty shells are 1077$\pm$35~K for the inner and 467$\pm$30~K for the outer one, 
with SD equal to 200 and 112~K, respectively. These temperatures are in excellent agreement with the values reported by 
Angeloni et al. (2010). Phillips (2007) also argued that the near-IR emission emitted from a  dusty shell is associated with 
silicate dust grains whose maximum temperature is of the order of 800~K, which is very close to the mean temperature between 
the two dust shells.

\begin{table*}
\centering
\caption{Number of positive and negative \ovi\ Raman-line Detections in the Milky Way, SMC, LMC, M31, and M33}
\label{table7}
\begin{tabular}{lllllllllll}
\hline 
Galaxy &  Total number$^{a}$ & Positive detections & Negative detections & \ovi\ Raman  & [Fe/H] & References \\
		&                        &                   &                      & (\%)         &        &  ([Fe/H]) \\    			
\hline 
Milky Way & 257  & 131 &108 & 55  & -0.11 & 1\\ 
SMC       &   9  &   9 &  0 & 100 & -0.99 & 2\\
LMC       &   9  &   4 &  3 & 57  & -0.60 & 3,4\\
M31       &  31  &  16 & 15 & 52  & -0.45 & 5\\
M33       &  12  &   5 &  7 & 42  & -0.11 & 6\\
\hline
\end{tabular}
\medskip{}
\begin{flushleft}
$a$Total number of optical spectra examined in this work.\\
References. 1. Sadler et al. (1996), 2. Dobbie et al. (2014), 3. Cole et al. (2000), 4. Salaris \& Girardi (2005), 5. Kalirai et al. (2006), 6. Gregersen et al. (2015)\\
\end{flushleft}
\end{table*}

\subsection{\DD-type SySts}

The typical SED profile of \DD-type SySts displays a nearly flat profile (Fig.~\ref{SEDsAll}). Both signatures from 
the cool giants and the dusty shells are present, and three BB models are required to reproduce all of their SED profiles, one for the cool 
companion and two for the inner and outer dust shells. Because of the flat profiles, we do not present an SED peak 
histogram for \DD-types. However, one can see that the dusty shells in \DD-type SySts show two distinct peaks between 2 and 10~$\mu$m.

Only 10 \DD-type SySts with a G/K spectral-type red giant are known (see Tables~\ref{table2} and \ref{table3new} in Appendix~A). 
According to our classification, only five of them show the typical SED profile of \DD-type SySts. The remaining 
are classified as a D-type (V417 Cen) and as S$+$IR-types (WRAY 15-157, Hen 3-1591, StHA~190 and SMP LMC 88). The classification of V417~Cen as a 
D-type SySt may not be correct despite the good fitting. V417~Cen shows a barely noticeable infrared excess in the {\it J}-band which 
could be attributed to the presence of a hotter red giant. In this case, a third BB model would be necessary to fit the SED profile to give 
the typical SED profile of \DD-types. V417~Cen has been classified as a \DD-type SySt with a G spectral-type companion (Gromadzki et al. 2011).  
Regarding the last four S$+$IR-type SySts, a fast rotating G spectral-type companion has been found in WRAY~15-157 (Zamanov et al. 2008) 
and StHa~190 (Smith, Pereira \& Cunha 2001), the optical spectra of SMP LMC 88 resembles that of a K giant (ilkiewicz et al. 2018) 
while Hen 3-1591 has a controversial classification of S or \DD-type (Bel2000).  

Six more SySts are found to have a typical flat SED profile, and most of them have been previously classified as D-type. 
One of them (GH Gem) has been previously incorrectly classified as an S-type, and our SED analysis reveals a \DD-type SySt. This agrees with 
the classification of the red giant as K3III by Munari et al. (2007). K5-33 displays a nearly flat profile, indicating a \DD-type 
classification, but its dust emission does not allow us to get an estimation of the temperature of the red giant. A D-type classification 
for K5-33 cannot be ruled out (Miszalski et al. 2013). Overall, 11 \DD-type SySts are listed in our census (3.5\%, Table~\ref{table3new}) 
based on their SED profiles. Whether the cold companion in a SySt exhibits characteristics of a cold G/K star, 
it does not mean a priori that it is a \DD-type since it may also be a yellow S-type (e.g. LT Del, LAMOST~J12280490-014825.7, StHa~63, 
see also Baella et al. 2016). \DD-type SySts have a G/K spectral-type companion with a strong IR excess (Allen 1982), which makes 
them to resemble PNe.

\section{\ovi\ Raman-scattered $\lambda\lambda$6830,7088 lines in symbiotic stars}

An updated census of the \ovi\ $\lambda$ 6830 Raman-scattered line in SySts is required in order to determine the total occurrence of 
these lines in SySts. We performed a systematic search for optical spectra of SySts in all of the works listed in Table~\ref{table3new} as 
well as in the following compilations by Blair et al. (1983), Allen (1984), Acker, Lundstrom and Stenholm (1988), Medina-Tanco \& Steiner (1995), 
Muerset, Schild \& Vogel (1996), Mikolajewska, Acker, Stenholm (1997), and Guti\'{e}rrez-Moreno Moreno \& Costa (1999). Whether 
or not the \ovi\ $\lambda$6830 Raman-scattered line is detected is given in Table~\ref{table3new}. For this work, we gathered information only for 
the $\lambda$6830 line since it is brighter by a factor of between 2 and 10 compared to the $\lambda$7088 line (Allen 1980, Schmid et al. 
1999) due to the different column densities of atomic hydrogen.

We gathered information from 298 optical spectra of known and candidate SySts. We end up with 165 positive 
confirmations of the \ovi\ Raman-scattered line, which corresponds to 55\% of the whole sample. This value is very close 
to the percentage calculated by Allen (1980). There are a number of SySts with spectra from different epochs, and 
the \ovi\ line is not always detected. The intensity of the \ovi\ line can significantly change during that time. 
A newly formed dust shell around the cool companion can absorb the \ovi\ 1032~\AA\ photons, resulting
in the depletion of \ovi\ $\lambda$6830 Raman-scattered photons (e.g. V1016~Cyg, Arkhipova et al. 2016) or the presence of an optically 
thick disk around the WD (e.g. Skopal 2006). 

The significant increase of new SySts in nearby galaxies ($\sim$350\%) implies that the aforementioned 
percentage derived from the whole sample should be taken cautiously, due to the different metallicities in galaxies. 
The percentage of SySts with the \ovi\ Raman-scattered line in their spectrum is determined separately for the Milky Way, 
SMC, LMC, M31, and M33 galaxies. In Table~\ref{table7}, we list the number of the positive and negative confirmations 
of the \ovi\ Raman-scattered line as well as their percentages, which range from 42\% to 100\%. 
Systematic surveys of SySts in nearby galaxies are required in order to increase the sample of SySts in other galaxies.

\subsection{\ovi\ Raman-scattered versus nebular lines}

The detection of the \ovi\ Raman-scattered line implies a very hot and luminous WD able to ionize the circumstellar envelope.
Hence, the recombination \heliumb\ $\lambda$4686 line (ionization potential, I.P.=54.4~eV) should also be detected 
when the \ovi\ Raman-scattered line (I.P.=113.9~eV) is detected. Indeed, we find that in all of the SySts in which the \ovi\ Raman-scattered line 
is detected, the \heliumb\ $\lambda$4686 line is also present, while the opposite is not true. The correlation 
between the two lines has been verified using the equivalent widths (e.g. Leedj\"{a}rv 2004, Leedj\"{a}rv et al. 2016). There are 
three cases in our sample for which both lines are not simultaneously detected (M31 SySt-23, Mikolajewska, Caldwell 
\& Shara 2014; QS Nor and Th 3-29, Acker, Lundstrom \& Stenholm 1988). We thus argue that the detection of the \ovi\ Raman-scattered 
line in these three sources is dubious.  

\begin{figure}
\begin{center}
\includegraphics[scale=0.425]{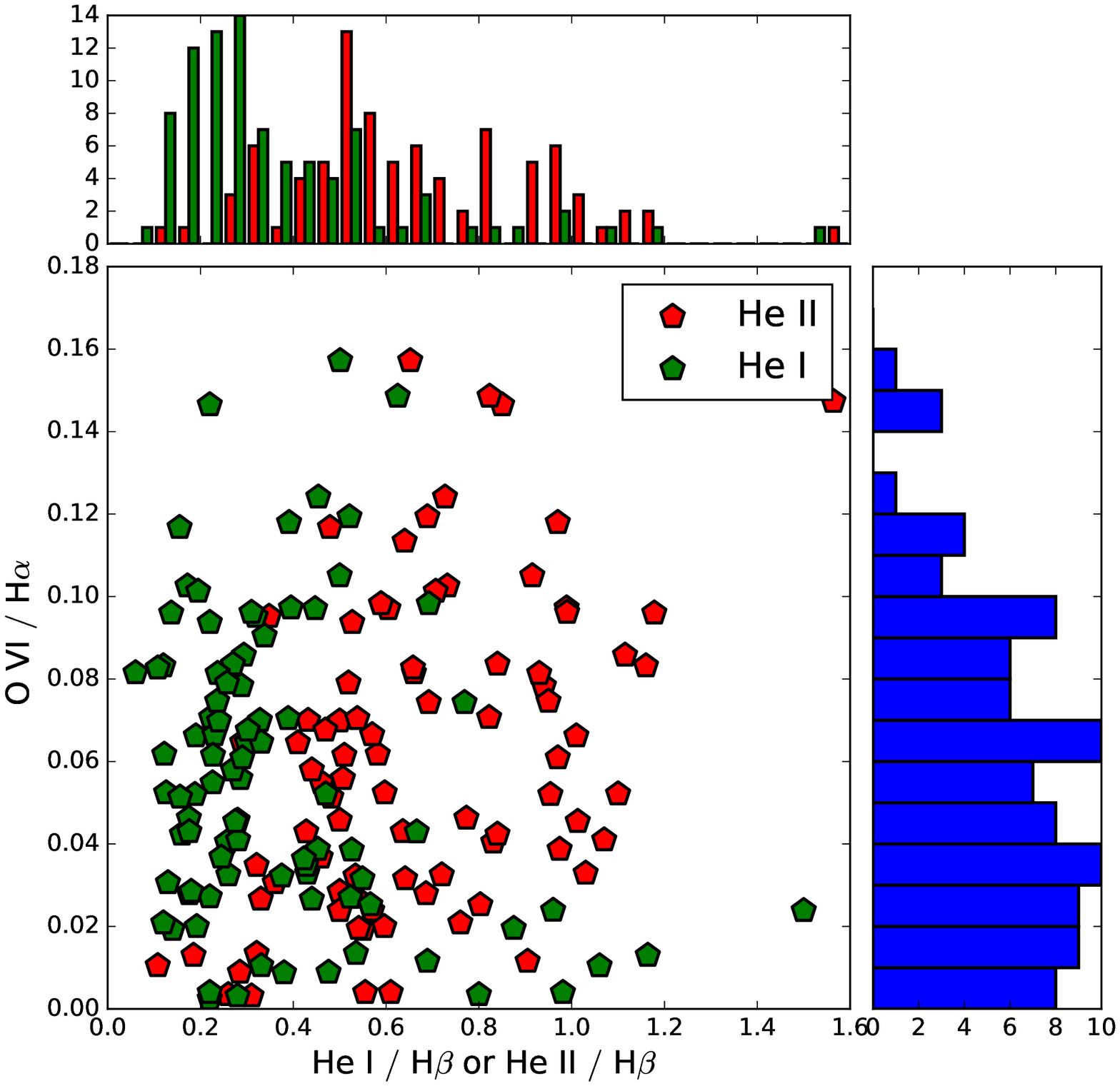}
\caption{\ovi\ $\lambda$6830/\ha\ vs. \helium\ $\lambda$5876/\hbeta\ or \heliumb\ $\lambda$4686/\hbeta\ line ratio
plot for a number of Galactic and extragalactic SySts.} 
\label{figovi}
\end{center}
\end{figure}

The average value of the \ovi\ $\lambda$6830/\ha\ line ratio is determined for all Galactic SySts (113) as well as for the SySts in 
M33 and M31 (43) with available emission-lines fluxes. The total average \ovi\ $\lambda$6830/\ha\ line ratio is 0.06 with SD=0.04, 
while for the Milky Way, M31, and M33, it is 0.06 (SD=0.04), 0.05 (SD=0.03), and 0.04 (SD=0.04), respectively. Figure~\ref{figovi} 
displays the \ovi\ $\lambda$6830/\ha\ versus \helium\ $\lambda$5876/\hbeta\ and \heliumb\ $\lambda$4686/\hbeta\ plot for all 
of these SySts (see also Table~\ref{table7new}). All the line ratios are estimated using integrated fluxes gathered from several 
studies in the literature. It can be seen that the \heliumb\ $\lambda$4686/\hbeta\ line ratio displays a lower threshold of 
$\sim$0.3-0.4, while the \helium\ $\lambda$5876/\hbeta\ line ratio has an upper threshold of $\sim$0.5-0.6.

The average values of the \heliumb\ $\lambda$4686/\hbeta\ and \helium\ $\lambda$5876/\hbeta\ line ratios for SySts with and 
without the \ovi\ Raman-scattered line in different galaxies are also estimated (see Table~\ref{table7new}). 
In general, SySts with the \ovi\ Raman line detected exhibit systematically higher \heliumb\ $\lambda$4686/\hbeta\ ratio (0.66) 
compared to those SySts without the \ovi\ line detected (0.22), while the \helium\ $\lambda$5876/\hbeta\ ratio has comparable 
values regardless of the detection of the Raman line (0.43 and 0.38, respectively). 

This implies that the \ovi\ Raman emission is also followed by an increase of \heliumb\ $\lambda$4686/\hbeta\ ratio. In this case, 
the WD is neither luminous nor hot enough to emit a significant number of UV \ovi\ photons, and it results in a lower 
\heliumb\ $\lambda$4686/\hbeta\ line ratio and a low-excitation nebula. On the other hand, if the \ovi\ line is detected, the He gas is 
mainly doubly ionized (high \heliumb\ $\lambda$4686/\hbeta) and the \helium\ $\lambda$5876 line becomes weaker 
(low \helium\ $\lambda$5876/\hbeta). The large scatter of the point in Figure~\ref{figovi} reflects the complex relation between 
the effective temperature of the hot companion, the mass-loss rates of the stars, the density of atomic hydrogen, and the 
efficiency of Raman scattering.

\begin{table*}
\centering
\caption{Emission-line ratios of SySts$^{\dag}$ }
\label{table7new}
\begin{tabular}{lllllllllll}
\hline 
Galaxy &  \ovi\ $\lambda$6830/\ha\ & \multicolumn{2}{c}{ \heliumb\ $\lambda$4686/\hbeta\ } & \multicolumn{2}{c}{ \helium\ $\lambda$5876/\hbeta\ } & \multicolumn{2}{c}{\heliumb\ $\lambda$4686/\oxygeniii\ $\lambda$5007 }\\
		&                          & \ovi\ (\cmark) &  \ovi\ (\xmark)                                   &\ovi\ (\cmark) &  \ovi\ (\xmark) &    \ovi\ (\cmark) &  \ovi\ (\xmark)  \\    			
\hline 
All       & 0.06 (0.04)  & 0.66 (0.26) & 0.22 (0.15) & 0.43 (0.33) & 0.38 (0.23) & 1.45 (1.49) & 1.21 (1.74) \\ 
Milky Way & 0.06 (0.04)  & 0.65 (0.26) & 0.21 (0.16) & 0.46 (0.34) & 0.40 (0.25) & 1.15 (1.24) & 1.42 (1.93) \\ 
M31       & 0.05 (0.03)  & 0.68 (0.27) & 0.25 (0.10) & 0.26 (0.22) & 0.34 (0.15) & 1.47 (1.40) & 0.69 (0.83) \\ 
M33       & 0.04 (0.04)  & 0.81 (0.16) & 0.24 (0.08) & 0.35 (0.15) & 0.27 (0.22) & 3.78 (1.12) & 0.73 (0.83) \\ 
\hline
\end{tabular}
\medskip{}
\begin{flushleft}
$^{\dag}$ The number in parentheses corresponds to the standard deviations\\
\end{flushleft}
\end{table*}

Besides the previous line ratios, we also revise the \heliumb\ $\lambda$4686/\oxygeniii\ $\lambda$5007 ratio. Interestingly, this line 
ratio is found to decrease as a function of metallicity for SySts with the \ovi\ Raman-scattered line detected and increase for those 
without Raman emission (Table~\ref{table7new}). We have to mention that in the Milky Way, there are three SySts (AS~289, AS~316, and AS 327) 
with extremely high \heliumb\ $\lambda$4686/\oxygeniii\ $\lambda$5007 line ratio between 30 and 45, which we did not take into 
consideration in these calculations. 

Overall, all of these line ratios provide new constraints on the identification of the \ovi\ $\lambda$6830 Raman-scattered line. 
There is, for instance, a number of sources with a \helium\ $\lambda$5876/\hbeta\ line ratio higher than 0.8, which puts them into doubt. 
According to Figure~\ref{figovi}, it is clear that there is a minimum value for the \heliumb\ $\lambda$4686/\hbeta\ ratio 
and a maximum value for the \helium\ $\lambda$5876/\hbeta\ in case the \ovi\ 6830\AA\ Raman-scattered line is detected. 

\subsection{Perplexing objects with the \ovi\ Raman emission reported}
The \ovi\ $\lambda$6830 Raman-scattered emission has also been reported for a few non-SySt objects, such as young PNe 
(Arrieta \& Torres-Peimbert 2003; Sanchez-Contreras et al. 2008), NGC~6302 (Groves et al. 2002), 
NGC~7027 (Zhang et al. 2005) and one Be star, namely LHA~115~S-18 (Torres et al. 2012). In order to verify the 
presence of the Raman line in these objects, we re-examined their spectra, and below we discuss each object in more detail.

\subsubsection{NGC~6302}
Groves et al. (2002) reported the detection of the \ovi\ $\lambda\lambda$6830,7088 Raman-scattered lines in the planetary nebula 
NGC~6302. Its very hot central star (150-400~kK, Pottasch et al. 1996; Groves et al. 2002) 
can emit a significant number of UV \ovi\ photons. The large full-width-half-maximum (FWHM) of the $\lambda\lambda$6830,7088 
line features in conjunction with the detection of other Raman-scattered lines at 4331\AA\ and 4852\AA\ suggests a Raman-scattering 
origin. Moreover, the large list of high-excitation emission-lines as well as Fe lines are consistent with what it is found in SySts.
The IUE data indicate the presence of a cool G-type companion (Feibelman 2001). 

However, \kriptoiii\ $\lambda$6826.7, \helium\ $\lambda$6827.9, and/or \carboni\ $\lambda$6828.1 lines can lead to a possible 
misidentification (see Zhang et al. 2005; Sharpee et al. 2007).  NGC 6302 has a $\lambda$6830/\ha\ line ratio equal to 0.001, 
which is more than one order of magnitude lower than the average ratio as well as the \heliumb\ $\lambda$4686/\oxygeniii $\lambda$5007 ratio.  

Assuming $\rm{N_e}$=10000~cm$^{-3}$ and $\rm{T_e}$=15000~K, the theoretical value of the \helium\ $\lambda$6830/$\lambda$4471 
ratio is $\sim$1.1$\times$10$^{-3}$ (Smits 1991), which is almost 50 times lower than the observed value (0.055). This discrepancy 
can be explained if the \carboni\ $\lambda$6828.1 line is taken into account. Hence, we consider the \carboni\ $\lambda$6828.1 line as 
a possible contaminant. The detection of other line from the same upper level such as \carboni\ $\lambda$6656.6 with an 
intensity of 0.045 (Groves et al. 2002) supports our hypothesis. 

According to this analysis, we argue that NGC~6302 is unlikely to be a SySt, and this is consistent with its very low 
\heliumb\ $\lambda$4686/\oxygeniii $\lambda$5007 and \ovi\ $\lambda$6830/\ha\ line ratios. We claim that the identification of the 
$\lambda\lambda$6830,7088 features such as \ovi\ Raman-scattered lines is dubious.

\subsubsection{NGC~7027}
NGC~7027 is a young, high-excitation nebula with a very hot central star (Middlemass 1990; Zhang et al. 2005) in which 
a line feature centered at 6828\AA\ has been detected (Zhang et al. 2005). In contrast to Groves et al. (2002), Zhang et al. 
(2005) identified it as \kriptoiii\ $\lambda$6826.7. Besides the \kriptoiii\ line, a weaker and broader line feature is also 
present and blended with the former line. Zhang et al. (2005) identified this broad feature as an \ovi\ Raman-scattered 
line, unlike its previous identification as a [Si\,{\sc II}] line by P\'{e}quignot \& Baluteau (1994). 

Similar to NGC~6302, the aforementioned broad line is significantly weaker than \heliumb\ $\lambda$4686. Moreover, 
the line ratio of the two Raman-scattered lines, $\lambda$7088/$\lambda$6830, is found to be at least twice as low as the 
ratio found in SySts. According to this, the identification of that feature as an \ovi\ Raman line is questionable despite
the broad feature. The \helium\ $\lambda$6827.9 line is a possibility.  

Using the list of theoretical \helium\ lines for a range of $\rm{N_e}$ and $\rm{T_e}$ by Smits (1991), 
we get a rough estimate of its theoretical intensity (\helium\ $\lambda$6830/$\lambda$4471$\sim$1.1$\times$10$^{-3}$ for 
$\rm{N_e}$=10$^4$ cm$^{-3}$ and $\rm{T_e}$=15000~K; Smits 1991). The intensity of 
the $\lambda$6830 feature is approximately three times lower than the intensity of the \kriptoiii\ $\lambda$6826.7 line 
(0.044 relative to H$\beta$=100; see Table~A1 in Zhang et al. 2005). 
Therefore, its intensity is of order of 0.014, and that of the observed \helium\ $\lambda$6830/$\lambda$4471 ratio of order of 0.0046. 
This value is almost three times higher than the theoretical value. The \kriptoiii\ $\lambda$6826.7 line is not corrected for the 
contribution of the \carboni\ $\lambda$6828.1 line, which may result in a lower line ratio.

In conclusion, the feature detected at 6828\AA\ in NGC 7027 may have been mistakenly classified as the \ovi\ Raman line. 
The \helium\ line centered at 6827.9\AA\ seems a more probable identification. Looking carefully at its spectrum, one can see that 
other \helium\ line centered at, e.g. 8634\AA, and 8652\AA\ also show red wings similar to the $\lambda$6828 feature (Zhang et al. 2005).

The high-resolution spectrum of NGC~7027 by Sharpee et al. (2007) shows that the $\lambda$6830 feature is blended with 
the \helium\ $\lambda$6827.9 and \carboni\ $\lambda$6828.1 lines as well as with a number of telluric OH bands. 
After all these analyses, we conclude that the identification of the \ovi\ Raman-scattered lines in NGC~7027 is dubious.

\subsubsection{Young PNe}

Arrieta \& Torres-Peimbert (2003) and Sanchez-Contreras et al. (2008) also reported the detection of the \ovi\ $\lambda$6830 Raman line 
in a number of young PNe, such as M~2-9, IRAS~17395-0841, IC~4997, IRC~+10420, M3-60, M1-92, IRAS 08005-2356, Hen~3-1475, and 
IRAS~22036+5306. We re-examined the spectra of these nine objects, and we find that (i) the $\lambda$6830 feature is very weak compared to 
SySts and (ii) \ovi\ $\lambda$6830 and \heliumb\ $\lambda$4686 lines are not simultaneously detected. 
M2-9 is the only object where the \heliumb\ $\lambda$4686 line and \ironiv\ are detected. 

The central stars of all these young PNe (except from M2-9) are not hot enough to emit a significant number of UV \ovi\ photons. 
We, therefore, argue that the line feature at $\sim$6830\AA\ is not associated with the \ovi\ Raman-scattered line. The \kriptoiii\ $\lambda$6826.7, \helium\ $\lambda$6827.9 or \carboni\ $\lambda$6828.1 nebular lines are possible identifications. As for M2-9, a further study 
is required in order to verify whether the $\lambda$6830 feature is associated with Raman-scattering.

\subsubsection{LHA~115 S-18}
LHA~115 S-18 is a highly controversial object for which the scenarios of being a luminous blue variable star, a symbiotic star, 
a PN, a Cygni variable, or a B[e] star have been proposed (Torres et al. 2012, Clark et al. 2013, Maravelias et al 2014, 
see also Lepo PhD dissertation 2015 for a detailed analysis). 

The most interesting point of this likely massive star is the recent detection of the \ovi\ Raman-scattered doublet lines 
$\lambda\lambda$6830,7088 with a simultaneous detection of the \heliumb\ $\lambda$4686 line (Torres et al. 2012). 
Maravelias et al. (2014) argued that the observed variations in some \heliumb\ lines and in the OGLE-II data 
are not consistent with its previous B[e] classification. On the contrary, its optical spectrum shows a number of [Fe\,{\sc II}], and Balmer 
lines, with the latter exhibiting a P Cygni profile typical for B[e] stars. Generally, P Cygni profiles imply a mass eruption event with high 
mass-loss rates and velocities up to 300~\kms. This event is likely responsible for the high conversion efficiency of Raman scattering and the detection of the \ovi\ 6830 \AA\ line.

According to the classification scheme of B[e]-type stars proposed by Lamers et al. (1998), some SySts show the B[e] phenomena and are 
classified as SymB[e] stars. Moreover, Skopal (2017) pointed out the similarities between B[e] stars and SySts during the active or outburst 
phase.

Based on {\it XMM-Newton} and {\it Chandra} data, LHA~115 S-198 is confirmed as an X-ray source. However, the low signal-to-noise ratio does 
not allow a spectrum to be extracted. Its X-ray luminosity is of the order of 10$^{32-33}$ erg~s$^{-1}$, and it is comparable with the 
luminosity of beta-type SySts but less compared to the luminosity of alpha-type SySts (Luna et al. 2013). LHA~115 S-18 is very likely a binary 
system with a compact object and a supergiant. The available X-ray data do not permit to identify the nature of the compact companion, 
which it can be either a WD or a neutron star. PU~Vul is also classified as beta-type based on its X-ray spectrum and also presents 
the B[e] phenomena. 

We thus decided to insert LHA~115 S-18 into our census as a known SySt with the \ovi\ Raman line detected, but more studies are required. 
We also argue that LHA~115 S-18 may belong to the group of SymB[e] stars showing the B[e] characteristics during an active or outburst phase.

\begin{table*}
\centering
\caption{Characteristics of Different Types of SySts.}
\label{table10a}
\begin{tabular}{lccccccccc} 
\hline  
Type   & All & \% & Known & \%  & Peak       & T$_{\rm eff}^{\dag}$ & T$_{\rm dust}$ & \ovi\ & X-Ray\\
	   &(\#) &    &  (\#) &     & ($\mu$m)   &  ($\mu$m)     & (K)            & (\%)  &  (\#)\\
\hline  
S      & 263 & 64  & 238   & 74  & 0.83-1.7   & 3000-4000  &  $-$               & 61   &  33  \\ 
S$+$IR & 37  &  9  & 26    &  8  & 0.88-1.7   & 3000-3900  & 150-500            & 39   &   3  \\
D      & 60  & 15  & 42    & 13  & 2.1-4.1    &   $-$      & 200-400/700-1350   & 55   &   8  \\        
\DD    & 31  & 7.5 & 11    & 3.5 & \lq\lq flat"    & 3500-4400  & 150-350/550-1000   & 50   &  $-$ \\ 
No type& 19  & 4.5 &  6    & 1.5 &   $-$      & $-$        &  $-$               & $-$  &   2  \\
\hline
\end{tabular}
\medskip{}
\begin{flushleft}
\end{flushleft} 
\end{table*}

\section{X-Ray emission in SySts}
All SySts with X-ray emission detection are presented in Table~\ref{table3new}. Only 46 have been found to 
be either soft or hard X-ray emission sources (see Luna et al. 2013 and references therein; Wheatley et al. 2003; Mukai et al. 2016;
Nu\~{n}ez et al. 2016). According to the $\alpha, \beta, \gamma$, and $\delta$ classification scheme 
(Muerset, Wolff \& Jordan 1997; Luna et al. 2013; Joshi et al. 2015; Bozzo et al. 2018), seven SySts have been classified as $\alpha$-types, 
12 as $\beta$-types, nine as $\gamma$-types, eight as $\delta$-types, and eight $\beta\delta$-types. 

From these four types, only the $\alpha$- and $\beta$-type SySts show the shell-burning process with or without accretion phenomena, 
while the last two types show only accretion phenomena (e.g. SU Lyn; Mukai et al. 2016). This implies that only $\alpha$- and $\beta$-type SySts 
are able to emit emission-lines, which correspond to 50\% of the total X-ray SySts detected so far. Mukai et al. (2016) claimed that the 
true population of SySts may have been significantly underestimated and that of all the published SySt 
catalogs are biased as they are based only on the optical emission.

Given that the shell-burning process on the hot WD's surface of $\alpha$-type SySts induces the emission of supersoft X-rays, 
we probe for a possible link between X-ray and \ovi\ Raman emission in SySts. We find that 12 out of 46 (or 27\%) 
X-ray SySts emit the \ovi\ Raman-scattered line whereas 21 of them (45\%) do not (Table~\ref{table3new}). To our knowledge, 
there are no available spectroscopic data for the remaining 13 X-ray SySts (27\%). Moreover, 12 out of 33 (34\%) 
SySts without detectable X-ray emission do not exhibit the Raman-scattered \ovi\ line, whereas it is detected in 20 (60\%) of them 
(Table~\ref{table3new}).

In addition to that, we also explored the presence of the \ovi\ Raman-scattered line in different X-ray-type SySts 
($\alpha, \beta, \gamma$ and $\delta$). None of the $\gamma$ or $\delta$ types exhibit the \ovi\ Raman-scattered line, 
but only the $\alpha$ and $\beta$-types. Specifically, six out of seven $\alpha$-type SySts, and six out of 
ten $\beta$-type SySts present the \ovi\ line. The $\alpha$-type SySts are expected to be hot and luminous enough due to the shell 
burning process, and o emit a large number of UV \ovi\ photons that are eventually transformed to \ovi\ Raman-scattered photons. 
But this is not evident in $\beta$-type SySts, in which the origin of the soft X-ray emission is from colliding winds 
(Muerset, Wolff \& Jordan 1997, Luna et al. 2013). However, such strong winds able to produce soft X-ray emission are 
indicative of strong mass loss and high luminosity, which are likely associated with some shell-burning process, as in $\alpha$-types. 
This could explain the detection of the \ovi\ Raman line in $\beta$-type SySts. Ramsay et al. (2016) claimed that the $\beta$-type 
AG~Peg SySt (Muerset et al. 1997) shows evidence of a quasi-shell-burning process. 

The presence of the \ovi\ Raman-scattered line in SySts may be an indication of a shell-burning process in $\alpha$ and $\beta$-type SySts 
(see also Mukai et al. 2016). Further X-ray observations are required to verify this correlation.

\section{Conclusion}

In this paper, we presented a new compilation of SySts. The total number of known SySts has been increased by 70\%. 
For the Galactic and extragalactic SySts, the numbers have increased from 173 to 257 ($\sim$45\%) and 15 to 66 
($\sim$350\%), respectively. 
	
The SED profiles of 348 SySts (known and candidates) were constructed using the 2MASS and AllWISE photometric data. 
These SEDs profile were used to verify their classification in the S/D/\DD scheme: 74\% of the known SySts were 
classified as S-type, 13\% as D-type and only 3.5\% as \DD-type. A new classification was proposed for 22 SySts 
with no previous classification. 

S-type are clearly dominated by the emission of the cool companion. Their SEDs show a peak between 0.8 and 1.7~$\mu$m, which 
corresponds to an effective temperature of cool giants between 3000 and 4100~K. The majority of S-type have an M spectral-type 
companion. The effective temperature derived in this work can be considered reliable within the uncertainties.

A small number of SySts in the whole sample (27, or 8\%) was found to display an S-type SED profile with a significant 
infrared excess between 10 and 22$\mu$m. We decided to separate this group of objects and classify them as S$+$IR-type. The presence of 
a dusty disk around the WD or a tenuous dusty envelope with temperatures lower compared to those of D-types is a possible explanation 
for this excess. S$+$IR-type SySts are likely a transition phase from the S- to D-type, in which the cool companion has just entered 
into the early AGB phase.

D-type SySts were found to have a peak at the wavelength range from 2 to 4~$\mu$m, which corresponds to a dust temperature 
between 700 and 1400~K. Several D-type SySts show the presence of two dusty shells, with the second one being colder 
between 200 and 400~K. Regarding the \DD-type SySts, their SEDs reveal the characteristics of cool companions and dusty shells 
resulting in a nearly flat profile, clearly distinct from the other three types. 
The overall characteristics of the four different types of SySts are listed in Table~\ref{table10a}.
 
Flux variations with amplitudes of 2~mag between the 2MASS and {\it WISE} surveys do not significantly affect the SED fitting, the 
classification, and the resulting temperatures. Moreover, we found that poor fitting of the 2MASS data or {\it W1} and {\it W2}
indicate some variability among the data. In general, flux variations with amplitudes lower than 2-3~mag result in 
temperature variations of 400~K.

Geometrical distances of 193 SySts, using {\it Gaia} DR2, were also presented. The vertical distances of known SySts 
from the Galactic disk showed that S-type belong to the Galactic thick and thin disks, while the S$+$IR- and D-type belong mainly 
to the Galactic thin disk.

Finally, a new census of the \ovi\ $\lambda$6830 Raman-scattered line in SySts was presented.
We found 165 cases or 55\% of the sample, in which the \ovi\ $\lambda$6830 Raman-scattered line is detected. 
No preference for the \ovi\ Raman-scattered line was found among the different types of SySts. 

Exploring the \ovi\ $\lambda$6830/\ha, \heliumb\ $\lambda$4686/\hbeta, \helium\ $\lambda$5876/\hbeta\ and 
\heliumb\ $\lambda$4686/\oxygeniii\ $\lambda$5007 line ratios, we came up with some additional criteria that can be used in order to 
identify any feature centered at 6830\AA\ as a Raman-scattered line. According to this analysis, we were able to confirm or reject the 
detection of the \ovi\ $\lambda$6830 Raman-scattered line reported in non-SySts. For most of the cases, we concluded that the line 
feature at 6830\AA\ does not correspond to the Raman-scattered line but probably to the  \kriptoiii\ $\lambda$6826.7, 
\helium\ $\lambda$6827.9 and \carboni\ $\lambda$6828.1 lines. Only two objects (M2-9 and LHA~115~S-18) show strong indications for a 
positive identification as an \ovi\ Raman-scattered line. 

Possible links between the \ovi\ Raman-scattered line and X-ray emission were also explored. From all known X-ray SySts (46), only 
12 of them emit the \ovi\ Raman-scattered line and 21 do not. There are 13 more X-ray SySts without available optical data that 
merit further observations. Moreover, only $\alpha$ and $\beta$-type X-ray SySts were found to show the \ovi\ Raman-scattered line. 
This may indicate a link between the mechanism responsible for the production of X-ray emission (shell-burning) and Raman-scattering, 
but further investigation is required.

In the future, more effort is needed to search for new SySts in the Milky Way and nearby galaxies. An extensive search in archive 
data such as the VPHAS+ survey needs to be done in order to find the hidden SySt population in the Galactic plane and bulge. 
X-ray observations of more SySts are also required in order to understand better the mechanisms of X-ray emission and their 
correlation with optical emission.

\acknowledgments
The authors thank the anonymous referee for the thorough revision and insightful comments and suggestions. 
S.A. and M.L.L.-F. acknowledge the support of CNPq, Conselho Nacional de Desenvolvimento Cient\'ifico e Tecnol\'ogico - Brazil 
(grant 300336/2016-0 and 248503/2013-8, respectively). G.R.L. acknowledges support from Universidad de Guadalajara, 
CONACyT, PRODEP, and SEP (Mexico). L.G.R. is supported by NWO funding toward the Allegro group at Leiden University. 
Authors also thank Romano Corradi for helpful discussions. This publication made use of data from the Two Micron All-Sky Survey, which is a 
joint project of the University of Massachusetts and the Infrared Processing and Analysis Center/California Institute of Technology, 
funded by the NASA and the National Science Foundation, data products from the {\it Wide-field Infrared Survey Explorer}, which is a 
joint project of the University of California, Los Angeles, and the Jet Propulsion Laboratory/California Institute of Technology, funded 
by the National Aeronautics and Space Administration. This work has made use of data from the European Space Agency (ESA) mission Gaia 
(https://www.cosmos.esa.int/gaia), processed by the Gaia Data Processing and Analysis Consortium 
(DPAC; https://www.cosmos.esa.int/web/gaia/dpac/consortium). Funding for the DPAC has been provided by national institutions, 
in particular the institutions participating in the Gaia Multilateral Agreement. This publication made also use of many 
software packages in {\sc Python}. \software{:Matplotlib (Hunter 2007), NumPy (van der Walt et al. 2011), SciPy (Jones et al. 2001), 
AstroPy Python (Astropy Collaboration et al. 2013; Muna et al. 2016), R (R Development Core Team 2008)}.

\appendix

\section{Lists of SySts, New Discoveries, 2MASS and AllWISE Data, and Their Physical Parameters}

\begin{longrotatetable}
\begin{deluxetable*}{llll}
\tablecaption{\label{table1} List of new SySt (Known or Candidate) Discoveries and Their References.}
\tablehead{
\colhead{Known} & \colhead{Candidate} & \colhead{Environment} & \colhead{References}
}
\startdata
$-$         & 2$^{a}$         & Galactic     & Corradi (1995), Weidmann \& Gamen (2011)\\
188         & 30$^{b}$        & Catalog      & Bel2000 and references therein\\
$-$         & 2$^c$           & Galactic     & Schmeja \& Kimeswenger (2001), Corradi et al. (2011b)\\
$-$         & 1               & Galactic     & Feibelman (2001), Groves et al. (2002)\\
2$^d$       & 2$^d$           & Galactic     & Van Eck \& Jorissen (2002)\\ 
4           & $-$             & Galactic     & Munari \& Zwitter (2002), Downes \& Keyes (1988)\\
1           & $-$             & Galactic     & Pereira et al. (2002) \\ 
1$^{e}$     & $-$             & Galactic     & Wheatley et al.(2003) \\
\enddata
\tablecomments{$^a$ Corradi (1995) refers to three objects, Hen 2-25, Th 2-B and 19W32. 19W32 is included in the Belczy\'{n}ski et al. catalog 
but not the other two. Here, we refer to the discovery of the other two candidate Systs.\\
$^b$ Table~2 in Bel2000 lists 28 candidate, whereas the correct number is 30 (see Belczy\'{n}ski's online VizieR catalog).\\
$^c$ M2-9, Mz~3.\\
$^d$ These discoveries were made based on the \ha\ line profiles. These four objects belong to the extrinsic s-type stars.\\
$^e$ The X-ray properties of this object are consistent with a symbiotic binary, ruling out its previous classifications as an AM Her-type cataclysmic variable.\\
$^f$ These are SySt X-ray binaries (SyXB; Luna et al. 2013 and references therein).\\
$^g$ Based on the IPHAS photometric constraints in Corradi et al. (2008).\\
$^h$ Miszalski et al. (2009) presented a preliminary list of 13 SySts and 11 candidate SySts. A second list is presented in 2013, with 20 SySts and 15 candidates (Miszalski et al. 2013). By cross matching the two studies, the exact number of SySts discovered in each study are given in the table. The second study provides a different but more robust classification. \\
$^i$ Two objects were classified as candidate SySts by Viironen et al. (2009). However, one of them (IPHASX~J190438.7+021424) was excluded from the list of SySt candidates by Weidmann \& Gamen (2011).\\
$^j$ Towards the Galactic Bulge.\\
$^k$ Two more objects are also classified as candidates, but they were first reported as candidates by Miszalski et al. (2011; [RP2006]~883 and [RP2006]~264).\\
$^l$ Sh 2-71.\\
$^{m}$ Seven previously classified candidate SySts have been confirmed.}
\end{deluxetable*}
\end{longrotatetable}

\begin{longrotatetable}
\begin{deluxetable*}{lcccccccccc} 
\tablecaption{\label{table2} Known and candidate symbiotic stars}
\tablehead{
\colhead{Name} & \colhead{R.A.} & \colhead{Decl.} & \colhead{\it J} & \colhead{\it H} & \colhead{K$_{\rm s}$} & \colhead{\it W1} & \colhead{\it W2} & 
\colhead{\it W3} & \colhead{\it W4} \\
\colhead{} & \colhead{J2000} & \colhead{J2000} & \colhead{(mag)} & \colhead{(mag)} & \colhead{(mag)} & \colhead{(mag)} & \colhead{(mag)} & 
\colhead{(mag)} & \colhead{(mag)}
} 
\startdata
IC10 SySt-1$^{4}$ & 00 20 33.59 & +59 18 45.9 						& - & - & - & 13.77$\pm$0.03 & 13.53$\pm$0.04 & 12.68$^{\dag}$ & 9.36 	\\
SMC1$^{1,2}$ (SMC SySt-1) & 00 29 10.86 & -74 57 39.8 				& 13.65$\pm$0.02 & 12.85$\pm$0.02 & 12.64$\pm$0.03 & 12.41$\pm$0.02 & 12.43 $\pm$0.02 & 11.88 $\pm$0.22 & 9.07				\\
Lin9$^{13}$ (SMC SySt-10) & 00 30 07.40 & -73 37 19.1 				& 13.29$\pm$0.03 & 12.44$\pm$0.03 & 12.27$\pm$0.03 & 12.14$\pm$0.02 & 12.17 $\pm$0.02 & 12.15 $\pm$0.26 & 8.90					\\
M31 SySt-1$^{16}$ & 00 38 46.16 & +40 07 17.0 						& - & - & - & - & - & - & - 				\\
M31 SySt-2$^{16}$ & 00 38 57.19 & +40 31 32.2 						& - & - & - & - & - & - & - 			\\
NGC185 SySt-1$^{21}$ & 00 39 00.36 & +48 19 26.6 					& - & - & - & 13.47$\pm$0.03 & 13.45$\pm$0.04 & 12.68 & 9.28 \\
M31 SySt-4$^{16}$ & 00 39 32.01 & +40 12 24.7 						& - & - & - & - & - & - & - 			\\
M31 SySt-5$^{16}$ & 00 39 51.85 & +40 21 53.8 						& - & - & - & - & - & - & - 				\\
\enddata
\end{deluxetable*}
\end{longrotatetable}

\begin{longrotatetable}
\begin{deluxetable*}{lccccccccccc} 
\tablecaption{\label{table3new} New Classification of SySts, Stellar and Dust Temperatures Derived from 
Blackbody Fitting, Stellar Temperatures of Red Giants Corrected for the BB Approximation, $\lambda$ Peak of SEDs, 
Effective Temperature from {\it Gaia} DR2, Distances of SySts Based on the Parallaxes Derived from {\it Gaia} DR2, 
and Information on the Detection of the OVI~$\lambda$6830 Raman-scattered Line and X-Ray Emission.}
\tablehead{
\colhead{Name} & \colhead{Old} & \colhead{New} & \colhead{T$_{\rm BB}$} & \colhead{T$_{\rm dust}$} & \colhead{T$_{\rm cool}$} & \colhead{$\lambda_{\rm peak}$} & \colhead{T$_{\rm Gaia}$ } & \colhead{D$_{\rm Gaia}$} & \colhead{\ovi} & \colhead{\ovi} & \colhead{X-ray$^{a}$} \\
\colhead{} & \colhead{Type} & \colhead{Type} & \colhead{(K)} & \colhead{(K)} & \colhead{(K)} & \colhead{($\mu$m)} & \colhead{(K)} & 
\colhead{(pc)} & \colhead{6830\AA} & \colhead{Refs.} & \colhead{Emission} 
} 
\startdata
IC10 SySt-1 				& ?   	& ? 	& $-$		   	& $-$  			 		 &  $-$ &				$-$ & $-$ &  $-$ & \xmark	& 25	&  \\
SMC1 (SMC SySt-1) 			& S  	& S 	& 3056$\pm$142 	& $-$ 					 & 3572 & 0.95$\pm$0.05 & $-$ & $-$  & \cmark	& 1,5	& \\
Lin9 (SMC SySt-10)		    & S  	& S 	& 3150$\pm$180 	& $-$				     & 3810 & 0.92$\pm$0.06 & $-$ & $-$ & \cmark	& 14	&\\
M31 SySt-1  				& S   	& ? 	& $-$ 			& $-$					 &	$-$	&				$-$ & $-$ & $-$ &\cmark	&17 	& \\
M31 SySt-2 					& S   	& ? 	& $-$ 			& $-$					 &	$-$	&				$-$ & $-$ & $-$ &\xmark	&17		& \\
NGC185 SySt-1 				& S   	& ? 	& $-$ 			& $-$					 &	$-$	&				$-$ & $-$ & $-$ &\cmark	& 10	& \\
M31 SySt-4 					& S   	& ? 	& $-$ 			& $-$			  		 &	$-$	&				$-$ & $-$ & $-$ &\cmark 	& 17	& \\
M31 SySt-5 					& S   	& ? 	& $-$ 			& $-$					 &	$-$	&	$-$			$-$ & $-$ &$-$ & \xmark	&17		&	\\
\enddata
\tablecomments{
$^{a}$ Muerset et al. (1997), Luna et al. (2013) and references therein.\\
\\
References. 1. Munari \& Zwitter (2002), 2. Blair et al. (1983), 3. Allen (1984), 4. Acker et al. (1988), 5. Muerset et al. (1996), 6. Downes \& Keyes (1988), 7. Kniazev et al. (2009), 8. Corradi \& Giammanco (2010), 9. Corradi et al. 2010b, 10. Gon\c{c}alves et al. 2012, 11. Oliveira et al. (2013),12. Miszalski et al. (2013), 13. Miszalski \& Mikolajewska (2014), 14. Miszalski et al. (2014), 15. Gon\c{c}alves et al. (2015), 16. Rodriguez-Flores et al. (2014), 17. Mikolajewska et al. (2014), 18. Li et al. (2015), 19. Guti\'{e}rrez-Moreno et al. (1999), 20. Medina Tanco \& Steiner (1995), 21. Corradi (1995), 22. Pereira et al. (2002), 23. Pereira \& Miranda (2005), 24. Masseti et al. (2006a), 25. Gon\c{c}alves et al. (2008), 26. Corradi et al. (2008), 27. Corradi et al. (2011a), 28. Baella et al. (2013), 29. Masetti et al. (2011), 30. Hajduk et al. (2014), 31. Baella et al. (2016), 32. Mikolajewska et al. (2017), 33. Cieslinski et al. (1994), 34. Torres et al. (2012), 35. Groves et al. (2002), 36. Macfarlane et al. (2017), 37. Skopal et al. (2017), 38. Remillard et al. (1992), 39. Mikolajewska et al. (1997), 40. Bond \& Kasliwal (2012), 41. Mennickent et al. (2008).}
\end{deluxetable*}
\end{longrotatetable}

\section{Exploring flux variations in the SED profiles of SySts}

Due to the large flux variations of SySts and the different epochs in which the 2MASS and {\it WISE} observations were performed, we had to explore 
how possible flux variations between the two surveys can alter the SED fittings. For this exercise, we used one S-type (V694~Mon) and one D-type 
(IPHAS~J205836.43+503307.28) SySt as test objects assuming four different maximum amplitude variations (mag$_{\rm{maximum}}$-mag$_{\rm{minimum}}$; Figures 11 and 12, panels a to d) and six different scenarios: (1) 2MASS in the maximum brightness and {\it WISE} in the minimum, 
(2) 2MASS in the minimum brightness and {\it WISE} in the maximum, (3) 2MASS in the maximum brightness and {\it WISE} without any variation (4) 
2MASS in the maximum brightness and {\it WISE} without any variation, (5) 2MASS without any variation and {\it WISE} in the minimum brightness, and 
(6) 2MASS without any variation and {\it WISE} in the minimum brightness. 

\begin{figure*}
\begin{center}

\includegraphics[scale=0.14]{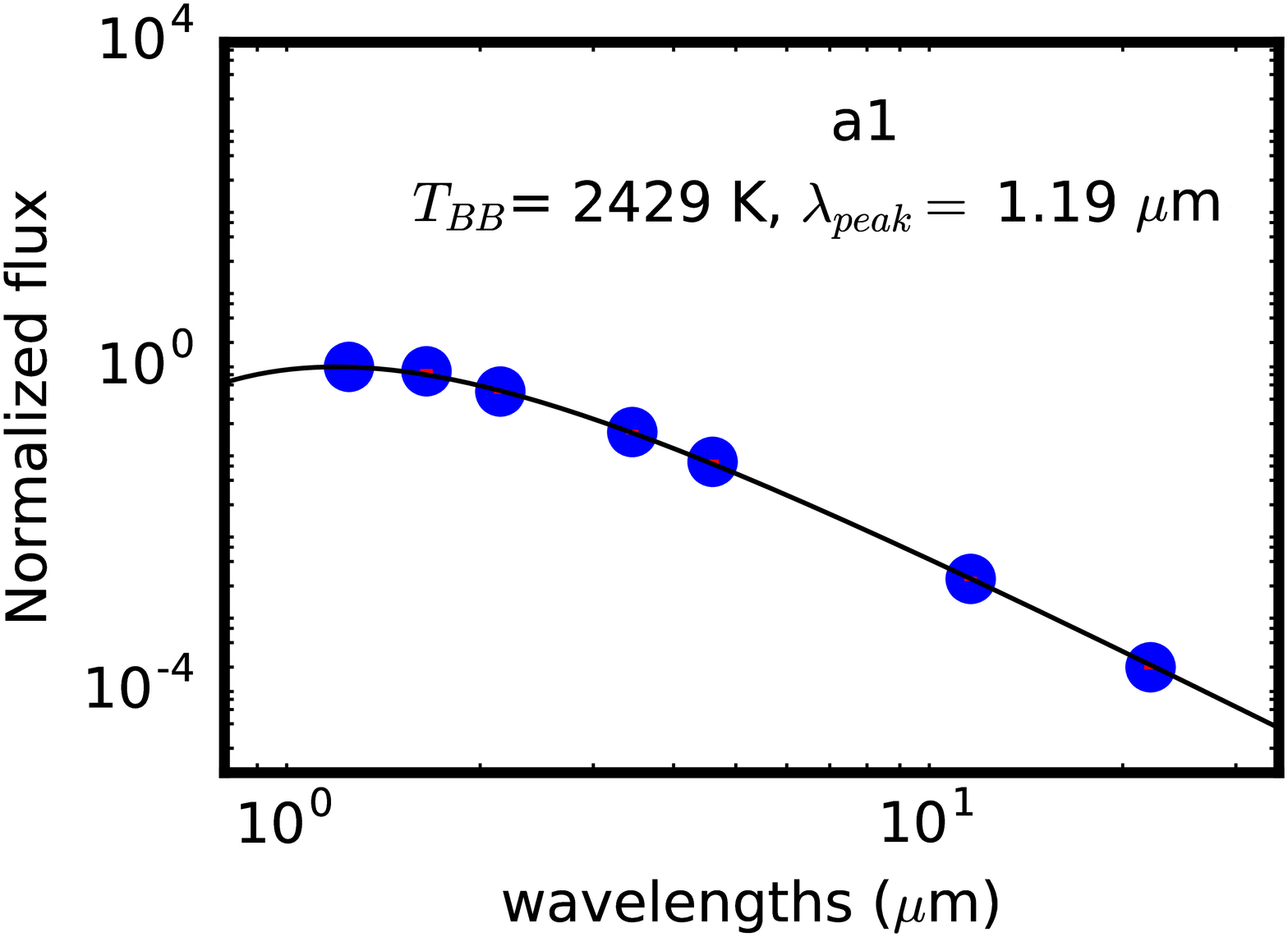}
\includegraphics[scale=0.14]{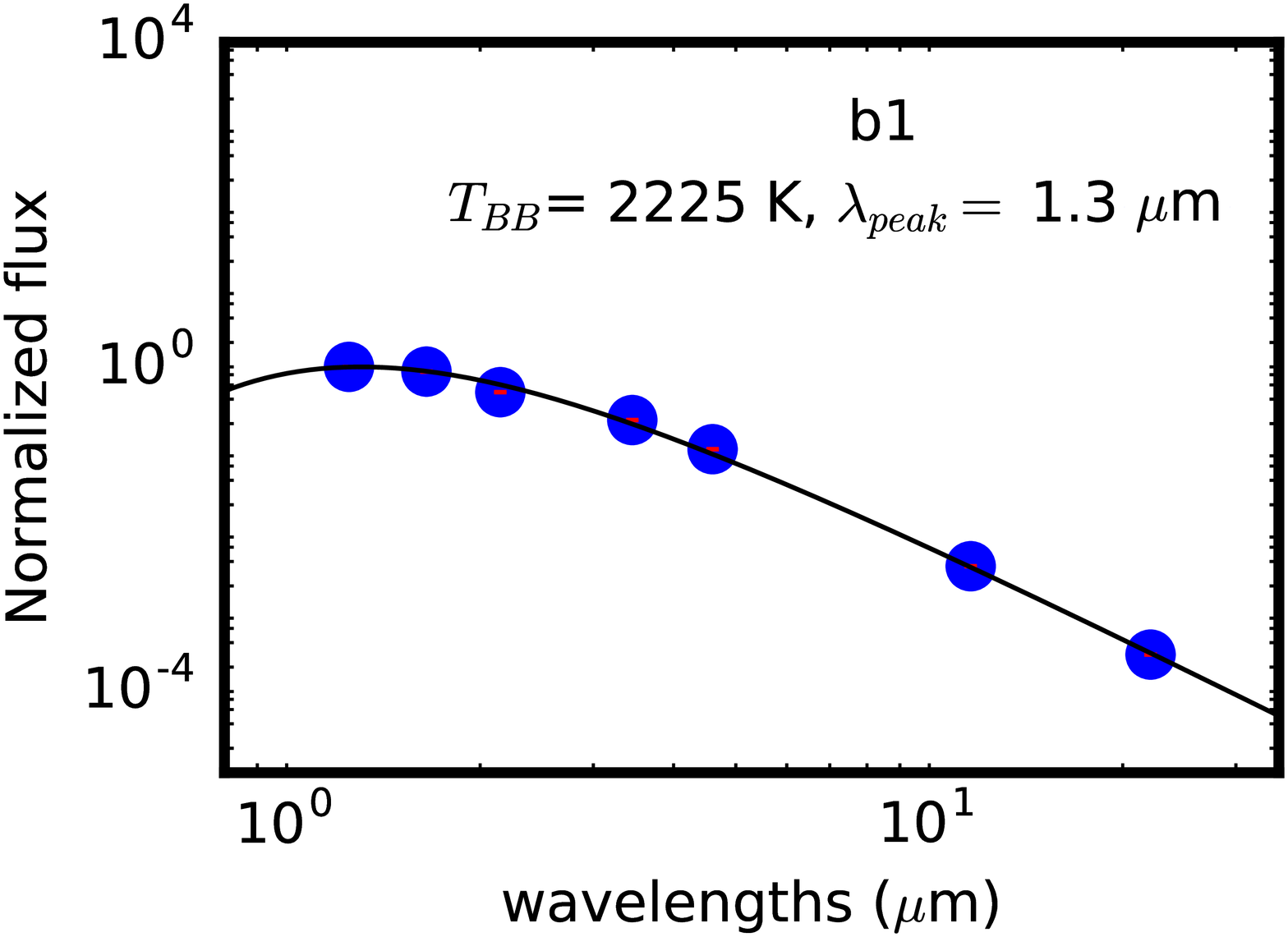}
\includegraphics[scale=0.14]{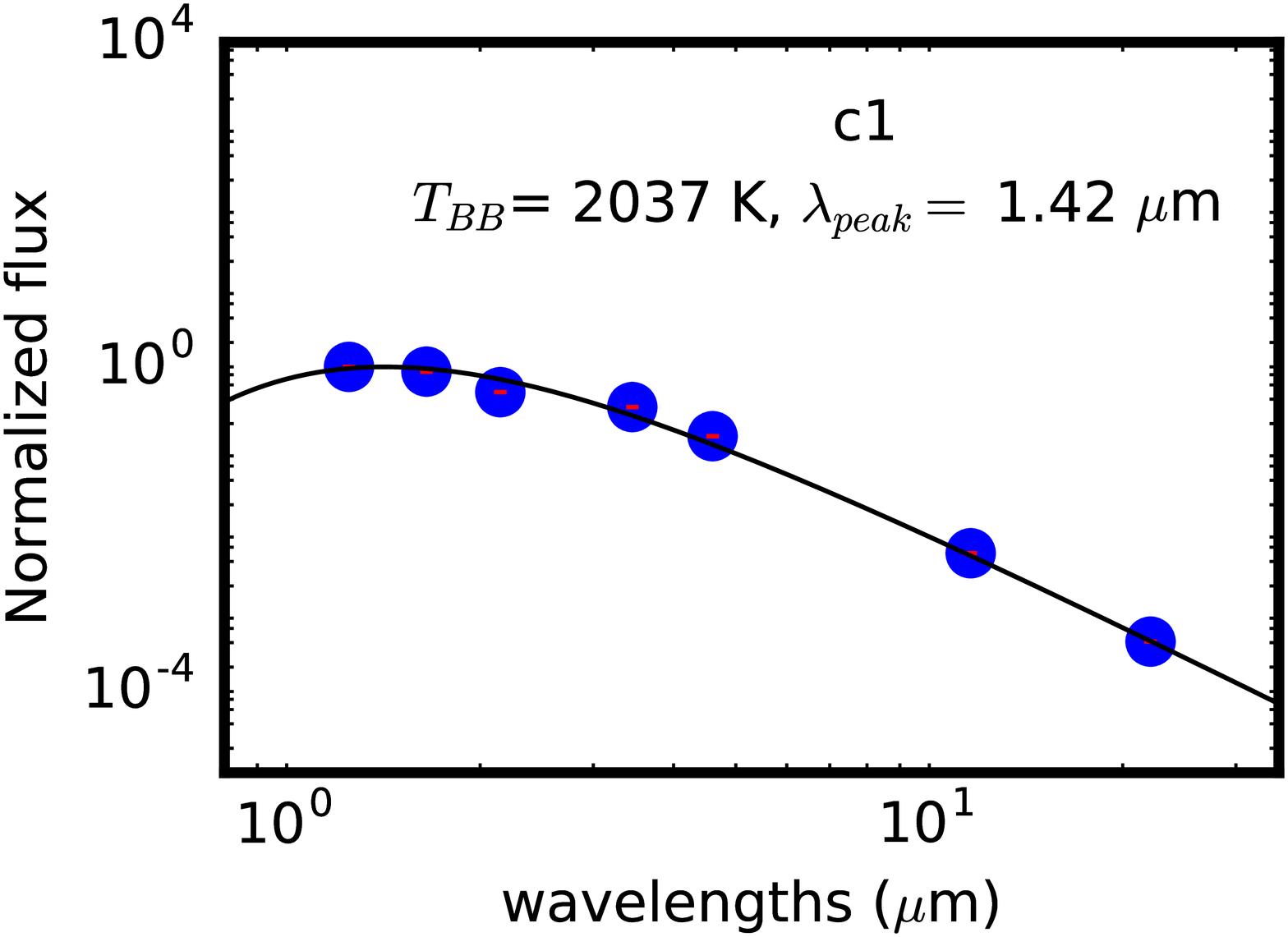}
\includegraphics[scale=0.14]{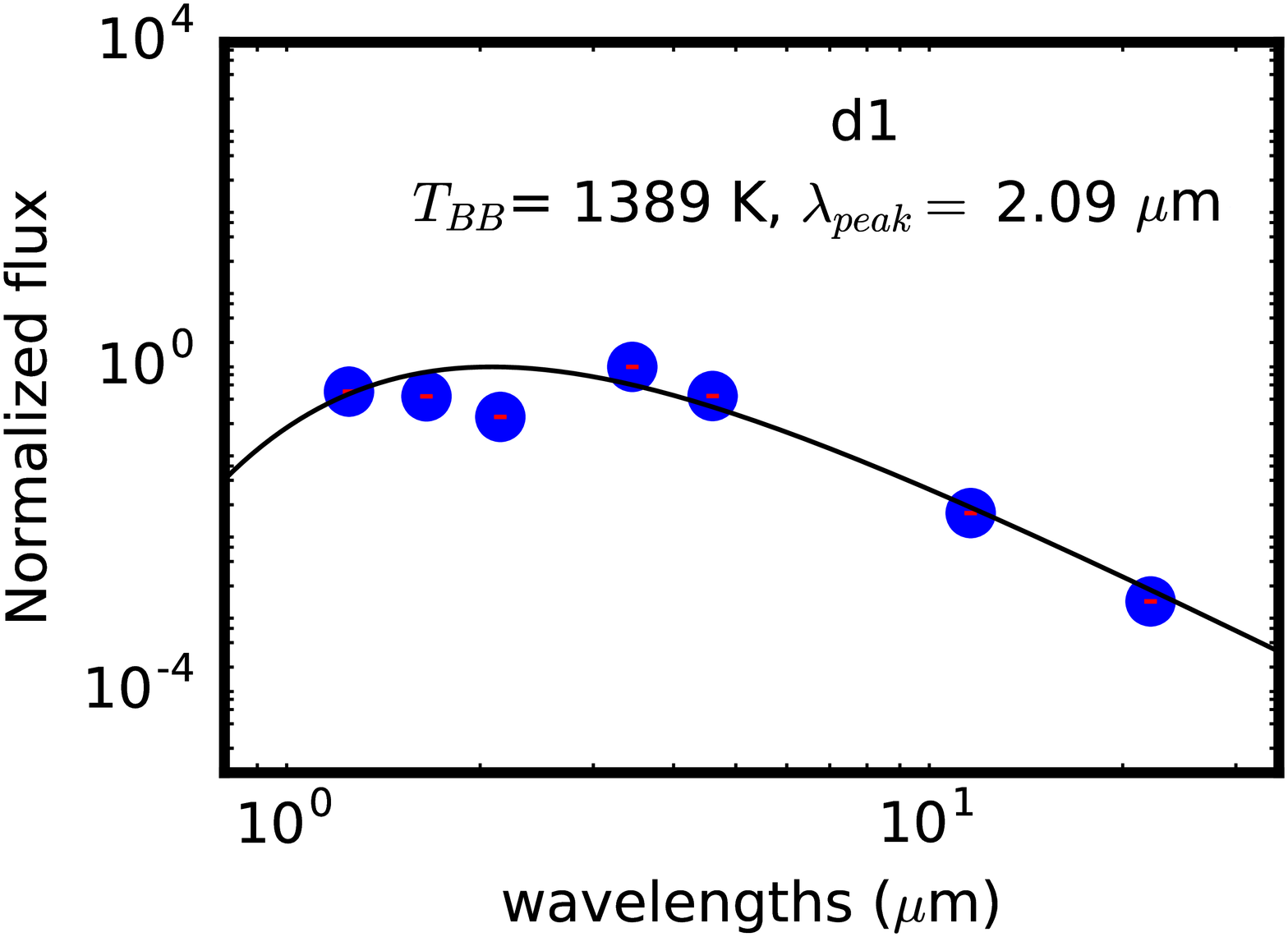}

\includegraphics[scale=0.14]{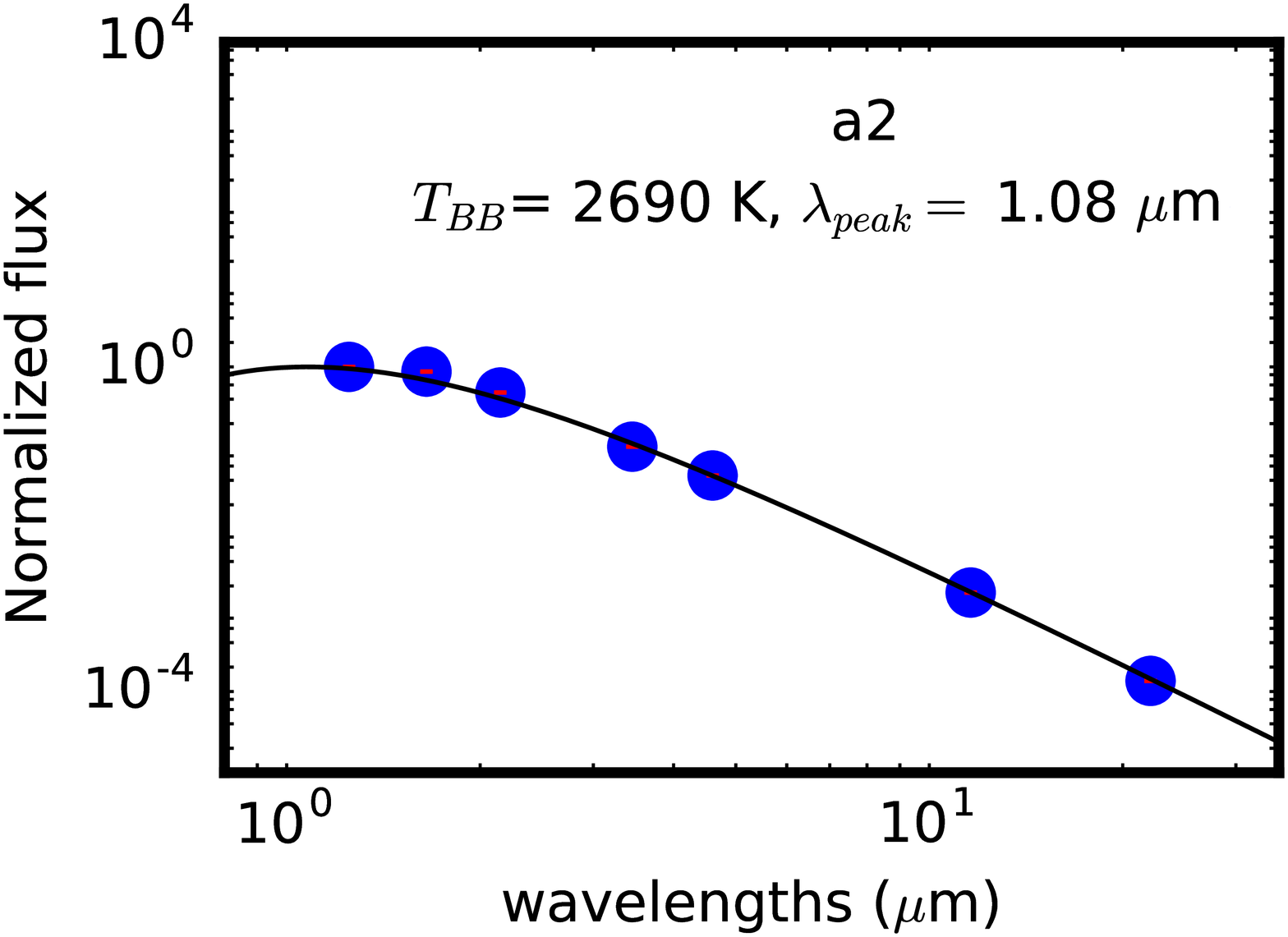}
\includegraphics[scale=0.14]{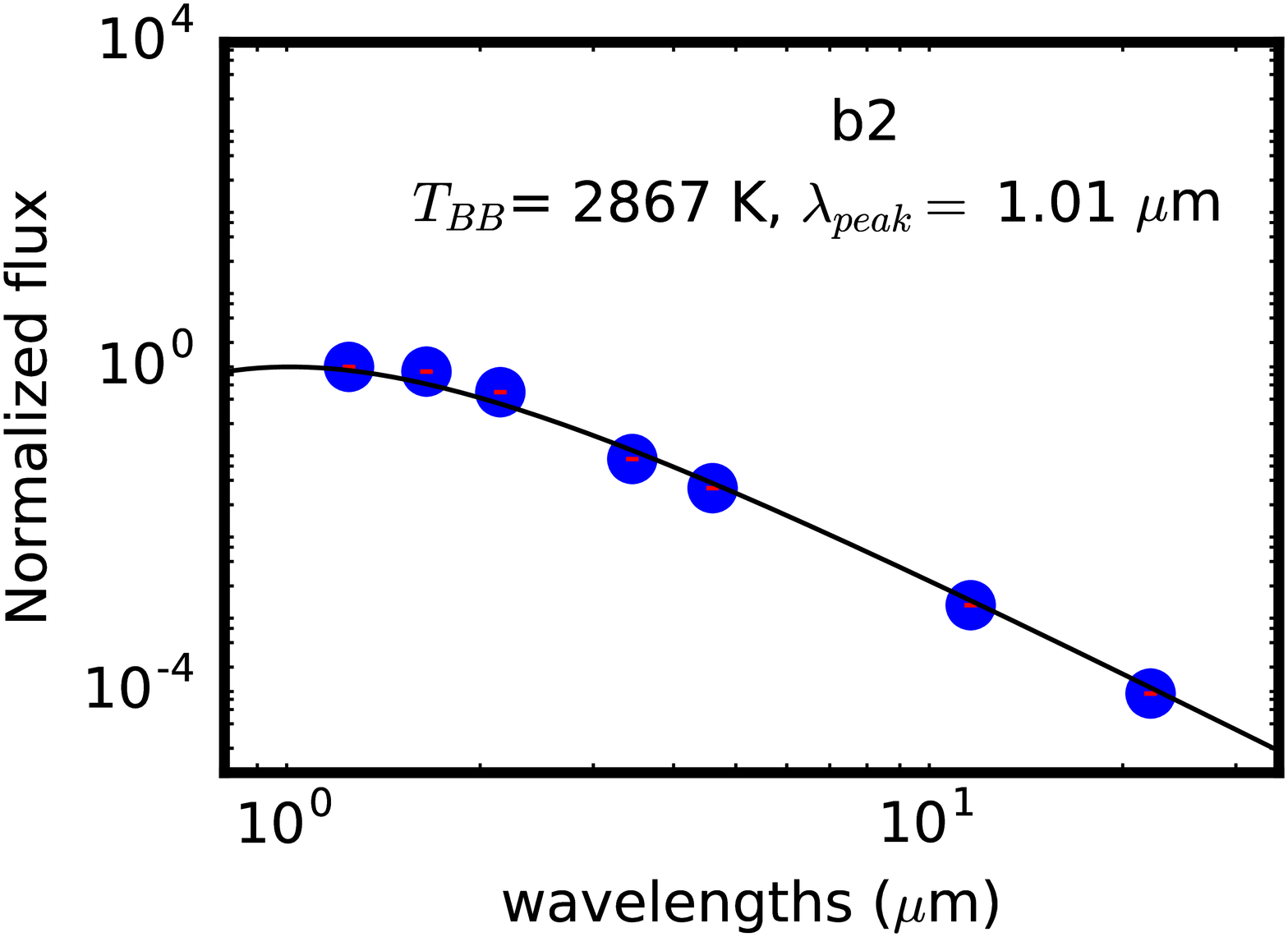}
\includegraphics[scale=0.14]{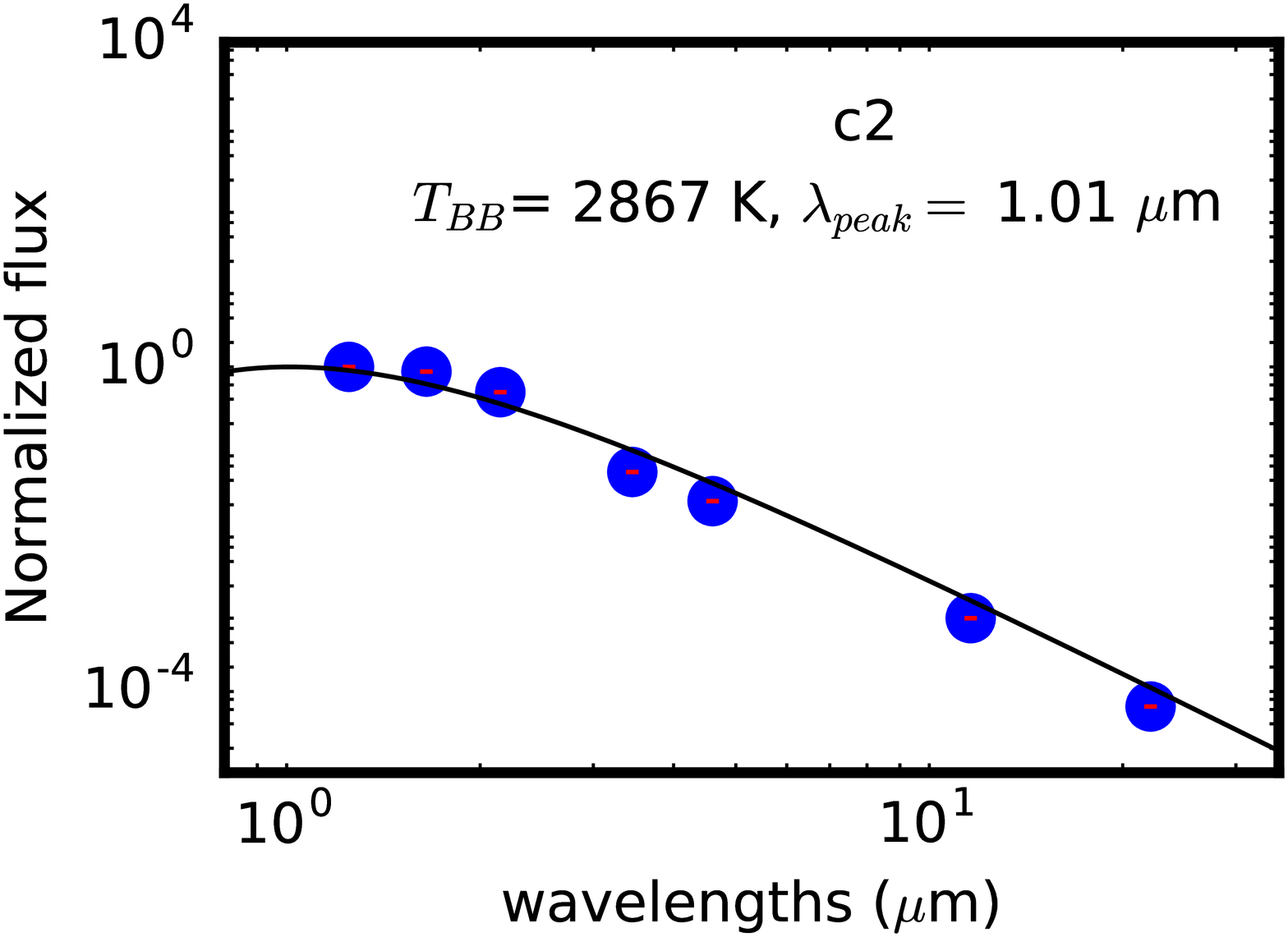}
\includegraphics[scale=0.14]{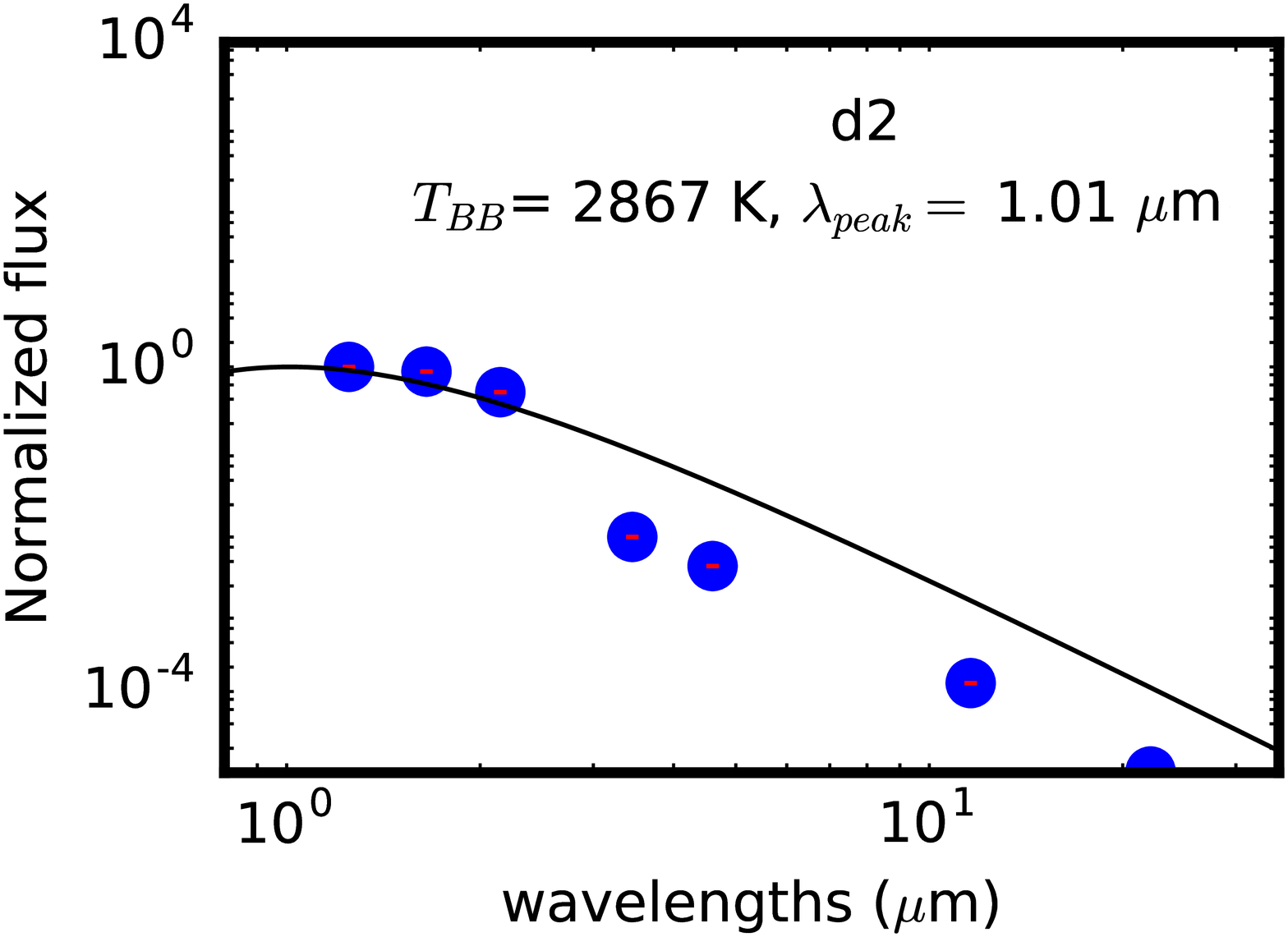}

\includegraphics[scale=0.14]{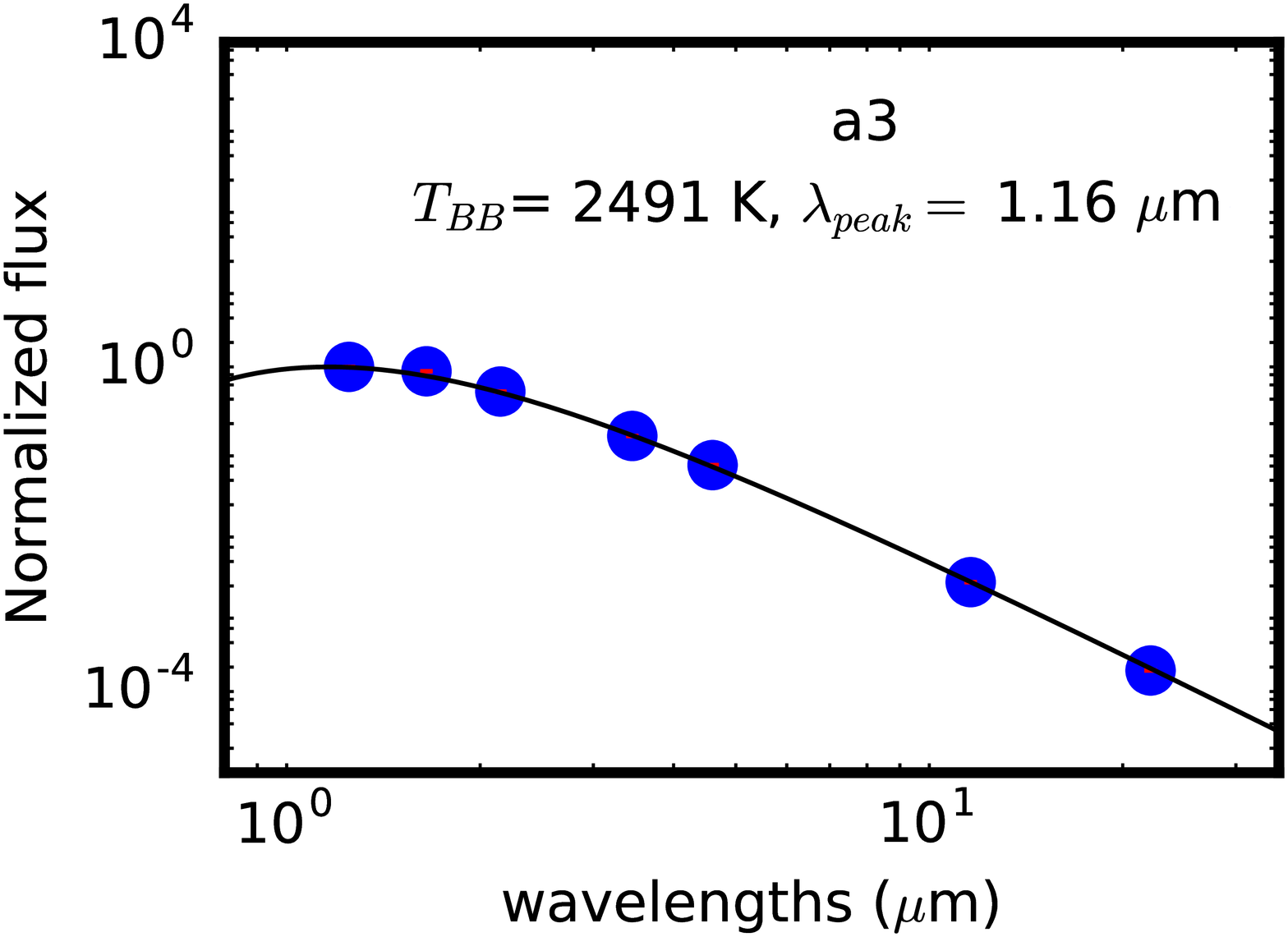}
\includegraphics[scale=0.14]{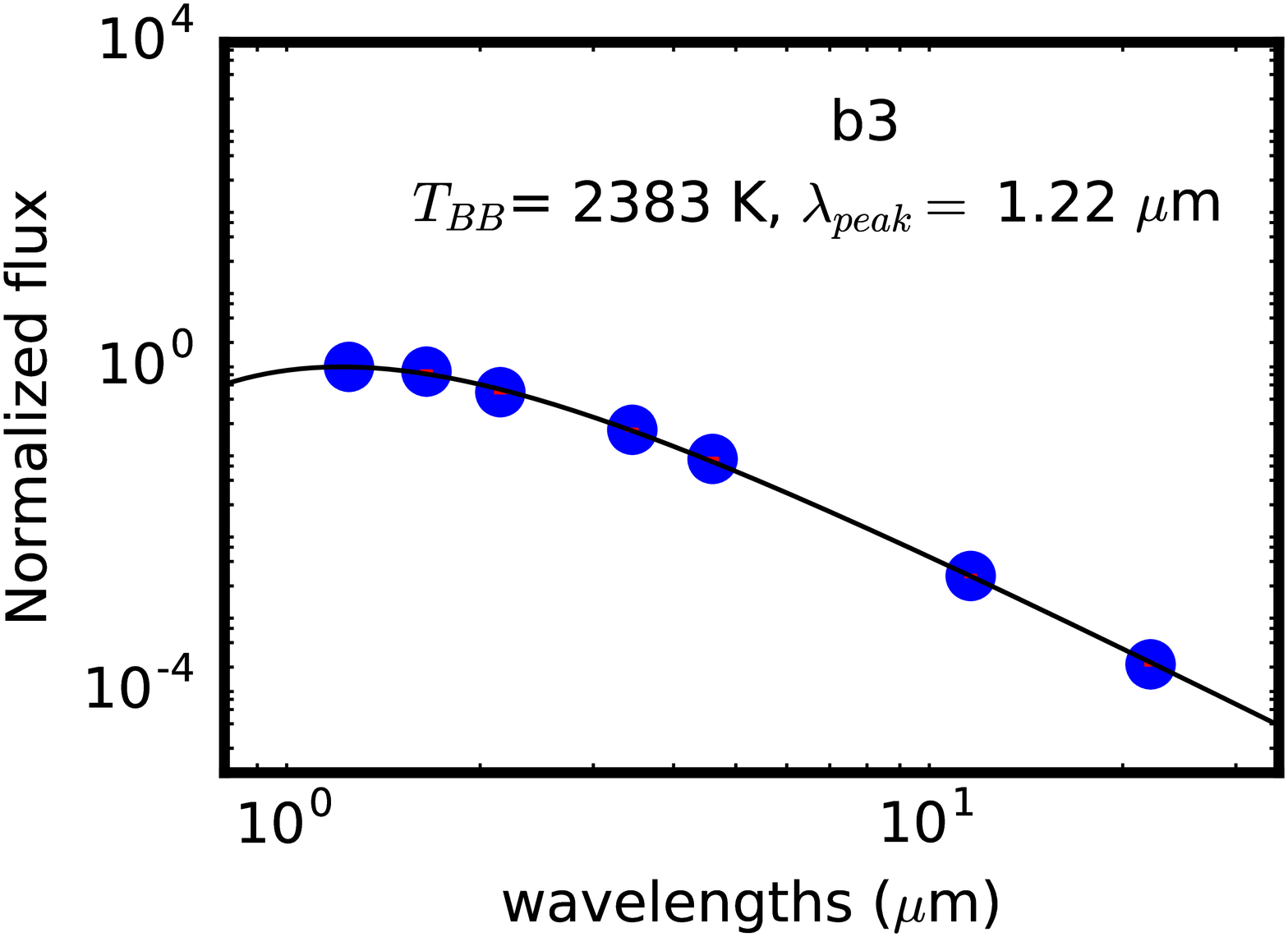}
\includegraphics[scale=0.14]{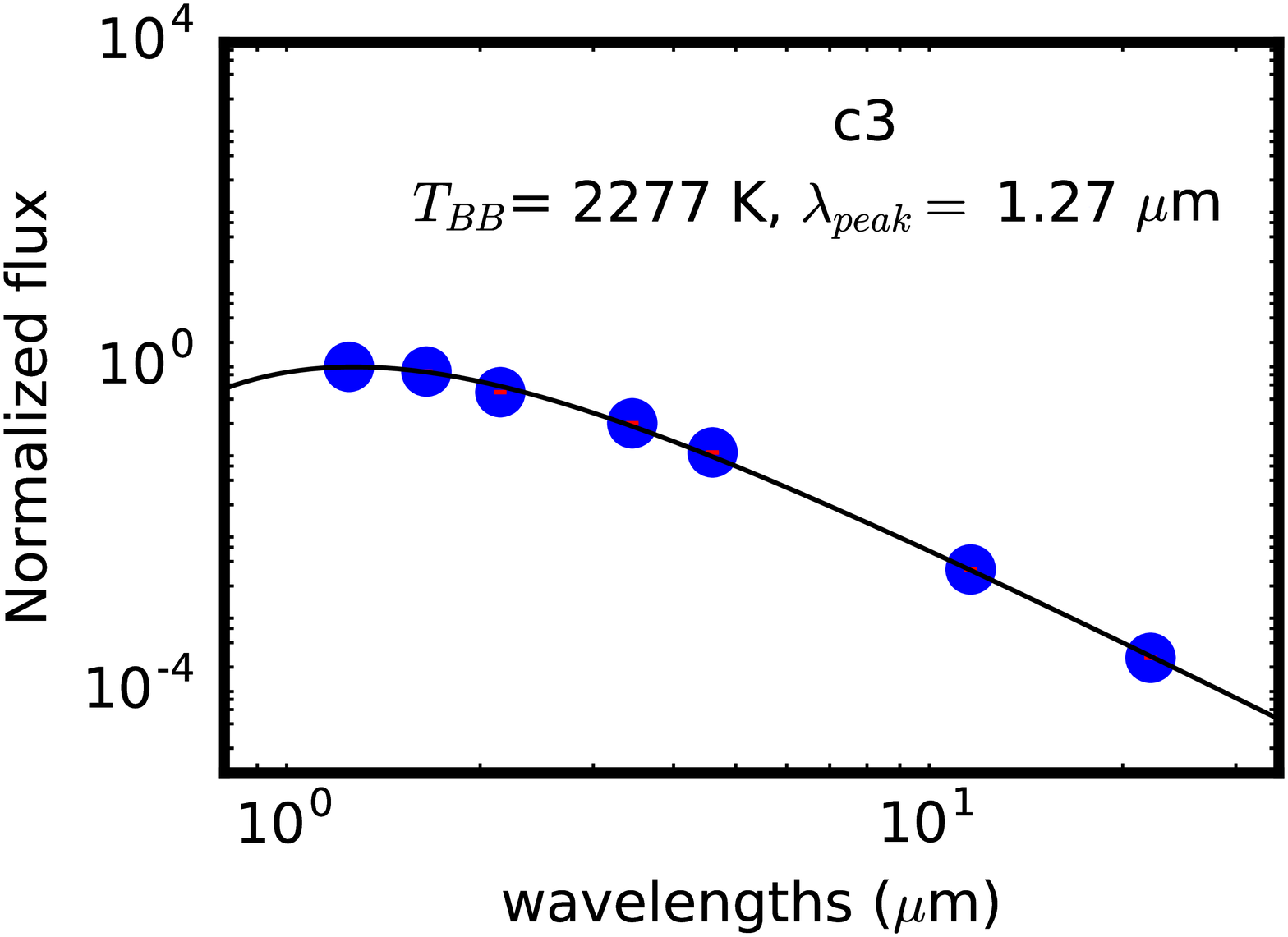}
\includegraphics[scale=0.14]{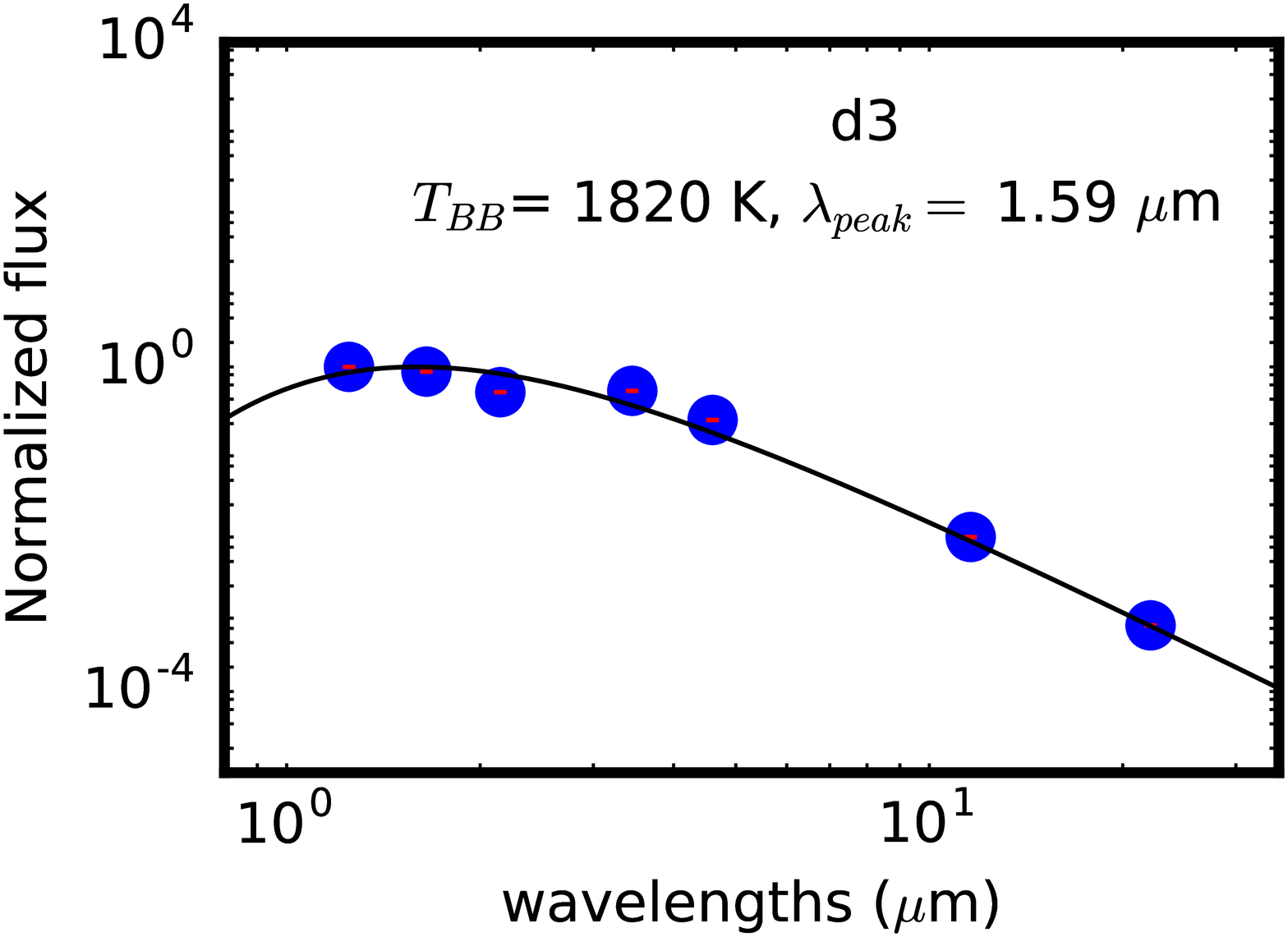}

\includegraphics[scale=0.14]{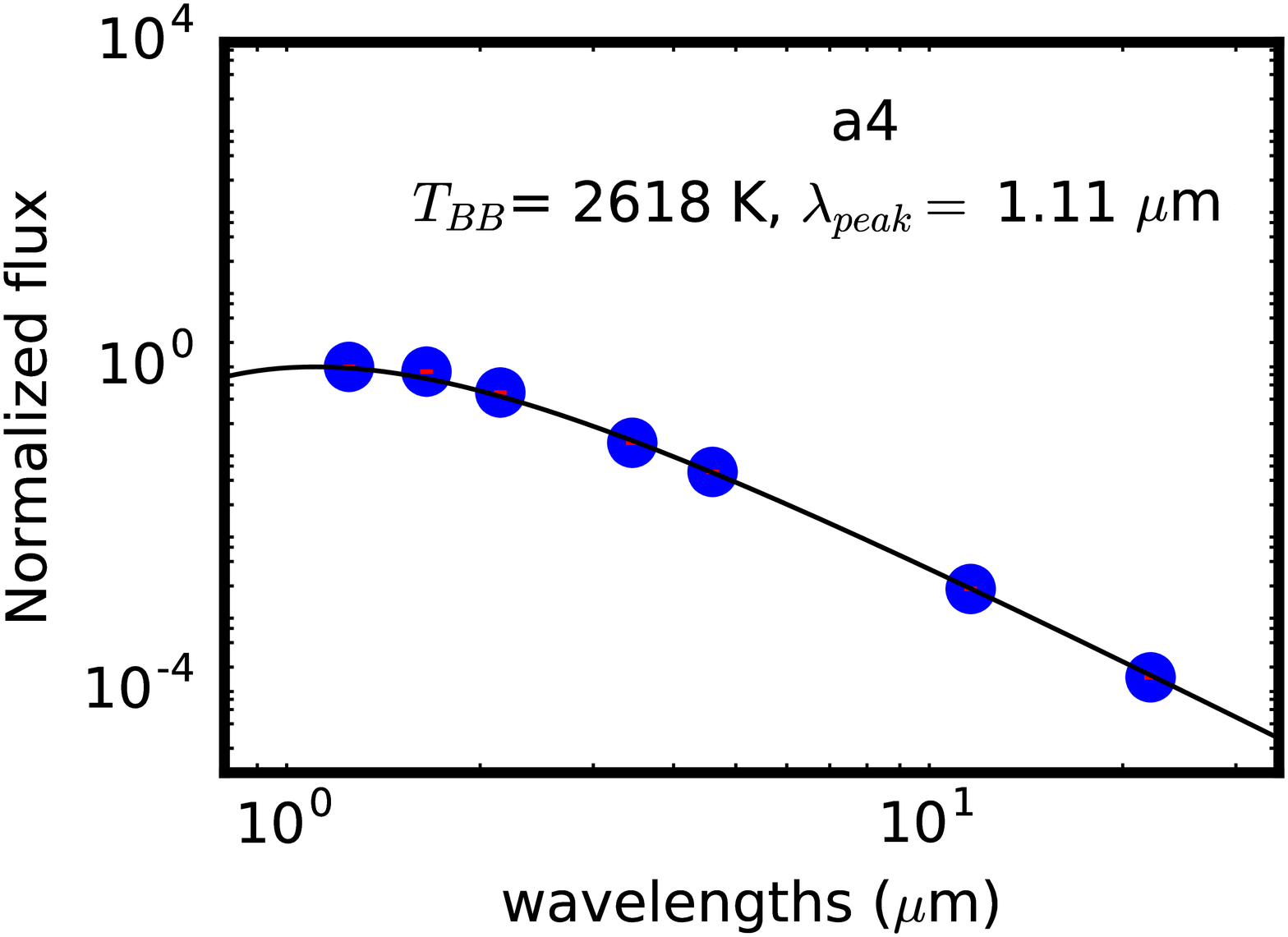}
\includegraphics[scale=0.14]{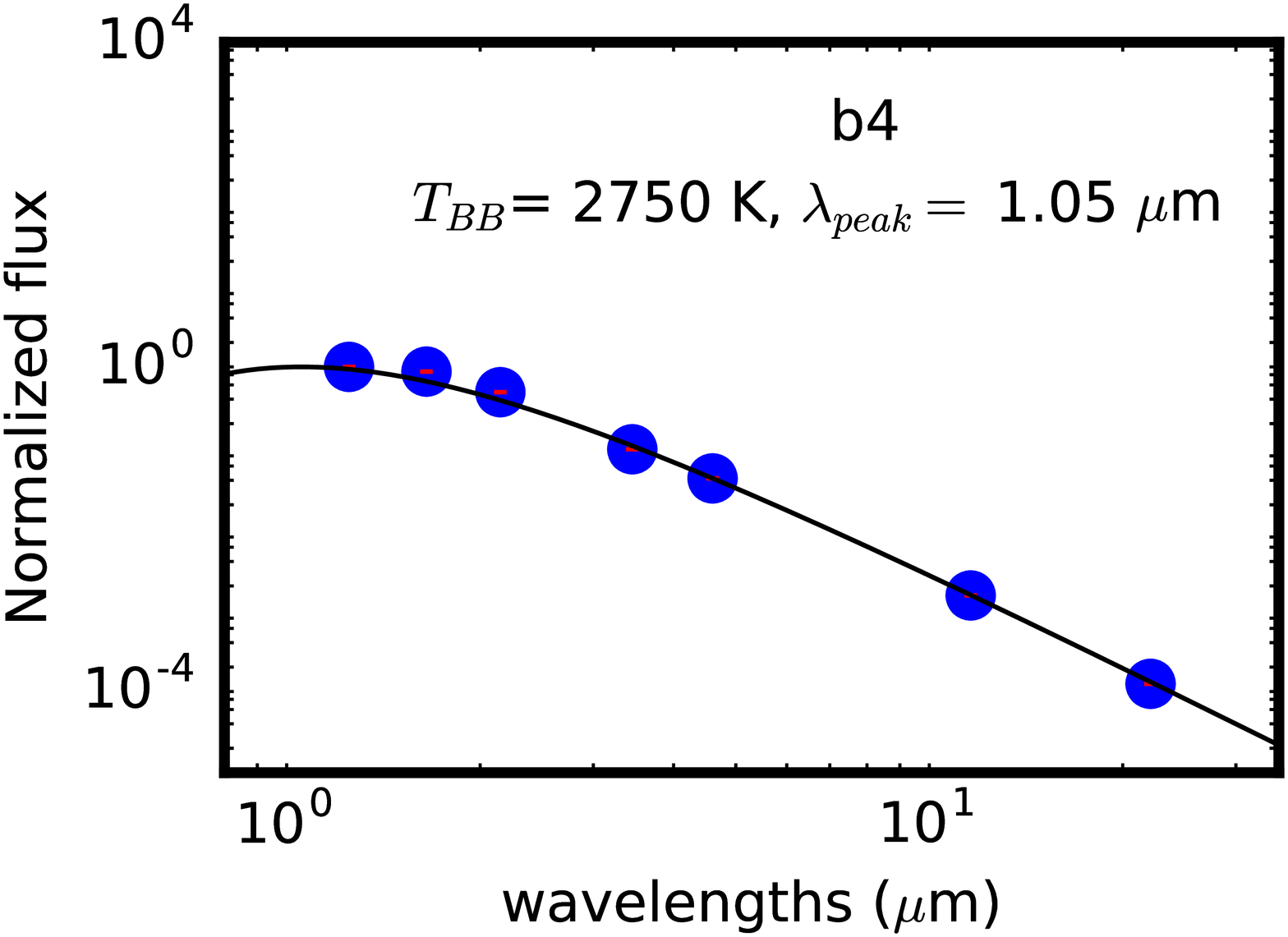}
\includegraphics[scale=0.14]{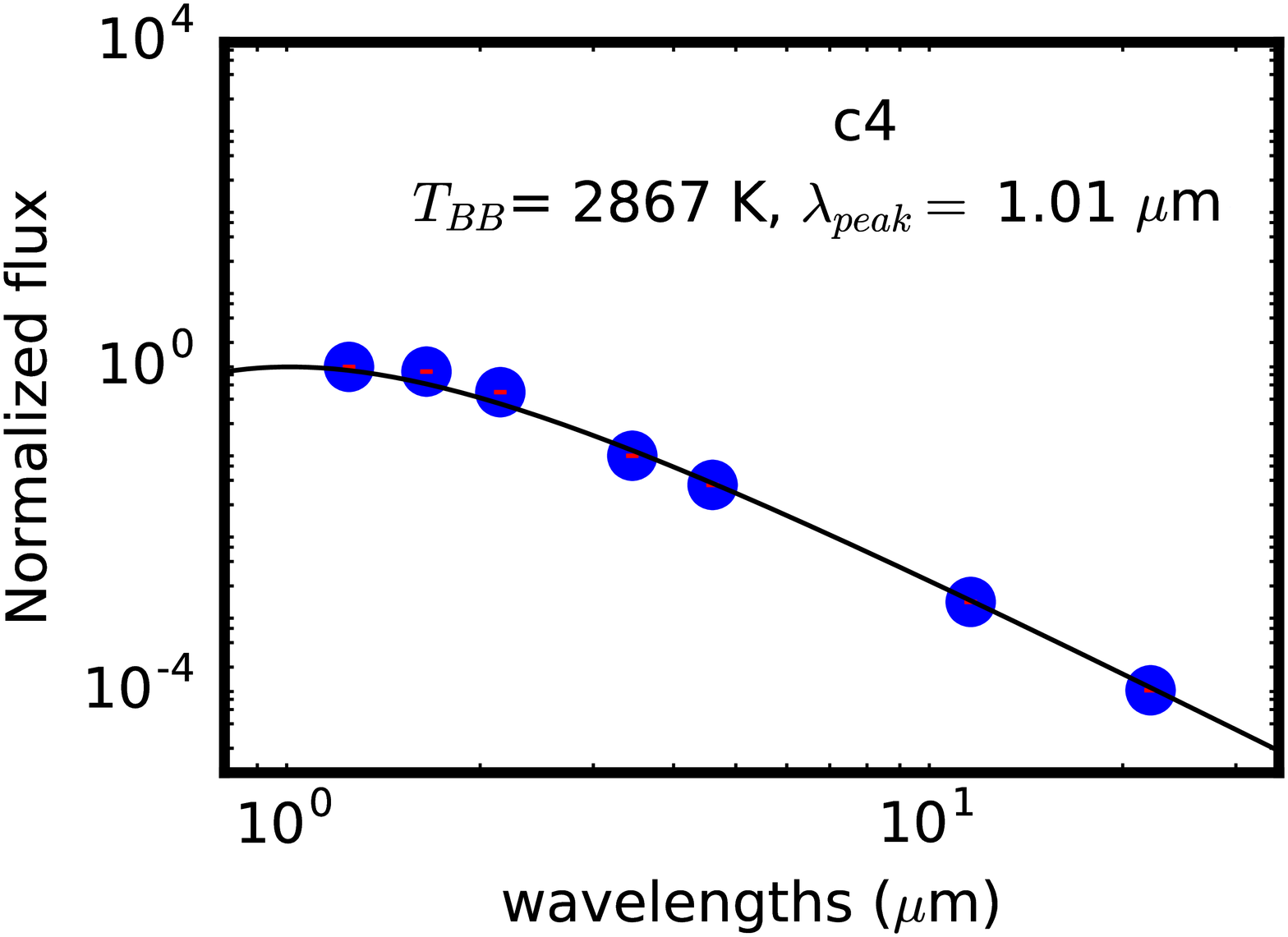}
\includegraphics[scale=0.14]{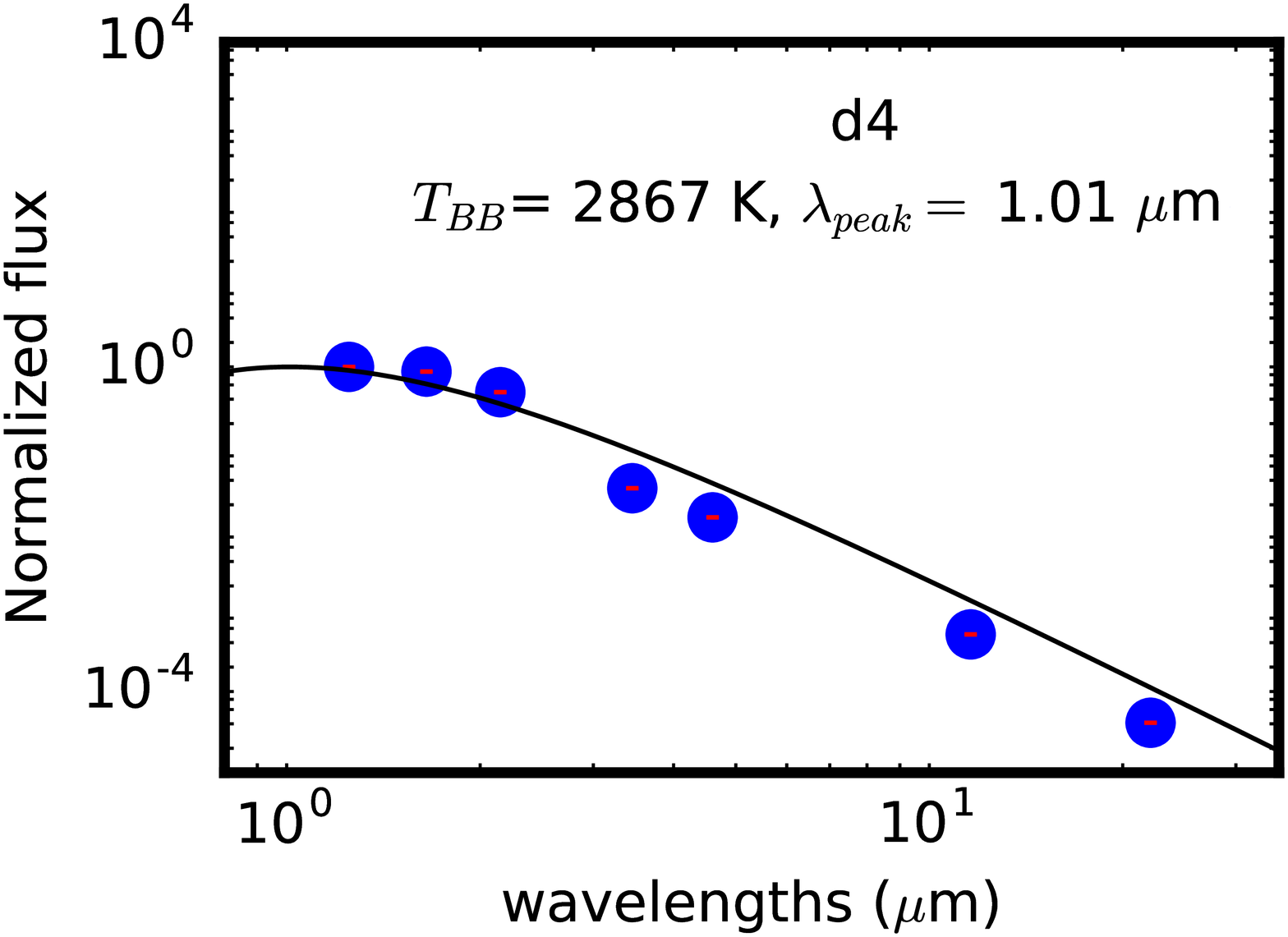}

\includegraphics[scale=0.14]{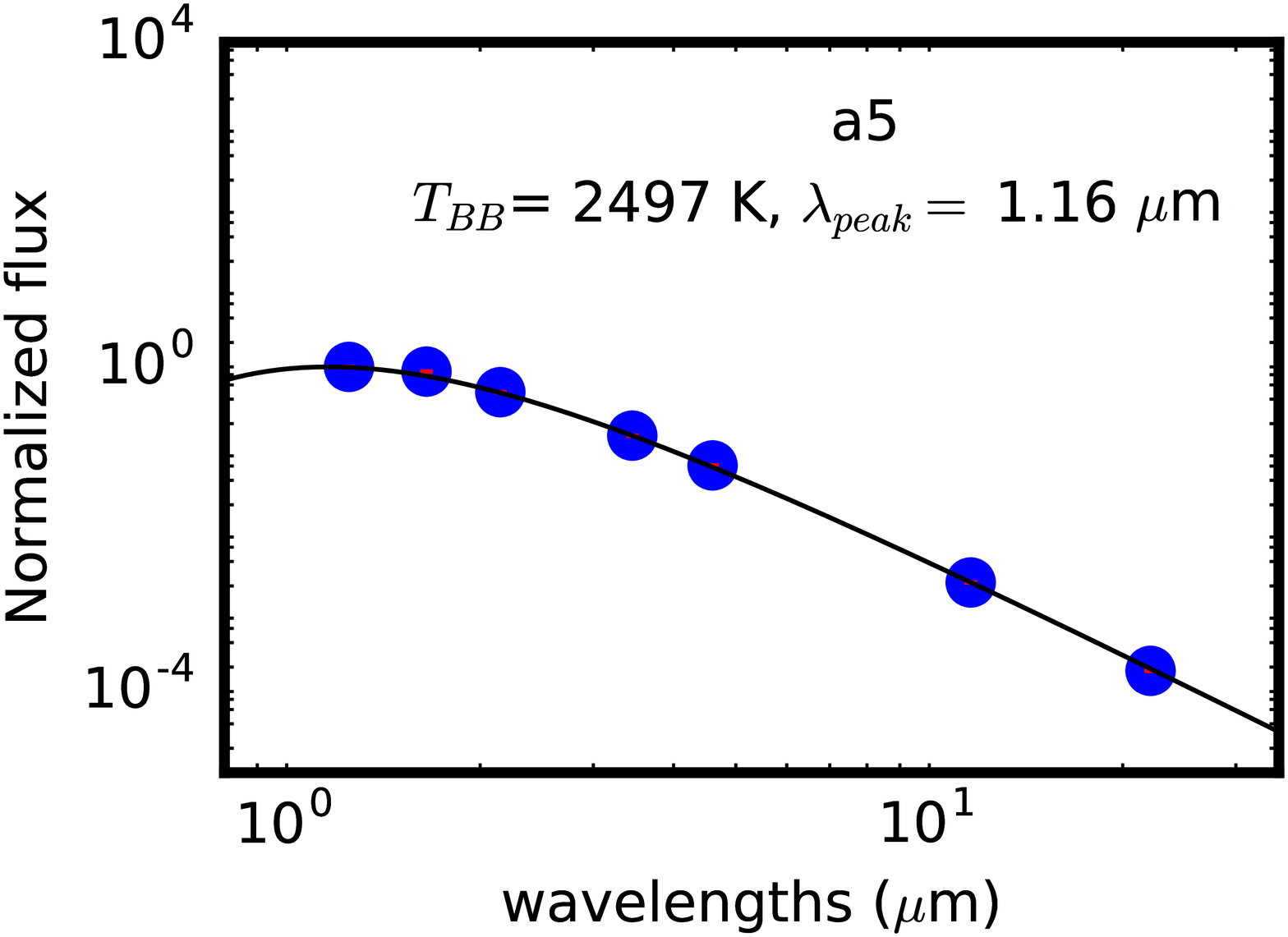}
\includegraphics[scale=0.14]{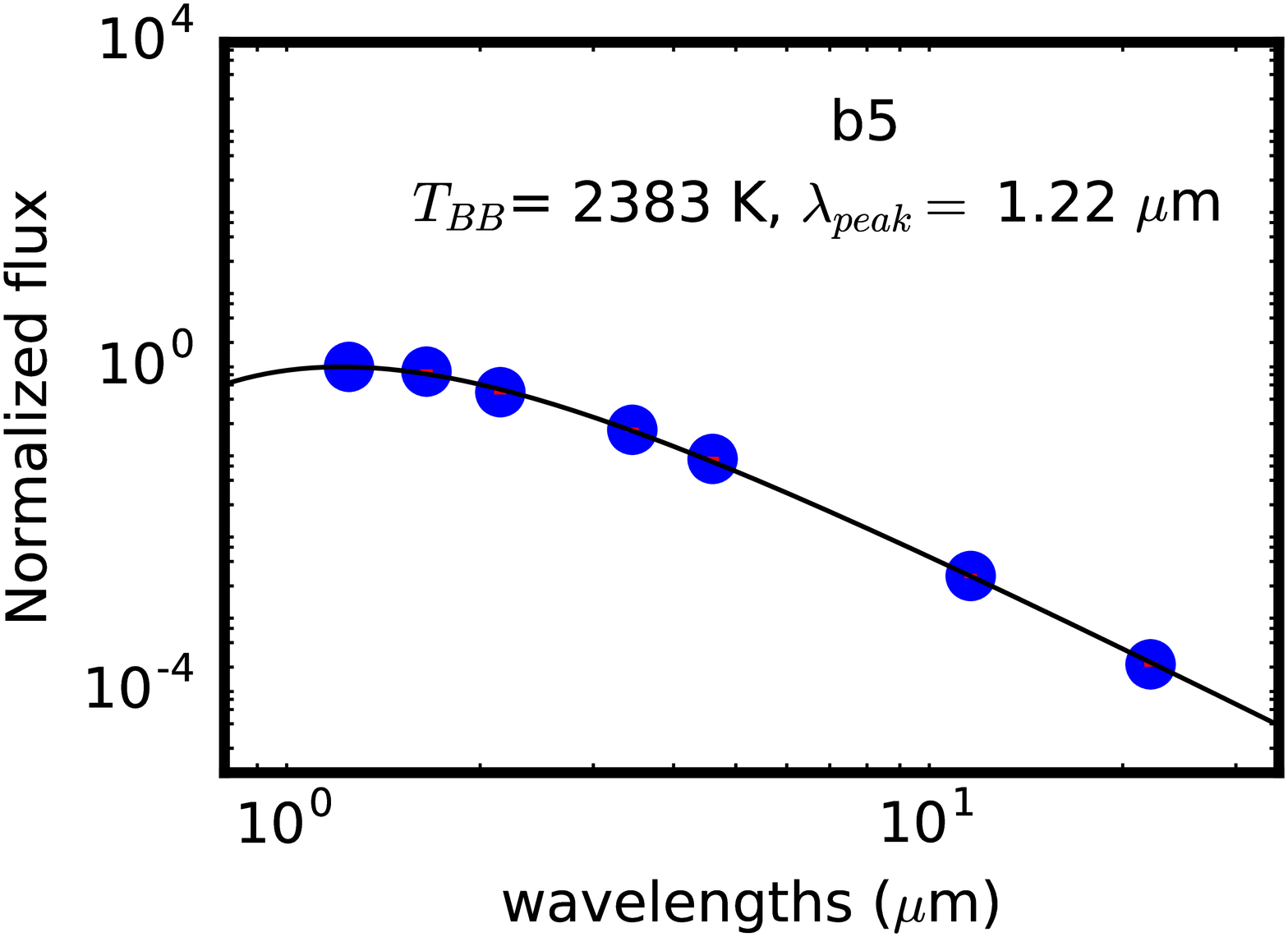}
\includegraphics[scale=0.14]{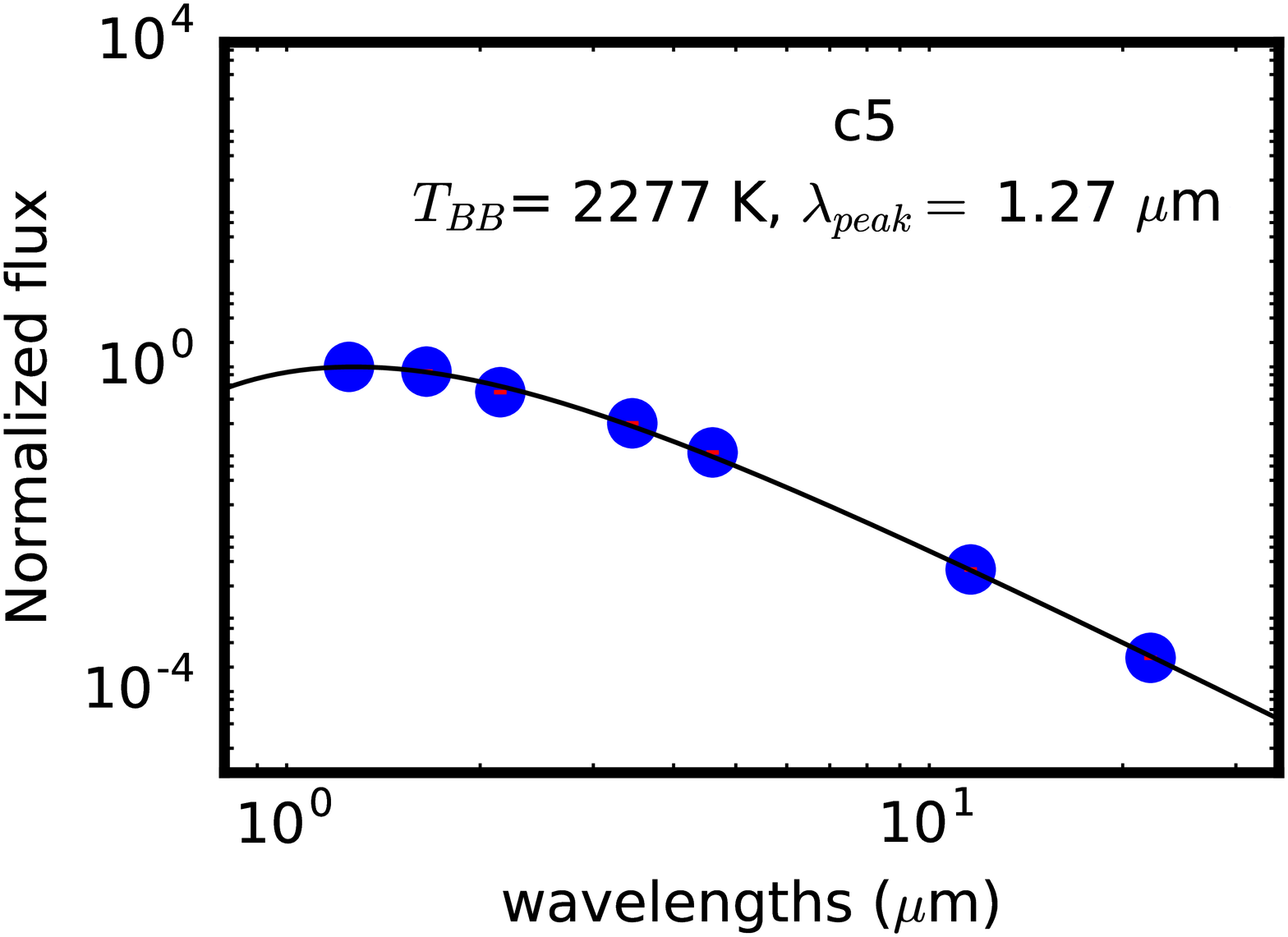}
\includegraphics[scale=0.14]{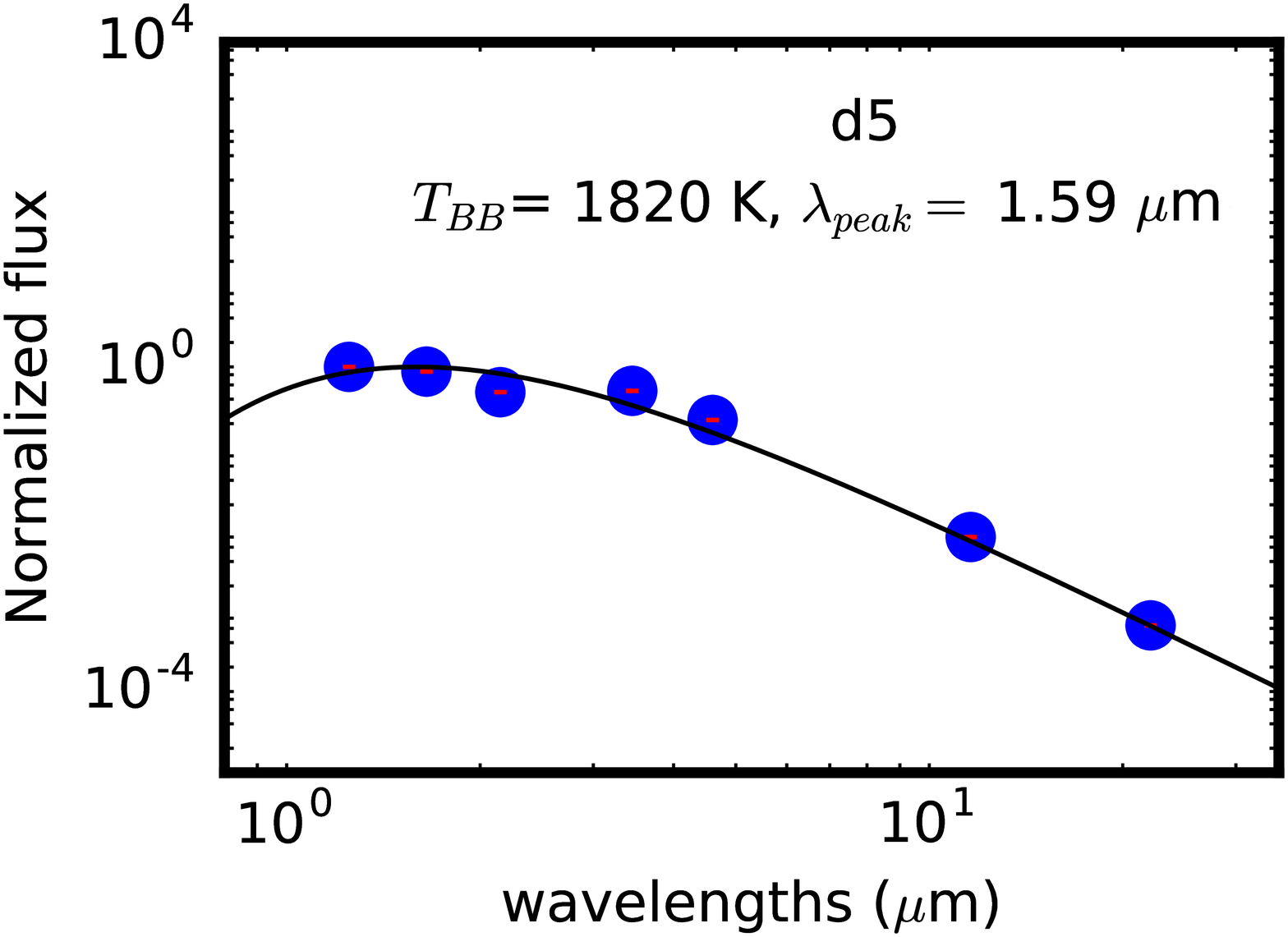}

\includegraphics[scale=0.14]{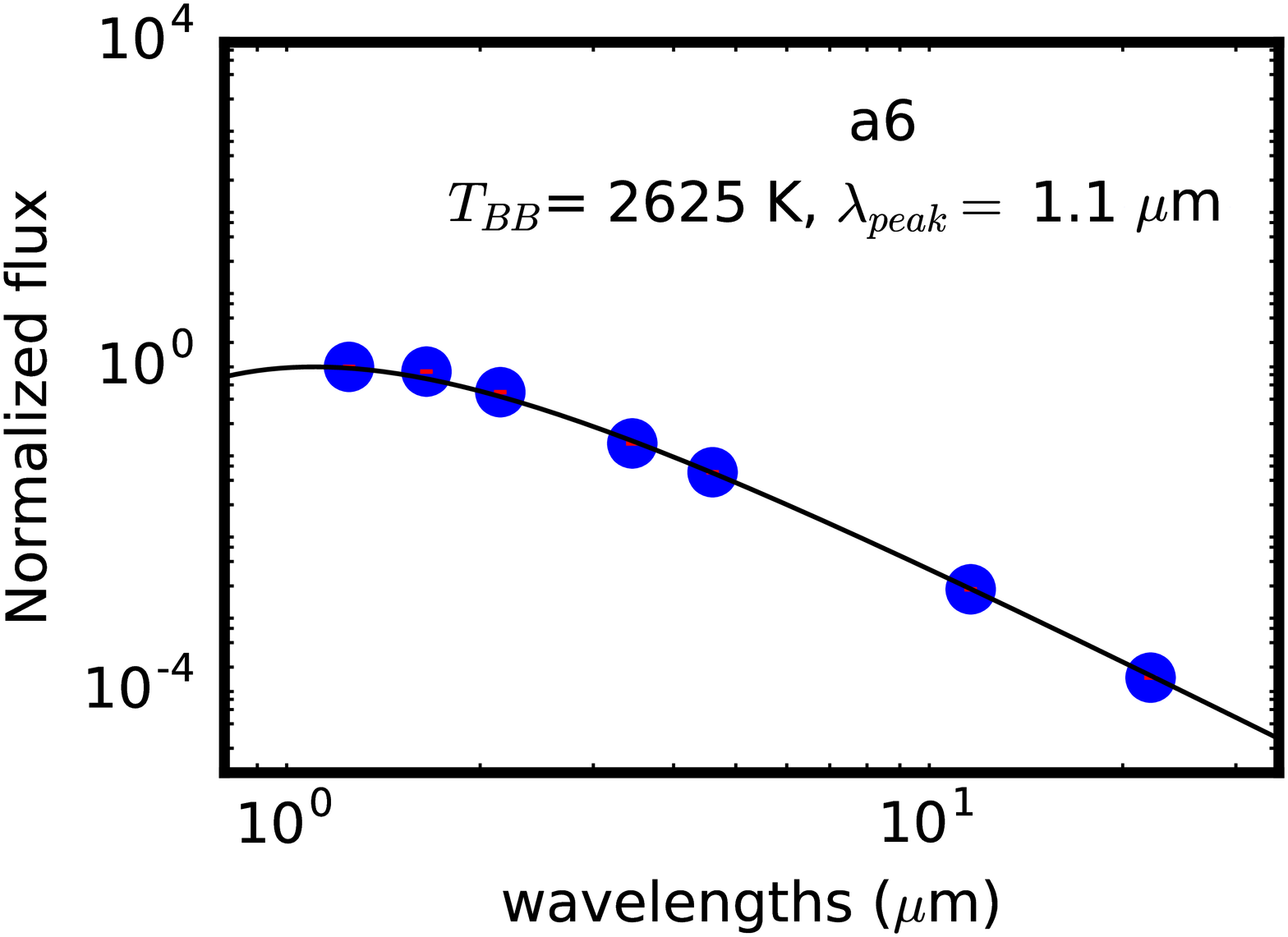}
\includegraphics[scale=0.14]{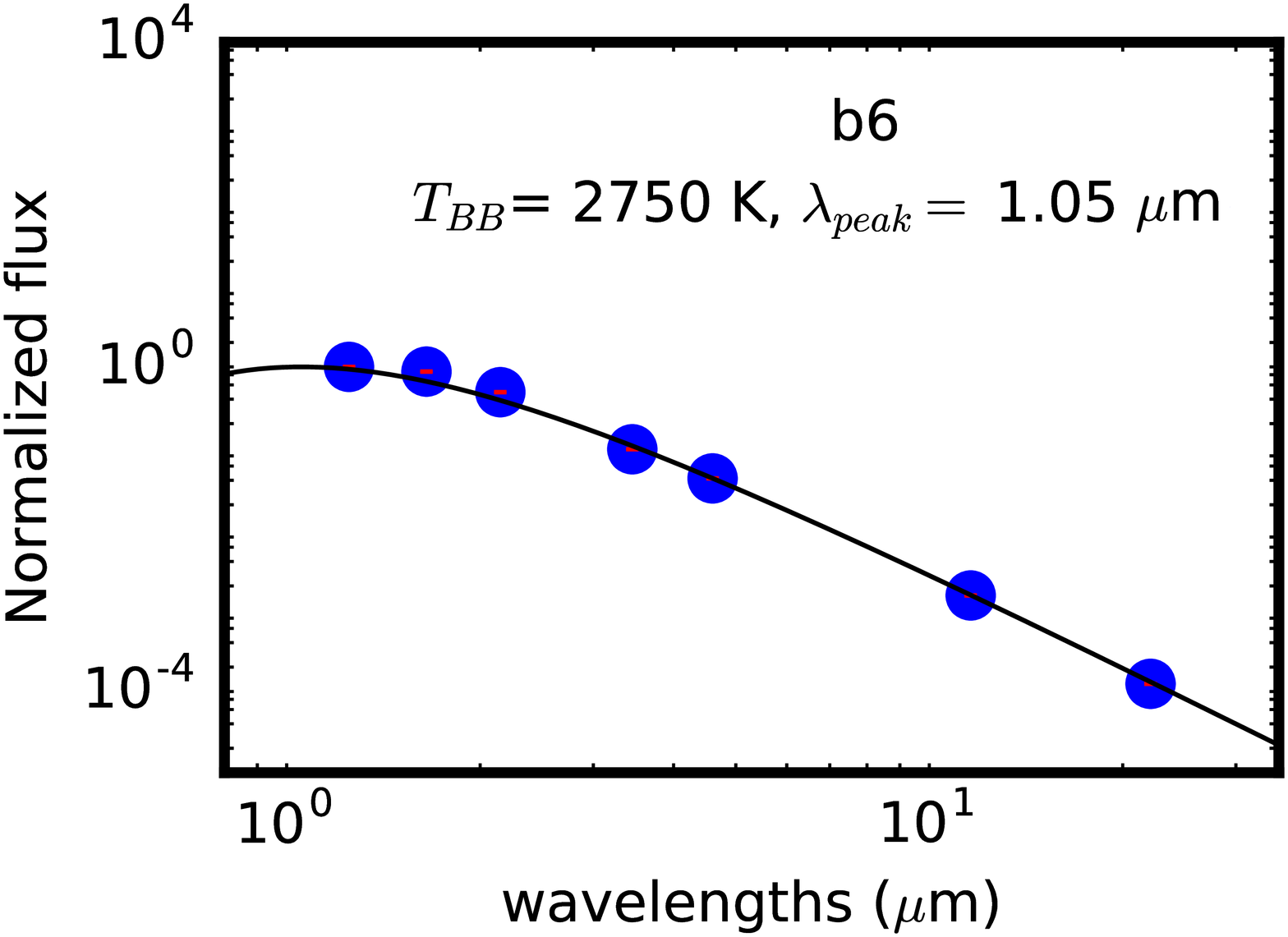}
\includegraphics[scale=0.14]{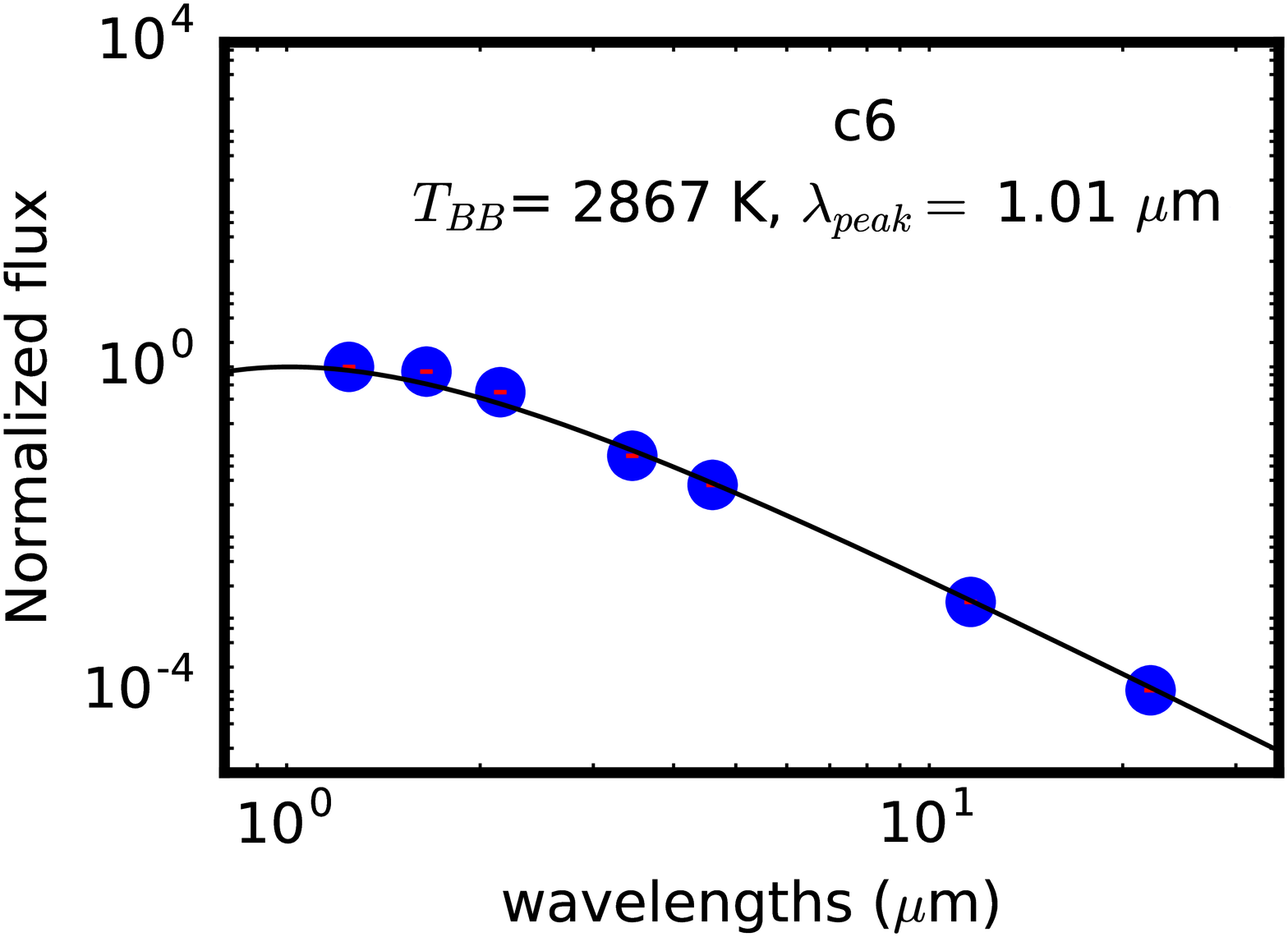}
\includegraphics[scale=0.14]{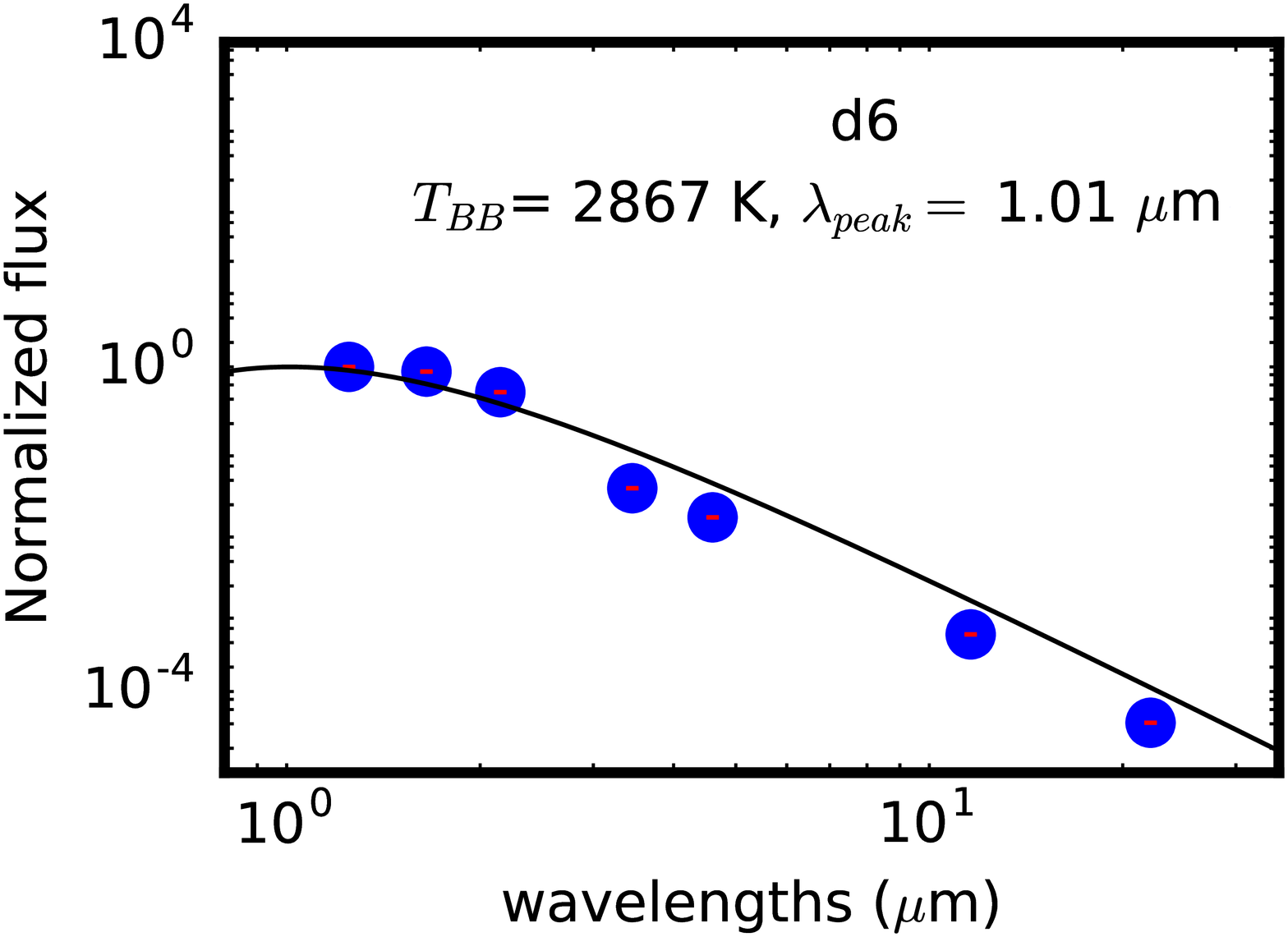}

\caption{SEDs profiles of V694~Mon for four cases of variation amplitude variations: (a) 0.2~mag (observed), (b) 0.6~mag, (c) 1~mag, and 
(d) 3~mag, assuming six different cases: (first row) 2MASS in the maximum brightness and {\it WISE} in the minimum, (second row) 
2MASS in the minimum brightness and {\it WISE} in the maximum, (third row) 2MASS in the maximum brightness and {\it WISE} without any 
variation (fourth row) 2MASS in the maximum brightness and {\it WISE} without any variation, (fifth row) 2MASS without any variation and {\it WISE} 
in the minimum brightness, and (sixth row) 2MASS without any variation and {\it WISE} in the minimum brightness.}
\label{fignew1}
\end{center}
\end{figure*}

For the S-type SySt, we assumed maximum amplitude variations in the {\it J} band of 0.2, 0.6, 1.0, and 3.0~mag. Figure~\ref{fignew1} 
illustrates the SEDs and BB fitting of each of the 24 examples. Apparently, the SED profile, as well as the classification of SySts,
does not change significantly due to possible flux variations between the two surveys. Only for an amplitude variation of 3~mag does
the SED fitting show some deviations from the BB fitting. In the case of a 1~mag amplitude variation, a poor fitting of the 
2MASS data or {\it W1} and {\it W2} bands can be seen in only two cases (c1 and c2). In general, the SED fitting for most of the cases is good,
and possible flux variations between the two surveys result in a temperature variation of 300-350~K.

\begin{figure*}
\begin{center}
\includegraphics[scale=0.14]{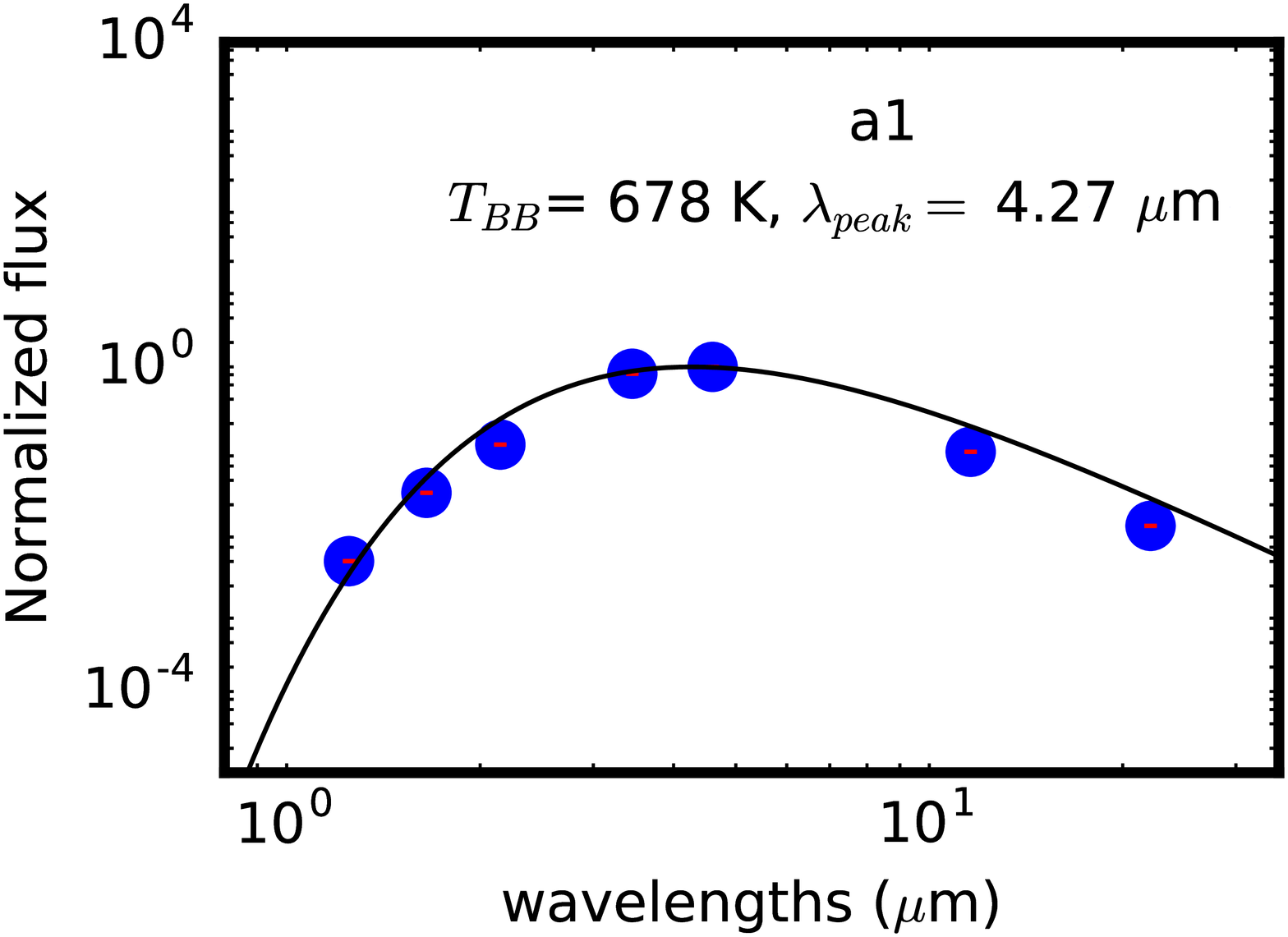}
\includegraphics[scale=0.14]{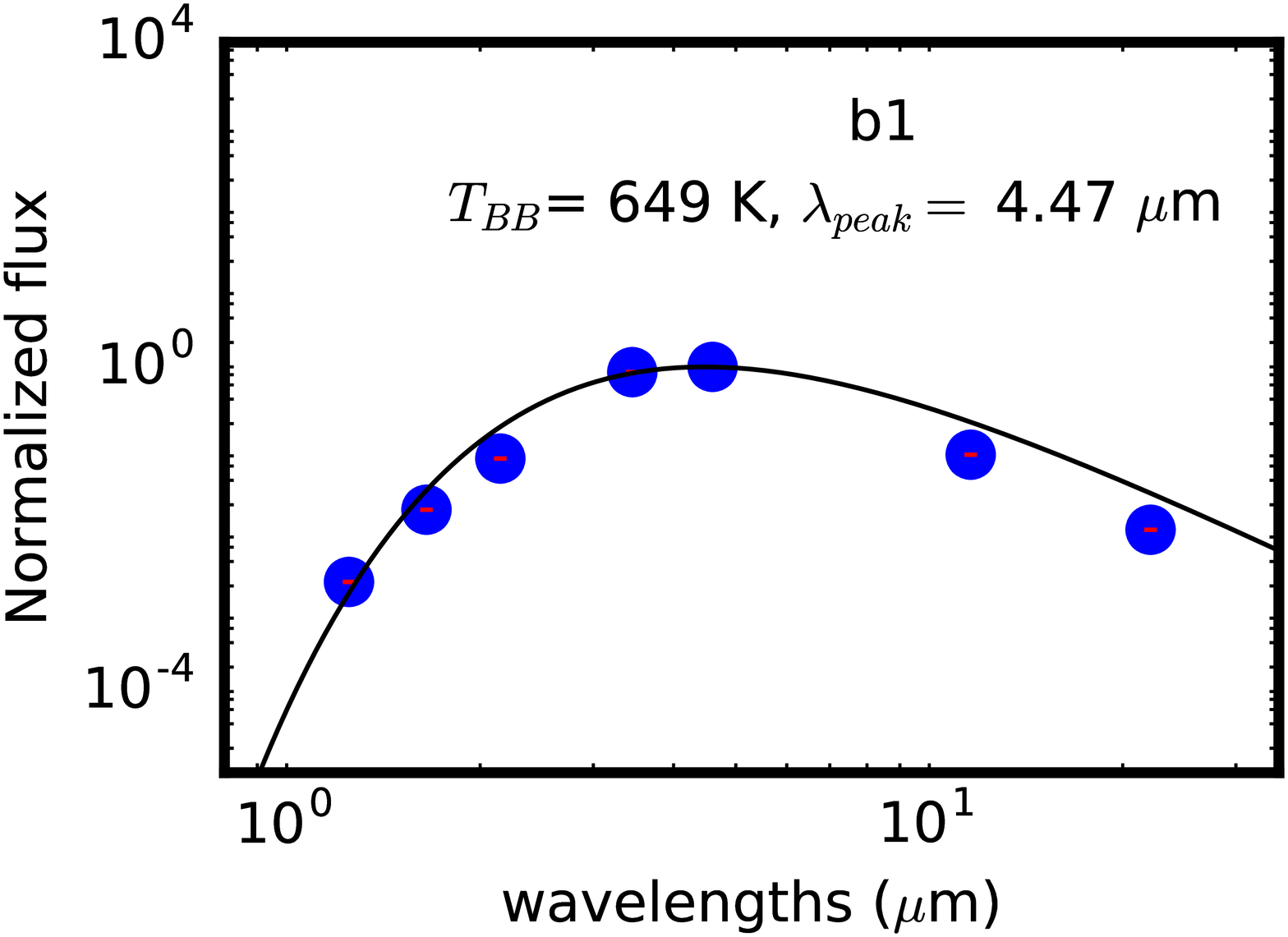}
\includegraphics[scale=0.14]{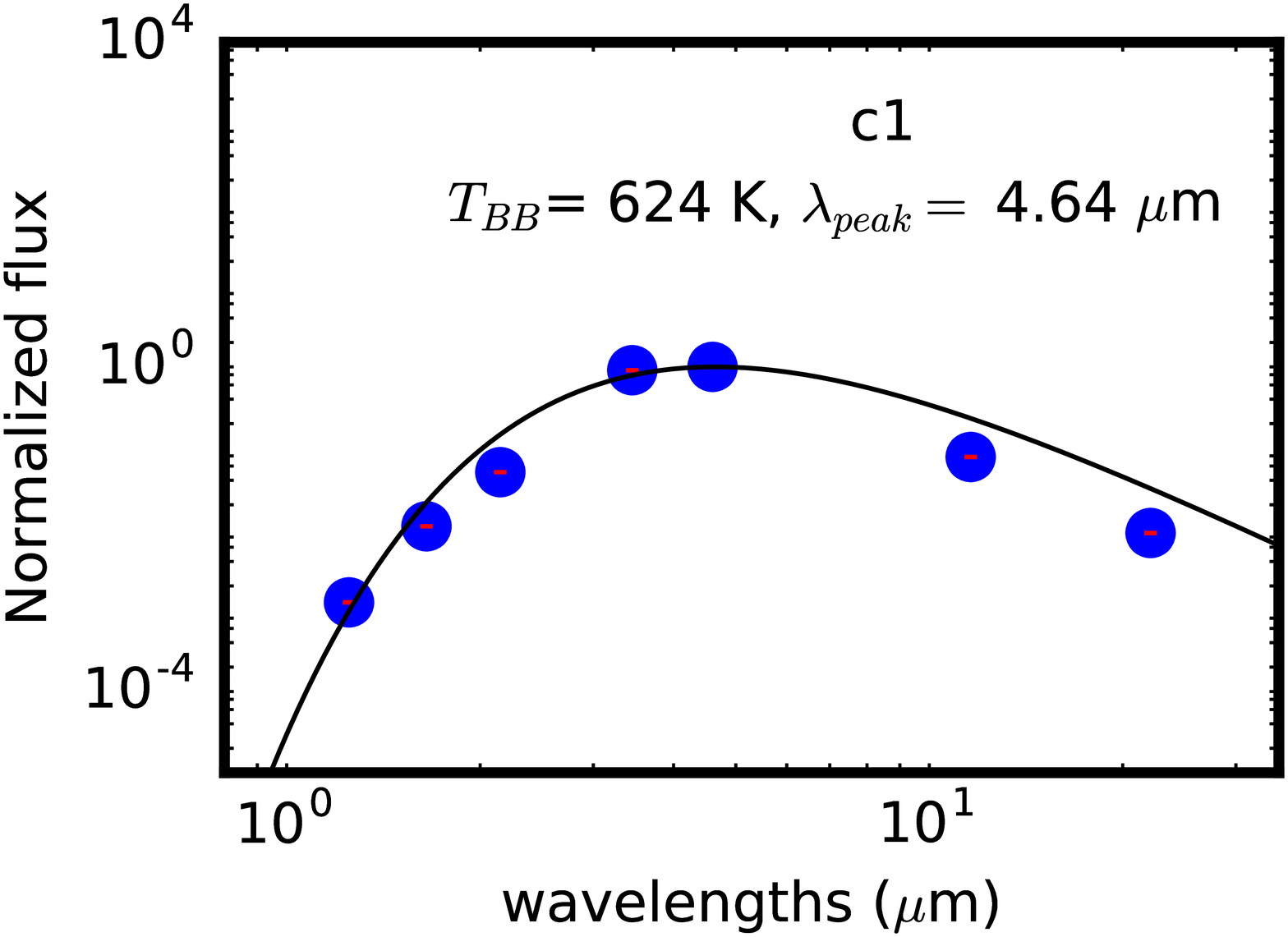}
\includegraphics[scale=0.14]{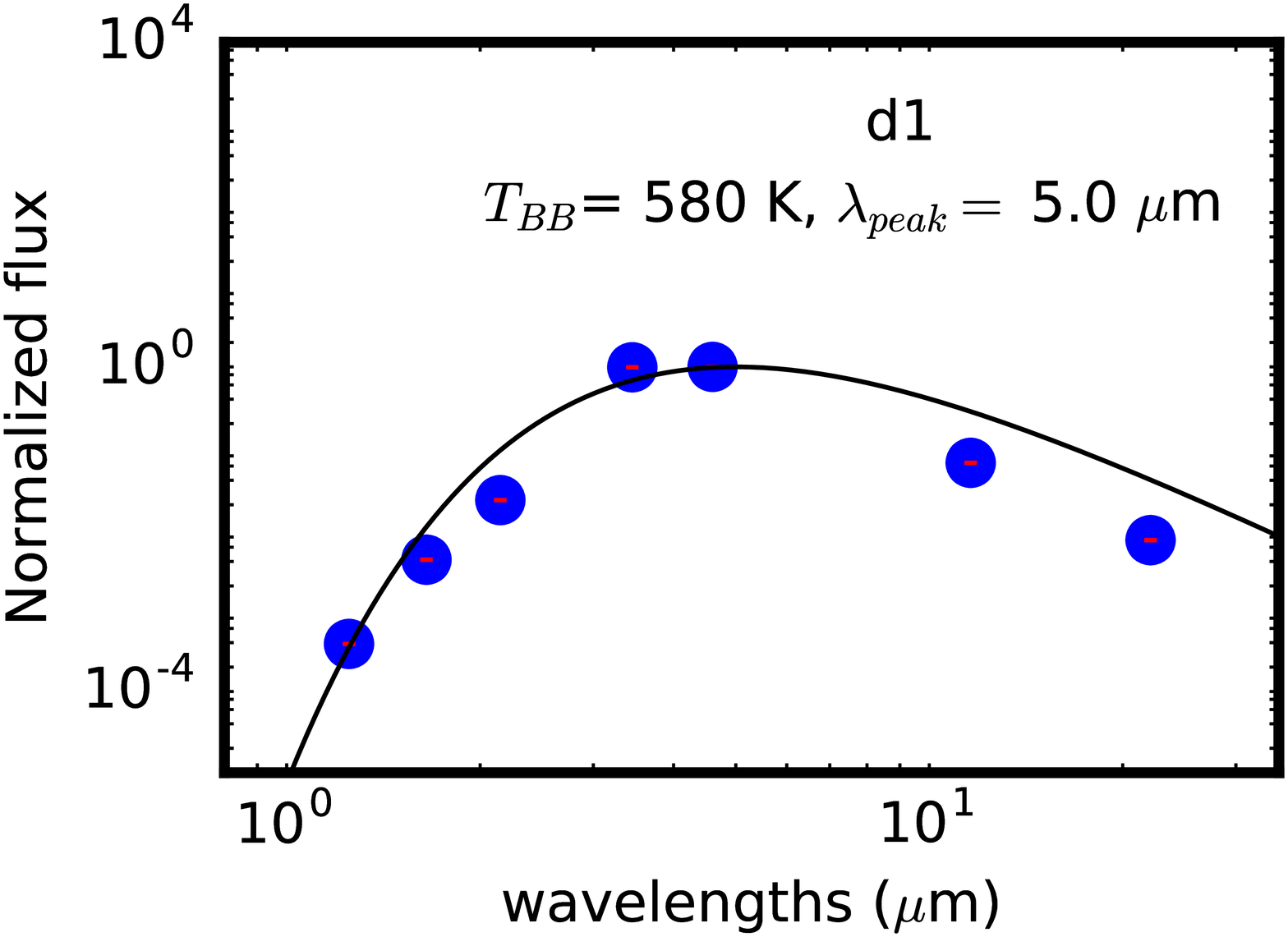}

\includegraphics[scale=0.14]{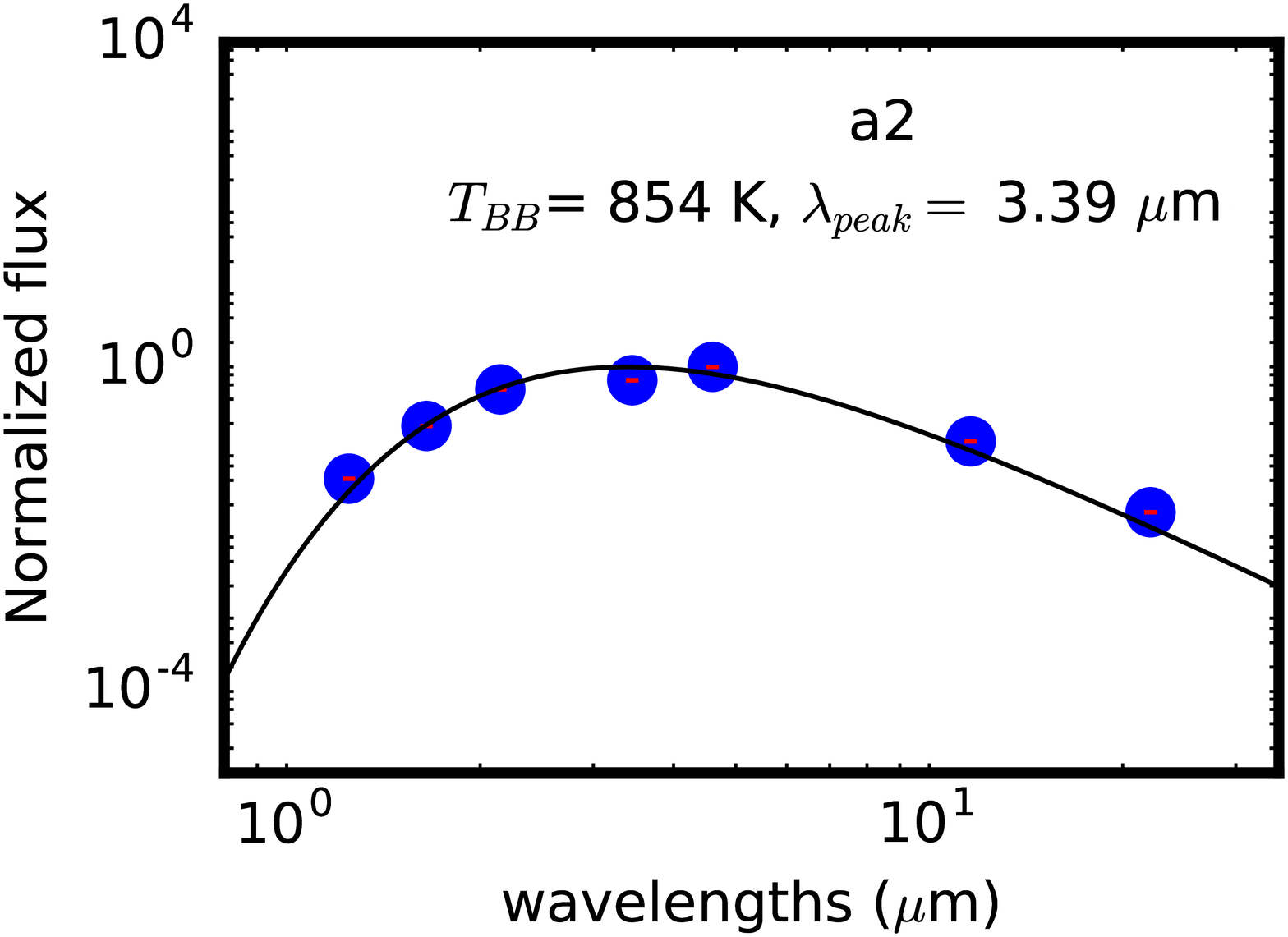}
\includegraphics[scale=0.14]{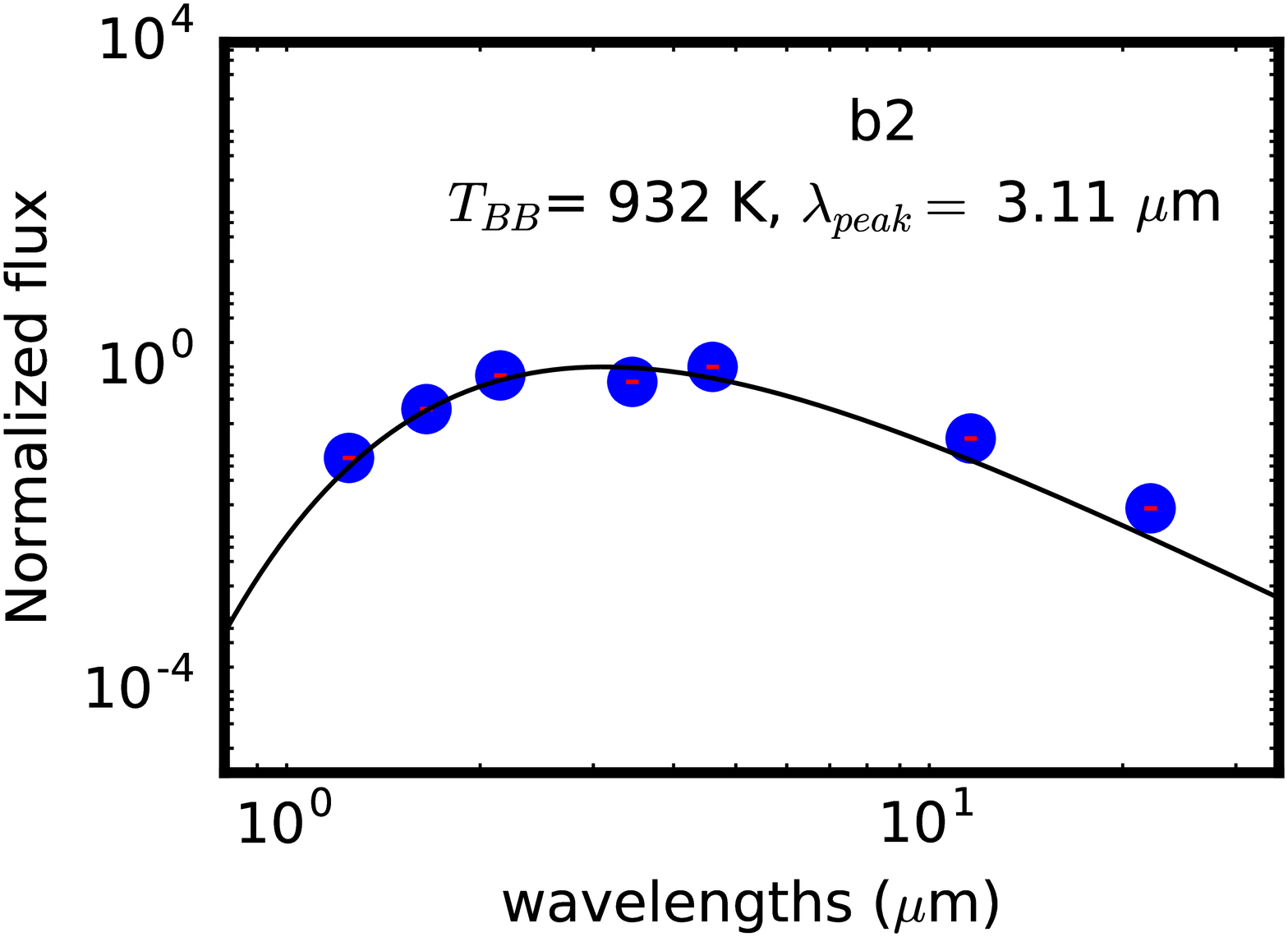}
\includegraphics[scale=0.14]{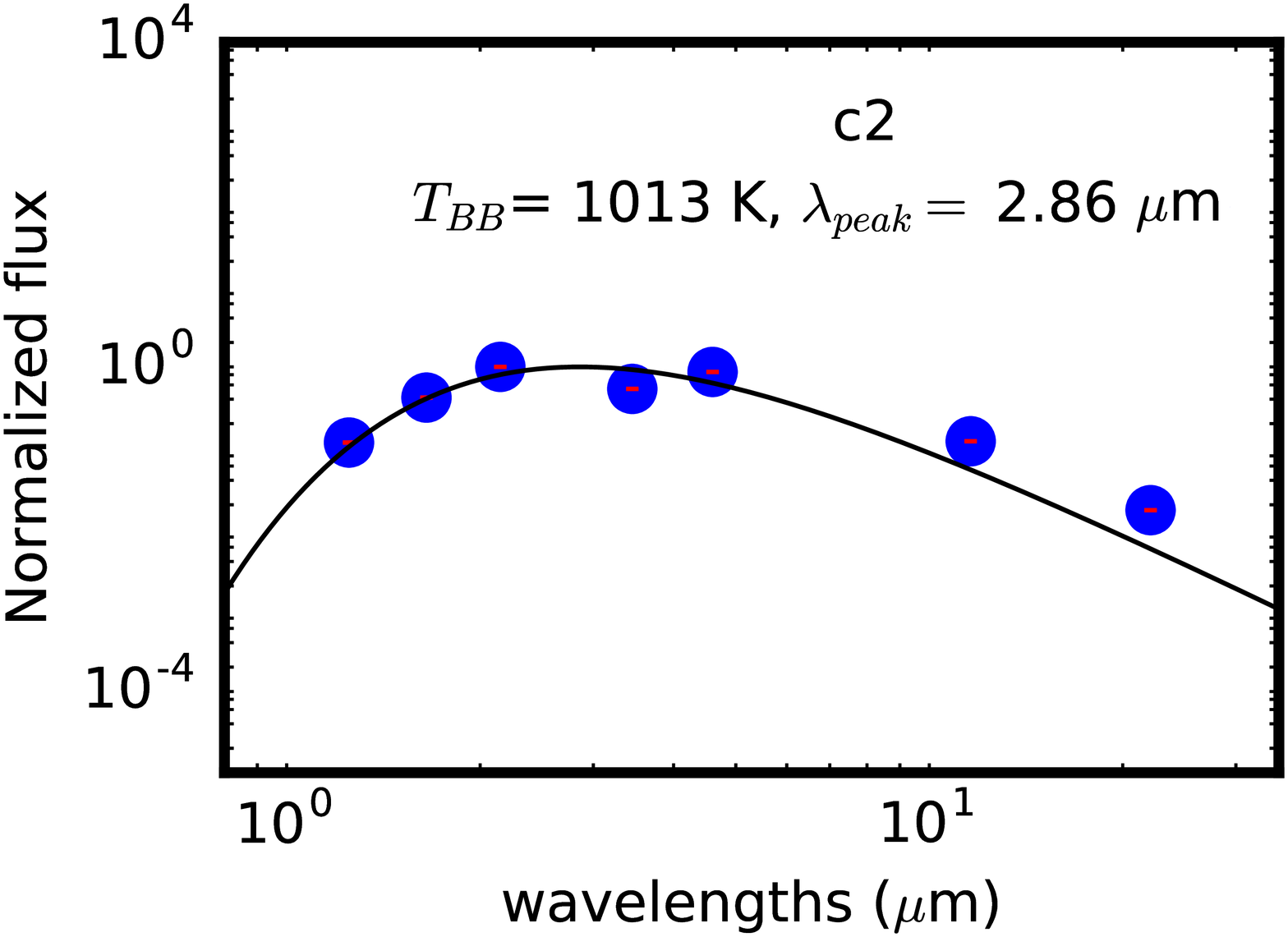}
\includegraphics[scale=0.14]{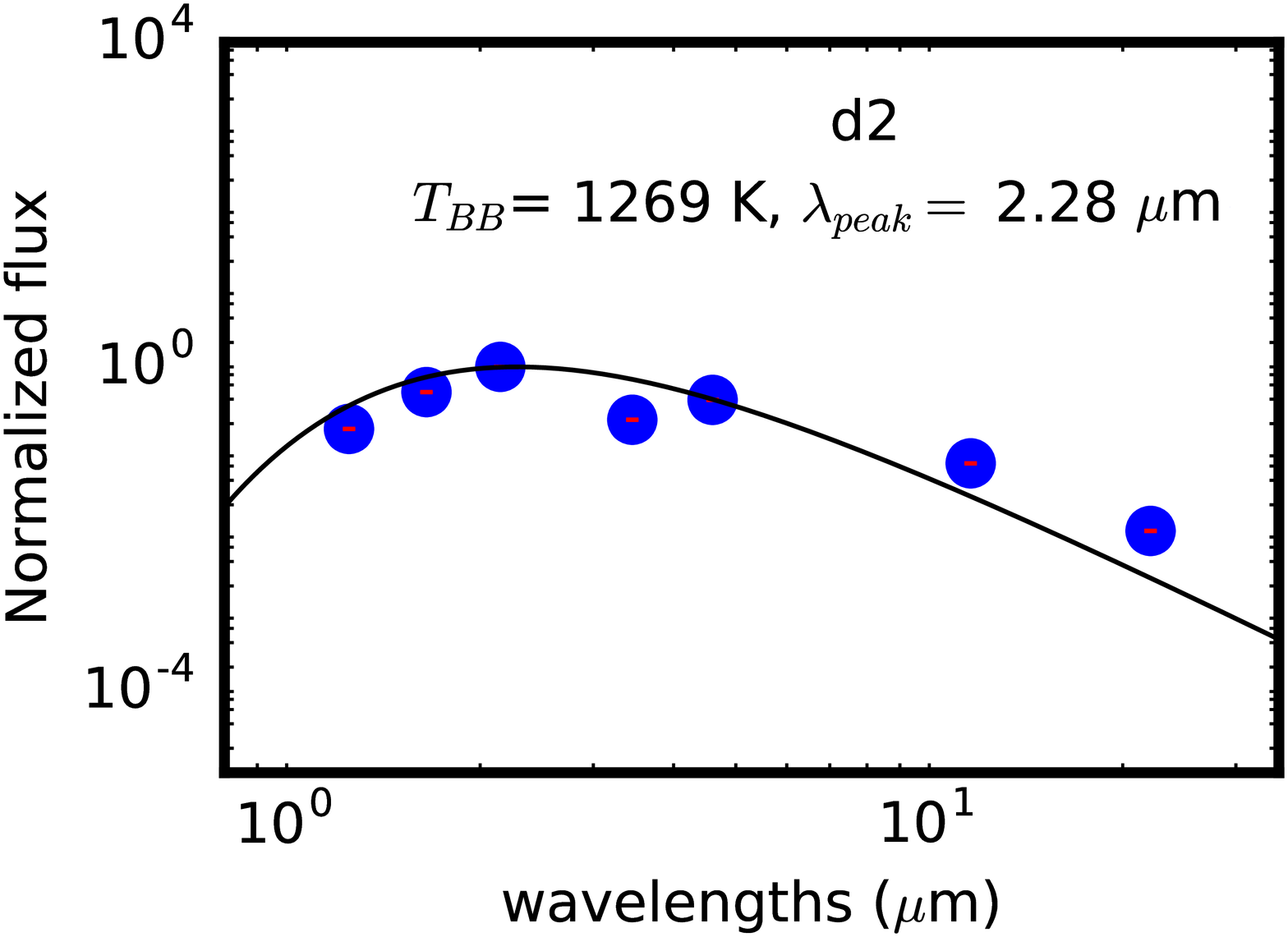}

\includegraphics[scale=0.14]{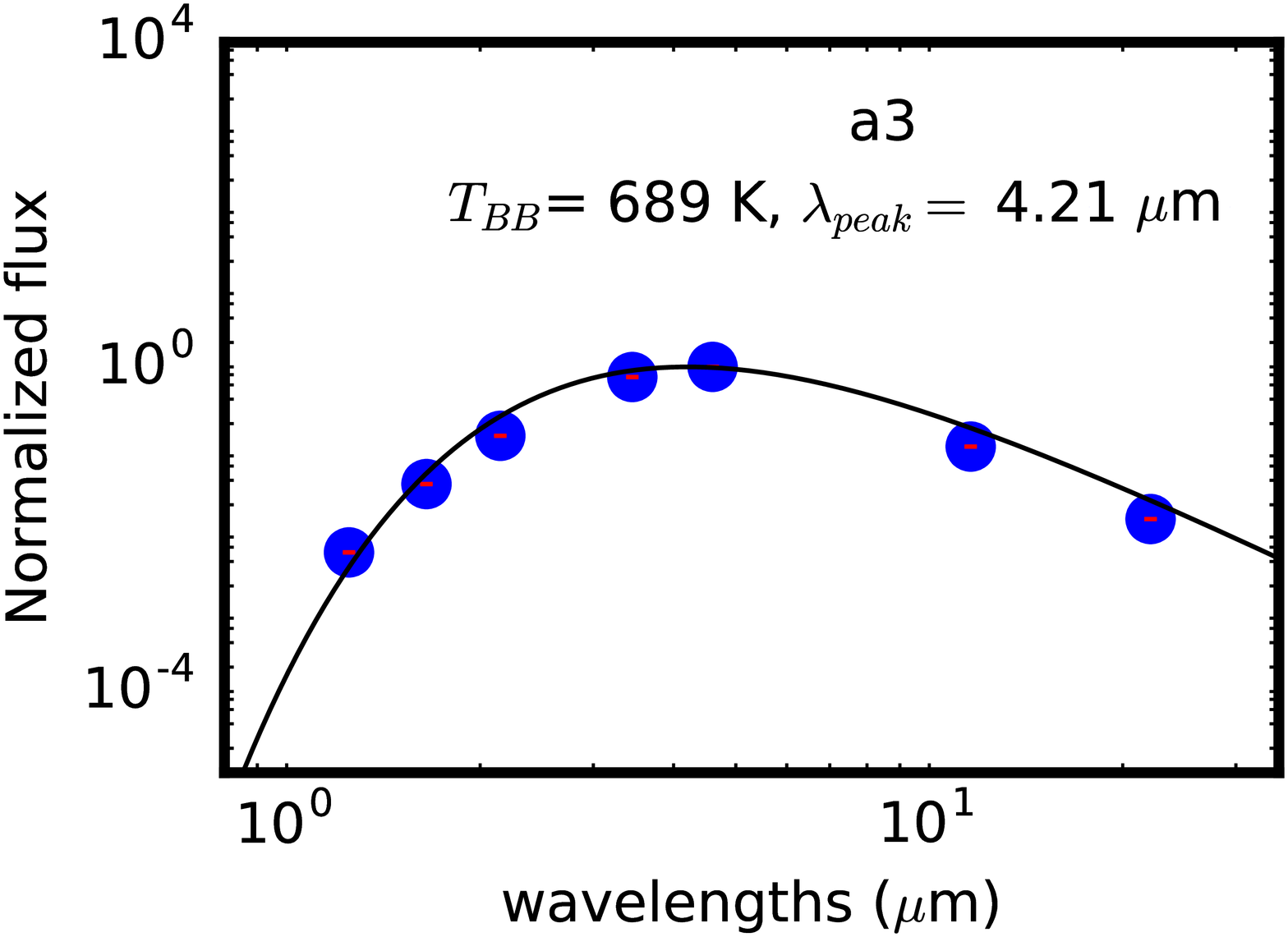}
\includegraphics[scale=0.14]{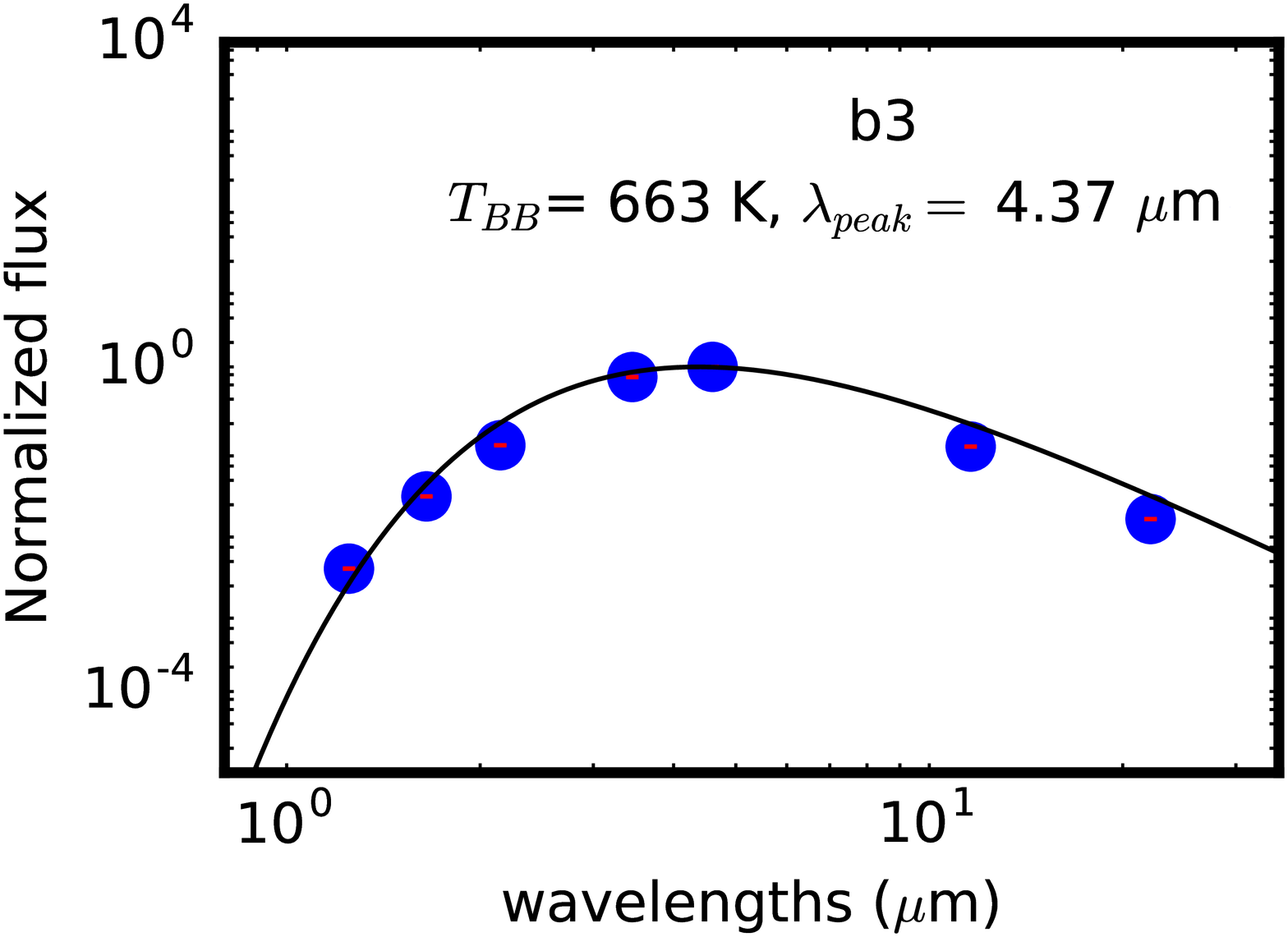}
\includegraphics[scale=0.14]{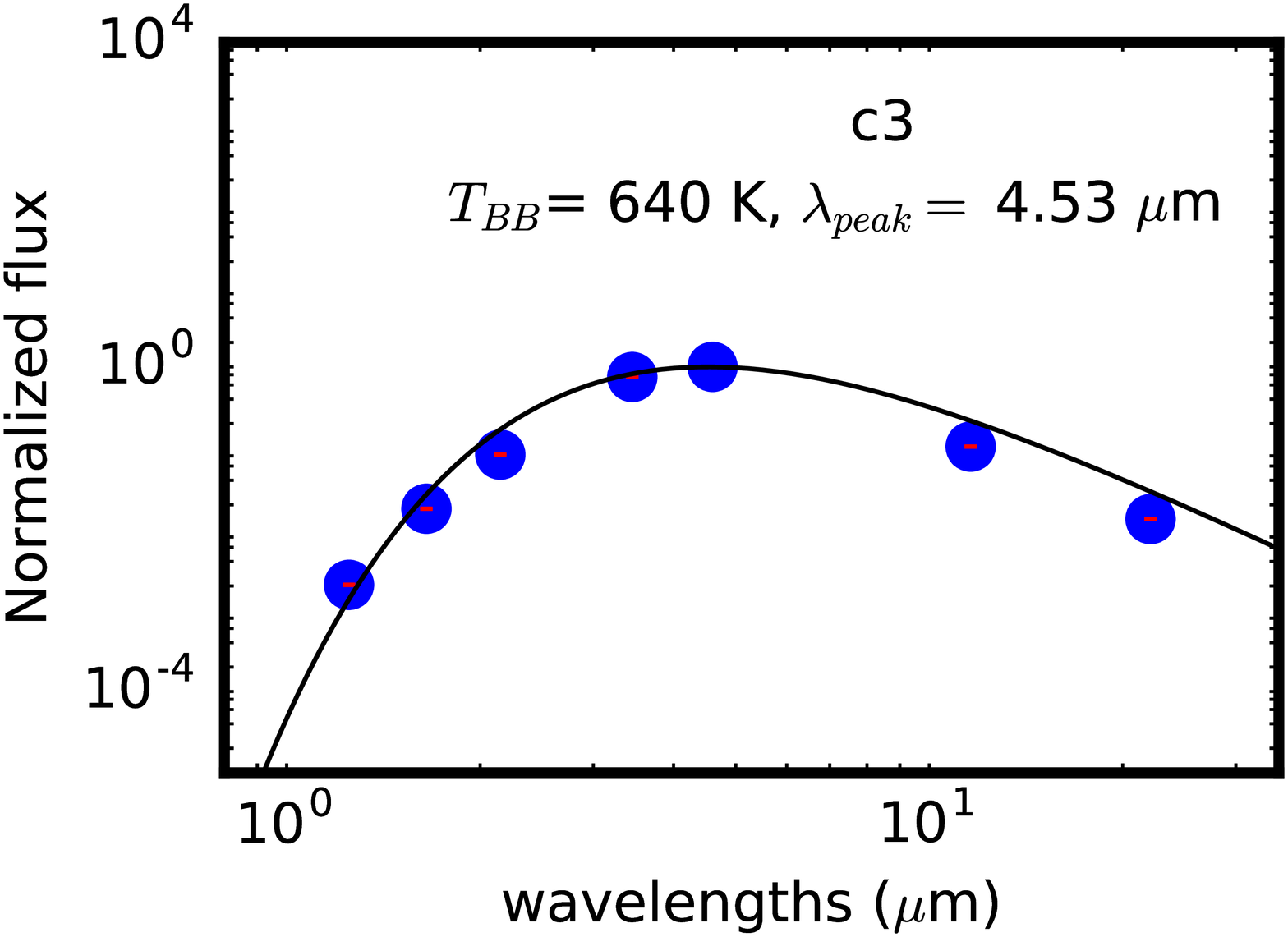}
\includegraphics[scale=0.14]{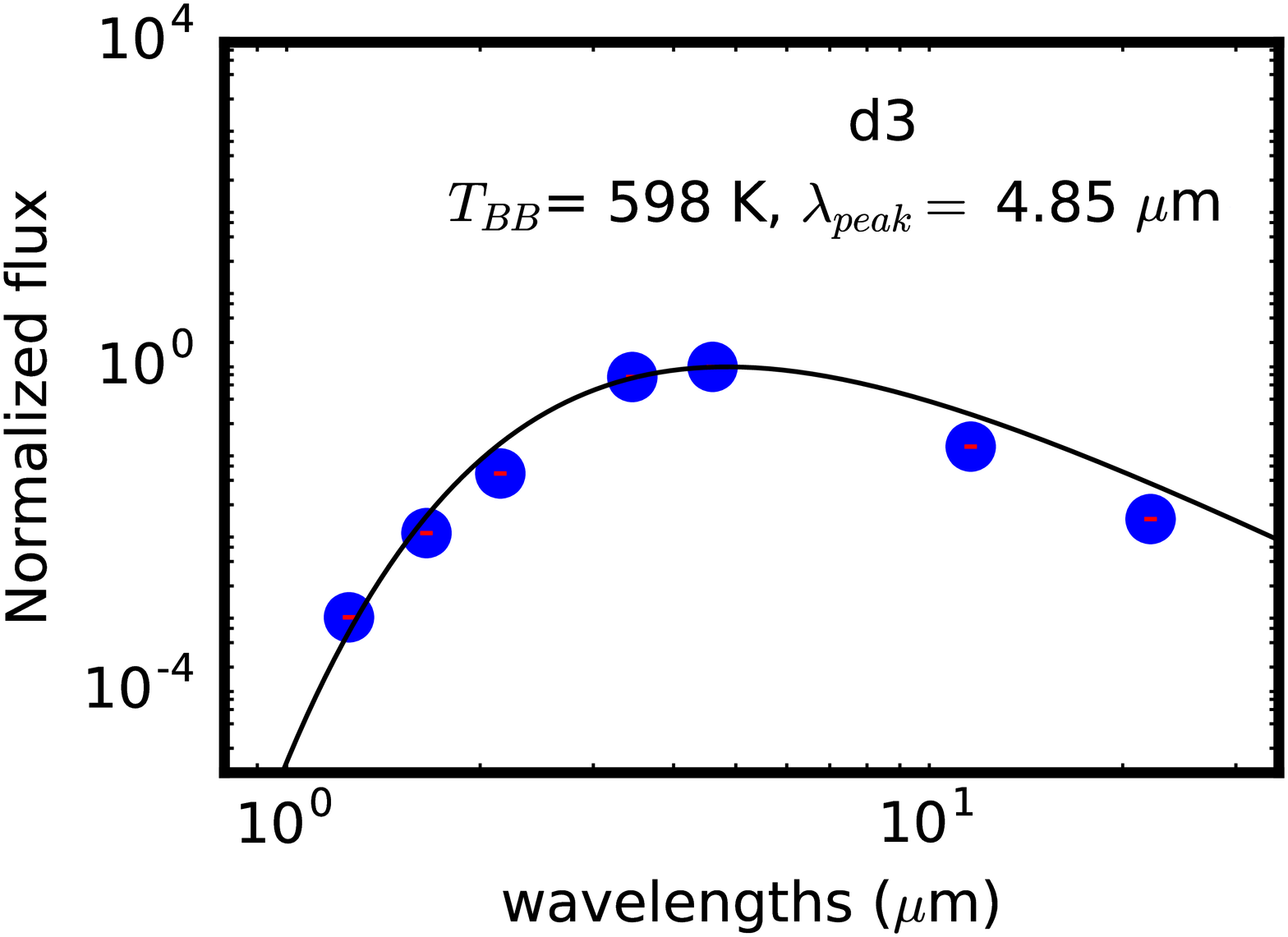}

\includegraphics[scale=0.14]{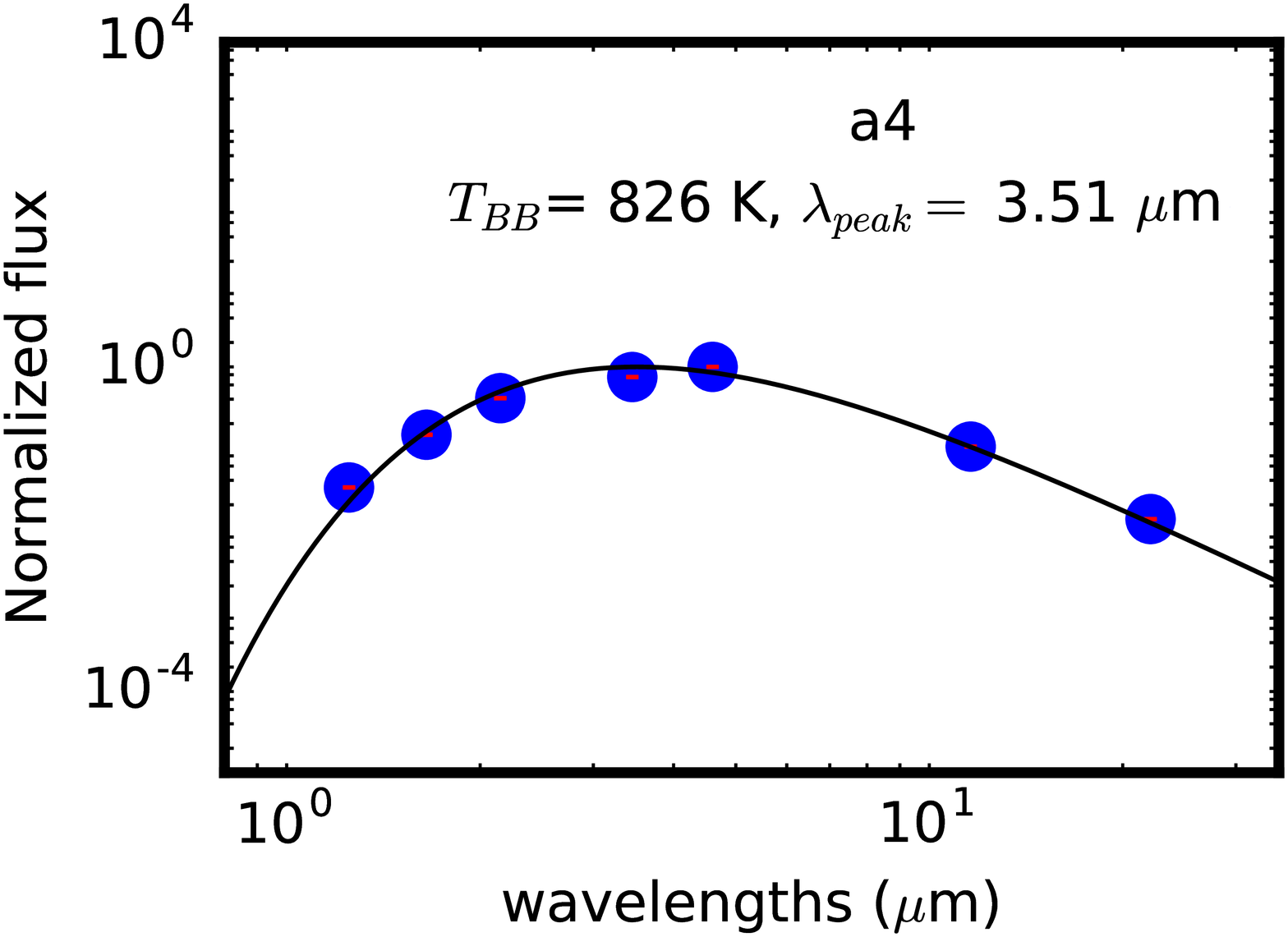}
\includegraphics[scale=0.14]{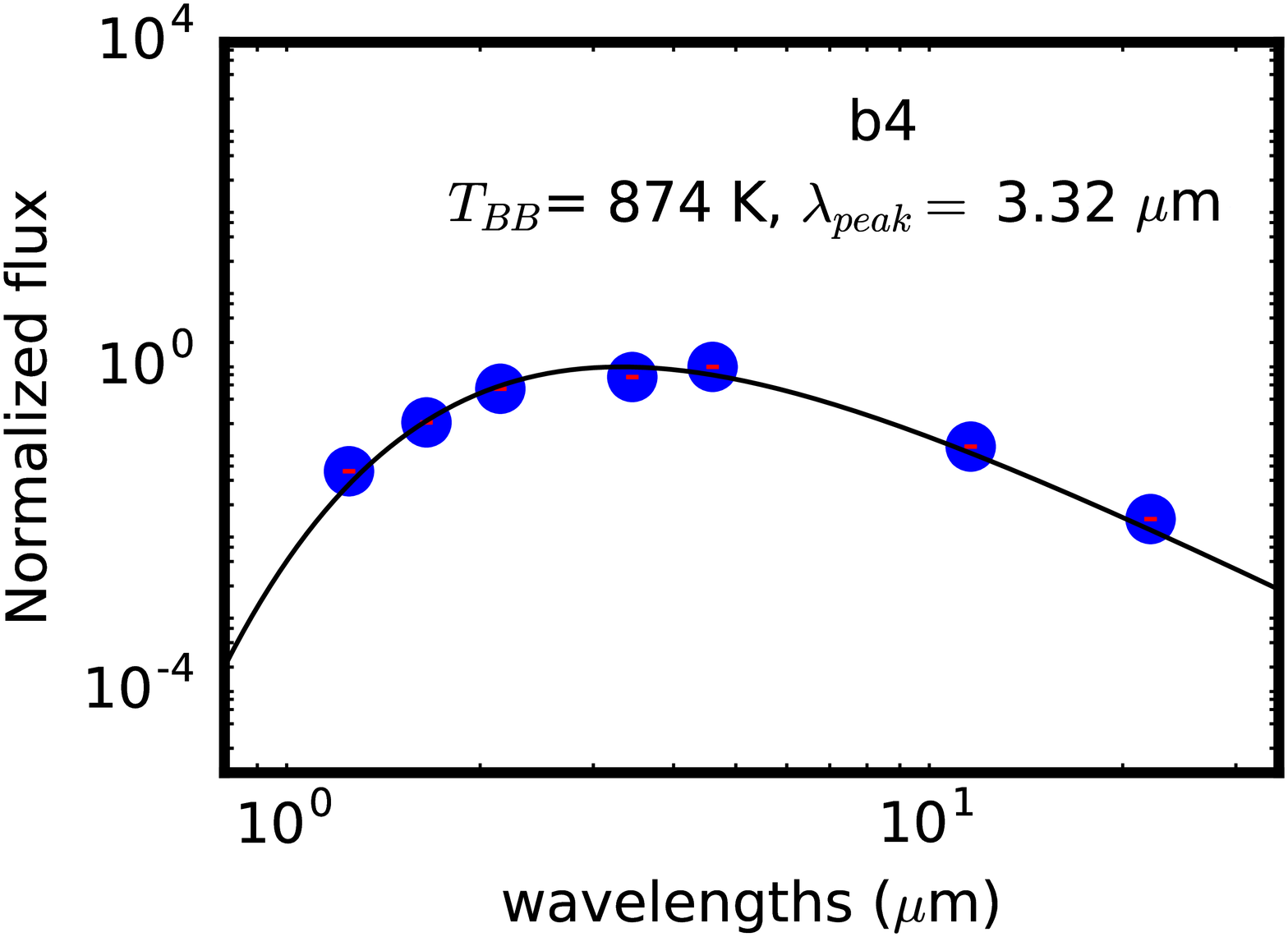}
\includegraphics[scale=0.14]{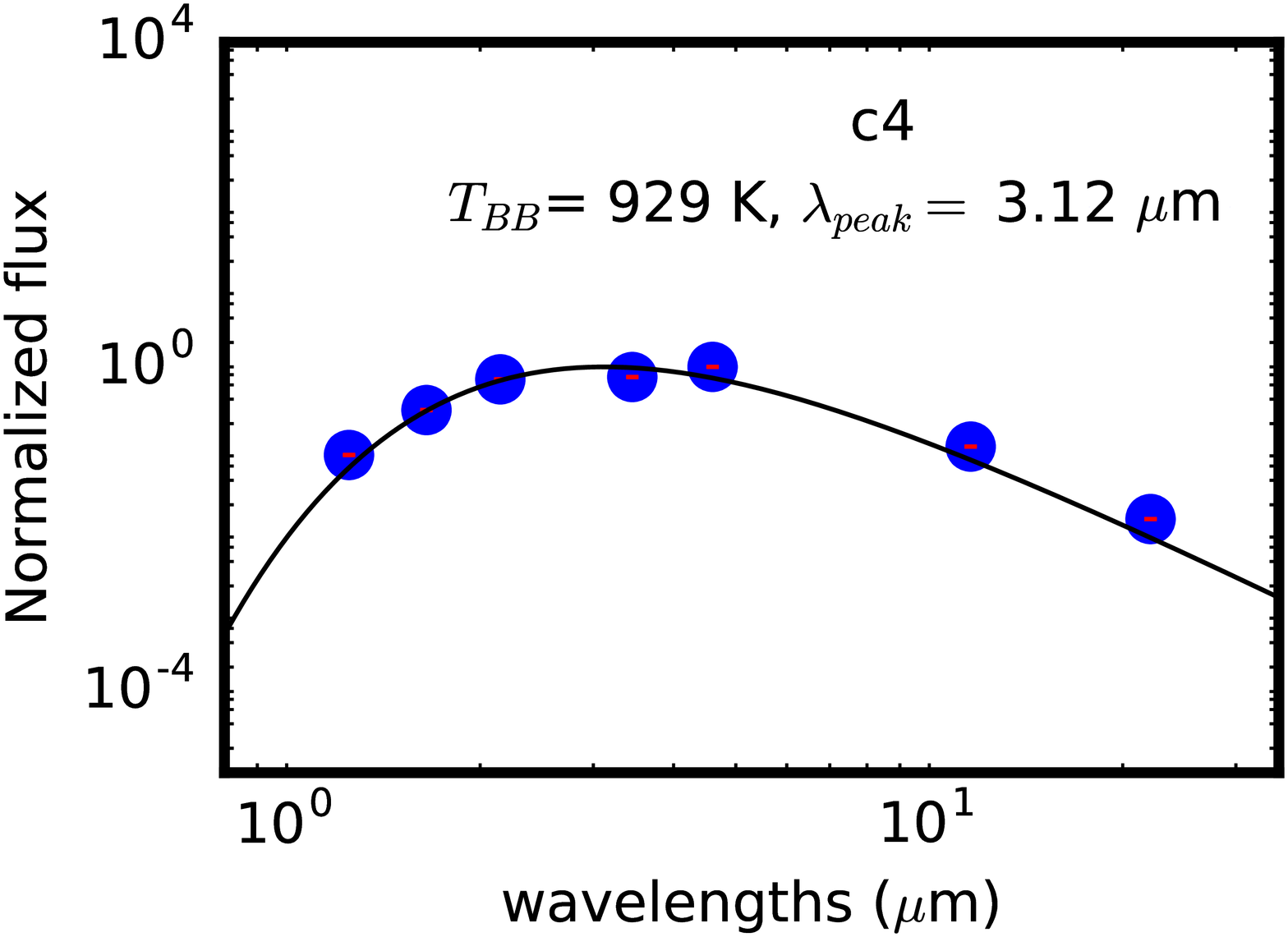}
\includegraphics[scale=0.14]{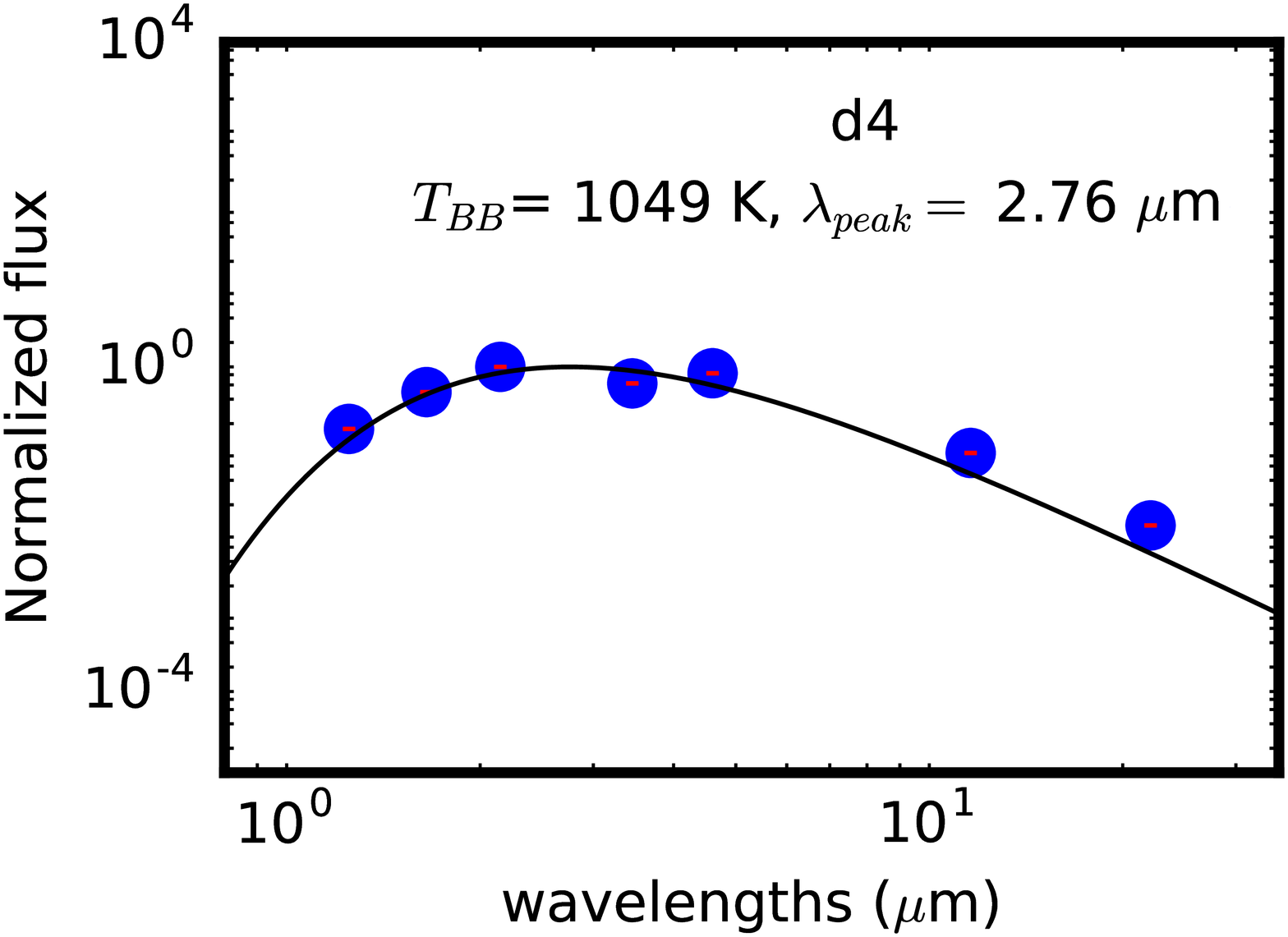}

\includegraphics[scale=0.14]{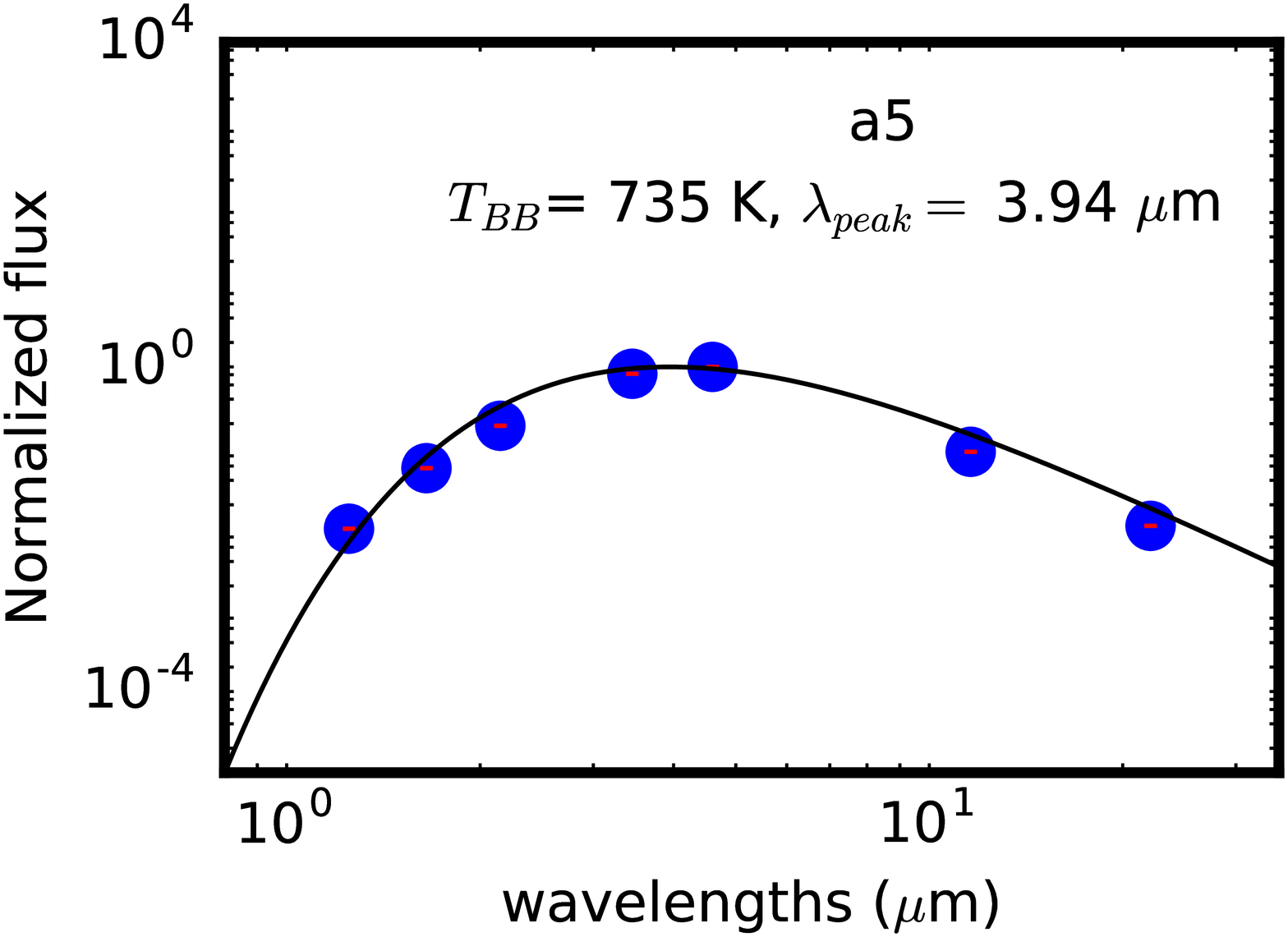}
\includegraphics[scale=0.14]{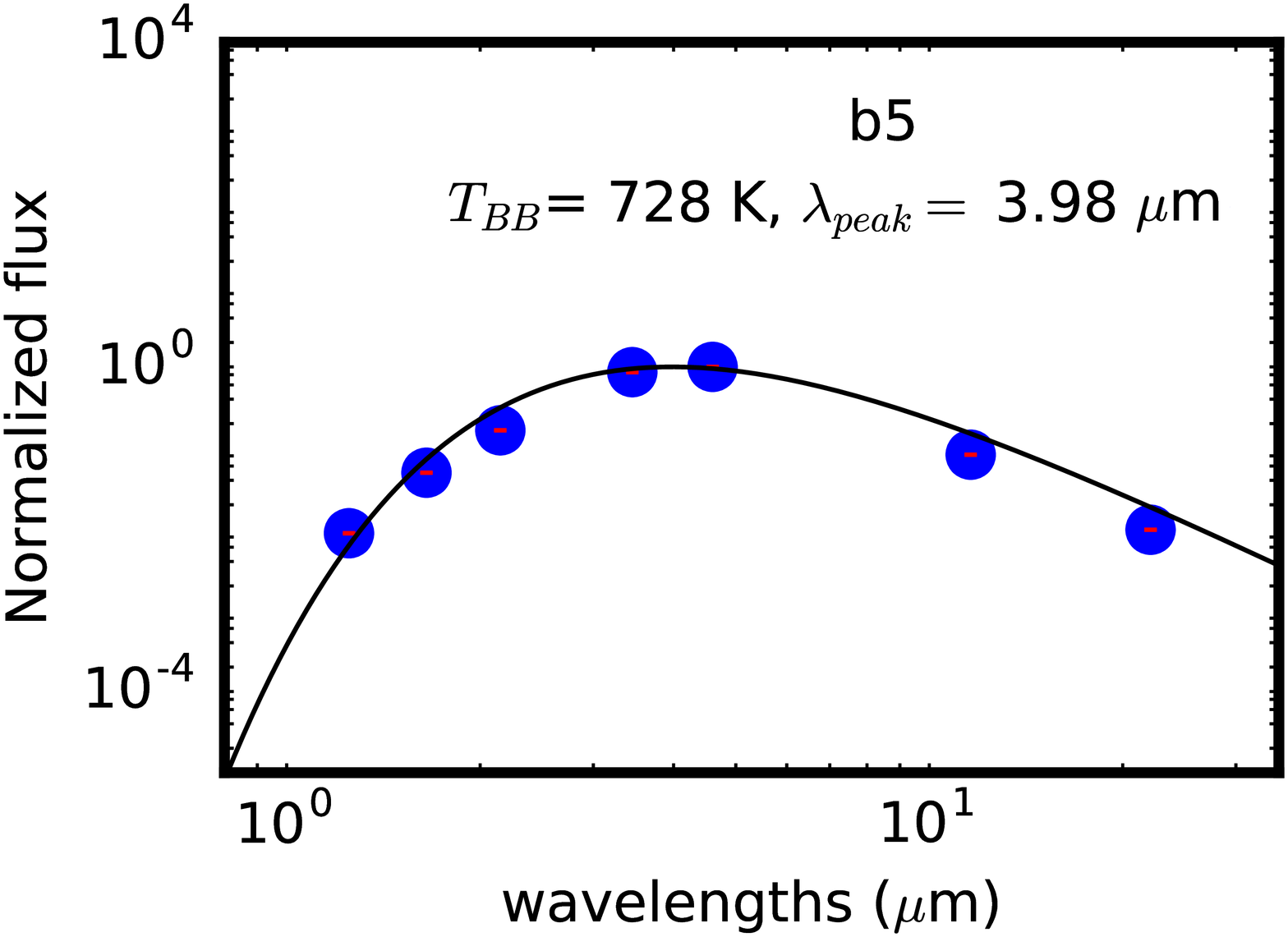}
\includegraphics[scale=0.14]{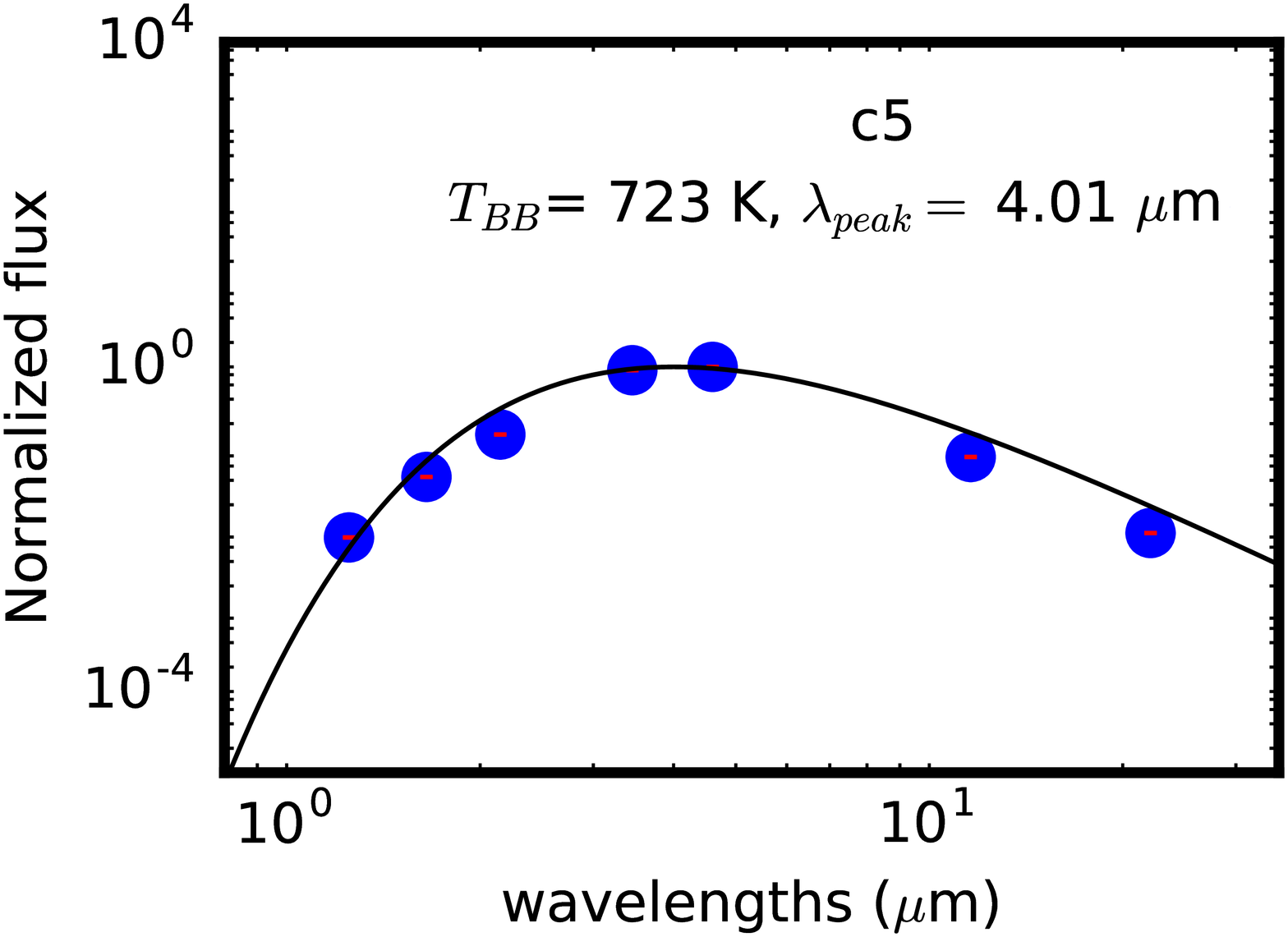}
\includegraphics[scale=0.14]{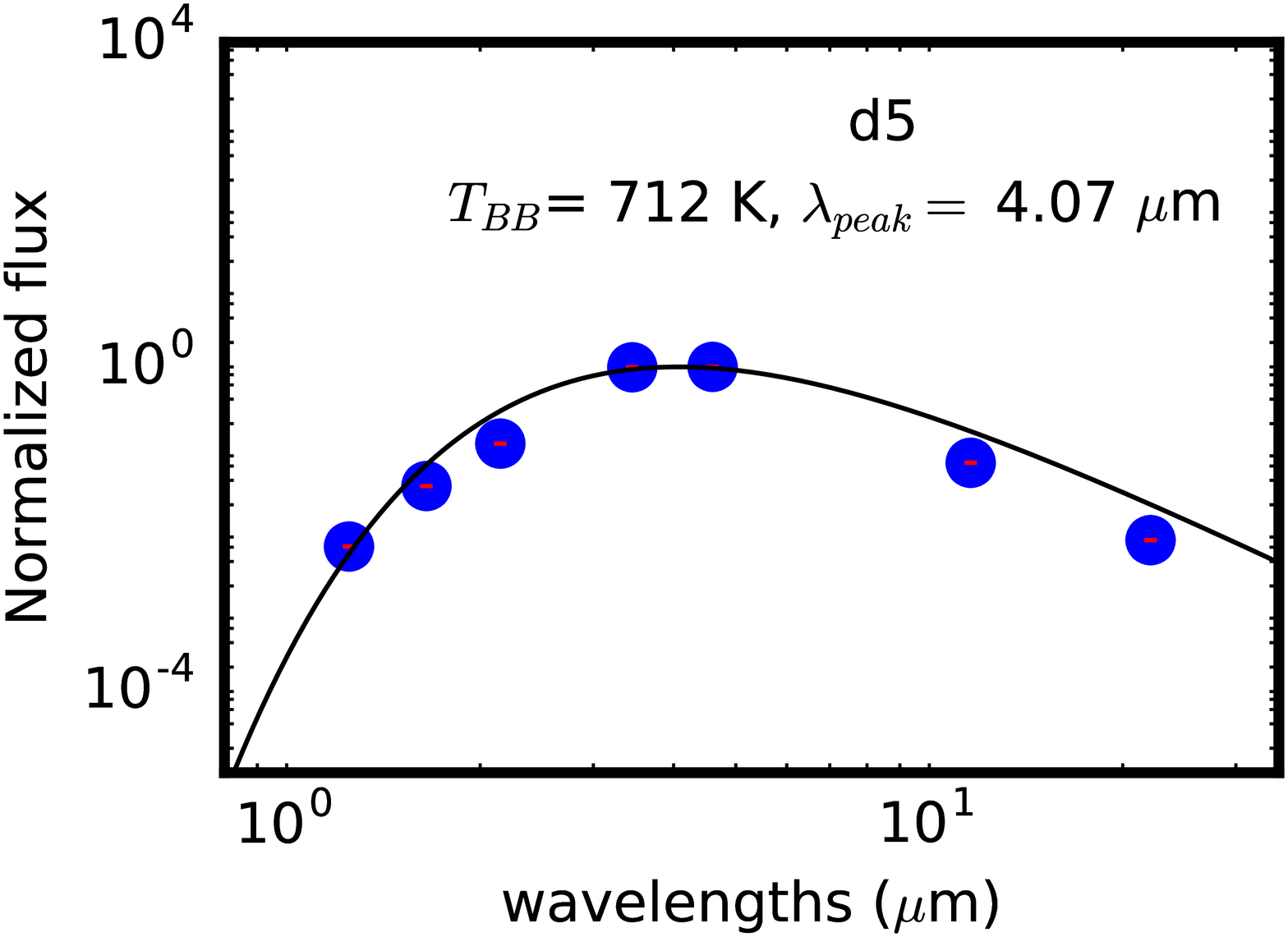}

\includegraphics[scale=0.14]{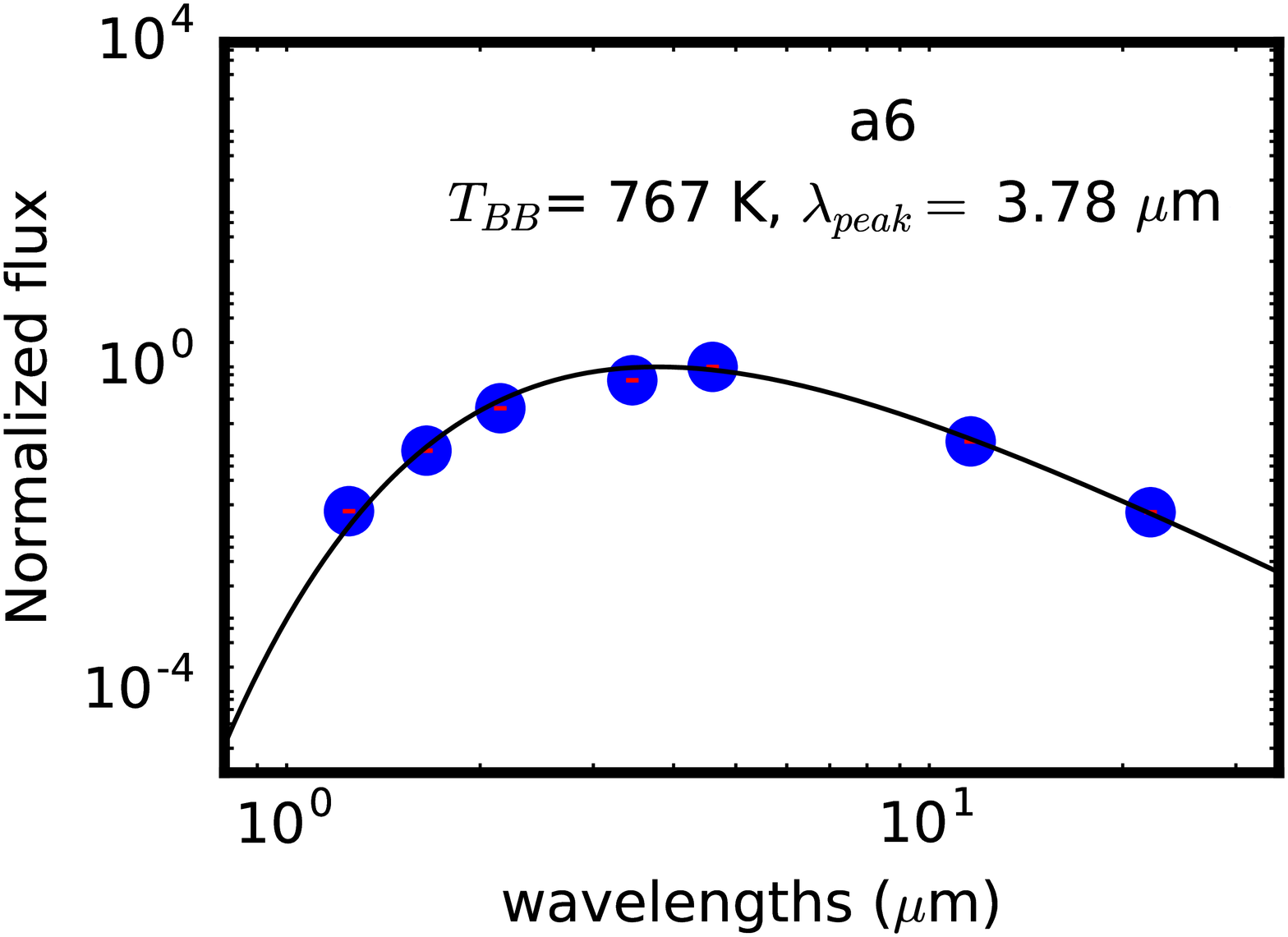}
\includegraphics[scale=0.14]{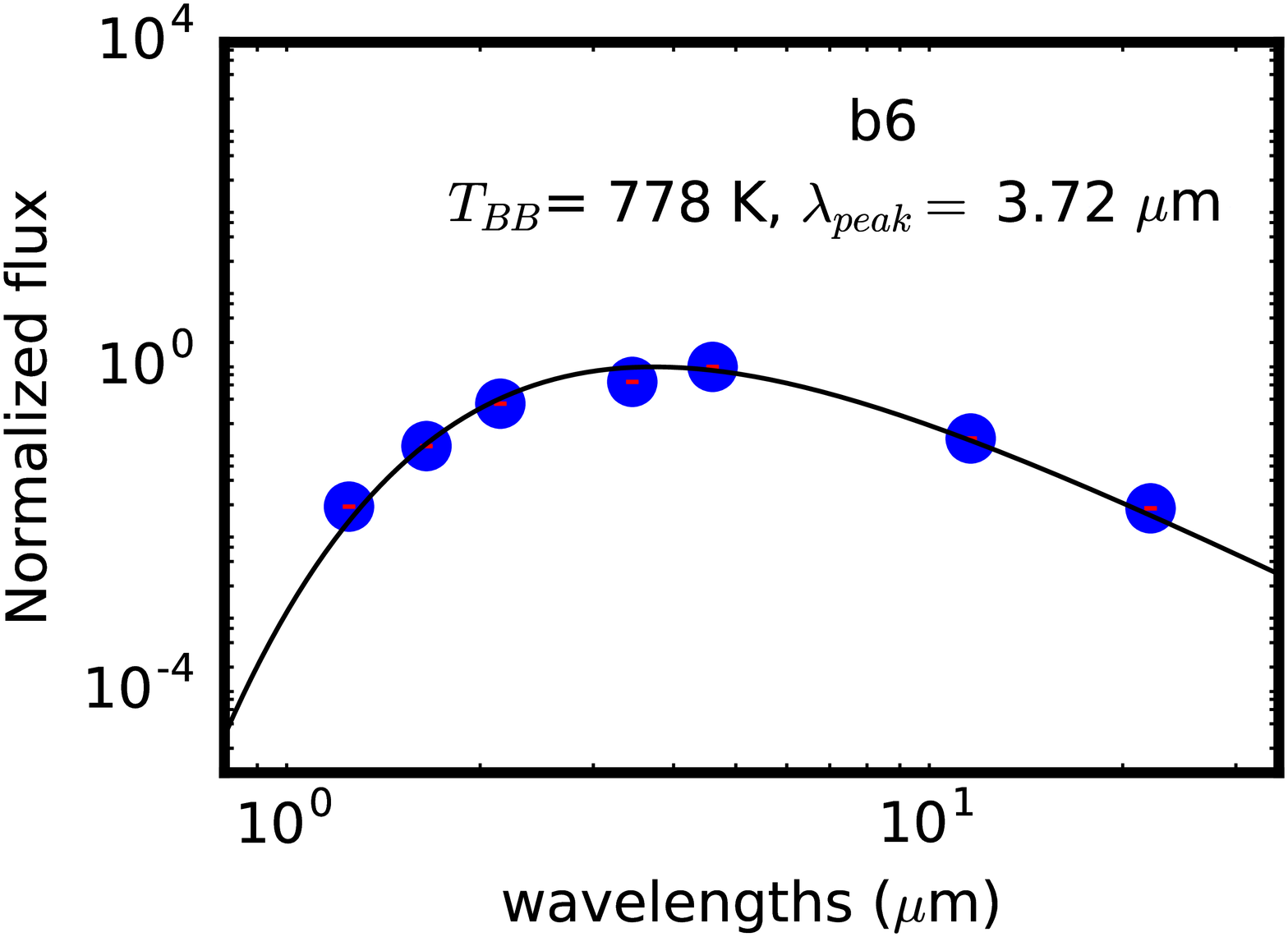}
\includegraphics[scale=0.14]{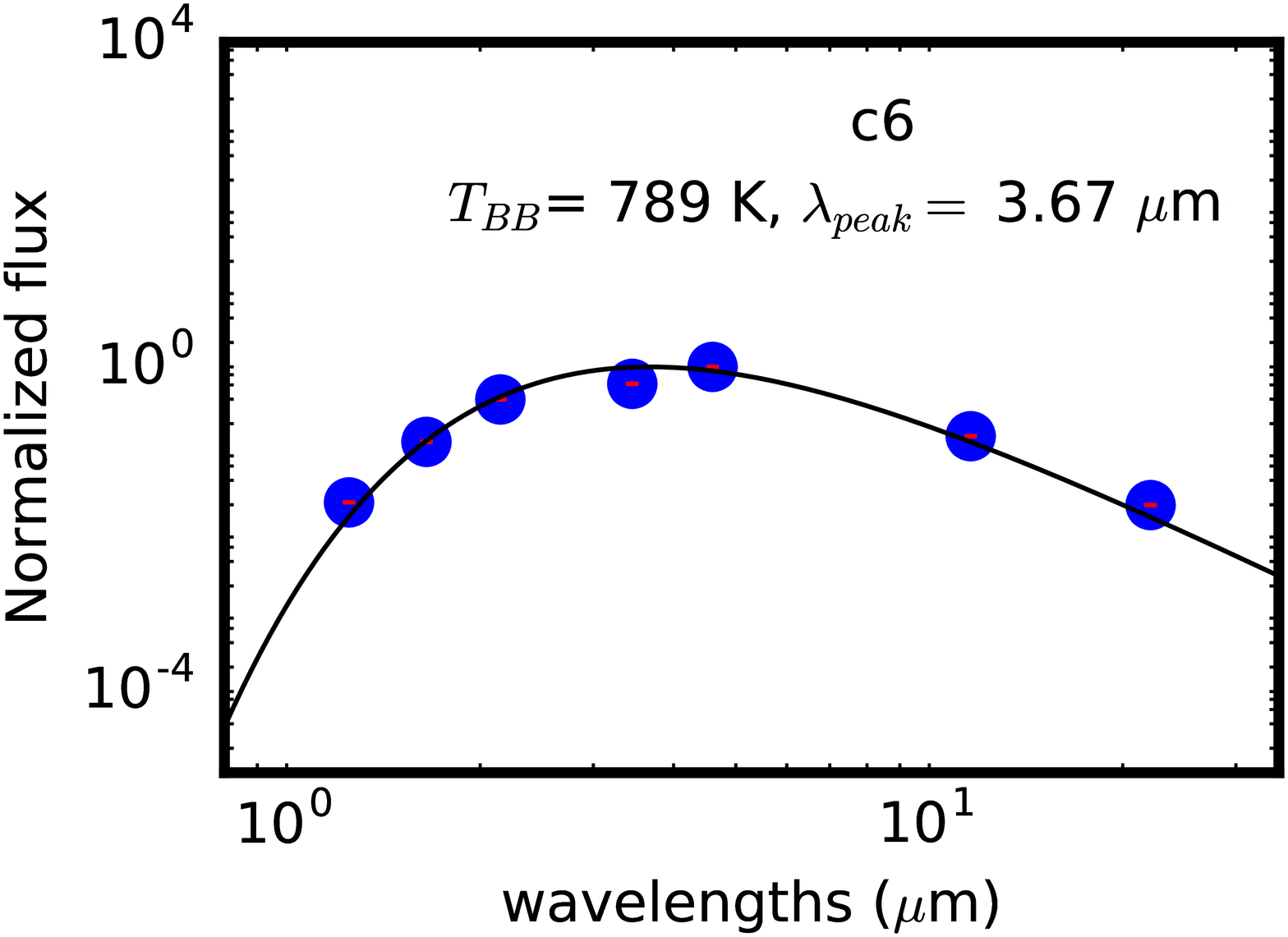}
\includegraphics[scale=0.14]{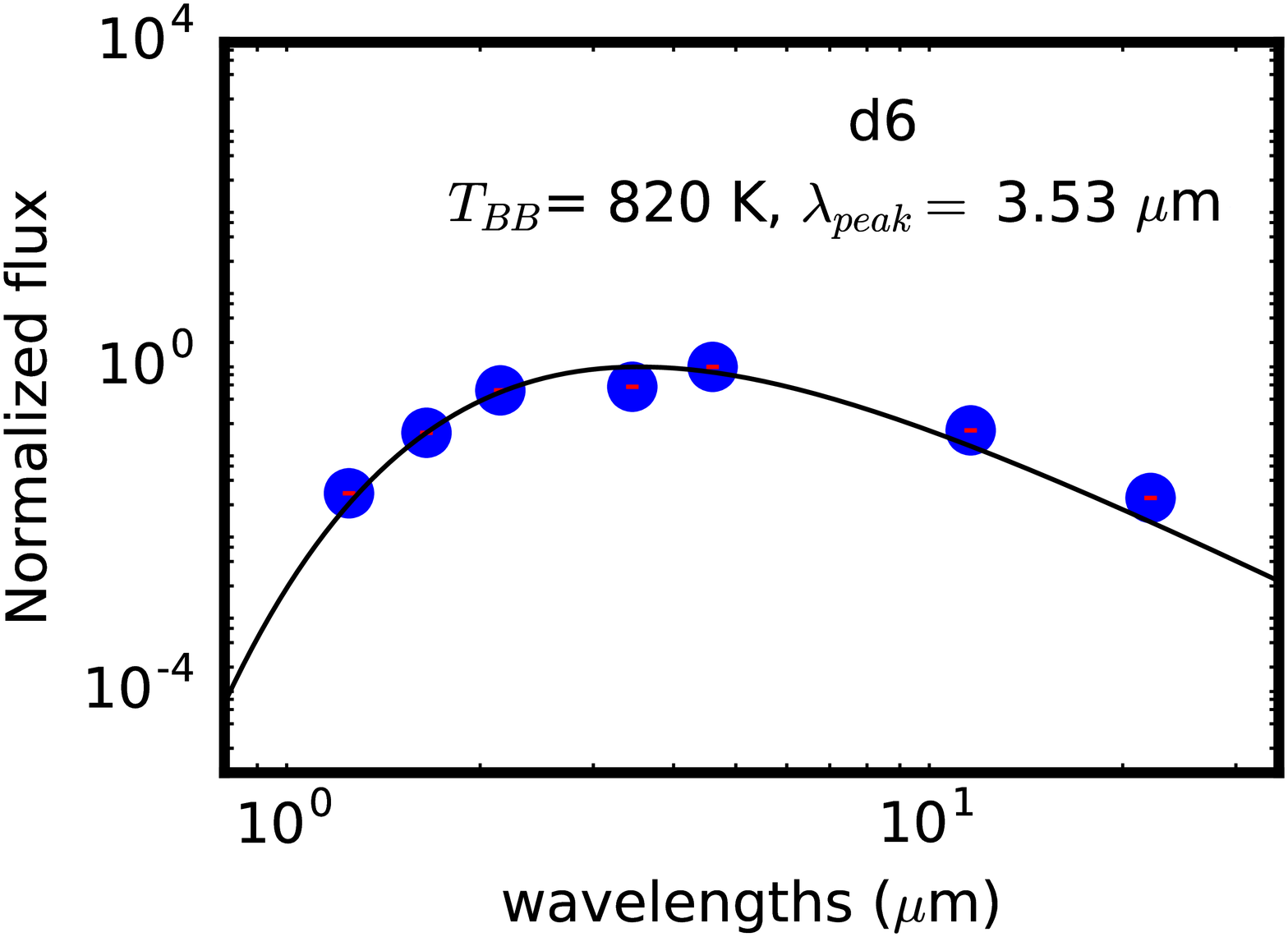}
\caption{The same as in Figure~\ref{fignew1} for the S-type SySt IPHAS~J205836.43+503307.28 and four cases of variation 
amplitude variations: (a) 2~mag (observed), (b) 3~mag, (c) 4~mag, and (d) 6~mag.}
\label{fignew2}
\end{center}
\end{figure*}

Regarding the D-type SySt, maximum amplitude variations of 2, 3, 4, and 6~mag in the {\it J}-band were considered. A poor fitting of the 
2MASS data (e.g., c1, d2, d4) or the {\it W1} and {\it W2} bands (e.g., b2, c2, d1, d2, d4, and d6) is an evidence of flux variation 
between the data of the two surveys (Fig.~\ref{fignew2}). Nevertheless, we can say that only amplitude variations higher than 3~mag result 
in significant deviations from the BB fitting. The resulting temperatures show a variation of 250~K. 

\begin{figure*}
\begin{center}
\includegraphics[scale=0.14]{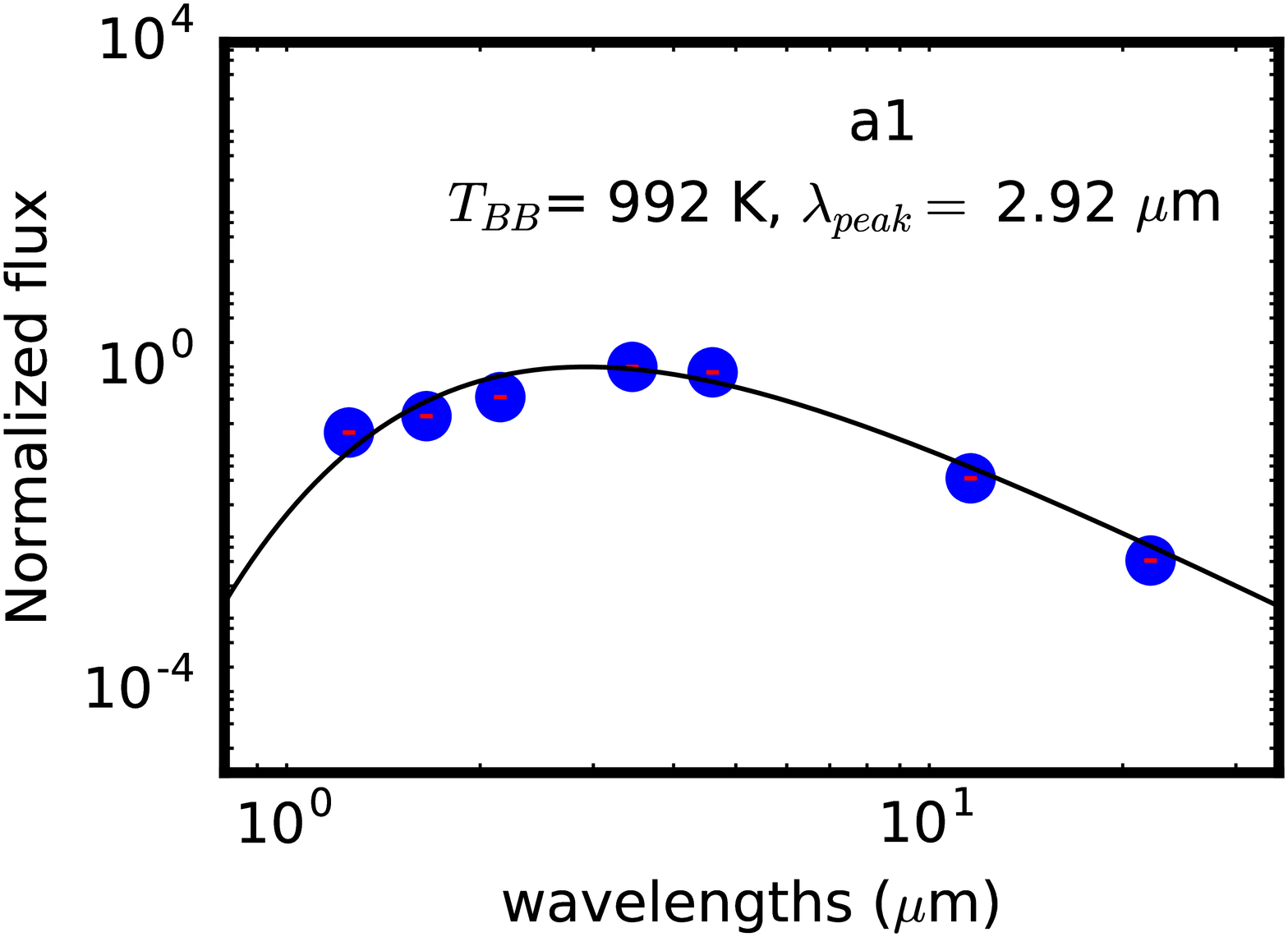}
\includegraphics[scale=0.14]{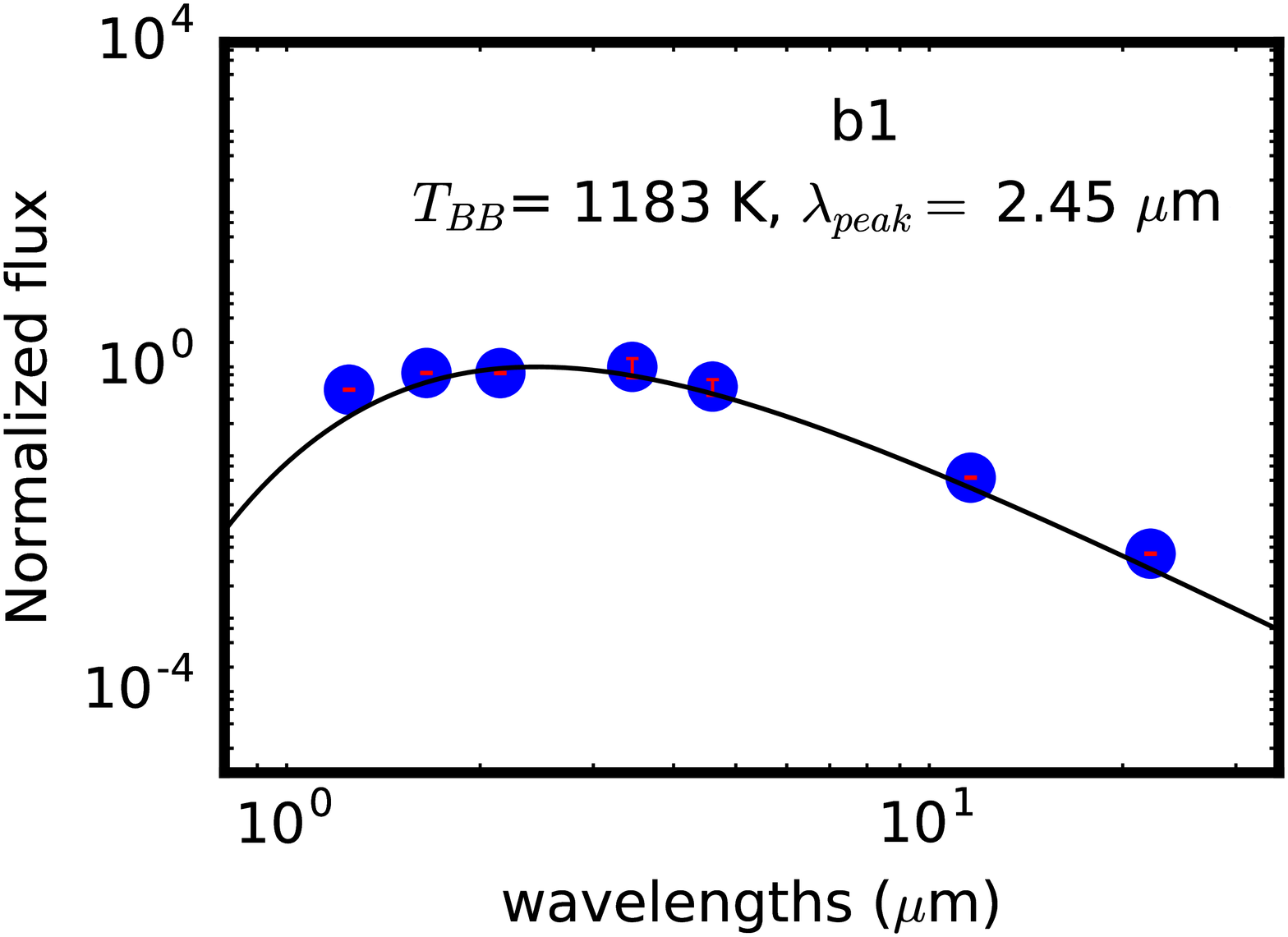}
\includegraphics[scale=0.14]{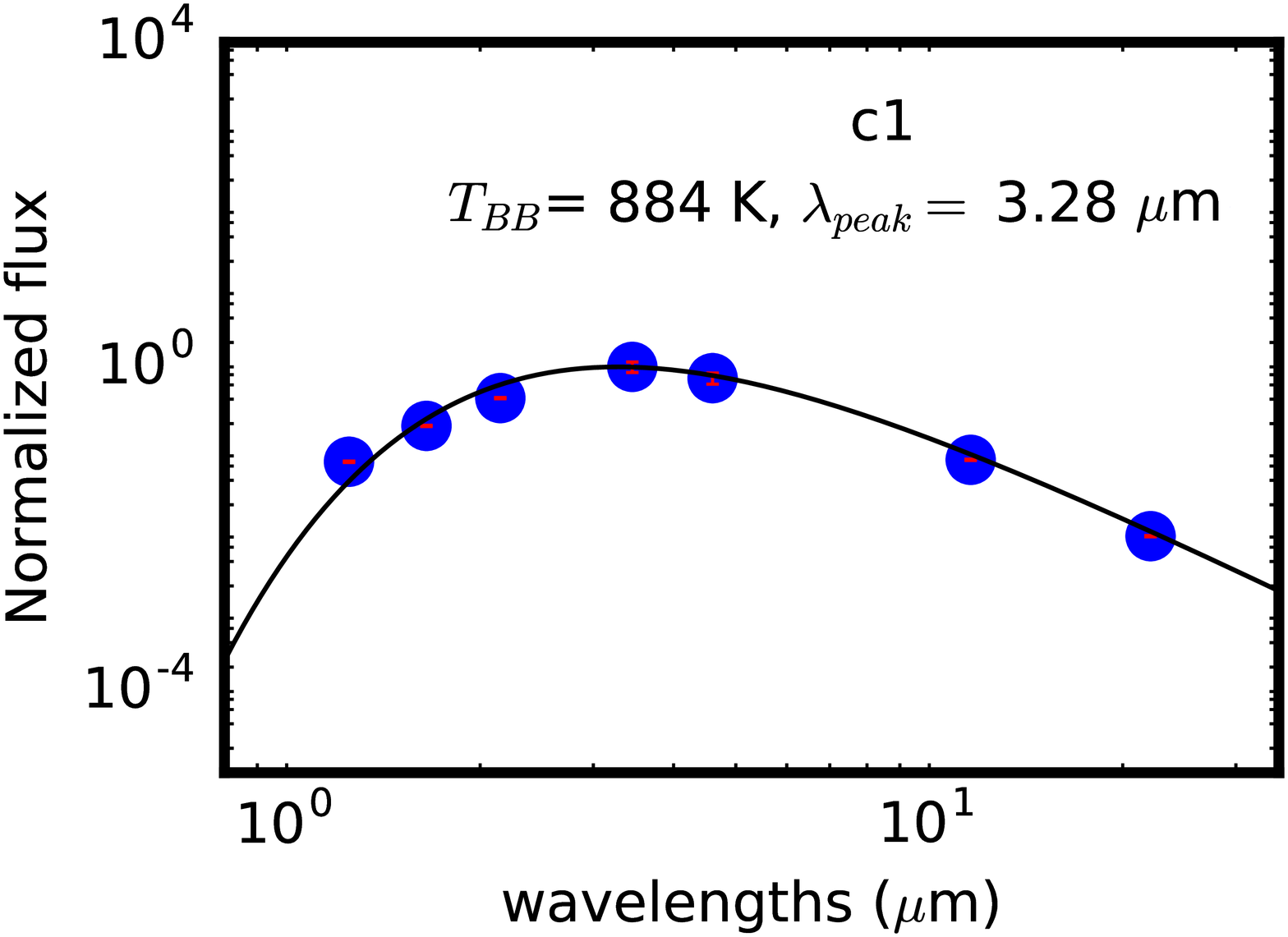}
\includegraphics[scale=0.14]{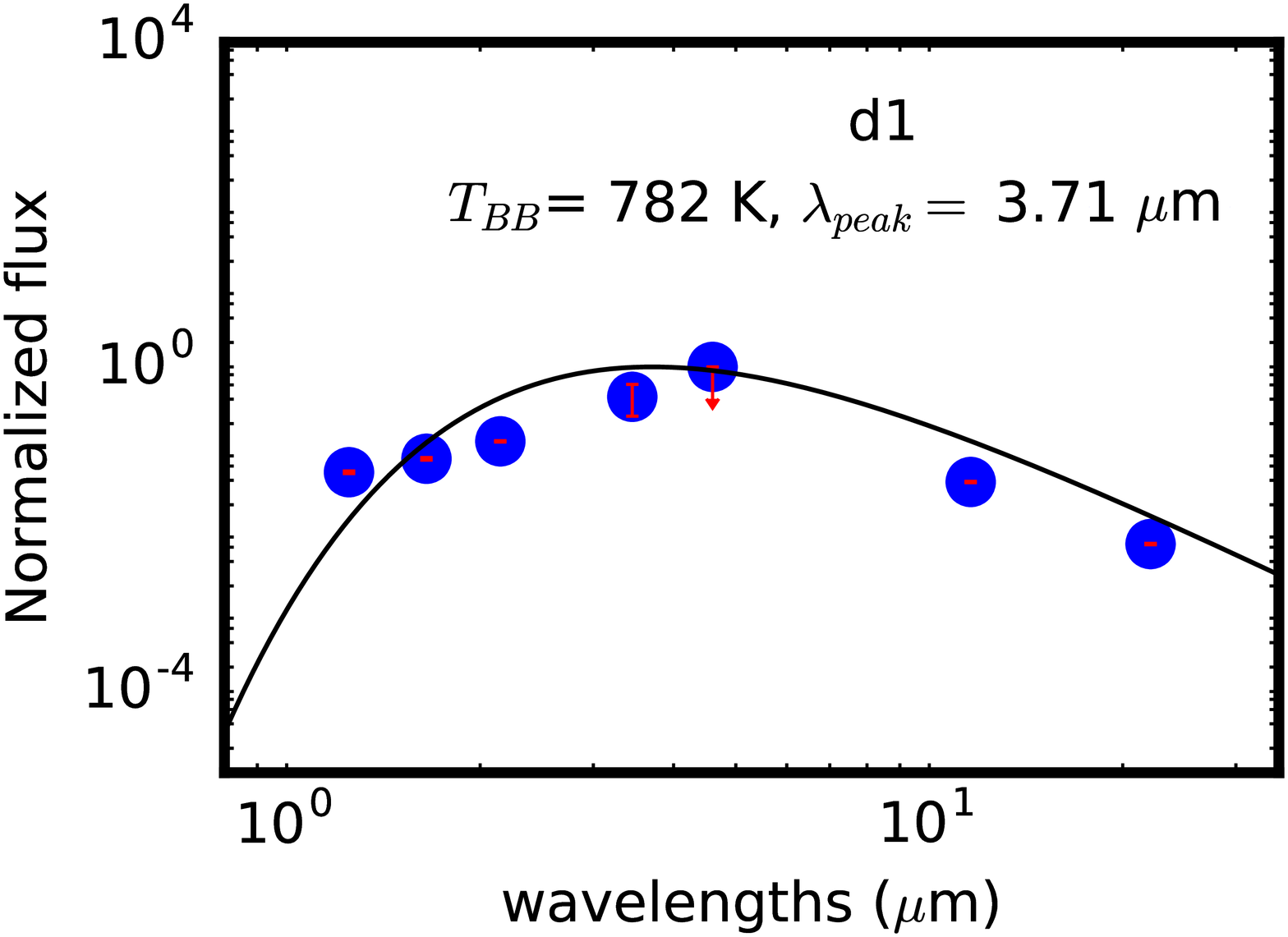}

\includegraphics[scale=0.14]{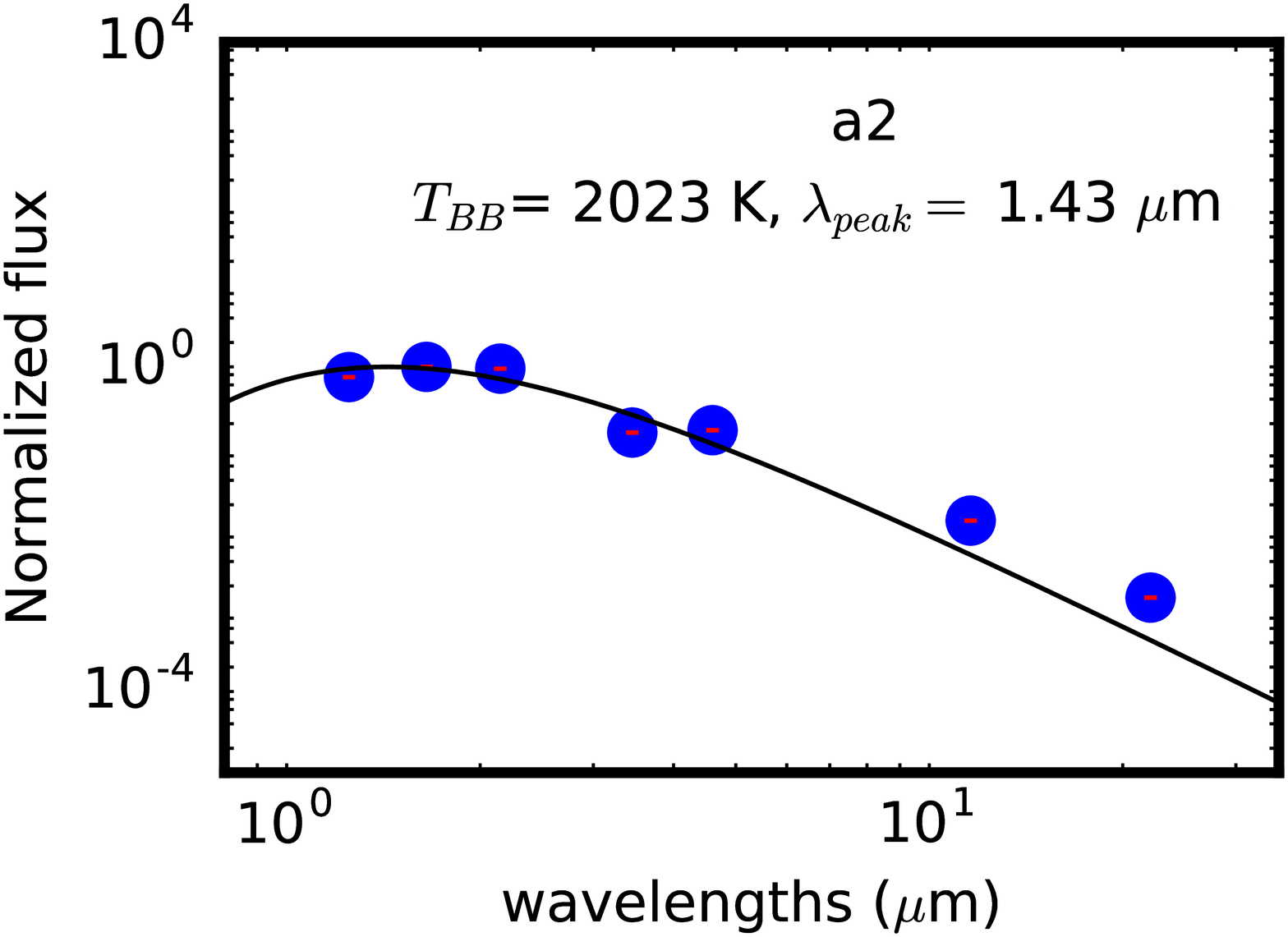}
\includegraphics[scale=0.14]{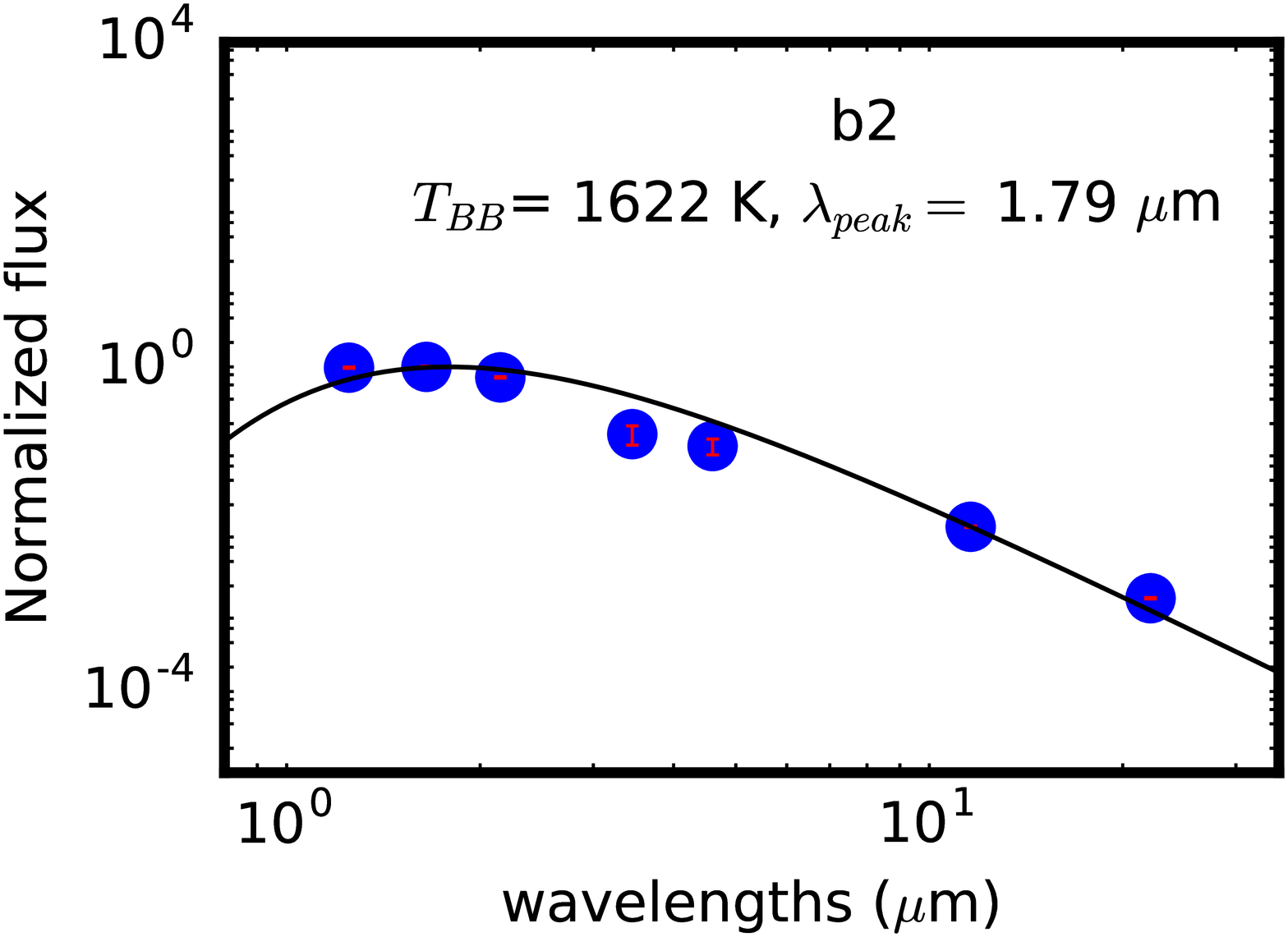}
\includegraphics[scale=0.14]{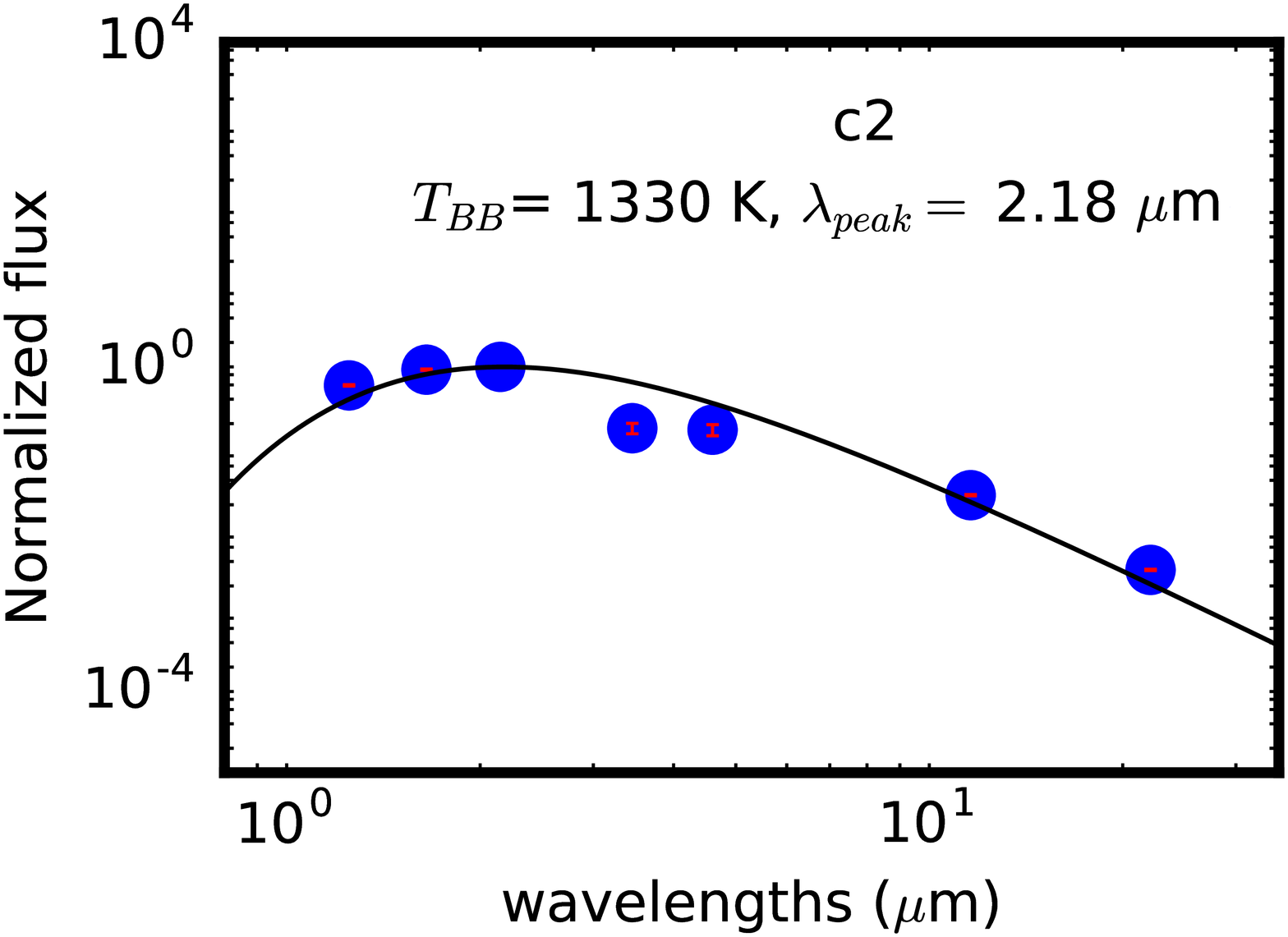}
\includegraphics[scale=0.14]{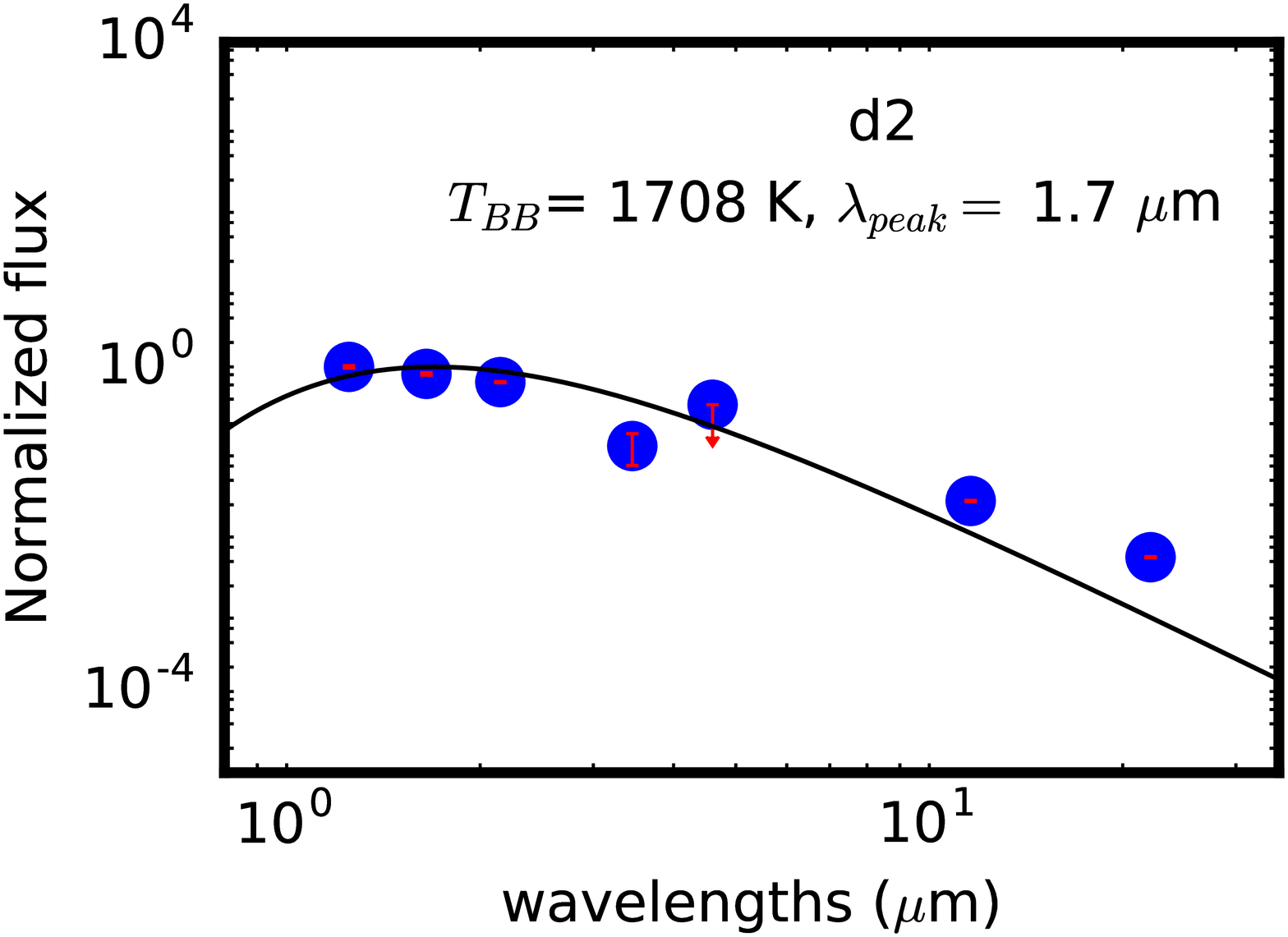}

\includegraphics[scale=0.14]{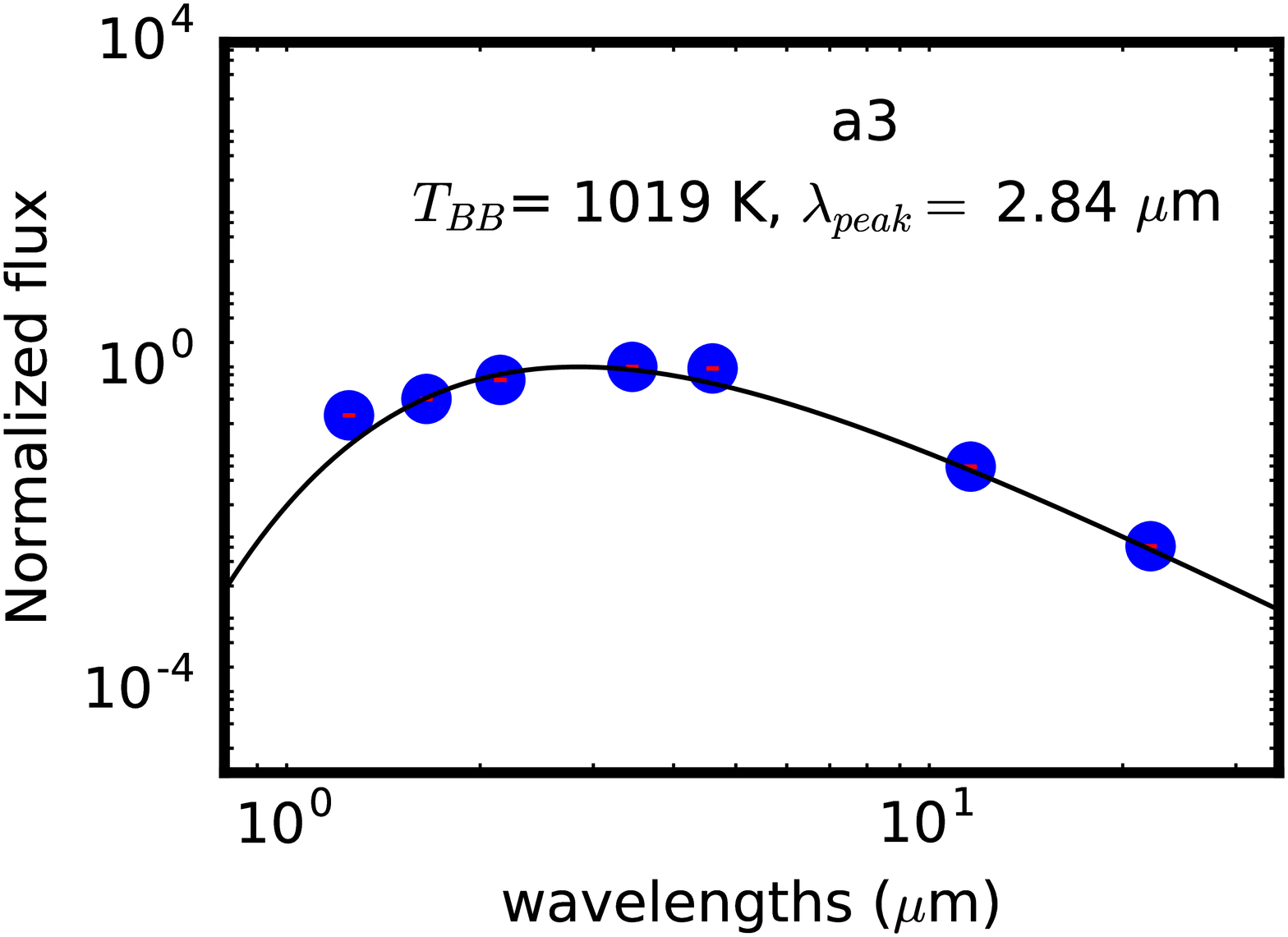}
\includegraphics[scale=0.14]{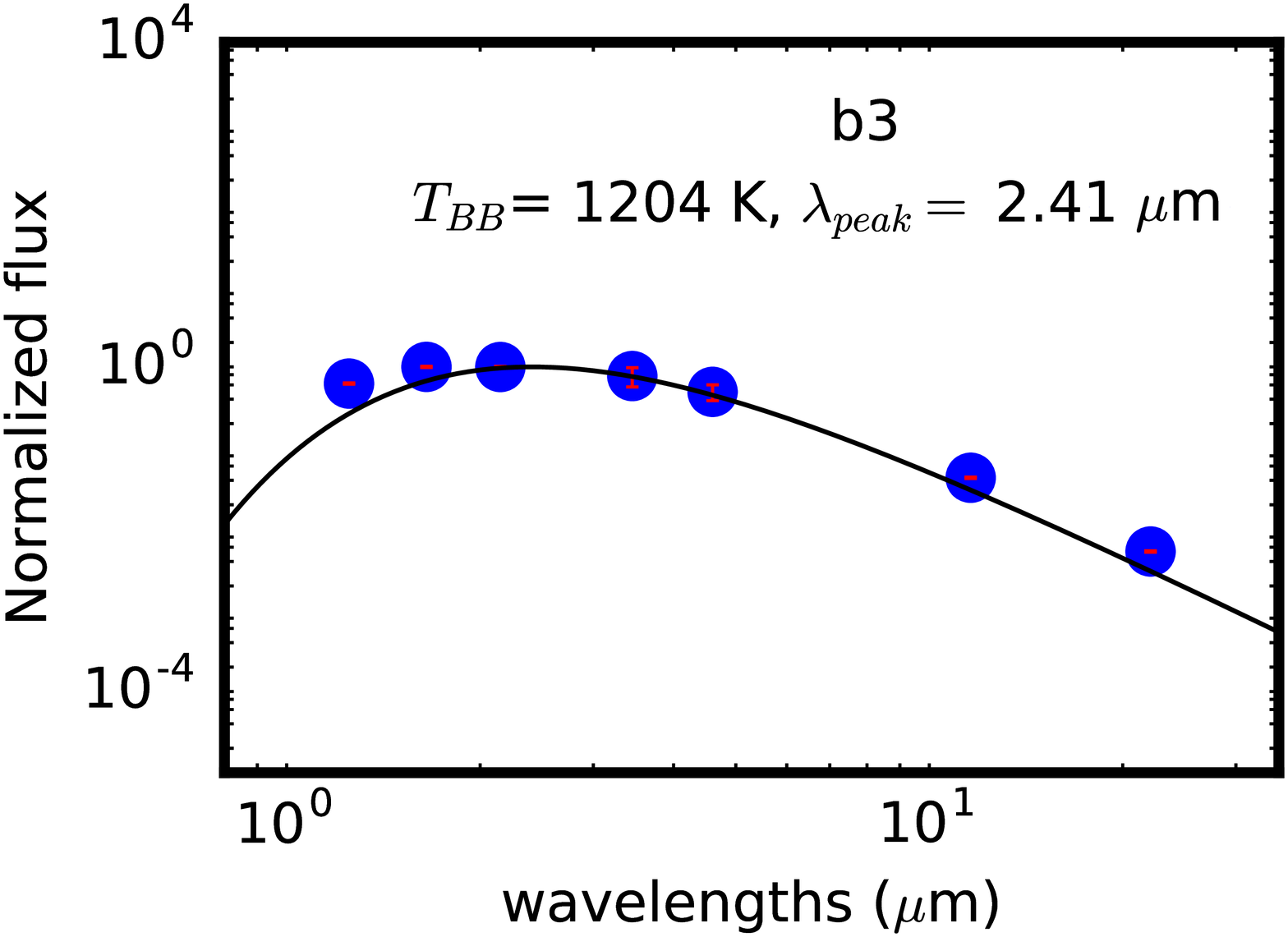}
\includegraphics[scale=0.14]{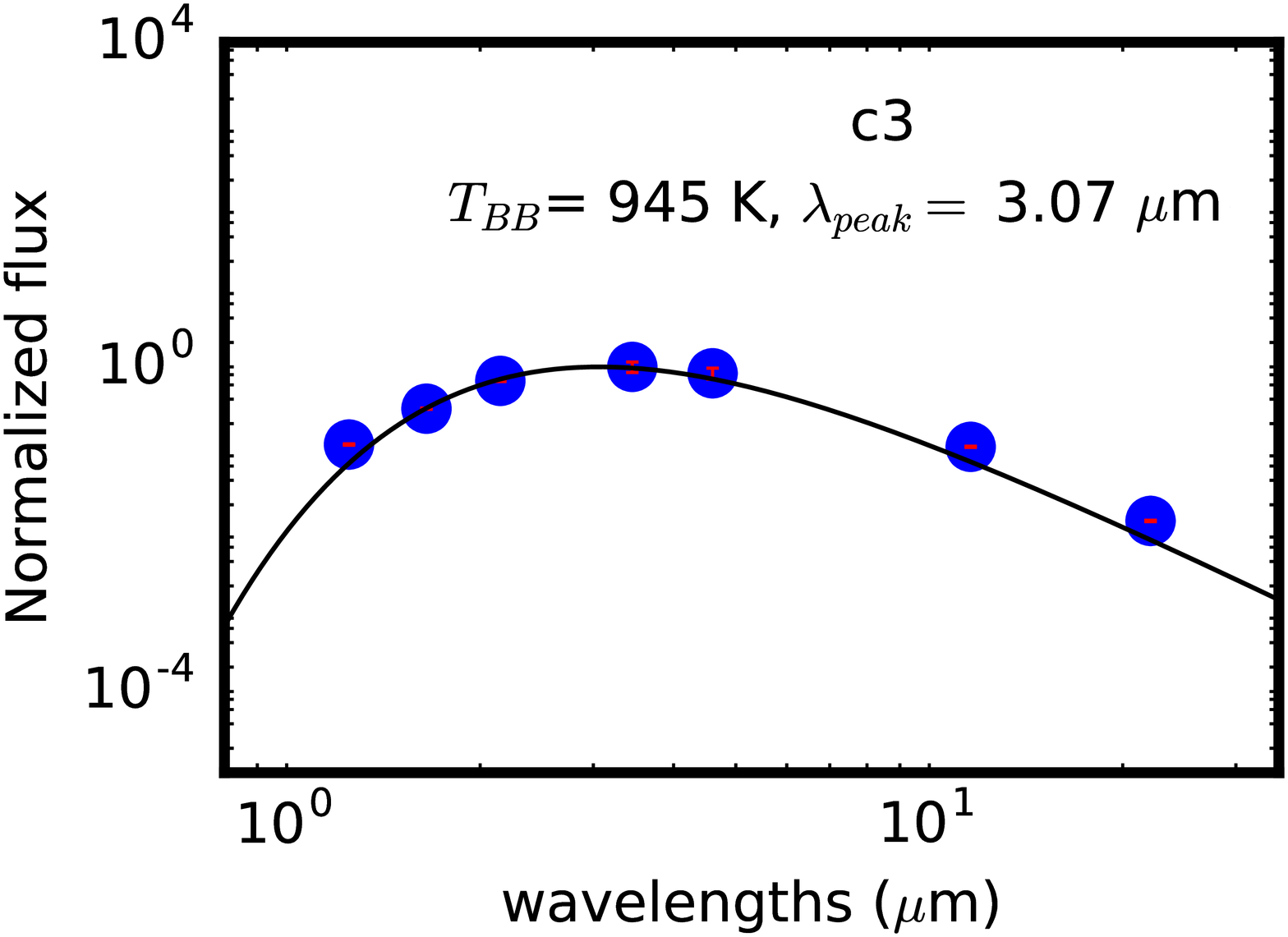}
\includegraphics[scale=0.14]{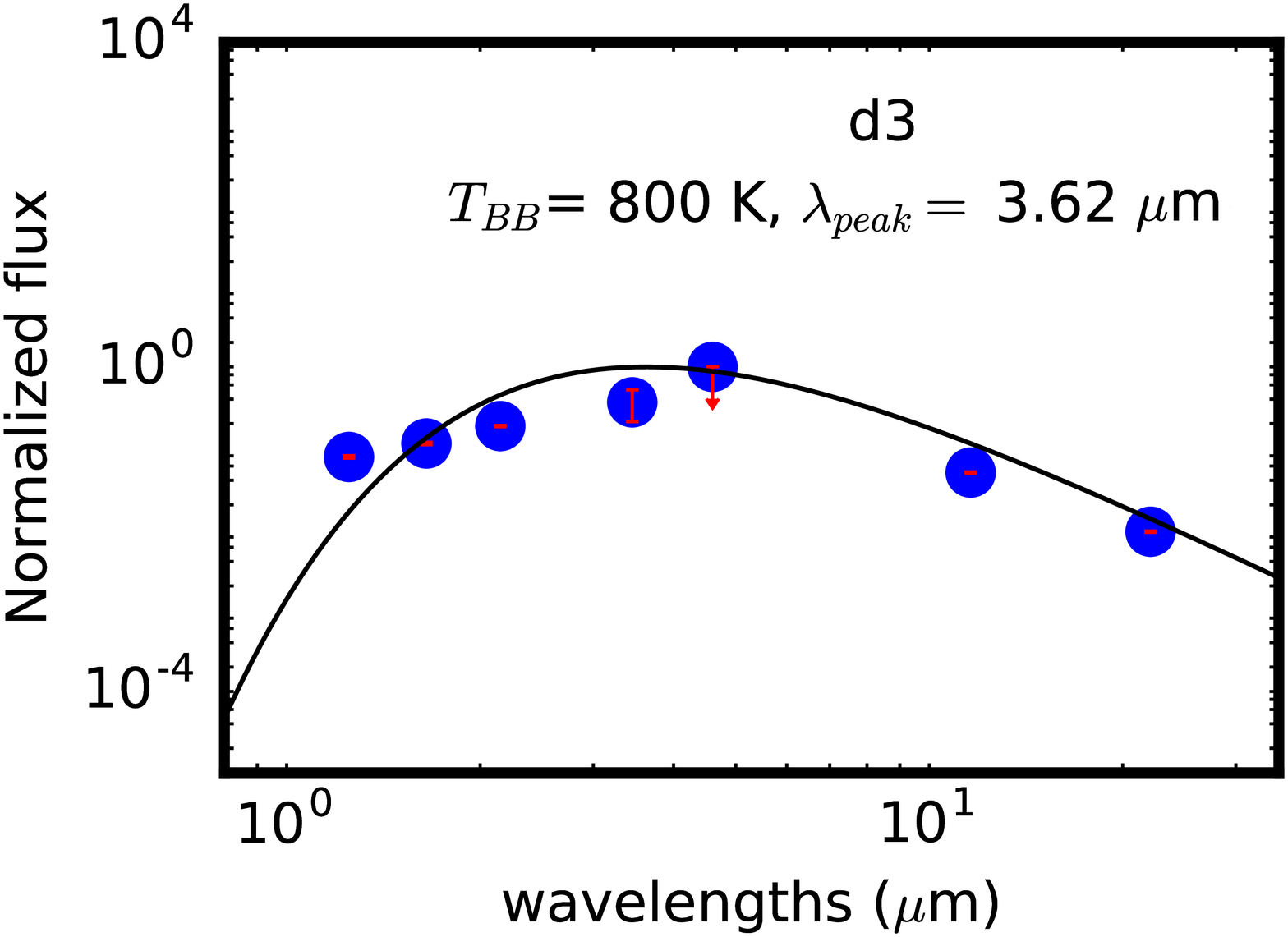}

\includegraphics[scale=0.14]{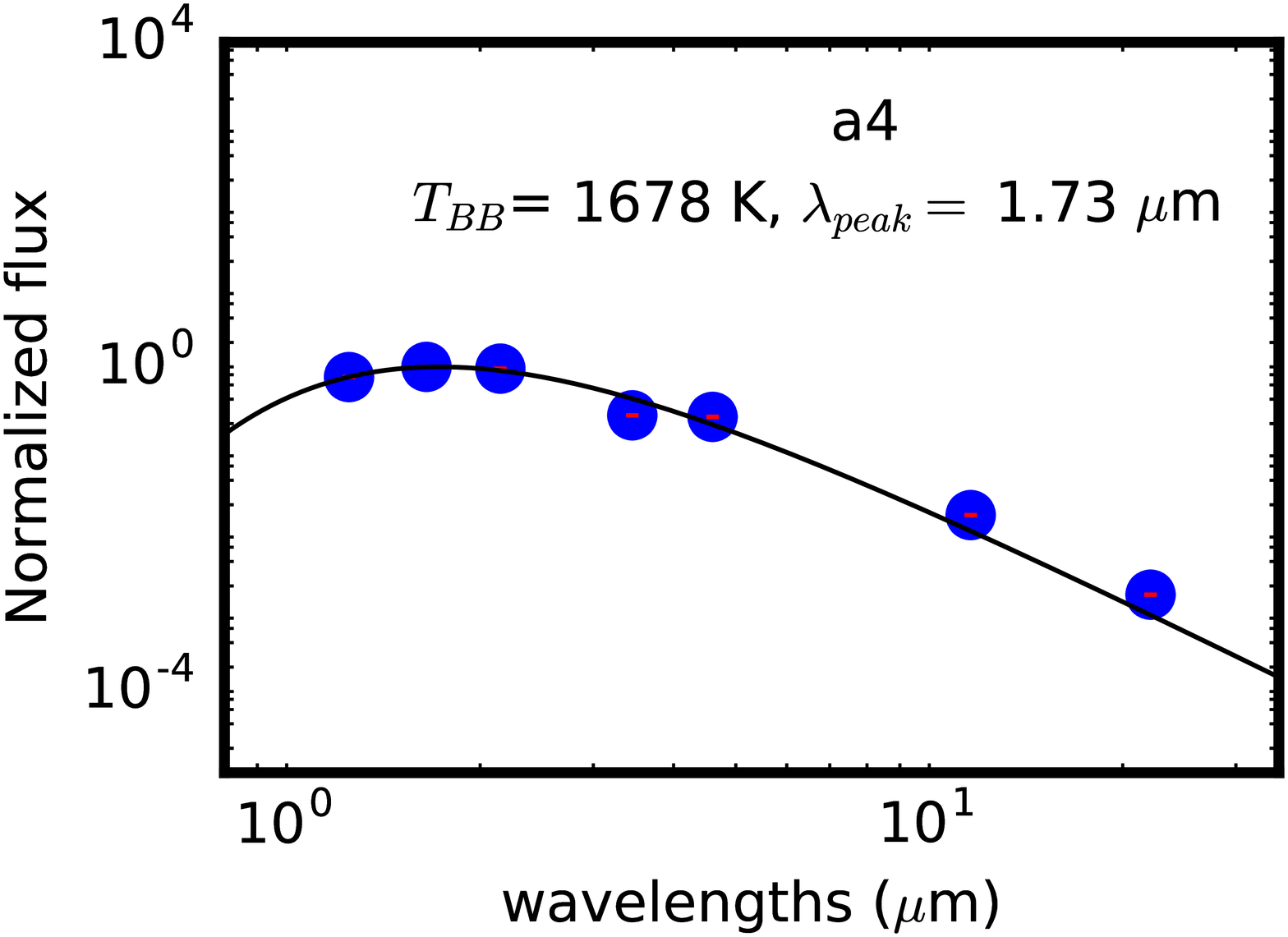}
\includegraphics[scale=0.14]{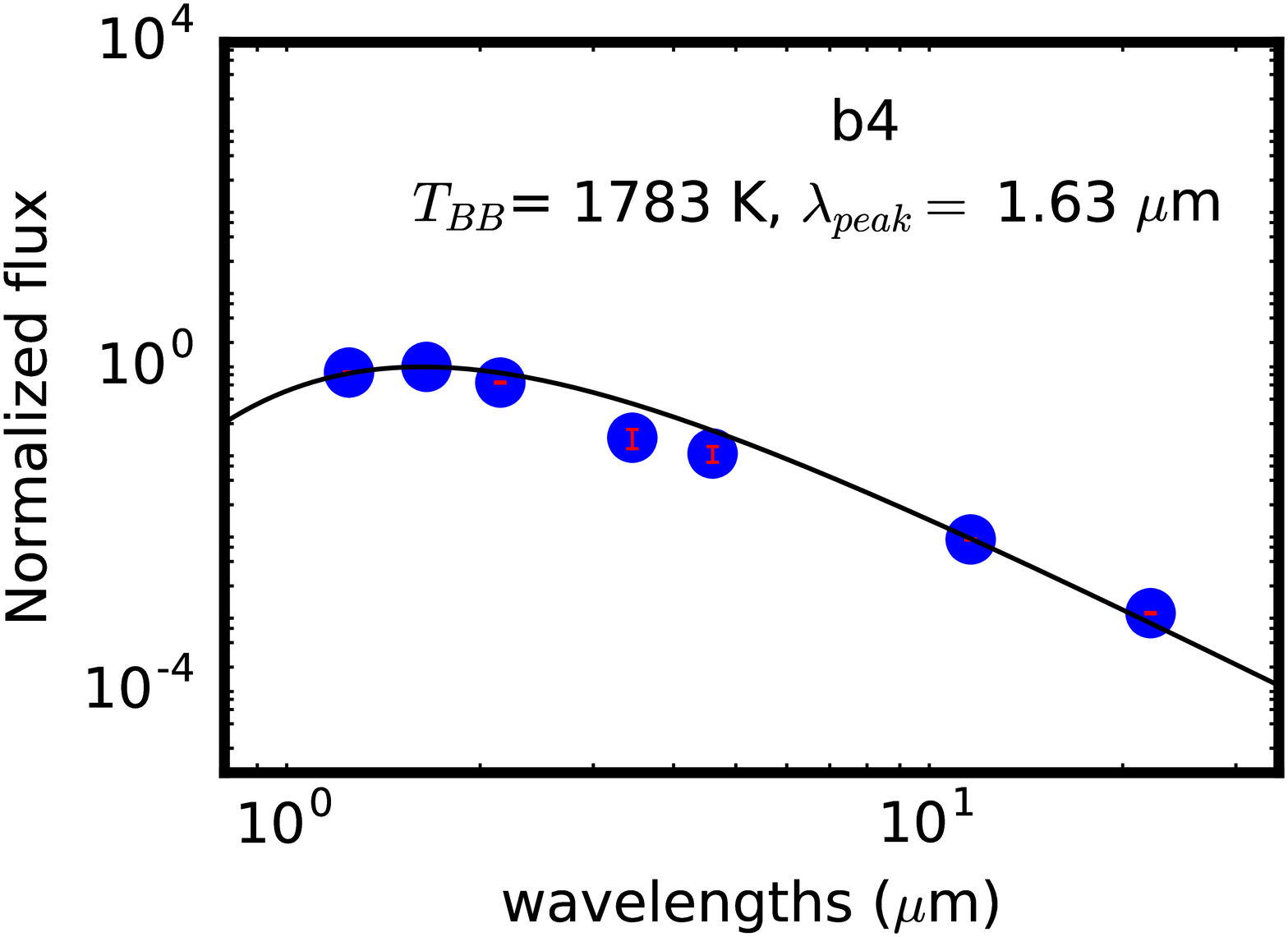}
\includegraphics[scale=0.14]{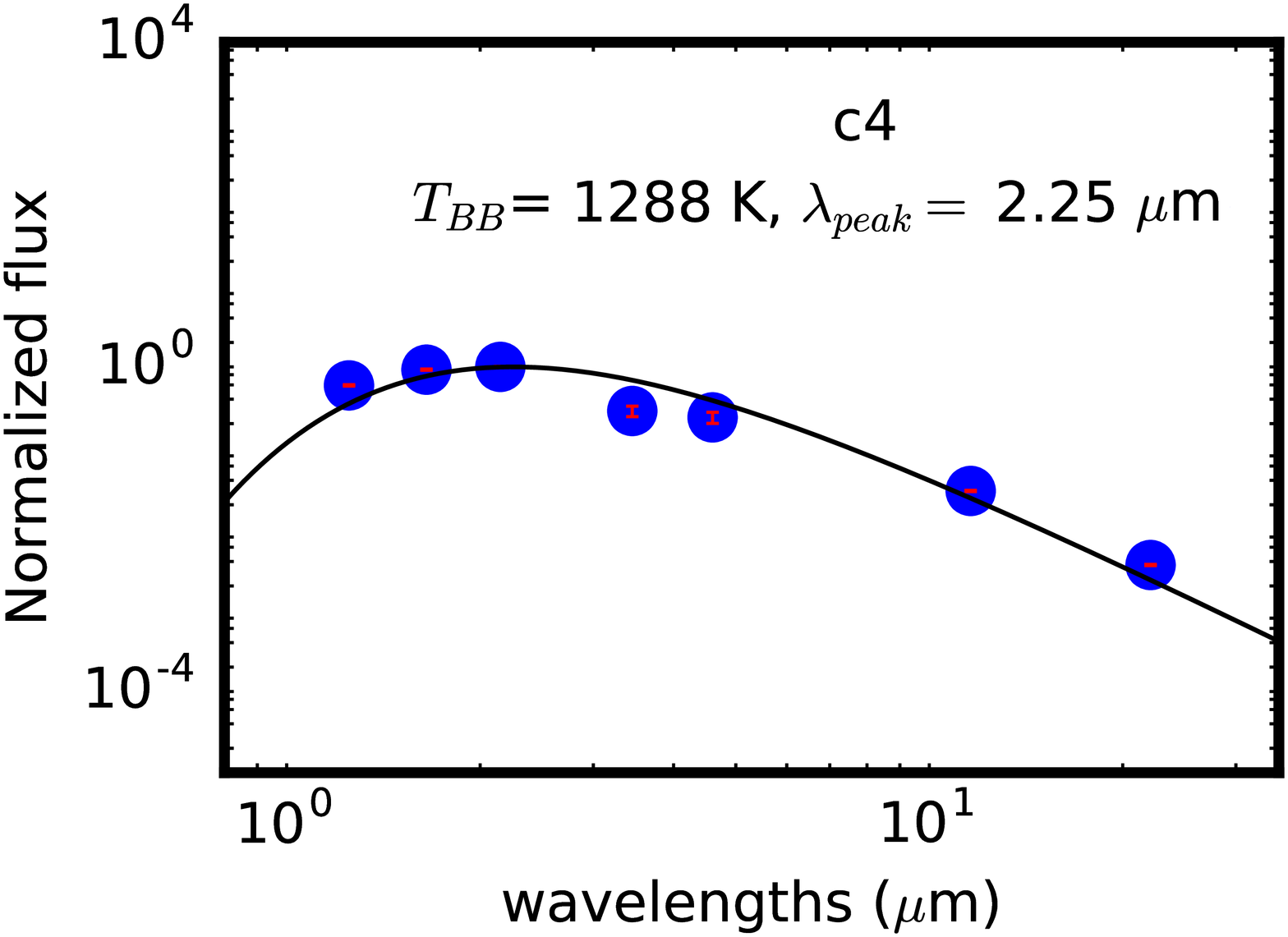}
\includegraphics[scale=0.14]{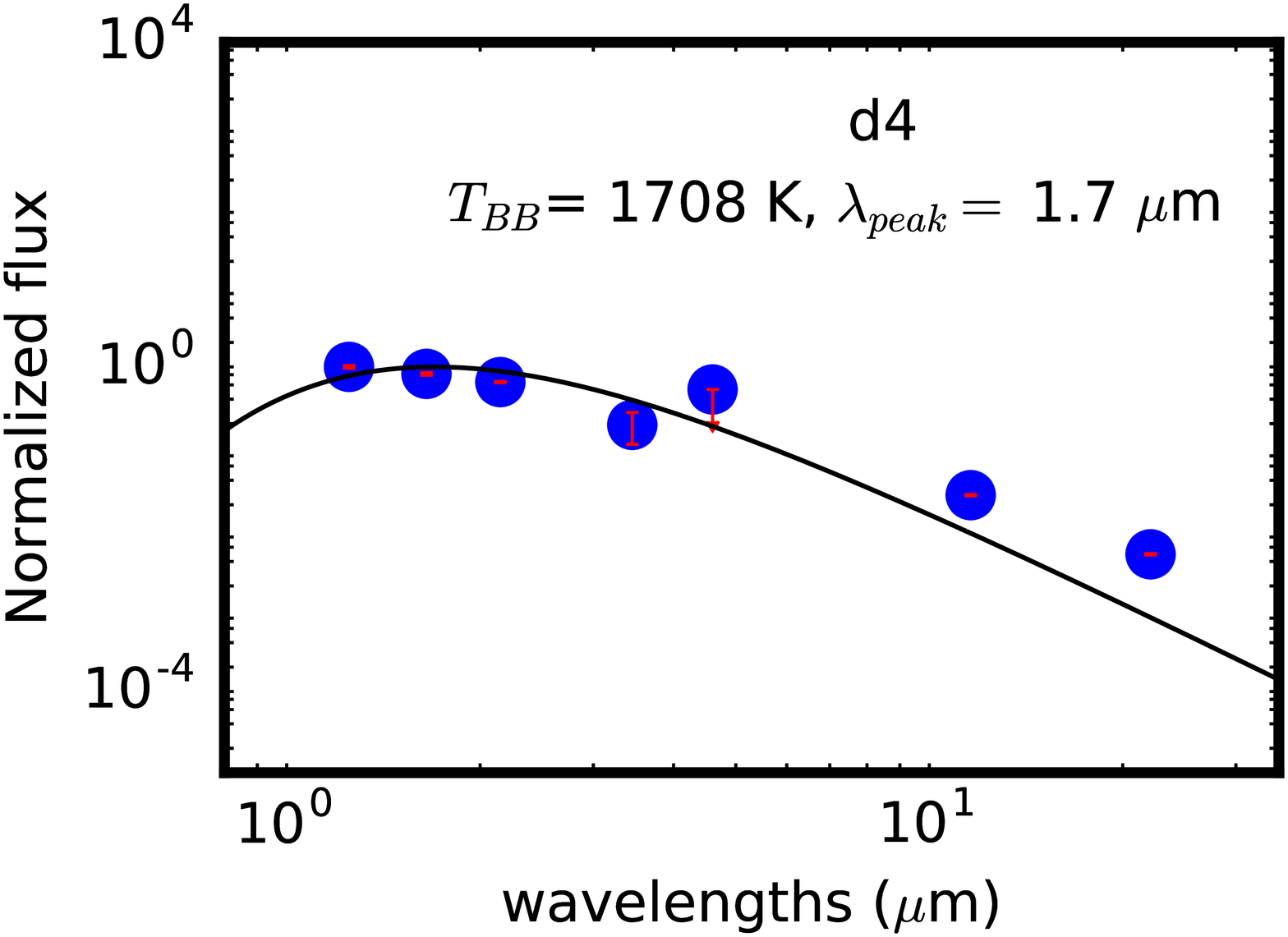}

\includegraphics[scale=0.14]{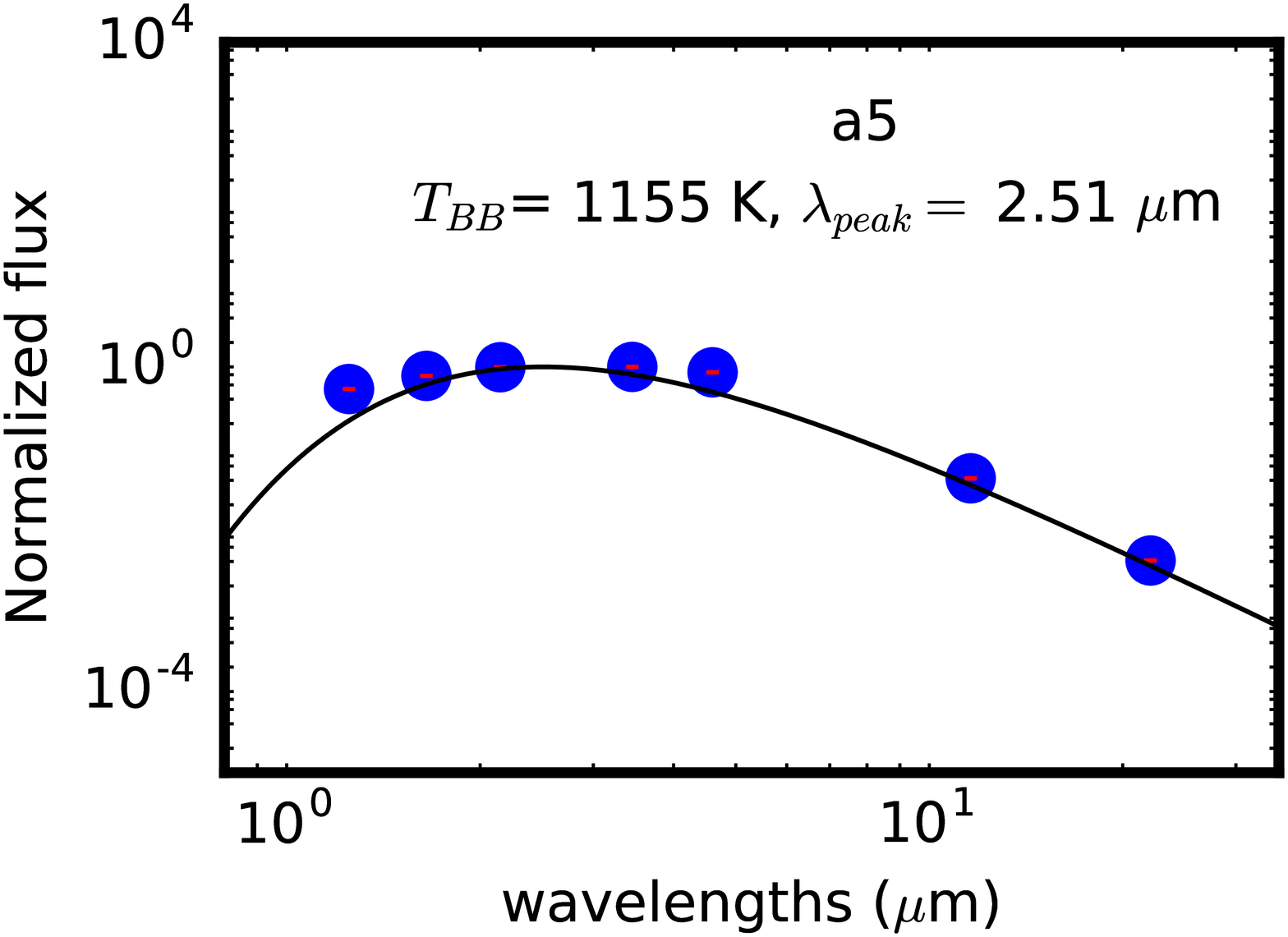}
\includegraphics[scale=0.14]{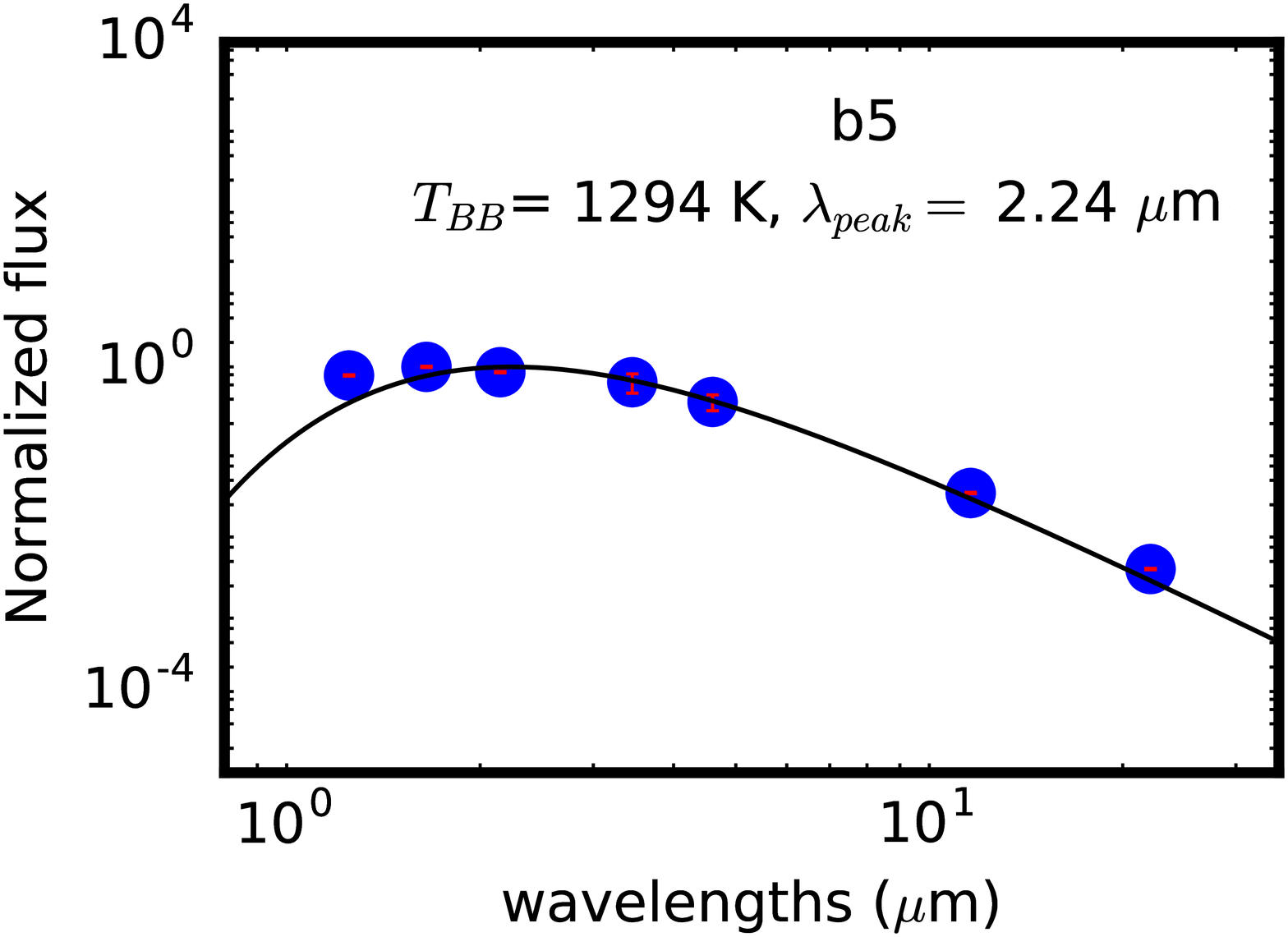}
\includegraphics[scale=0.14]{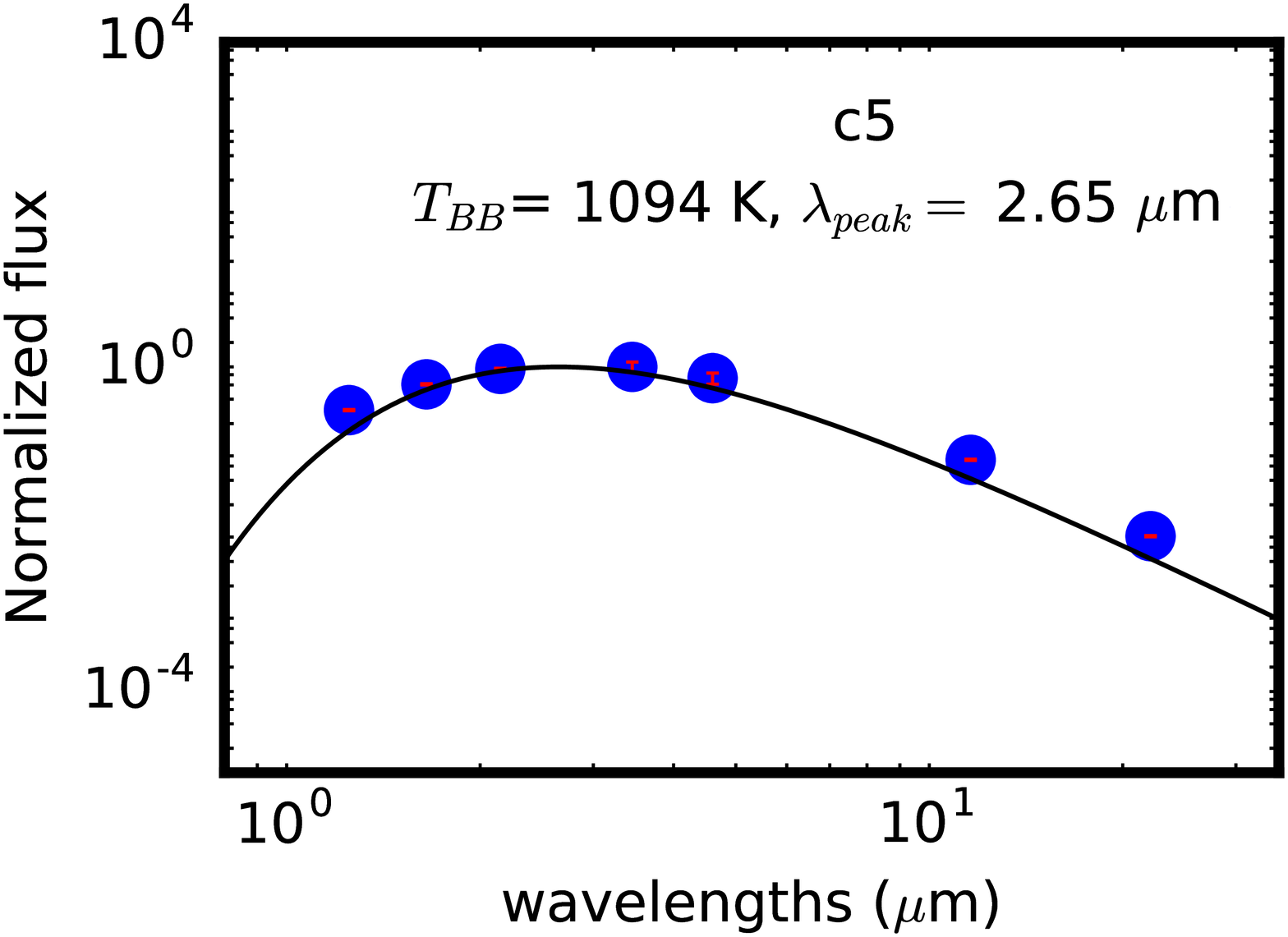}
\includegraphics[scale=0.14]{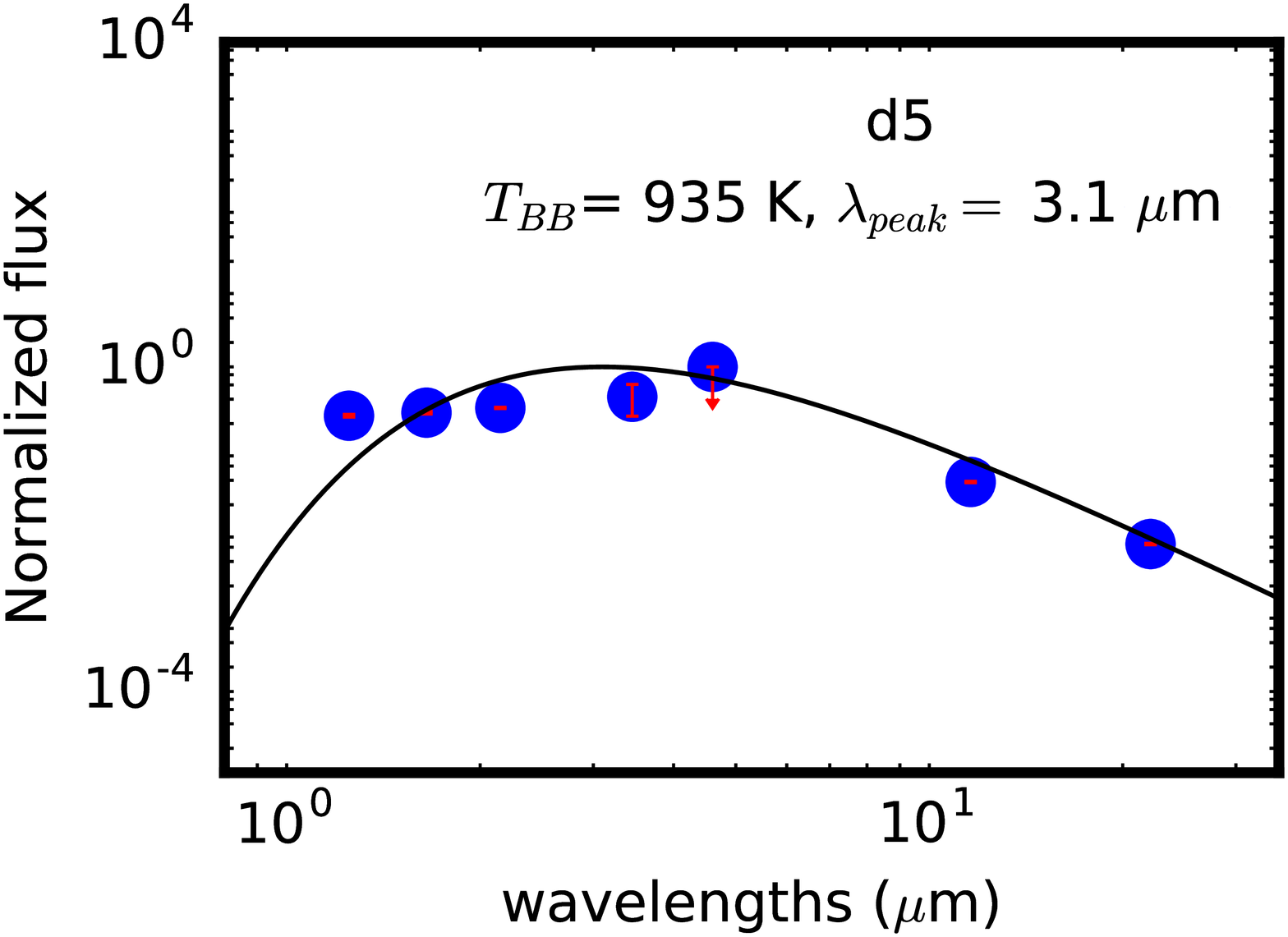}

\includegraphics[scale=0.14]{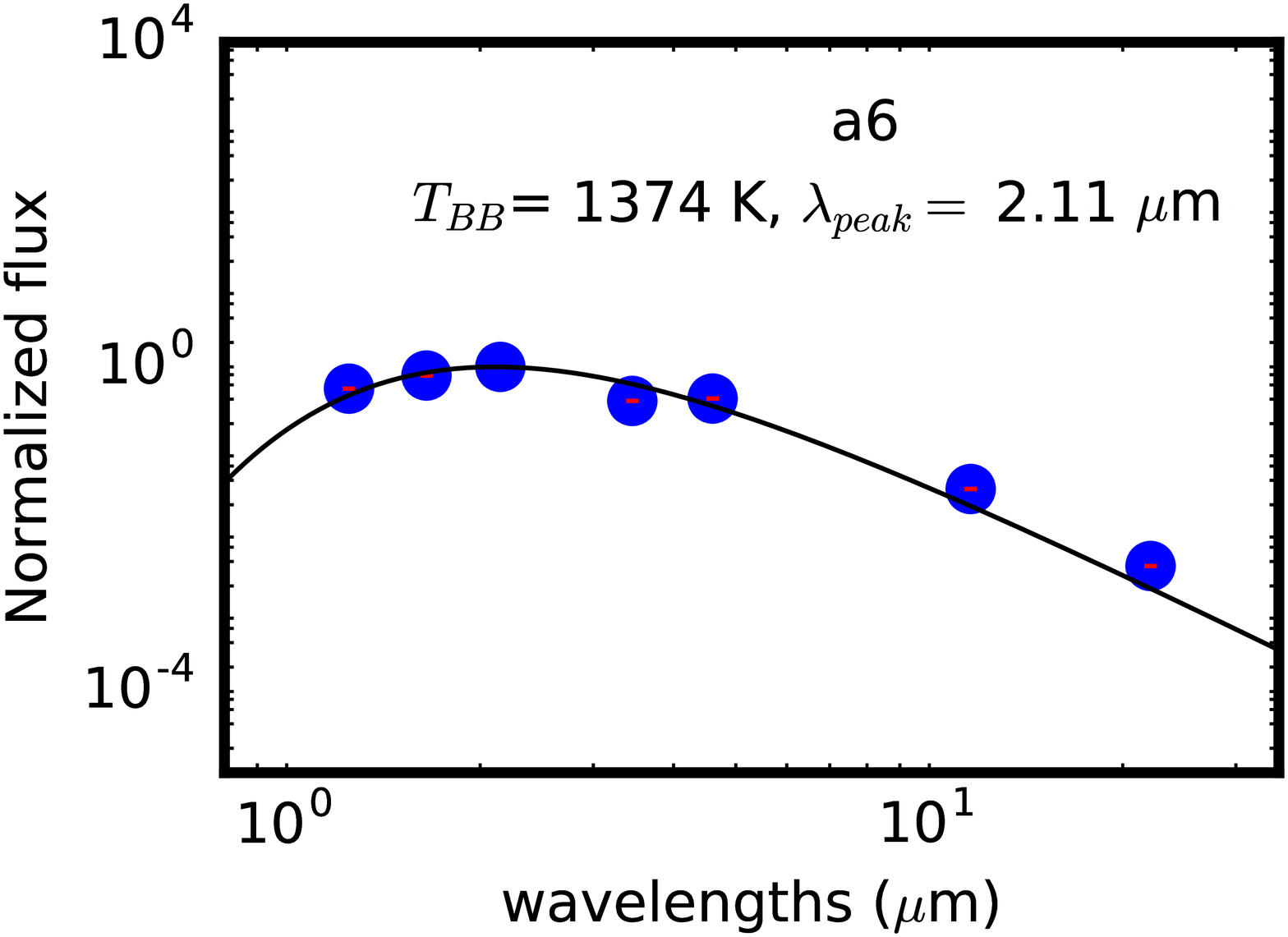}
\includegraphics[scale=0.14]{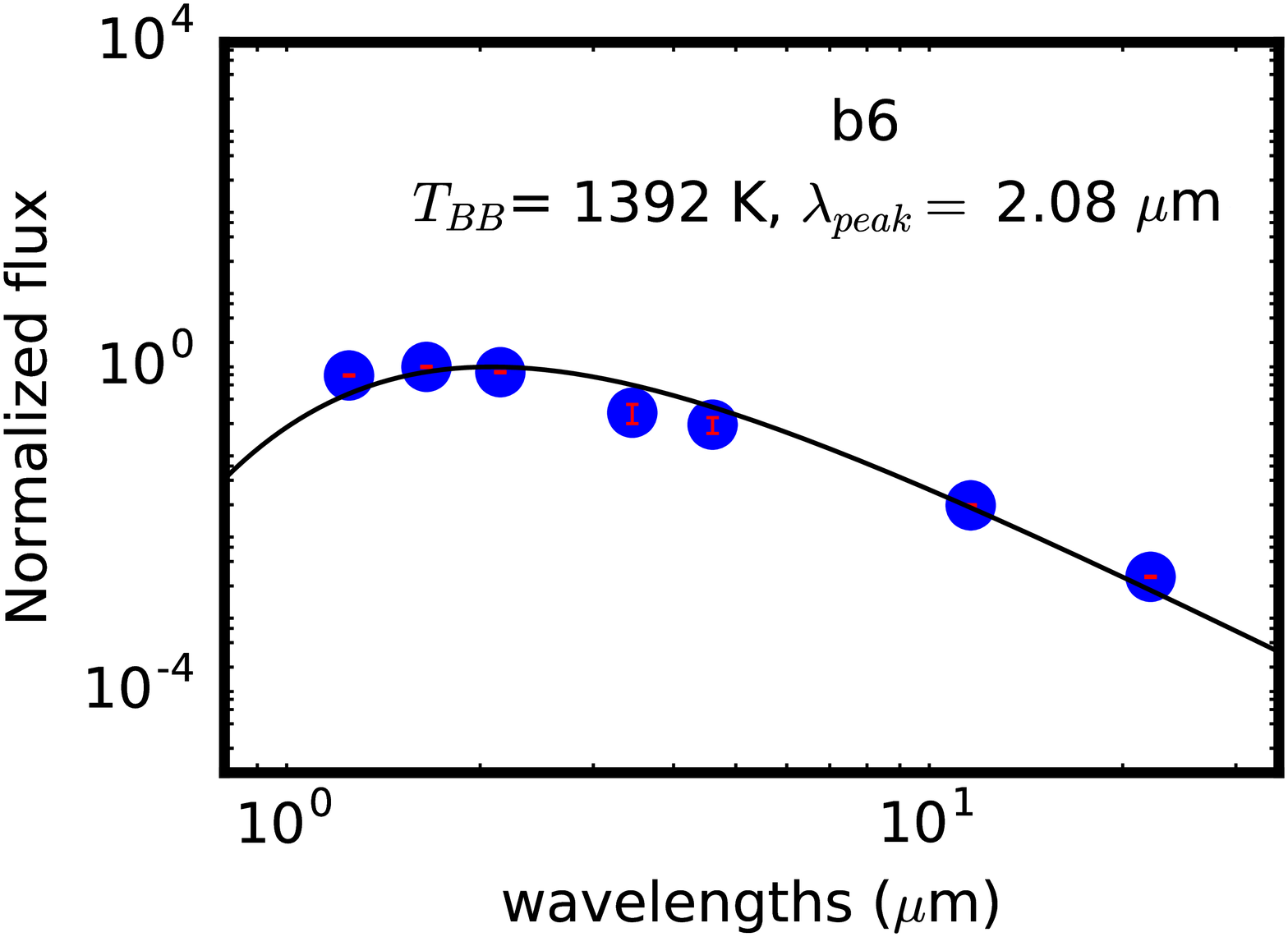}
\includegraphics[scale=0.14]{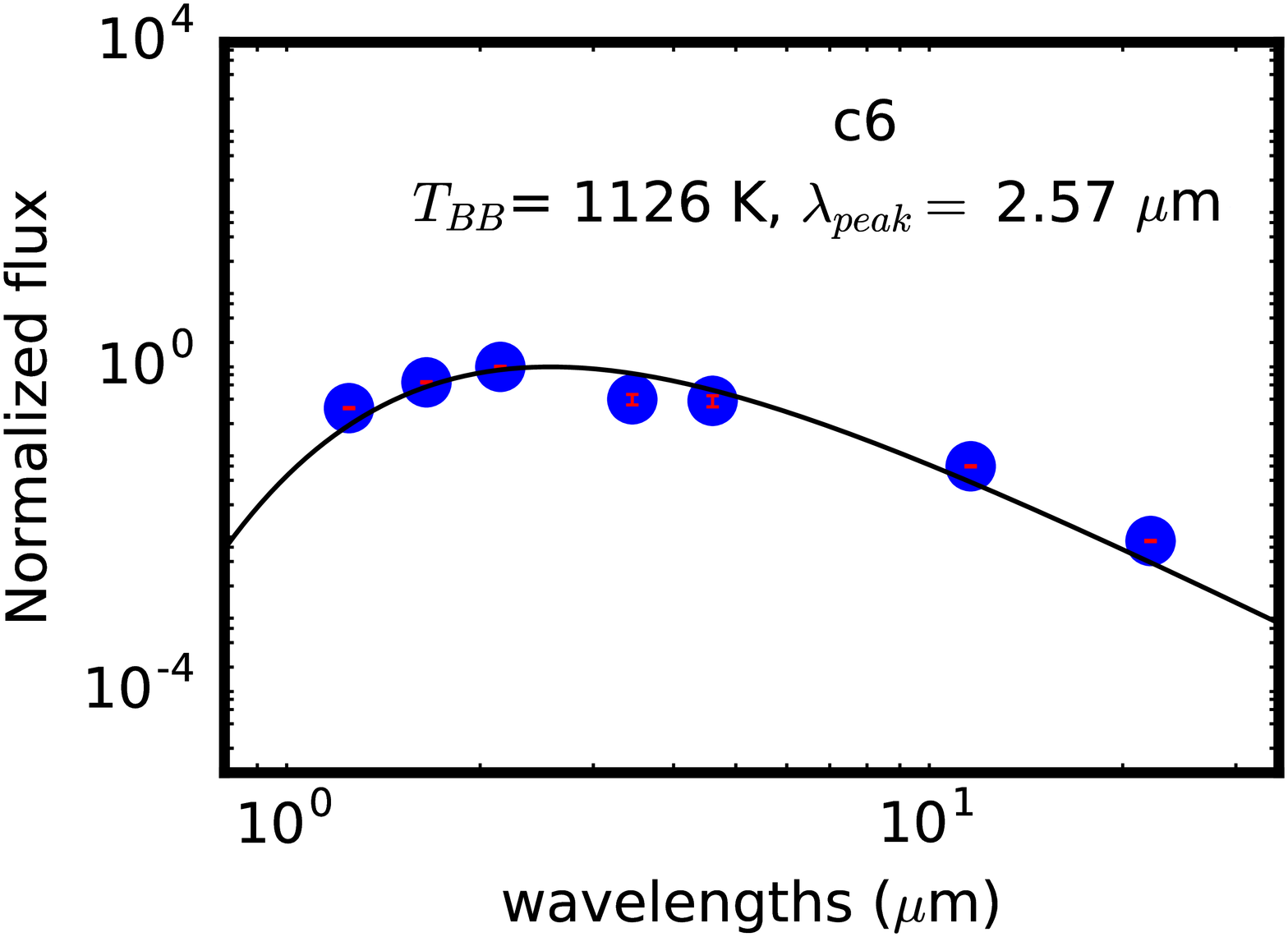}
\includegraphics[scale=0.14]{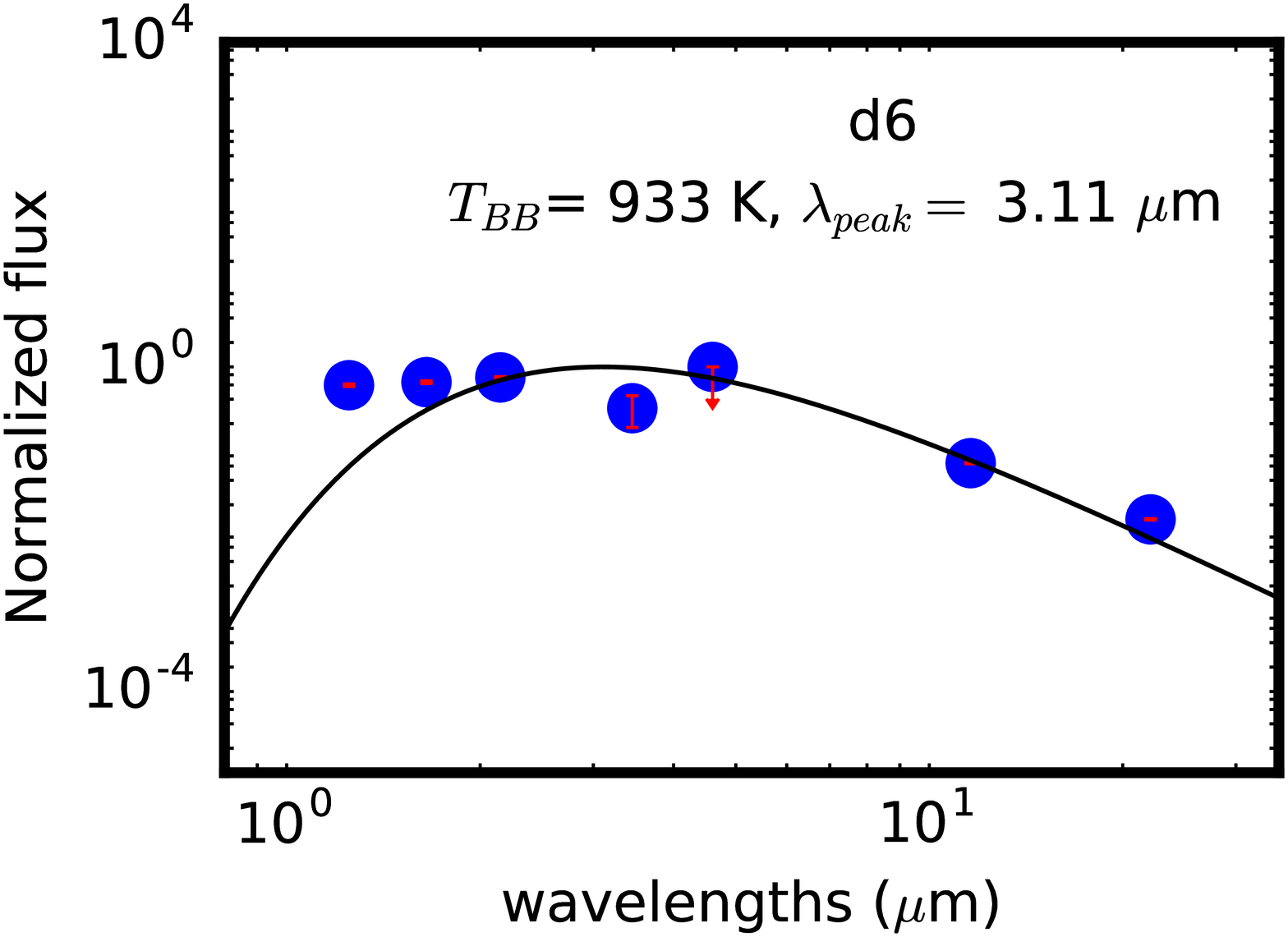}

\caption{The same as in Figure~\ref{fignew1} for four D-type SySts: (a) AS~210, (b)V347, (c) SS!73 38, and (d) RR~tel. 
The maximum amplitude variations of these SySts were obtained from Gromadzki et al. (2009).}
\label{fignew3}
\end{center}
\end{figure*}

Then, we applied the same exercise to four D-type SySts (a) AS~210, (b)V347, (c) SS~73 38, and (d) RR~tel, using their observed 
maximum amplitude variation in the {\it J}, {\it H}, {\it K} and {\it L} bands from Gromadzki et al. (2009). The amplitude variation for 
the {\it W2}, {\it W3} and {\it W4} bands were calculated by extrapolating a power law. The generated SED profiles are presented in Figure~\ref{fignew3}. The derived dust temperatures of these four sources show a variation of only of 350, 210, 400, and 230~K, respectively, 
if the poorly fitting SEDs are taken out of consideration (amplitude flux variation $>$3~mag). 

RR~Tel is a special case for which two BBs are required in order to fit the SED, and our dust temperature estimate agrees with 
the previous work from Jurkic \& Kotnik-Karuza (2012). Moreover, the upper limit of the {\it W2} magnitude makes this SySt quite more 
uncertain. In general, SySts with upper limit {\it W1} and {\it W2} magnitudes are more uncertain but there are fewer than 20 objects 
in our list. In these real cases, we verify that a poor fitting on the {\it W1} and {\it W2} bands (e.g., a2, a4, b2, b4, c2, c6, etc.) or 
the 2MASS bands (e.g. a1, a2, b3, b5, c2, d3, d5. etc.) is likely associated with flux variations between the data. 

All of these examples represent extreme cases given that the possibility that the 2MASS and {\it WISE} observations were performed during the 
maximum and minimum phases is not relatively high. Scrutinizing the SED profiles of all SySts in our list, we found that fewer than 20 
exhibit some evidence of variability among the 2MASS and {\it WISE} data.

Overall, the effective temperatures of cold giants or the dust temperatures are reliable within their uncertainties. The agreement of 
our temperature estimates with previous spectroscopic studies supports our approach. Flux variations in SySts results in temperature 
variations of the order of 250-350~K, which are comparable with their uncertainties. This analysis is not possible if a source has been 
observed during or close to an outburst event, which is true for the vast majority of the SySts in our list.

\end{document}